\documentclass[10pt]{article}

\usepackage{amssymb,amsmath,amsthm,mathrsfs} 
\usepackage{psfrag}
\usepackage{colordvi,color,pspicture}
\usepackage{natbib}          
\usepackage{graphicx}        

\newcommand{\Y}{\mathcal{Y}}
\newcommand{\X}{\mathcal{X}}
\newcommand{\U}{\mathcal{U}}
\newcommand{\I}{\mathbf{I}}

\newcommand{\mI}{\mathcal{I}}

\newcommand{\F}{\mathcal F}
\newcommand{\g}{\gamma}
\newcommand{\G}{\Gamma}

\newcommand{\V}{\mathcal{V}}
\newcommand{\A}{\mathcal{A}}
\newcommand{\R}{\mathbb{R}}
\newcommand{\mS}{\mathcal{S}}
\newcommand{\E}{\mathbf{E}}

\newcommand{\D}{\mathscr D}
\newcommand{\ve}{\varepsilon}

\newcommand{\al}{\alpha}
\newcommand{\be}{\beta}

\DeclareMathOperator{\argmin}{argmin}

\DeclareMathOperator{\BER}{BER}
\DeclareMathOperator{\BIN}{BIN}

\theoremstyle{plain}
\newtheorem{theorem}{Theorem}[section]
\newtheorem{lemma}[theorem]{Lemma}
\newtheorem{proposition}[theorem]{Proposition}
\newtheorem{corollary}[theorem]{Corollary}

\theoremstyle{definition}

\theoremstyle{remark}

\newtheorem{remark}[theorem]{Remark}

\begin{document}

\title{Domination Number of an Interval Catch Digraph Family and Its Use for Testing Uniformity}
\author{
Elvan Ceyhan\\
Department of Mathematics and Statistics,\\
Auburn University,\\
Auburn, AL\\
e-mail: ceyhan@auburn.edu
}

\date{\today}

\maketitle

{\bf short title:}
Domination Number of an ICD Family for Testing Uniformity

\begin{abstract}
\noindent
We consider a special type of interval catch digraph (ICD) family
for one-dimensional data in a randomized setting and propose its use for testing uniformity.
These ICDs are defined with an expansion and a centrality parameter,
hence we will refer to this ICD as parameterized ICD (PICD).
We derive the exact (and asymptotic) distribution of the domination number of
this PICD family when its vertices are from a uniform (and non-uniform) distribution in one dimension
for the entire range of the parameters;
thereby determine the parameters
for which the asymptotic distribution is non-degenerate.
We observe jumps
(from degeneracy to non-degeneracy or
from a non-degenerate distribution to another)
in the asymptotic distribution of the domination number at certain parameter combinations.
We use the domination number for testing uniformity of data in real line,
prove its consistency against certain alternatives,
and compare it with two commonly used tests and three recently proposed tests in literature and
also arc density of this ICD and of another ICD family in terms of size and power.
Based on our extensive Monte Carlo simulations,
we demonstrate that domination number of our PICD has higher power for certain
types of deviations from uniformity
compared to other tests.
\end{abstract}

\vspace{0.1 in}



\section{Introduction}
\label{sec:intro}
Graphs and digraphs for one dimensional points as vertices have been extensively studied and
have far-reaching applications despite their simplicity.
In this article, we introduce an interval catch digraph (ICD) family,
provide the distribution of its domination number for random vertices,
and employ the domination number in testing uniformity of one-dimensional data.
Interval graphs and digraphs have applications in many fields
such as chronological ordering of artifacts in archeology,
modeling traffic lights in transportation,
food web models in ecology, document localization, classification of RNA structures
 and so on (see \cite{roberts:1976}, \cite{drachenberg:masters-thesis}, \cite{arlazarov:2017}, and \cite{quadrini:2017}).
ICDs were introduced as a special type of interval digraphs and found applications in various fields
(see \cite{prisner:1989, prisner:1994} for a characterization and detailed discussion of ICDs).
The new digraph family we consider in this article is parameterized
by an expansion parameter and a centrality parameter.
We demonstrate that this digraph family is actually an ICD family,
hence it is referred to as parameterized ICD (PICD).
A \emph{digraph} is a directed graph with
vertex set $\V$ and arcs (directed edges) each of which is from one
vertex to another based on a binary relation.
The pair $(p,q) \in \V \times \V$ is an ordered pair
which stands for an \emph{arc} from vertex $p$ to vertex $q$ in $\V$.

The PICDs are closely related to the \emph{class cover problem} (CCP) of \cite{cannon:2000}
which is motivated by applications in statistical classification.
To properly describe the CCP problem,
let $(\Omega,d)$ be a metric space
with a dissimilarity function $d:\Omega \times \Omega \rightarrow \mathbb R$
such that $d(a,b)=d(b,a) \ge d(a,a)=0$
for all $a,b \in \Omega$.
Let $\X_n=\{X_1,X_2,\ldots,X_n\}$ and
$\Y_m=\{Y_1,Y_2,\ldots,Y_m\}$ be two sets of i.i.d. $\Omega$-valued random variables
from classes $\X$ and $\Y$,
with class-conditional distributions $F_X$ and $F_Y$, respectively.
We also assume that each $X_i$ is independent of each $Y_j$
and all $X_i \in \X_n$ and all $Y_j \in \Y_m$ are distinct with probability one,
and $(X_i,Y_j) \sim F_{X,Y}$
(i.e., $(X_i,Y_j)$ has joint distribution $F_{X,Y}$
with the marginal distributions $F_X$ for $X_i$ and $F_Y$ for $Y_j$).
The CCP for a target class refers to finding a collection of neighborhoods,
$\mathcal N$ around $X_i$, denoted $N(X_i) \in \mathcal N$, such that
(i) $\X_n \subseteq  \bigl(\cup_i N(X_i) \bigr)$ and (ii) $\Y_m \cap \bigl(\cup_i N(X_i) \bigr)=\emptyset$.
The neighborhood $N(X_i)$ is a subset of $\Omega$, containing $X_i$,
and is defined based on the dissimilarity $d$ (between $X_i$ and $\Y_m$).
A collection of neighborhoods satisfying both conditions is called a {\em class cover}.
Clearly, it follows by condition (i) that the set of all covering regions
(i.e., neighborhoods $N(X_i)$ around $X_i$) is a class cover;
however, the goal is to have a class cover for $\X_n$ that
has as few points as possible.
Thus, e.g. in statistical learning,
the classification will be less complex while most of the
relevant information being kept.
Hence, the CCP considered here is a \emph{minimum-cardinality class cover}.
One can convert the CCP to the graph theoretical problem of
finding dominating sets.
In particular,
our ICD is the digraph $D=(\V,\A)$
with vertex set $\V=\X_n$ and arc set $\A$
such that there is an arc $(X_i,X_j) \in \A$ iff $X_j \in N(X_i)$.
It is easy to see that
solving the CCP is equivalent to finding a minimum domination set of the corresponding PICD,
hence \emph{cardinality of a solution to CCP is equal to the domination number
of the associated digraph} (see \cite{marchette:2004}).
Hence the tool introduced in this article can be seen as a
parameterized extension to the original CCP problem of \cite{cannon:2000}.
That is, the cardinality of the smallest cover (i.e., the domination number) is investigated
when the cover(ing) regions, $N(X_i)$, depend on two parameters
and the distribution of this cardinality is based on $N(X_i)$ (hence the parameters) and $F_{X,Y}$.

Our PICDs are \emph{random digraphs}
(according to the digraph version of classification of \cite{beer:2011})
in which  each vertex corresponds to a data point
and arcs are defined in terms of some bivariate relation on the data,
and are also related to the class cover catch digraph (CCCD) introduced by \cite{priebe:2001}
who derived the exact distribution of its domination number for
uniform data from two classes in $\R$.
A CCCD consists of a vertex set in $\mathbb R^d$
and arcs $(u,v)$ if $v$ is inside the ball centered at $u$ with a radius based on spatial proximity of the points.
CCCDs were also extended to higher dimensions and
were demonstrated to be a competitive alternative to the existing methods in classification
(see \cite{devinney:2006} and references therein)
and to be robust to the class imbalance problem (\cite{ceyhan:jmlr-cccd-2016}).
Furthermore,
a CLT result for
CCCD based on one-dimensional data is proved (\cite{xiangCLT:2009}) and
the distribution of the domination number
of CCCDs is also derived for non-uniform data (\cite{ceyhan:dom-num-CCCD-NonUnif}).

We investigate the distribution of domination number of the PICDs
for data in $\Omega=\mathbb R$.
The domination in graphs has been studied extensively in recent decades
(see, e.g., \cite{hedetniemi:1990} and the references therein and \cite{henning:2013}),
and domination in digraphs has received comparatively less attention
but is also studied in literature (see, e.g., \cite{lee:1998}, \cite{niepel:2009} and \cite{hao:2017}).
We provide the exact and asymptotic distributions of the domination number of
PICDs with vertices from uniform (and non-uniform) one-dimensional distributions.
Some special cases and bounds for the domination number of PICDs
are handled first,
then the domination number is investigated
for uniform data in one interval (in $\R$) and the analysis is generalized to
uniform data in multiple intervals and to non-uniform data in one and multiple intervals.

We use domination number in testing uniformity of one-dimensional data.
Testing uniformity is important in its own right in numerous fields,
e.g., in assessing the quality of random number generators (\cite{lecuyer:2001})
and in chemical processes (\cite{fahidy:2013}).
Furthermore,
testing that data come from a particular distribution can be reduced to testing uniformity,
hence uniformity tests are of great importance for goodness-of-fit tests (see \cite{milosevic:2018}
and references therein).
Some graph theoretical tools are employed (although not so commonly) in two-sample testing (\cite{chen:2017}
and in testing uniformity;
for example,
\cite{jain:ICPR-2002} use minimum spanning trees
and \cite{ceyhan:revstat-2016} use the arc density of another family of ICDs for this purpose.
Moreover, \cite{ceyhan:metrika-2012} provide the probabilistic investigation of the arc density for the
PICD of this article, but it is not applied for uniformity testing previously.
In (\cite{ceyhan:dom-num-CCCD-NonUnif}), the distribution of the domination number of CCCDs is studied
when vertices are from a non-uniform one-dimensional distribution,
but the domination number of the PICD introduced here is not studied previously.
To the author's knowledge
\emph{domination number is not used in literature for testing uniformity}.
We compare the size and power performance of our test with
two well known competitors, namely Kolmogorov-Smirnov (KS) test
and Pearson's $\chi^2$ goodness-of-fit test,
and the arc density of PICDs and of another ICD family,
and also a uniformity test which is based on Too-Lin characterization
of the uniform distribution due to \cite{milosevic:2018},
and two entropy-based tests due to \cite{zamanzade:2015}.
We demonstrate that the test based on the domination number
has higher power for certain types of deviations from uniformity.
Furthermore, this article forms the foundation of the extensions of the methodology to higher dimensions.
The domination number has other applications, e.g.,
in testing spatial point patterns (see, e.g., \cite{ceyhan:dom-num-NPE-SPL})
and our results can help make the power comparisons
possible for a large family of alternative patterns in such a setting.
Some trivial proofs regarding PICDs are omitted,
while others
are mostly deferred to the Supplementary Materials Section.

We define the PICDs
and their domination number in Section \ref{sec:prop-edge-PICD},
provide the exact and asymptotic distributions of the domination number of PICDs
for uniform data in one interval in Section \ref{sec:gamma-dist-uniform},
discuss the distribution of the domination number for
data from a general distribution in Section \ref{sec:non-uniform}.
We extend these results to multiple intervals in Section \ref{sec:dist-multiple-intervals},
use domination number in testing uniformity in Section \ref{sec:app-test-unif},
prove consistency of the domination number tests under certain alternatives in Section \ref{sec:consistency},
and provide discussion and conclusions in Section \ref{sec:disc-conclusions}.

\section{A Parameterized Random Interval Catch Digraph Family}
\label{sec:prop-edge-PICD}

Let $N:\Omega \rightarrow \wp(\Omega)$ be a map
where $\wp(\Omega)$ represents the power set of $\Omega$.
Then the {\em proximity map}
$N(\cdot)$
associates with each point $x \in \Omega$
a {\em proximity region} $N(x) \subseteq \Omega$.
For $B \subseteq \Omega$, the \emph{$\G_1$-region} is the image of the map
$\G_1(\cdot,N):\wp(\Omega) \rightarrow \wp(\Omega)$
that associates the region $\G_1(B,N):=\{z \in \Omega: B \subseteq  N(z)\}$
with the set $B$.
For a point $x \in \Omega$,
for convenience,
we denote $\G_1(\{x\},N)$ as $\G_1(x,N)$.
Notice that while the proximity region is defined for one point,
a $\G_1$-region can be defined for a set of points.
The PICD has the vertex set $\V=\X_n$
and arc set $\A$ defined by $(X_i,X_j) \in \A$ iff $X_j \in N(X_i)$.

Although the above definition of the proximity region does not require multiple classes,
in this article, we will define proximity regions in a two-class setting
based on relative allocation of points from one class (say $\X$)\
with respect to points from the other class (say $\Y$).
We now get more specific and restrict our attention to $\Omega=\mathbb R$ and define $N$ explicitly.
Let $\Y_m$ consist of $m$ distinct points from class $\Y$
and $Y_{(i)}$ be the $i^{th}$ order statistic
(i.e., $i^{th}$ smallest value) of $\Y_m$ for $i=1,2,\ldots,m$
with the additional notation for $i \in \{0,m+1\}$ as
$$-\infty =: Y_{(0)}<Y_{(1)}< \ldots <Y_{(m)}< Y_{(m+1)}:=\infty.$$

\begin{figure} [h]
\centering
\rotatebox{0}{ \resizebox{2.5 in}{!}{ \includegraphics{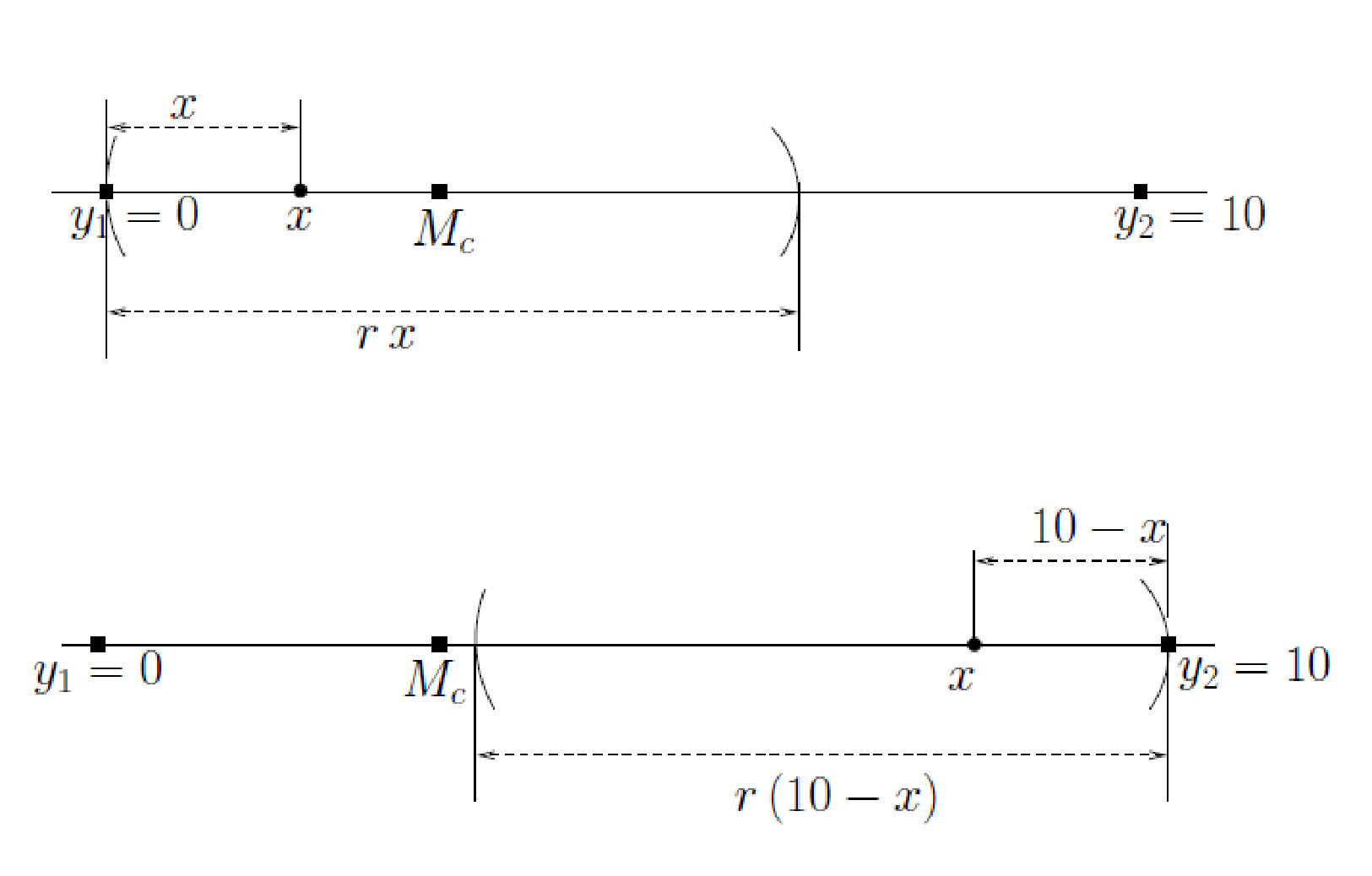}}}
\caption{
\label{fig:ProxMapDef1D}
Illustrations of the construction of the parameterized proximity region, $N(x,r,c)$
with $c \in (0,1/2)$
for $\Y_2=\{y_1,y_2\}$ with $y_1=0$ and $y_2=10$ (hence $M_c=10 c$)
and $x \in (0,M_c)$ (top) and $x \in (M_c,10)$ (bottom).}

\end{figure}

Then $Y_{(i)}$ values partition $\R$ into $(m+1)$ intervals
which is called the \emph{intervalization} of $\R$ by $\Y_m$.
Let also that
$\mI_i:=\left( Y_{(i)},Y_{(i+1)} \right)$ for $i \in \{0,1,2,\ldots,m\}$
and $M_{c,i}:=Y_{(i)}+c\left(Y_{(i+1)}-Y_{(i)}\right)$
(i.e., $M_{c,i} \in \mI_i$ such that $c \times 100$ \% of
length of $\mI_i$ is to the left of $M_{c,i}$).
We define the parameterized proximity region
with the expansion parameter $r \ge 1$ and
centrality parameter $c \in [0,1]$
for two one-dimensional data sets, $\X_n$ and $\Y_m$,
from classes $\X$ and $\Y$, respectively, as follows
(see also Figure \ref{fig:ProxMapDef1D}).
For $x \in \mI_i$ with $i \in \{1,2,\ldots,m-1\}$
(i.e. for $x$ in the middle intervals)
\begin{equation}
\label{eqn:NPEr-general-defn1}
N(x,r,c)=
\begin{cases}
\left( Y_{(i)}, \min\left(Y_{(i+1)},Y_{(i)}+r\,\left( x-Y_{(i)} \right)\right) \right)   & \text{if $x \in (Y_{(i)},M_{c,i})$,}\\
\left( \max\left(Y_{(i)},Y_{(i+1)}-r\left(Y_{(i+1)}-x\right)\right), Y_{(i+1)} \right)   & \text{if $x \in \left( M_{c,i},Y_{(i+1)} \right)$.}
\end{cases}
\end{equation}
Additionally,
for $x \in \mI_i$ with $i \in \{0,m\}$ (i.e. for $x$ in the end intervals)
\begin{equation}
\label{eqn:NPEr-general-defn2}
N(x,r,c)=
\begin{cases}
\left( Y_{(1)}-r\left( Y_{(1)}-x \right), Y_{(1)} \right)     & \text{if $x < Y_{(1)}$,}\\
\left( Y_{(m)}, Y_{(m)}+r\,\left( x-Y_{(m)} \right) \right) & \text{if $x > Y_{(m)}$.}
\end{cases}
\end{equation}

Notice that for $i \in \{0,m\}$,
the proximity region does not have a centrality parameter $c$.
For $x \in \Y_m$,
we define $N(x,r,c)=\{x\}$ for all $r \ge 1$ and $c \in [0,1]$.
If $x = M_{c,i}$, then in Equation \eqref{eqn:NPEr-general-defn1},
we arbitrarily assign $N(x,r,c)$ to be one of
the defining intervals.
For $c = 0$,
we have $\left( M_{c,i},Y_{(i+1)} \right)=\mI_i$
and
for $c = 1$,
we have $(Y_{(i)},M_{c,i})=\mI_i$.
So,
we set
$N(x,r,0):= \left( \max\left(Y_{(i)},Y_{(i+1)}-r\left(Y_{(i+1)}-x\right)\right), Y_{(i+1)} \right)$
and
$N(x,r,1):= \left( Y_{(i)}, \min \left(Y_{(i+1)},Y_{(i)}+r\,\left( x-Y_{(i)} \right)\right) \right)$.
For $r > 1$, we have $x \in N(x,r,c)$ for all $x \in \mI_i$.
Furthermore,
$\lim_{r \rightarrow \infty} N(x,r,c) = \mI_i$
for all $x \in \mI_i$,
so we define $N(x,\infty,c) = \mI_i$ for all such $x$.

The PICD has the vertex set $\X_n$ and arc set $\A$ defined by
$(X_i,X_j) \in \A$ iff $X_j \in N(X_i,r,c)$.
We denote such PICDs as $\mathbf D_{n,m}(F_{X,Y},r,c)$.
The randomness of the PICD lies in the fact that
the vertices are randomly generated from the distribution $F_X$
and proximity regions are random depending on $F_{X,Y}$,
but arcs $(X_i,X_j)$ are
deterministic functions of the random variable $X_j$ and the random set $N(X_i)$.
Notice that although $N$ depends on $\Y_m$,
we omit $\Y_m$ for brevity in notation of proximity region $N(x,r,c)$.

\subsection{Relation of PICDs with other Graph Families}
\label{sec:PICD-wrt-othergraphs}
Interval graphs are a special type of intersection graphs,
which have emerged from a problem in genetics called Benzer problem
(see \cite{roberts:1976} for details)
and they have been extensively studied in graph theory
since their introduction (\cite{drachenberg:masters-thesis} and \cite{francis:2018}).
On the other hand,
interval digraphs have recently gained attention
after their introduction in \cite{sen:1989} (see, e.g., \cite{das:2016}).
Let $\V$ be a set of $n$ index points in some arbitrary space;
for simplicity take $\V=\{1,2,\ldots,n\}$.
Consider a set of ``source" intervals $S_v$ and a set of ``target" intervals $T_v$ in $\mathbb R$
associated with $v \in \V$.
The family of ordered pairs of these intervals
$(S_v,U_v)_{v \in \V}$ such that $U_v \in S_v$ for each $v$
is called a \emph{nest representation} (\cite{prisner:1994}).
The digraph $D=(\V,\A)$ is called an \emph{interval nest digraph},
if there exists a nest representation with the index set $\V$
such that $(i,j) \in \A$ iff $S_i \cap U_j \ne \emptyset$.
\emph{Interval catch digraphs} (ICDs)
are interval nest digraphs with each $T_v$ containing just one element (\cite{prisner:1994}).
In fact,
for catch digraphs the nest representation
constitutes a family of sets with points (or \emph{pointed sets}) $(S_v,p_v)_{v \in \V}$
where each set $S_v$ is associated with a base point $p_v \in S_v$.
Then $D=(\V,\A)$ is a catch digraph
with $(i,j) \in \A$ iff $p_j \in S_i$.
Such a catch digraph is called an \emph{interval catch digraph},
if there is a totally ordered set $(T,\le)$
such that $D$ is the catch digraph of a family of pointed intervals in $T$.
Here, $I \subset T$ is an interval if,
for all $x,y,z \in T$, $x \le y \le z$ and $x,z \in I$ imply that $y \in I$.
For finite ICDs, $T$ can always be taken as the real line
(see, e.g., \cite{prisner:1989} who also provides a characterization of ICDs).

The PICDs are closely related to the {\em proximity graphs} of \cite{jaromczyk:1992}
and might be considered as one-dimensional versions of
proportional-edge proximity catch digraphs of \cite{ceyhan:dom-num-NPE-SPL}.
Furthermore, when $r=2$ and $c=1/2$ (i.e., $M_{c,i}=\left( Y_{(i)}+Y_{(i+1)} \right)/2$)
we have $N(x,r,c)=B(x,r(x))$
where $B(x,r(x))$ is the ball centered at $x$ with radius $r(x)=d(x,\Y_m)=\min_{y \in \Y_m}d(x,y)$.
The region $N(x,2,1/2)$ corresponds to the proximity region
which gives rise to the CCCD of \cite{priebe:2001}.
Note also that, $N(x,r,c)$ can be viewed as a \emph{homothetic transformation
(enlargement)} with $r \ge 1$ applied on a translation of the region $N(x,1,c)$.
Furthermore, this transformation is also an \emph{affine similarity transformation}.
Since $(\mathbb R,\le)$ is a total order,
by the characterization theorem of \cite{maehara:1984},
our random digraph is clearly an interval catch digraph,
since there exists a total order ``$\le$" on $\X_n \subset \mathbb R$
such that for $x<y<z \in \X_n$,
$(x,z) \in \A$ implies $(x,y) \in \A$
and $(z,x) \in \A$ implies $(z,y) \in \A$.
Our ICD is based on two parameters,
so we call it \emph{parameterized interval catch digraph} (PICD).

\subsection{Domination Number of PICDs}
\label{sec:domination-number-picds}
In a digraph $D=(\V,\A)$ of order $|\V|=n$, a vertex $u$ {\em dominates}
itself and all vertices of the form $\{v:\,(u,v) \in \A\}$.
A {\em dominating set}, $S_D$, for the digraph $D$ is a subset of
$\V$ such that each vertex $v \in \V$ is dominated by a vertex in $S_D$.
A {\em minimum dominating set},  $S^*_D$, is a dominating set of minimum cardinality;
and the {\em domination number}, denoted $\g(D)$, is defined as $\g(D):=|S^*_D|$,
where $|\cdot|$ stands for set cardinality (\cite{west:2001}).
\cite{chartrand:1999} distinguish domination in digraphs as out- and in-domination
and provide definitions for out- and in-domination numbers for digraphs.
Domination in this article refers to the \emph{out-domination} in PICDs.
If a minimum dominating set consists of only one vertex,
we call that vertex a {\em dominating vertex}.
Clearly, the vertex set $\V$ itself is always a dominating set,
so we have $\g(D) \le n$ in general,
and $1 \le \g(D) < n$ for nontrivial digraphs.

Let
$$\F\left( \R^d \right):=\{F_{X,Y} \text{ on } \R^d \text { with } (X,Y) \sim F_{X,Y}, ~
\text{ and random variables X and Y do not collide}\}.$$
That is, if $\X_n$ and $\Y_m$ are two samples from $F_X$ and $F_Y$, respectively
with $(X,Y)\sim F_{X,Y}$ and the marginal distributions of $X$ and $Y$ are $F_X$ and $F_Y$, respectively.
Furthermore,
``no collision of $X$ and $Y$" condition is equivalent to $P(X_i=Y_j)=0$ for all $i=1,\ldots,n$ and $j=1,\ldots,m$.
Notice that if $F_{X,Y}$ continuous,
then $F_{X,Y} \in \F\left( \R^d \right)$ follows.
Furthermore, if the probability distributions $F_X$ and $F_Y$ respectively have probability measures $\mathcal M_X$ and $\mathcal M_Y$ which are non-atomic,
then the associated joint distribution would be in $\F\left( \R^d \right)$ as well.
If $\mathcal M_Y$ contains an atom,
$Y_j$ points might collide,
but without loss of generality
we can assume that there are $m$ distinct $\Y$ points.
We restrict our attention to one dimensional data (i.e., $d=1$),
so we consider the random digraph for which
$\X_n$ and $\Y_m$ are samples from $F_X$ and $F_Y$, respectively,
with the joint distribution of $X,Y$ being $F_{X,Y} \in \F\left( \R \right)$.
We focus on the random variable $\g(\mathbf D_{n,m}(F_{X,Y},r,c))$,
the domination number of the digraph $\mathbf D_{n,m}(F_{X,Y},r,c)$.
To make the notation simpler,
we will use $\g_{{}_{n,m}}(F_{X,Y},r,c)$ instead of $\g(\mathbf D_{n,m}(F_{X,Y},r,c))$.
For $n \ge 1$ and $m \ge 1$,
it is immediate to see that $1 \le \g_{{}_{n,m}}(F_{X,Y},r,c) \le n$.

Let $\X_{[i]}:=\X_n \cap \mI_i$,
and $\Y_{[i]}:=\left\{Y_{(i)},Y_{(i+1)}\right\}$ for $i=0,1,2,\ldots,m$.
This yields a disconnected digraph with subdigraphs
each of which might be null or itself disconnected.
Let $\mathbf D_{[i]}$ be the component of $\mathbf D_{n,m}(F_{X,Y},r,c)$
induced by $\X_{[i]}$ for $i=0,1,2,\ldots,m$,
$n_i:=\left|\X_{[i]}\right|$ (provided that $n_i>0$),
and $F_i$ be the density $F_X$ restricted to $\mI_i$
(note that $\mI_i$ is also random here),
and
$\g_{{}_{[i]}}(F_i,r,c)$ be the domination number of $\mathbf D_{[i]}$.
Let also that $M_{c,i} \in \mI_i$ be the internal point that divides
the interval $\mI_i$ in ratios $c/(1-c)$
(i.e., length of the subinterval to the left of $M_{c,i}$ is
$c \times 100$ \% of the length of $\mI_i$).
Then $\g_{{}_{n,m}}(F_{X,Y},r,c)=\sum_{i=0}^m \g_{{}_{[i]}}(F_i,r,c)$.

\textbf{A Summary of Results in this article is as follows:}
\begin{itemize}
\item
In the middle intervals (i.e., for $i=1,2,\ldots,m-1$),
we show that $\g_{{}_{[i]}}(F_i,r,c)-1$ has a Bernoulli distribution
with the parameter depending on $F_{X,Y}$.
In the end intervals (i.e., $i \in \{0,m\}$)
where the domination number
$\g_{{}_{[i]}}(F_i,r,c)$ is $\I(n_i>0)$.

\item
Conditional on $\Y_m$ (i.e., $\Y_m$ is given),
randomness in the digraph (hence in the domination number)
stem from $F_X$.
So if $\Y_m$ is given,
we write the corresponding domination number as $\g_{{}_{n,m}}(F_X,r,c)$.
In this case,
we modify our notations as $\mathbf D_{n,m}(F,r,c)$ and $\g_{{}_{n,m}}(F,r,c)$
for the PICD and the associated domination number, where $F=F_X$.
\begin{itemize}
\item[(i)] Then we show that $\g_{{}_{n,2}}(F,r,c)$ is scale invariant for $\Y_2=\{a,b\}$,
$F=\U(a,b)$ with $-\infty<a<b<\infty$,
where $\U(a,b)$ stands for uniform distribution on $(a,b)$,
hence (without loss of generality) we can consider $\U(0,1)$.
\item[(ii)] We find the exact (and hence the asymptotic) distribution
of $\g_{{}_{n,2}}(\U,r,c)$ for $r \ge 1,c \in [0,1]$ (which is the most general case for these parameters).
\item[(iii)] We extend the result in (ii) by considering the
general non-uniform $F$ satisfying mild regularity conditions,
thereby find the asymptotic distribution of $\g_{{}_{n,2}}(F,r,c)$.
\item[(iv)] Finally,
we provide the more general form (in terms of $n$ and $m$) of $\g_{{}_{n,m}}(F,r,c)$ by considering general $m$
(i.e., $m>2$)
and find the asymptotic distribution of $\g_{{}_{n,m}}(F,r,c)$.
\end{itemize}
\item Domination number is employed as a test statistic for testing uniformity of one-dimensional data,
is consistent and
exhibits a good performance for certain types of alternatives.
\end{itemize}

\subsection{Special Cases for the Distribution of $\g_{n,m}(F_{X,Y},r,c)$}
\label{sec:special-cases-dom-numb-Dnm}
We study the simpler random variable $\g_{{}_{[i]}}(F_i,r,c)$ first.
The following lemma follows trivially.

\begin{lemma}
\label{lem:end-intervals}
For $i \in \{ 0,m \}$,
we have $\g_{{}_{[i]}}(F_i,r,c)=\I(n_i >0)$ for all $r \ge 1$.
For $i=1,2,3,\ldots,(m-1)$,
if $n_i =1$, then $\g_{{}_{[i]}}(F_i,r,c)=1$.
\end{lemma}

Let $\G_1\left( B,r,c \right)$ be the $\G_1$-region
for set $B$ associated with the proximity map $N(\cdot,r,c)$.

\begin{lemma}
\label{lem:G1-region-in-Ii}
The $\G_1$-region for $\X_{[i]}$ in $\mI_i$ with $r \ge 1$ and $c \in [0,1]$
is
$$\G_1\left( \X_{[i]},r,c \right) =
\Biggl(\frac{\max\,\left( \X_{[i]} \right)+Y_{(i)}(r-1)}{r},M_{c,i} \Biggr] \bigcup
\Biggl[M_{c,i}, \frac{\min\left( \X_{[i]} \right)+Y_{(i+1)}(r-1)}{r}\Biggr)$$
with the understanding that the intervals $(a,b)$, $(a,b]$, and $[a,b)$ are empty if $a \ge b$.
\end{lemma}

Notice that if $\X_{[i]} \cap \G_1\left( \X_{[i]},r,c \right) \not=\emptyset$,
we have $\g_{{}_{[i]}}(F_i,r,c)=1$,
hence the name \emph{$\G_1$-region} and the notation $\G_1(\cdot)$.
For $i=1,2,3,\ldots,(m-1)$ and $n_i > 1$,
we prove that $\g_{{}_{[i]}}(F_i,r,c) = 1$ or $2$
with distribution dependent probabilities.
Hence,
to find the distribution of $\g_{{}_{[i]}}(F_i,r,c)$,
it suffices to find the probability of $\g_{{}_{[i]}}(F_i,r,c)$ is 1 or 2.
For computational convenience, we employ the latter in our calculations henceforth
and denote it as  $p(F_i,r,c):=P\bigl( \g_{{}_{[i]}}(F_i,r,c)=2 \bigr)=
P\left( \X_{[i]} \cap \G_1\left( \X_{[i]},r,c \right) =\emptyset \right)$.

Furthermore,
let $\BER(p)$ and $\BIN(n',p)$, respectively, denote the Bernoulli and Binomial distributions
where $p$ is the probability of success with $p \in [0,1]$ and $n'>0$ is the number of trials.

\begin{lemma}
\label{lem:gamma 1 or 2}
For $i=1,2,3,\ldots,(m-1)$,
let the support of $F_i$ have positive Lebesgue measure.
Then for $n_i > 1$, $r \in (1,\infty)$, and $c \in (0,1)$,
we have $\g_{{}_{[i]}}(F_i,r,c)-1 \sim \BER\left( p(F_i,r,c) \right)$.
Furthermore,
$\g_{{}_{1,2}}(F_i,r,c)=1$ for all $r \ge 1$ and $c \in [0,1]$;
$\g_{{}_{[i]}}(F_i,r,0)=\g_{{}_{[i]}}(F_i,r,1)=1$ for all $n_i \ge 1$ and $ r \ge 1$;
and
$\g_{{}_{[i]}}(F_i,\infty,c)=1$ for all $n_i \ge 1$ and $c \in [0,1]$.
\end{lemma}

The probability $p(F_i,r,c)$
depends on the distribution $F_{X,Y}$ and the interval $\G_1\left( \X_{[i]},r,c \right)$,
which, if known, will make the computation of $p(F_i,r,c)$ possible.
We can bound the domination number with some crude bounds in this general case
(see the Supplementary Materials Section).

Based on Proposition \ref{prop:gamma-Dnm-r=1-M},
we have
$P\left(\g_{{}_{[i]}}(F_i,1,c)=1\right)=
P\left(\X_{[i]} \subset \left( Y_{(i)},M_{c,i} \right)\right)+
P\left(\X_{[i]} \subset \left( M_{c,i},Y_{(i+1)} \right)\right)$
and
$P\left(\g_{{}_{[i]}}(F_i,1,c)=2\right)=
P\left(\X_{[i]} \cap \left( Y_{(i)},M_{c,i} \right) \not= \emptyset,
\X_{[i]} \cap \left( M_{c,i},Y_{(i+1)} \right) \not= \emptyset \right)$.

\begin{remark}
\textbf{Restrictions on the Joint and Marginal Distributions for the Rest of the Article:}
The only restriction we imposed on $F_{X,Y}$ thus far was that
$P(X=Y)=0$ and collisions were not allowed (i.e., $P(X_i=Y_j)=0$ for all $i=1,\ldots,n$ and $j=1,\ldots,m$).
Note that $\X_n$ and $\Y_m$ need not be independent of each other;
collisions would be avoided if $X$ has a continuous distribution.
But in general $X$ and $Y$ can both be continuous, discrete or mixed.
Although we define in this very general setting,
\emph{in the rest of the article we will condition on a realization of $\Y_m$.}
Henceforth for brevity in notation,
we write $F=F_X$ and $\mathcal M=\mathcal M_X$
and we also assume that $\X_n$ is a random sample from $F$
(i.e., $X_j \stackrel{iid}{\sim}F$ for $j=1,\ldots,n$).
For $X_j \stackrel{iid}{\sim} F$,
with the additional assumption that
support $\mS(F_i) \subseteq \mI_i$
and $F$ is absolutely continuous around $M_{c,i}$ and around the end points of $\mI_i$,
it follows that the special cases in the construction
of $N(\cdot,r,c)$ ---
$X$ falls at $M_{c,i}$ or the end points of $\mI_i$ ---
occurs with probability zero.
Notice that $X_j$ having a nondegenerate one-dimensional
probability density function (pdf) $f$
which is continuous around $M_{c,i}$ and around the end points of $\mI_i$
is a special case of this (additional) assumption.
Furthermore, for such an $F$,
the region $N(X_i,r,c)$ is an interval a.s.
$\square$
\end{remark}

The results so far have been straightforward so far.
The more interesting cases are presented in the subsequent sections.

\section{The Distribution of the Domination Number of PICDs for Uniform Data in One Interval}
\label{sec:gamma-dist-uniform}
We first consider the simplest case of $m=2$ with
$\Y_2=\{y_1,y_2\}$ with $-\infty<y_1<y_2<\infty$
and $\X_n =\{X_1,X_2,\ldots,X_n\}$ a random sample from $\U(y_1,y_2)$,
we have the PICD with vertices from $F=\U(y_1,y_2)$.
The special case of $m=2$ is important in deriving the distribution of the domination number
in the general case of $m>2$,
because the domination number in multiple interval case is the sum of the domination numbers for the intervals.
We denote such digraphs as $\mathbf D_{n,2}(\U(y_1,y_2),r,c)$
and provide the exact distribution of their domination
number for the entire range of $r$ and $c$.
Let $\g_{{}_{n,2}}(\U(y_1,y_2),r,c)$
be the domination number of the PICD based on $N(\cdot,r,c)$ and $\X_n$
and $p_n(\U(y_1,y_2),r,c):=P\left(\g_{{}_{n,2}}(\U(y_1,y_2),r,c)=2\right)$,
and  $p(\U(y_1,y_2),r,c):=\lim_{n\rightarrow \infty}p_n(\U(y_1,y_2),r,c)$.
We first present a ``scale invariance" result for $\g_{{}_{n,2}}(\U(y_1,y_2),r,c)$.

\begin{theorem}
\label{thm:scale-inv-NYr}
(Scale Invariance Property)
Suppose $\X_n$ is a random sample from $\U(y_1,y_2)$
with $-\infty<y_1<y_2<\infty$.
Then for any $r \in [1,\infty]$ the distribution of $\g_{{}_{n,2}}(\U(y_1,y_2),r,c)$ is
independent of $\Y_2$ and hence independent of the support interval $(y_1,y_2)$.
\end{theorem}

\noindent {\bf Proof:}
Let $\X_n$ be a random sample from $\U(y_1,y_2)$ distribution.
Any $\U(y_1,y_2)$ random variable can be transformed into a $\U(0,1)$
random variable by the transformation $\phi(x)=(x-y_1)/(y_2-y_1)$,
which maps intervals $(t_1,t_2) \subseteq (y_1,y_2)$ to
intervals $\bigl( \phi(t_1),\phi(t_2) \bigr) \subseteq (0,1)$.
That is,
if $X \sim \U(y_1,y_2)$,
then we have $\phi(X) \sim \U(0,1)$
and
$P_1(X \in (t_1,t_2))=P_2\left(\phi(X) \in \bigl( \phi(t_1),\phi(t_2) \bigr)\right)$
for all $(t_1,t_2) \subseteq (y_1,y_2)$
where $P_1$ is the probability measure with respect to $\U(y_1,y_2)$
and $P_2$ is with respect to $\U(0,1)$.
So,
the distribution of $\g_{{}_{n,2}}(\U(y_1,y_2),r,c)$
does not depend on the support interval $(y_1,y_2)$,
i.e.,
it is scale invariant.
$\blacksquare$

Note that scale invariance of $\g_{{}_{n,2}}(F,\infty,c)$ follows trivially
for all $\X_n$ from any $F$ with support in $(y_1,y_2)$,
since for $r=\infty$, we have $\g_{{}_{n,2}}(F,\infty,c)=1$  a.s.
for all $n>1$ and $c \in (0,1)$.
The scale invariance of $\g_{{}_{1,2}}(F,r,c)$ holds
for all $r \ge 1$ and $c \in [0,1]$,
and scale invariance of $\g_{{}_{n,2}}(F,r,c)$ with $c \in \{0,1\}$ holds
for all $n \ge 1$ and $r \ge 1$ as well.
The scale invariance property in Theorem \ref{thm:scale-inv-NYr}
will \emph{simplify the notation and calculations in
our subsequent analysis of $\g_{{}_{n,2}}(\U(y_1,y_2),r,c)$ by allowing us to consider the special case
of the unit interval, $(0,1)$}.
Hence we drop the interval end points $y_1$ and $y_2$ in our notation
and write $\g_{{}_{n,2}}(\U,r,c)$ and $p_u(r,c,n)$,
and $p_u(r,c)$ for $p_n(\U,r,c)$ and $p(\U,r,c)$ henceforth
when vertices are from uniform distribution.
Then the proximity region for $x \in (0,1)$
with parameters $r \ge 1$ and $c \in [0,1]$ simplifies to
\begin{equation}
\label{eqn:NPEr-(0,1)-defn1}
N(x,r,c)=
\begin{cases}
(0, \min(1,r\,x)) & \text{if $x \in (0,c)$,}\\
(\max(0,1-r(1-x)), 1)     & \text{if $x \in (c,1)$}
\end{cases}
\end{equation}
with the comments below Equation \eqref{eqn:NPEr-general-defn2} applying to $N(x,r,c)$ as well.

\begin{remark}
\label{rem:gam2-computation}
Given $X_{(1)}=x_1$ and $X_{(n)}=x_n$,
let $\G_1(\X_n,r,c)=(\delta_1,\delta_2)$.
Then
the probability of $\g_{{}_{n,2}}(F,r,c)=2$ (i.e., the quantity $p_{{}_n}(F,r,c)$) is
$\displaystyle ( 1-[F(\delta_2)-F(\delta_1)]/[F(x_n)-F(x_1)])^{(n-2)}$
provided that $\delta_1 <\delta_2$ (i.e. $\G_1(\X_n,r,c)\not= \emptyset$);
if $\G_1(\X_n,r,c) = \emptyset$,r
then we would have $\g_{{}_{n,2}}(F,r,c)=2$.
That is,
$P(\g_{{}_{n,2}}(F,r,c)=2)=
P(\g_{{}_{n,2}}(F,r,c)=2,\;\G_1(\X_n,r,c) \not= \emptyset)+
P(\g_{{}_{n,2}}(F,r,c)=2,\;\G_1(\X_n,r,c) = \emptyset)$.
Then
\begin{equation}
\label{eqn:Pg2-int-first}
P(\g_{{}_{n,2}}(F,r,c)=2,\;\G_1(\X_n,r,c) \not= \emptyset)=
\int\int_{\mS_1}f_{1n}(x_1,x_n)\left(1-\frac{F(\delta_2)-F(\delta_1)}{F(x_n)-F(x_1)}\right)^{(n-2)}\,dx_n dx_1
\end{equation}
where $\mS_1=\{0<x_1<x_n<1:(x_1,x_n) \not\in \G_1(\X_n,r,c),\text{ and }\G_1(\X_n,r,c) \not= \emptyset\}$
and $f_{1n}(x_1,x_n)=n(n-1)f(x_1)f(x_n)\bigl(F(x_n)-F(x_1)\bigr)^{(n-2)}\I(0<x_1<x_n<1)$
is the joint pdf of $X_{(1)},X_{(n)}$.
The integral in \eqref{eqn:Pg2-int-first} becomes
\begin{equation}
\label{eqn:Pg2-integral-G1nonempty-U}
P(\g_{{}_{n,2}}(F,r,c)=2,\;\G_1(\X_n,r,c) \not= \emptyset)=
\int\int_{\mS_1}H(x_1,x_n)\,dx_n dx_1,
\end{equation}
where
\begin{equation}
\label{eqn:integrand}
H(x_1,x_n):=n\,(n-1)f(x_1)f(x_n)\bigl(F(x_n)-F(x_1)+F\left(\delta_1 \right)-F\left( \delta_2 \right)\bigr)^{n-2}.
\end{equation}

If $\G_1(\X_n,r,c) = \emptyset$,
then $\g_{{}_{n,2}}(F,r,c)=2$.
So
\begin{equation}
\label{eqn:Pg2-integral-G1empty}
P(\g_{{}_{n,2}}(F,r,c)=2,\;\G_1(\X_n,r,c)= \emptyset)=
P(\G_1(\X_n,r,c)= \emptyset)=
\int\int_{\mS_2}f_{1n}(x_1,x_n)\,dx_n dx_1
\end{equation}
where $\mS_2=\{0<x_1<x_n<1:\G_1(\X_n,r,c) = \emptyset\}$.
$\square$
\end{remark}

\subsection{Exact Distribution of $\g_{n,2}(\U,r,c)$}
\label{sec:r-and-M}
We first consider the case of $\U(y_1,y_2)$ data
with $r \ge 1$ and $c \in [0,1]$ and $n=1,2,\dots$.
That is,
we derive the distribution of $\g_{{}_{n,2}}(\U,r,c)$ for the entire range of the parameters $r$ and $c$.
For $r \ge 1$ and $c \in (0,1)$,
the $\G_1$-region is $\G_1(\X_n,r,c)=(X_{(n)}/r,c] \cup [c,(X_{(1)}+r-1)/r)$
where $(X_{(n)}/r,c]$ or $[c,(X_{(1)}+r-1)/r)$ or both could be empty.
\begin{theorem}
\label{thm:r and M}
\textbf{(Main Result 1)}
Let $\X_n$ be a random sample from $\U(y_1,y_2)$ distribution
with $n \ge 1$, $r \ge 1$, and $c \in (0,1)$.
Then we have
$$\g_{{}_{n,2}}(\U,r,c)-1 \sim \BER(p_u(r,c,n))$$
with
\begin{equation*}
p_u(r,c,n) =
\left\lbrace \begin{array}{ll}
       p_{u,a}(r,c,n)  & \text{for $c \in \big[\left(3-\sqrt{5}\right)/2,1/2\big]$,}\\
       p_{u,b}(r,c,n)  & \text{for $c \in \big[1/4,\left(3-\sqrt{5}\right)/2\big)$,}\\
       p_{u,c}(r,c,n)  & \text{for $c \in (0,1/4)$,}\\
\end{array} \right.
\end{equation*}
where explicit forms of $p_{u,a}(r,c,n)$, $p_{u,b}(r,c,n)$, and $p_{u,c}(r,c,n)$ are provided in Section \ref{sec:explicit-forms} in the Supplementary Materials.
By symmetry,
for $c \in \big(1/2,\left(\sqrt{5}-1\right)/2\big]$,
we have
$p_u(r,c,n) =p_{u,a}(r,1-c,n)$,
for $c \in \big(\left(\sqrt{5}-1\right)/2),3/4\big]$,
$p_u(r,c,n) =p_{u,b}(r,1-c,n)$,
and for $c \in (3/4,1)$,
$p_u(r,c,n) =p_{u,c}(r,1-c,n)$
with the understanding that the transformation $c \to 1-c$ is also applied
in the interval endpoints in the piecewise definitions of
$p_{u,a}(r,c,n)$, $p_{u,b}(r,c,n)$ and $p_{u,c}(r,c,n)$, respectively.

Furthermore, we have
$\g_{{}_{n,2}}(\U,r,0)=\g_{{}_{n,2}}(\U,r,1)=1$ for all $n \ge 1$.
\end{theorem}

Some remarks are in order for Main Result 1.
The partitioning of $c \in (0,1/2)$
as $c \in (0,1/4)$,
$c \in \big[1/4,\left(3-\sqrt{5}\right)/2\big)$, and
$c \in \big[\left(3-\sqrt{5}\right)/2,1/2\big)$
is due to the relative positions of $1/(1-c)$ and $(1-c)/c$
and the restrictions arising from various cases in the probability computations
(see the Supplementary Materials Section).
For example, for $c \in \left(\left(3-\sqrt{5}\right)/2,1/2\right)$, we have $1/(1-c) > (1-c)/c$
and
for $c \in \left(0,\left(3-\sqrt{5}\right)/2\right)$,
we have $1/(1-c) < (1-c)/c$.

\begin{figure}
\begin{center}
\psfrag{r}{\huge{$r$}}
\psfrag{c}{\huge{$c$}}
\rotatebox{0}{ \resizebox{2.5 in}{!}{ \includegraphics{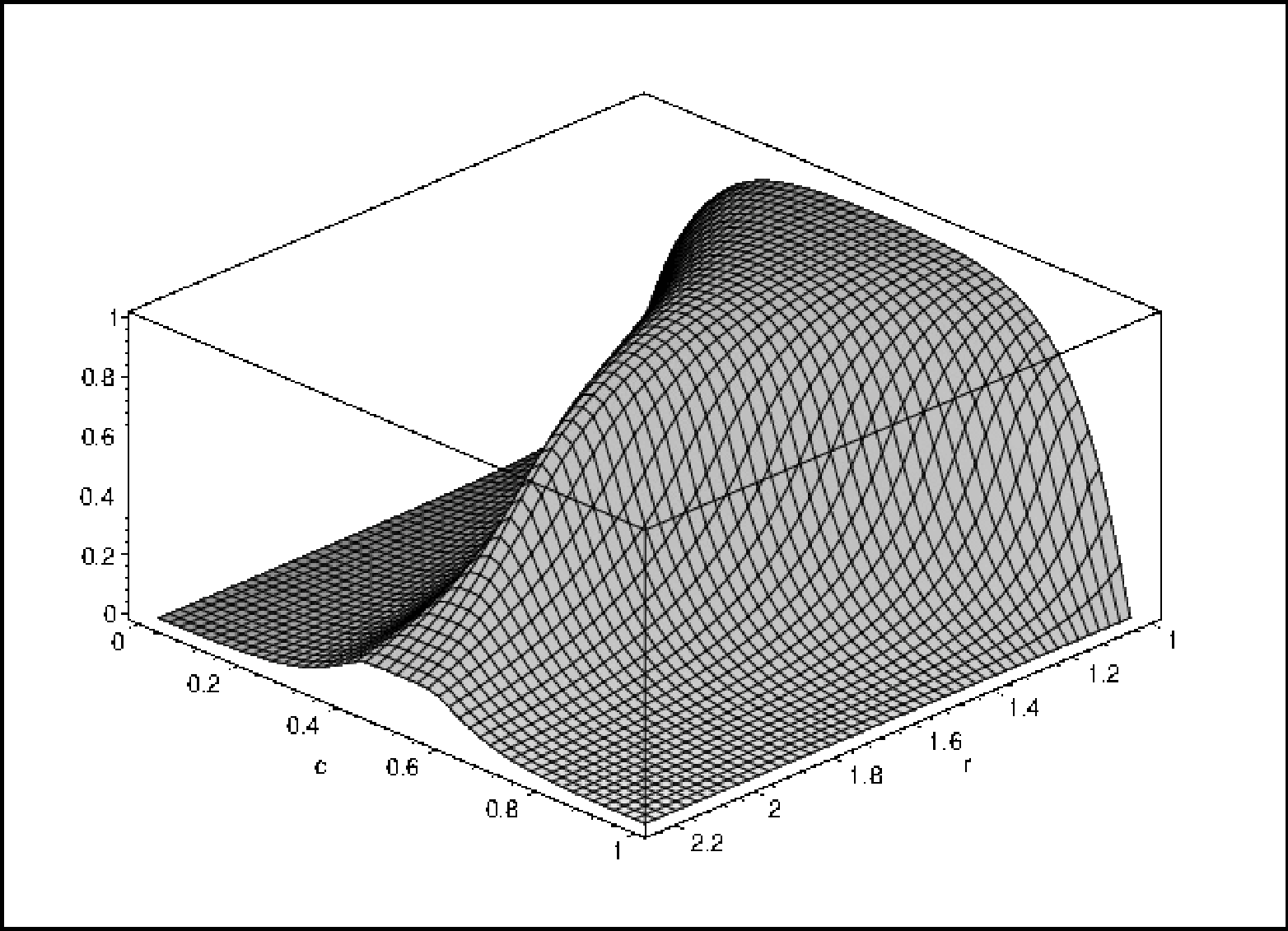}}}
\rotatebox{0}{ \resizebox{2.5 in}{!}{ \includegraphics{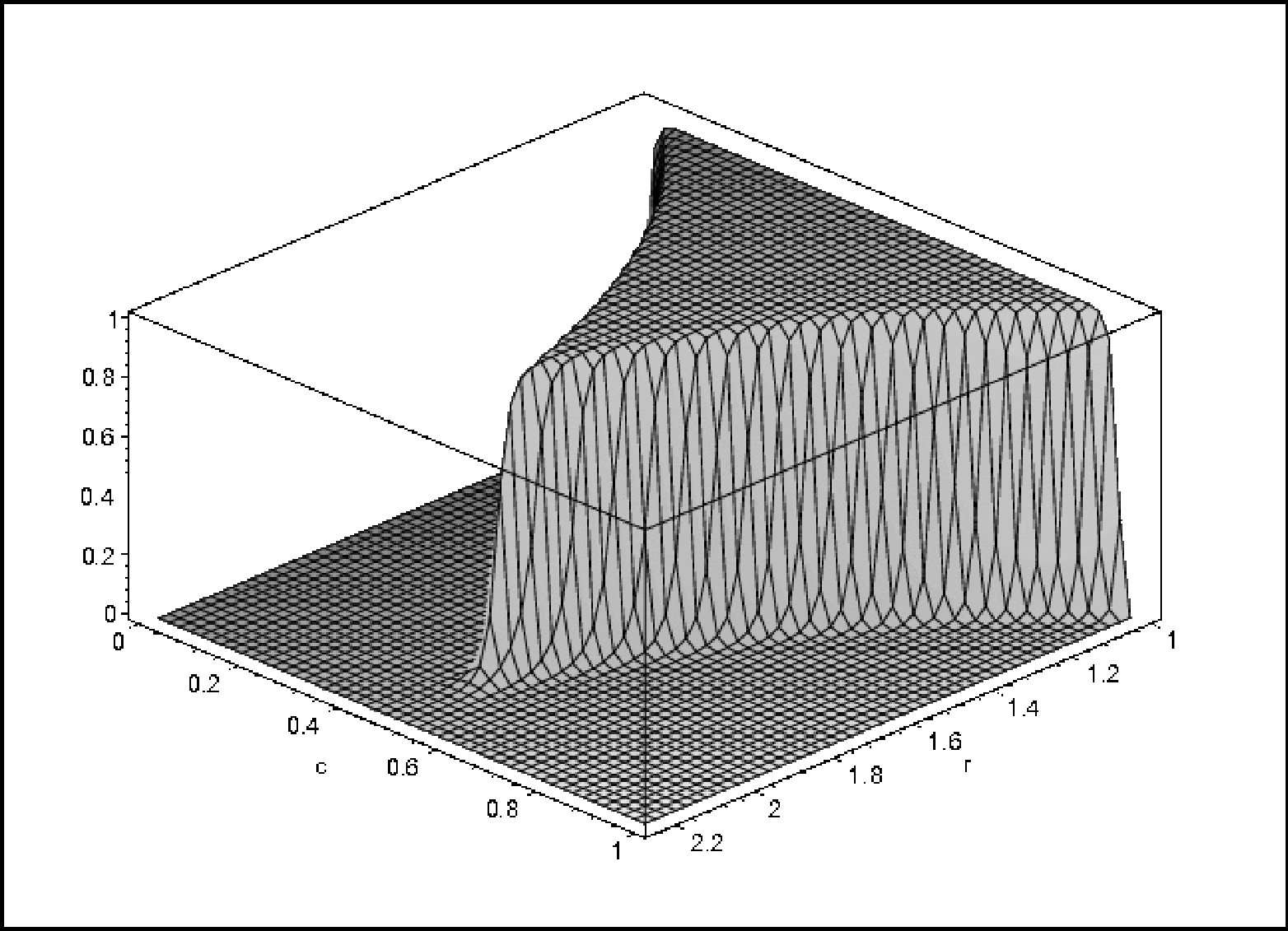}}}
\end{center}
\caption{
\label{fig:Prgam=2-withrcn}
Surface plots of $p_u(r,c,n)$ with $n = 10$ (left) and $n=100$ (right).
}
\end{figure}

We present the (three-dimensional) surface plots of $p_u(r,c,n)$ for $n =10$ and $n=100$ in Figure \ref{fig:Prgam=2-withrcn}.
As expected $\lim_{r \rightarrow 1} p_u(r,c,n)=0$.
For finite $n \ge 1$,
the probability $p_u(r,c,n)$
is continuous in $(r,c) \in \{(r,c) \in \R^2: r \ge 1, 0 \le c \le 1\}$.
For fixed $c \in (0,1)$ and fixed $n$,
$p_u(r,c,n)$ is decreasing as $r$ is increasing,
while
for fixed $r \in (1,\infty)$ and fixed $n$,
$p_u(r,c,n)$ is increasing as $c$ is approaching to 1/2.
In particular,
as $(r,c) \rightarrow (2,1/2)$
the distribution of $\g_{{}_{n,2}}(\U,r,c)-1$ converges to $\BER(p_u(2,1/2,n))$,
where $p_u(2,1/2,n)=4/9-(16/9) \, 4^{-n}$ as in \cite{priebe:2001}.
In the special cases of $c=1/2$ or $r=2$ or $(r,c)=(2,1/2)$,
the probability $p_u(r,c,n)$ reduces to much simpler forms.
See Section \ref{sec:special-cases} in the Supplementary Materials.

\subsubsection{Asymptotic Distribution of $\g_{n,2}(\U,r,c)$}
\label{sec:asy-dist-r-and-c}

\begin{theorem}
\label{thm:r and M-asy}
\textbf{(Main Result 2)}
For the PICD, $\mathbf D_{n,2}(\U,r,c)$, with $c \in (0,1)$ and $r^*=1/\max(c,1-c)$,
the domination number $\g_{{}_{n,2}}(\U,r,c)$ has the following
asymptotic distribution.
As $n \rightarrow \infty$, for $c \in (0,1)$,
\begin{equation}
\label{eqn:asy-unif-rM}
\g_{{}_{n,2}}(\U,r,c)-1 \stackrel{\mathcal L}{\to}
\left\lbrace \begin{array}{ll}
       0,           & \text{for $r> r^*$,}\\
       \BER(p_r),           & \text{for $r = r^*$,}\\
       1,           & \text{for $1 \le r < r^*$.}\\
\end{array} \right.
\end{equation}
where
\begin{equation}
\label{eqn:p_r}
p_r=
\left\lbrace \begin{array}{ll}
       \frac{r^*}{r^*+1},           & \text{for $c \ne 1/2$,}\\
       \frac{4}{9},           & \text{for $c =1/2$,}
       \end{array} \right.
\end{equation}
\end{theorem}

\begin{figure}[ht]
\centering
\rotatebox{-90}{ \resizebox{2.5 in}{!}{ \includegraphics{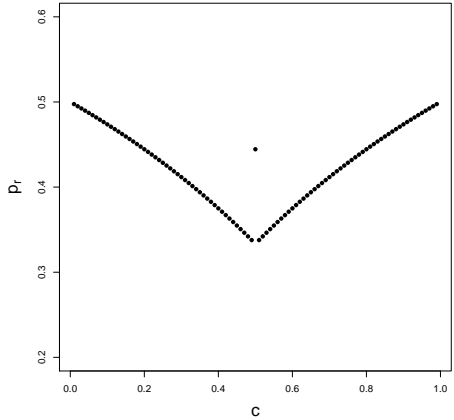}}}
\caption{
\label{fig:prob-p_r}
Plot of the limiting probability $p_r:=\lim_{n\rightarrow \infty} \g_{{}_{n,2}}(\U,r,c)$
for $r=r^*=1/\max(c,1-c)$ (see also Equation \eqref{eqn:p_r}).}
\end{figure}

Notice the interesting behavior of the asymptotic distribution
of $\g_{{}_{n,2}}(\U,r,c)$ around $r=r^*$ for any given $c \in (0,1)$.
The asymptotic distribution is non-degenerate only for $r = r^*$.
For $r>r^*$,
$\lim_{n \rightarrow \infty}\g_{{}_{n,2}}(\U,r,c) = 1$ w.p. 1,
and
for $1 \le r < r^* $, $\lim_{n \rightarrow \infty}\g_{{}_{n,2}}(\U,r,1/2) = 2$ w.p. 1.
The critical value $r=r^*$ corresponds to
$c=(r-1)/r$, if $c \in(0,1/2)$ (i.e., $r^*=1/(1-c)$)
and
$c=1/r$, if $c \in(1/2,1)$ (i.e., $r^*=1/c$)
and $r=r^*$ only possible for $r \in (1,2)$.
The probability $p_u(r,c)$ is continuous in $r$ and $c$ for $r \ne r^*$
and there is a jump (hence discontinuity) in the probability $p_u(r,c)$
at $r=r^*$,
since $p_u(r^*,c)=r^*/(r^*+1)$ for $c \ne 1/2$
(see also Figure \ref{fig:prob-p_r}).
Therefore, given a centrality parameter $c \in (0,1)$,
we can choose the expansion parameter $r$
for which the asymptotic distribution is non-degenerate,
and vice versa.
There is yet another interesting behavior of the asymptotic distribution
around $(r,c)=(2,1/2)$.
The probability $p_u(r^*,c)$
has jumps at $(r,c)=(r^*,c)$ for $r \in [1,2]$
with $p_u(r^*,c)=r^*/(r^*+1)$ for $c \ne 1/2$.
That is,
for fixed $(r,c) \in S$,
$\lim_{n \rightarrow \infty}p_u(r^*,c,n)=r^*/(r^*+1)$ for $c \ne 1/2$.
Letting $(r,c) \rightarrow (2,1/2)$,
we get
$p_u(r^*,c) \rightarrow 2/3$,
but
$p_u(2,1/2)=4/9$.
Hence for $(r,c) \ne (2,1/2)$
the distribution of $\g_{{}_{n,2}}(\U,r^*,c)-1$
converges to $\BER(r^*/(r^*+1))$,
but the distribution of
$\g_{{}_{n,2}}(\U,2,1/2)-1$
converges to $\BER(4/9)$ as $n \rightarrow \infty$
(rather than $\BER(2/3)$).
In other words,
$p_u(r^*,c)$ has another jump at $(r,c)=(2,1/2)$.
This interesting behavior occurs due to the symmetry around $c=1/2$.
Because for $c \in (0,1/2)$,
with $r=1/(1-c)$,
for sufficiently large $n$,
a point $X_i$ in $(c,1)$ can dominate all the points in $\X_n$
(implying $\g_{{}_{n,2}}(\U,1/(1-c),c)=1$),
but no point in $(0,c)$ can dominate all points a.s.
Likewise,
for $c \in (1/2,1)$ with $r=1/c$,
for sufficiently large $n$,
a point $X_i$ in $(0,c)$ can dominate all the points in $\X_n$
(implying $\g_{{}_{n,2}}(\U,1/c,c)=1$),
but no point in $(c,1)$ can dominate all points a.s.
However,
for $c=1/2$ and $r=2$,
for sufficiently large $n$,
points to the left or right of $c$
can dominate all other points in $\X_n$.

\section{Distribution of $\g_{n,2}(F,r,c)$}
\label{sec:non-uniform}
We now relax the assumption of uniformity for the vertices of our PICD (i.e., for $\X$ points).
Let $\F(y_1,y_2)$ be a family of continuous distributions
with support in $\mS_F \subseteq (y_1,y_2)$.
Consider a distribution function $F \in \F(y_1,y_2)$.
For simplicity, assume $y_1=0$ and $y_2=1$.
Let $\X_n$ be a random sample from $F$,
$\Gamma_1$-region $\G_1(\X_n,r,c)=(\delta_1,\delta_2)$,
and $p_{{}_n}(F,r,c):=P(\g_{{}_{n,2}}(F,r,c)=2)$,
$p(F,r,c):=\lim_{n \rightarrow \infty}P(\g_{{}_{n,2}}(F,r,c)=2)$.
The exact and asymptotic distributions of $\g_{{}_{n,2}}(F,r,c)-1$
are $\BER\left(p_{{}_n}(F,r,c)\right)$ and $\BER\left(p(F,r,c)\right)$, respectively.
That is, for finite $n > 1$, $r \in [1,\infty)$, and $c \in (0,1)$,
we have
\begin{equation}
\g_{{}_{n,2}}(F,r,c)=  \left\lbrace \begin{array}{ll}
       1           & \text{w.p. $1-p_{{}_n}(F,r,c)$},\\
       2           & \text{w.p. $p_{{}_n}(F,r,c)$}.
\end{array} \right.
\end{equation}
Moreover,
$\g_{{}_{1,2}}(F,r,c)=1$ for all $r \ge 1$ and $c \in [0,1]$,
$\g_{{}_{n,2}}(F,r,0)=\g_{{}_{n,2}}(F,r,1)=1$ for all $n \ge 1$ and $r \ge 1$,
$\g_{{}_{n,2}}(F,\infty,c)=1$ for all $n \ge 1$ and $c \in [0,1]$,
and
$\g_{{}_{n,2}}(F,1,c)=k_4$ for all $n \ge 1$ and $c \in (0,1)$
where $k_4$ is as in Proposition \ref{prop:gamma-Dnm-r=1-M} with $m=2$.
The asymptotic distribution is similar with
$p_{{}_n}(F,r,c)$ being replaced with $p(F,r,c)$.
The special cases are similar in the asymptotics
with the exception that
$p(F,1,c)=1$ for all $c \in (0,1)$.
The finite sample mean and variance of $\g_{{}_{n,2}}(F,r,c)-1$ are
$p_{{}_n}(F,r,c)$ and $p_{{}_n}(F,r,c)\,(1-p_{{}_n}(F,r,c))$, respectively;
and similarly the asymptotic mean and variance of $\g_{{}_{n,2}}(F,r,c)-1$ are
$p(F,r,c)$ and $p(F,r,c)\,(1-p(F,r,c))$, respectively.

For $\Y_2=\{y_1,y_2\} \subset \R$ with $-\infty<y_1<y_2<\infty$,
a quick investigation shows that, by Lemma \ref{lem:G1-region-in-Ii},
the $\G_1$-region is $\G_1(\X_n,r,c)= \Bigl(\frac{X_{(n)}+y_1(r-1)}{r},M_c\Bigr] \cup
\Bigl[M_c, \frac{X_{(1)}+y_2(r-1)}{r}\Bigr)$.
Notice that for a given $c \in [0,1]$,
the corresponding $M_c \in [y_1,y_2]$
is $M_c=y_1+c(y_2-y_1)$.
Let $F$ be a continuous distribution with support $\mS(F)\subseteq (0,1)$.
The simplest of such distributions is $\U(0,1)$,
which yields the simplest exact distribution for $\g_{{}_{n,2}}(F,r,c)$
with $(r,c)=(2,1/2)$.
If $X \sim F$, then by probability integral transform, $F(X) \sim \U(0,1)$.
So for any continuous $F$,
we can construct a proximity map depending on $F$ for which
the distribution of the domination number of the
associated digraph has the same distribution
as that of $\g_{{}_{n,2}}(\U,r,c)$,
which is explicated in the below proposition
whose proof is provided in the Supplementary Materials Section.

\begin{proposition}
\label{prop:NF vs NPE}
Let $X_i \stackrel{iid}{\sim} F$ which is an absolutely continuous
distribution with support $\mS(F)=(0,1)$ and let $\X_n:=\{X_1,X_2,\ldots,X_n\}$.
Define the proximity map $N_F(x,r,c):=F^{-1}(N(F(x),r,c))$.
That is,
\begin{equation}
\label{eqn:N-F-(y1,y2)}
N_F(x,r,c)=
\begin{cases}
(0, \min(1,F^{-1}(r\,F(x)) )) & \text{if $x \in (0,F^{-1}(c))$,}\\
(\max(0,F^{-1}(1-r(1-F(x))), 1)     & \text{if $x \in (F^{-1}(c),1)$.}
\end{cases}
\end{equation}
Then the domination number of the digraph based on $N_F$, $\X_n$, and $\Y_2=\{0,1\}$
has the same distribution as $\g_{{}_{n,2}}(\U,r,c)$.
\end{proposition}
\noindent

The result in Proposition \ref{prop:NF vs NPE}
can easily be generalized for a distribution $F$ with $\mathcal S(F)=(a,b)$
with finite $a<b$.
For $X \sim F$,
the transformed random variable $W=\frac{X-a}{b-a}$
would have cdf
$F_W(w)=F_X(a+w(b-a))$ which has support $\mathcal S(F_W)=(0,1)$.
Then one can apply Proposition \ref{prop:NF vs NPE}
to $W_i \stackrel{iid}{\sim}F_W$.
There is also a stochastic ordering between $\g_{{}_{n,2}}(F,r,c)$ and $\g_{{}_{n,2}}(\U,r,c)$
provided that $F$ satisfies some regularity conditions,
which are provided in Proposition \ref{prop:stoch-order} in the Supplementary Materials Section.
We can also find the exact distribution of $\g_{{}_{n,2}}(F,r,c)$
for $F$ whose pdf is piecewise constant
with support in $(0,1)$, see Remark \ref{rem:exact-dist-gam} in the Supplementary Materials Section for more details.

Recall the PICD, $\mathbf D_{n,m}(F,r,c)$.
We denote the digraph which is obtained
in the special case of $\Y_2=\{y_1,y_2\}$ and support of $F_X$ in $(y_1,y_2)$
as $\mathbf D_{n,2}(F,r,c)$.
Below, we provide asymptotic results
pertaining to the distribution of domination number of such digraphs.

\subsection{Asymptotic Distribution of $\g_{n,2}(F,r,c)$}
\label{sec:asy-dist-generalF}
Although the exact distribution of $\g_{{}_{n,2}}(F,r,c)$ may not be analytically available
in a simple closed form for $F$
whose density is not piecewise constant,
the asymptotic distribution of $\g_{{}_{n,2}}(F,r,c)$ is available for larger families of distributions.
First, we present the asymptotic distribution of $\g_{{}_{n,2}}(F,r,c)$ for
$\mathbf D_{n,2}(F,r,c)$ with $\Y_2=\{y_1,y_2\} \subset \R$ with $-\infty<y_1<y_2<\infty$
for general $F$ with support $\mS(F) \subseteq (y_1,y_2)$.
Then we will extend this to the case with $\Y_m \subset \R$ with $m>2$.

Let $c \in (0,1/2]$ and $r \in (1,2]$.
Then for $(r,c)=(1/(1-c),c)$,
we define the family of distributions
\begin{equation*}
\F_1\bigl(y_1,y_2\bigr) :=
\Bigl \{\text{$F$ :
$(y_1,y_1+\ve) \cup \bigl( M_c,M_c+\ve \bigr) \subseteq \mS(F)\subseteq(y_1,y_2)$
for some $\ve \in (0,c)$ with $c=(0,1/2]$} \Bigr\}.
\end{equation*}
Similarly,
let $c \in [1/2,1)$ and $r \in (1,2]$.
Then for $(r,c)=(1/c,c)$,
we define
\begin{equation*}
\F_2\bigl(y_1,y_2\bigr) :=
\Bigl \{\text{$F$ :
$(y_2-\ve,y_2)\cup \bigl(M_c-\ve,M_c\bigr) \subseteq \mS(F)\subseteq(y_1,y_2)$
for some $\ve \in (0,1-c)$ with $c=[1/2,1)$} \Bigr\}.
\end{equation*}

Let $k^{th}$ order right (directed) derivative at $x$ be defined as
$f^{(k)}(x^+):=\lim_{h \rightarrow 0^+}\frac{f^{(k-1)}(x+h)-f^{(k-1)}(x)}{h}$
for all $k \ge 1$ and the right limit at $u$ be defined as $f(u^+):=\lim_{h \rightarrow 0^+}f(u+h)$.
Let the left derivatives and limits be defined similarly with $+$'s being replaced by $-$'s.

\begin{theorem}
\label{thm:kth-order-gen-r,c}
\textbf{(Main Result 3)}
Suppose $\Y_2=\{y_1,y_2\} \subset \R$ with $-\infty < y_1 < y_2<\infty$,
$\X_n=\{X_1,X_2,\ldots,X_n\}$ with $X_i \stackrel {iid}{\sim} F$
with $\mS(F) \subseteq (y_1,y_2)$,
and $c \in (0,1)$ and $r^*=1/\max(c,1-c)$.
Let $\mathbf D_{n,2}(F,r,c)$ be the PICD based on $\X_n$ and $\Y_2$.
\begin{itemize}
\item[(i)]
Then for $n>1$, $r \in (1,\infty)$,
we have $\g_{{}_{n,2}}(F,r^*,c)-1 \sim \BER\bigl(p_{{}_n}(F,r^*,c)\bigr)$.
Note also that $\g_{{}_{1,2}}(F,r,c)=1$ for all $r \ge 1$ and $c \in [0,1]$;
for $r=1$,
we have $\g_{{}_{n,2}}(F,1,0)=\g_{{}_{n,2}}(F,1,1)=1$ for all $n \ge 1$ and
for $r = \infty$,
we have $\g_{{}_{n,2}}(F,\infty,c)=1$ for all $n \ge 1$ and $c \in [0,1]$.

\item[(ii)]
Suppose
$c \in (0,1/2)$ and $r =r^* =1/(1-c)$,
$F \in \F_1(y_1,y_2)$ with pdf $f$,
and $k \ge 0$ is the smallest integer for which
$F(\cdot)$ has continuous right derivatives up to order $(k+1)$ at $y_1$,
$M_c$, and
$f^{(k)}(y_1^+)+r^{-(k+1)}\,f^{(k)}\left( M_c^+ \right) \not= 0$
and $f^{(i)}(y_1^+)=f^{(i)}\left( M_c^+ \right)=0$ for all $i=0,1,2,\ldots,(k-1)$
and suppose also that $F(\cdot)$ has a continuous left derivative at $y_2$.
Then for bounded $f^{(k)}(\cdot)$,
we have the following limit
$$p(F,1/(1-c),c) =
\lim_{n \rightarrow \infty}p_{{}_n}(F,1/(1-c),c) =
\frac{f^{(k)}(y_1^+)}
{f^{(k)}(y_1^+)+(1-c)^{(k+1)}\,
f^{(k)}\left( M_c^+ \right)}.$$

\item[(iii)]
Suppose
$c \in (1/2,1)$ and $r =r^*=1/c$,
$F \in \F_2(y_1,y_2)$ with pdf $f$,
and $\ell \ge 0$ is the smallest integer for which
$F(\cdot)$ has continuous left derivatives up to order $(\ell+1)$ at $y_2$, and $M_c$,
and
$f^{(\ell)}(y_2^-)+r^{-(\ell+1)}\,f^{(\ell)}\left( M_c^- \right) \not= 0$
and $f^{(i)}(y_2^-)=f^{(i)}\left( M_c^- \right)=0$ for all $i=0,1,2,\ldots,(\ell-1)$
and suppose also that $F(\cdot)$ has a continuous right derivative at $y_1$.
Then for bounded $f^{(\ell)}(\cdot)$,
we have the following limit
$$
p(F,1/c,c) =
\lim_{n \rightarrow \infty}p_{{}_n}(F,1/c,c) =
\frac{f^{(\ell)}(y_2^-)}
{f^{(\ell)}(y_2^-)+c^{(\ell+1)}\,f^{(\ell)}\left( M_c^- \right)}.$$

\item[(iv)]
Suppose $(M_c-\ve,M_c+\ve) \cup (y_1,y_1+\ve) \cup (y_2-\ve,y_2) \subset \mathcal S(F)$
for some $\ve > 0$,
then
\begin{equation*}
p(F,r,c) =
\left\lbrace \begin{array}{ll}
       1           & \text{if $r > r^*$},\\
       0           & \text{if $r < r^*$ }.
\end{array} \right.
\end{equation*}
\end{itemize}
\end{theorem}

The asymptotic distribution of  $\g_{{}_{n,2}}(F,r,c)$ for $r=2$ and $c=1/2$
is provided in Theorem \ref{thm:kth-order-gen} in the Supplementary Materials Section.

In Theorem \ref{thm:kth-order-gen-r,c} parts (ii) and (iii),
we assume that $f^{(k)}(\cdot)$ and $f^{(\ell)}(\cdot)$
are bounded on $(y_1,y_2)$, respectively.
The extension to the unbounded derivatives is provided in Remark \ref{rem:unbounded} in the Supplementary Materials Section.
The rates of convergence in  Theorem \ref{thm:kth-order-gen-r,c} parts (ii) and (iii) depend on $f$
and are provided in Remark \ref{rem:rate-of-conv} in the Supplementary Materials Section.
The conditions of the Theorems \ref{thm:kth-order-gen-r,c} and \ref{thm:kth-order-gen}
might seem a bit esoteric.
However, most of the well known functions that are scaled
and properly transformed to be pdf of some random variable
with support in $(y_1,y_2)$ satisfy the conditions
for some $k$ or $\ell$,
hence one can compute the corresponding limiting probability $p(F,r^*,c)$.

\textbf{Examples:}
(a)
With $F=\U(y_1,y_2)$, in Theorem \ref{thm:kth-order-gen-r,c} (ii),
we have
$k=0$ and $f(y_1^+)=f( M_c^+ )=1/(y_2-y_1)$,
and
in Theorem \ref{thm:kth-order-gen-r,c} (iii), we have
$\ell=0$ and $f(y_2^-)=f\left( M_c^- \right)=1/(y_2-y_1)$.
Then $\lim _{n \rightarrow \infty}p_n(\U,r^*,c)=r^*/(r^*+1)$ for $c \ne 1/2$,
which agrees with the result given in Equation \eqref{eqn:asy-unif-rM}
and $\lim _{n \rightarrow \infty}p_u(2,1/2,n)=4/9$. $\square$

(b)
For $F$ with pdf $f(x)=\bigl( x+1/2 \bigr)\,\I\bigl( 0 <x<1 \bigr)$,
we have $k=0$, $f(0^+)=1/2$, and $f\left( c^+ \right)=c+1/2$ in Theorem \ref{thm:kth-order-gen-r,c} (ii).
Then $p(F,1/(1-c),c)=\frac{1}{2+c-2 c^2}$ for $c \ne 1/2$.
In Theorem \ref{thm:kth-order-gen-r,c} (iii),
we have $\ell=0$,  $f(1^-)=3/2$ and $f\left( c^- \right)=c+1/2$,
then $p(F,1/c,c)=\frac{3}{3+c+2 c^2}$ for $c \ne 1/2$.
Based on Theorem \ref{thm:kth-order-gen},
$p(F,2,1/2)=3/8$. $\square$

(c)
For $F$ with pdf
$f(x)=(\pi/2)|\sin(2 \pi x)|\I(0 < x < 1)=
(\pi/2)(\sin(2 \pi x)\I(0 < x \le 1/2)-\sin(2 \pi x)\I(1/2 < x <1 ))$,
we have $k=0$, $f(0^+)=0$,
and $f\left( c^+ \right)=(\pi/2)(\sin(2 \pi c))$ in Theorem \ref{thm:kth-order-gen-r,c} (ii).
Then $p(F,1/(1-c),c)=0$ for $c \ne 1/2$.
As for Theorem \ref{thm:kth-order-gen-r,c} (iii),
we have $\ell=0$,
$f(1^-)=0$ and $f\left( c^- \right)=-(\pi/2)(\sin(2 \pi c)$.
Then $p(F,1/c,c)=0$ for $c \ne 1/2$.
Moreover,
by Theorem \ref{thm:kth-order-gen},
$p(F,2,1/2)=0$ as well. $\square$

For more examples, see Supplementary Materials Section.
In Theorem \ref{thm:kth-order-gen-r,c} (ii),
if we have $f^{(k)}(0^+)=f^{(k)} \left( c^+ \right)$,
then
$\lim_{n \rightarrow \infty}p_{{}_n}(F,1/(1-c),c)=\frac{1}{1+(1-c)^{(k+1)}}.$
In particular, if $k=0$, then
$\lim_{n \rightarrow \infty}p_{{}_n}(F,1/(1-c),c)=1/(2-c)$.
Hence
$\g_{{}_{n,2}}(F,1/(1-c),c)$ and $\g_{{}_{n,2}}(\U,1/(1-c),c)$
would have the same limiting distribution.
Likewise,
in Theorem \ref{thm:kth-order-gen-r,c} (iii),
if we have $f^{(\ell)}(1^-)=f^{(\ell)} \left( c^- \right)$,
then
$\lim_{n \rightarrow \infty}p_{{}_n}(F,1/c,c)=\frac{1}{1+c^{(\ell+1)}}.$
In particular, if $\ell=0$, then
$\lim_{n \rightarrow \infty}p_{{}_n}(F,1/c,c)=1/(1+c)$.
Hence
$\g_{{}_{n,2}}(F,1/c,c)$ and $\g_{{}_{n,2}}(\U,1/c,c)$
would have the same limiting distribution.

\section{Distribution of $\g_{n,m}(F_{X,Y},r,c)$}
\label{sec:dist-multiple-intervals}
We now consider the more challenging case of $m>2$.
For $\omega_1<\omega_2$ in $\R$, define the family of distributions
$$
\mathscr H(\R):=\bigl \{ F_{X,Y}:\;(X_i,Y_i) \sim F_{X,Y} \text{ with support }
\mS(F_{X,Y})=(\omega_1,\omega_2)^2 \subsetneq \R^2,\;\;X_i \sim F_X \text{ and } Y_i \stackrel{iid}{\sim}F_Y \bigr\}.
$$
We provide the exact distribution of $\g_{{}_{n,m}}(F_{X,Y},r,c)$ for
the PICD, $\mathbf D_{n,m}(F_{X,Y},r,c)$, with $F_{X,Y} \in \mathscr H(\R)$ in Theorem \ref{thm:general-Dnm} in the Supplementary Materials Section.

This exact distribution for finite $n$ and $m$ has a simpler form
when $\X$ and $\Y$ points are both uniformly distributed in a bounded interval in $\R$.
Define $\mathscr U(\R)$ as follows
$$
\mathscr U(\R):=\bigl \{ F_{X,Y}: \text{ $X$ and $Y$ are independent}
\;X_i \stackrel{iid}{\sim} \U(\omega_1,\omega_2) \text{ and } Y_i \stackrel{iid}{\sim}\U(\omega_1,\omega_2),
\text{ with } -\infty <\omega_1<\omega_2<\infty \bigr\}.
$$
Clearly, $\mathscr U(\R) \subsetneq \mathscr H(\R)$.
Then we have Corollary \ref{cor:uniform-Dnm} to Theorem \ref{thm:general-Dnm} (see the Supplementary Materials Section).

For $n,m < \infty$, the expected value of domination number is
\begin{equation}
\label{eqn:expected-gamma-Dnm}
\E[\g_{{}_{n,m}}(F_{X,Y},r,c)]=P\left(X_{(1)}<Y_{(1)}\right)+P\left(X_{(n)} > Y_{(m)}\right)+
\sum_{i=1}^{m-1}\sum_{k=1}^n\,P(N_i=k)\,\E[\g_{{}_{[i]}}(F_i,r,c)]
\end{equation}
see Supplementary Materials Section for details and its limit as $n \rightarrow \infty$.

\begin{theorem}
\label{thm:asy-general-Dnm}
\textbf{(Main Result 4)}
Let $\mathbf D_{n,m}(F_{X,Y},r,c)$ be the PICD with $F_{X,Y} \in \mathscr H(\R)$.
Then
\begin{itemize}
\item[(i)]
for fixed $n<\infty$, $\lim_{m \rightarrow \infty}\g_{{}_{n,m}}(F_{X,Y},r,c)=n$ a.s. for all $r \ge 1$ and $c \in [0,1]$.
\item[] For fixed $m<\infty$, and
\item[(ii)]
for $r=1$ and $c \in (0,1)$,
$\lim_{n \rightarrow \infty}P(\g_{{}_{n,m}}(F_{X,Y},1,c)=2m)=1$
and
$\lim_{n \rightarrow \infty}P(\g_{{}_{n,m}}(F_{X,Y},1,0)=m+1)=
\lim_{n \rightarrow \infty}P(\g_{{}_{n,m}}(F_{X,Y},1,1)=m+1)=1$,
\item[(iii)]
for $r>2$ and $c \in (0,1)$, $\lim_{n \rightarrow \infty}P(\g_{{}_{n,m}}(F_{X,Y},r,c)=m+1)=1$,
\item[(iv)]
for $r=2$,
if $c \not= 1/2$,
then $\lim_{n \rightarrow \infty}P(\g_{{}_{n,m}}(F_{X,Y},2,c)=m+1)=1$;\\
if $c = 1/2$,
then $\lim_{n \rightarrow \infty}\g_{{}_{n,m}}(F_{X,Y},2,1/2)\stackrel{d}{=}
m+1+\sum_{i=1}^m B_i \text{ with } B_i \sim \BER(p(F_i,2,1/2))$,
\item[(v)]
for $r \in [1,2)$,
if $r \not= r^*=1/\max(c,1-c)$,
then $\lim_{n \rightarrow \infty} \g_{{}_{n,m}}(F_{X,Y},r,c)$ is degenerate;
otherwise,
it is non-degenerate.
That is, for $r \in [1,2)$,
as $n \rightarrow \infty$,
\begin{equation}
\label{eqn:asy-unif-rM-mult-int}
\g_{{}_{n,m}}(F_{X,Y},r,c) \stackrel{\mathcal L}{\to}
\left\lbrace \begin{array}{ll}
       m+1,           & \text{for $r > r^*$,}\\
       m+1+\sum_{i=1}^m B_i,  & \text{for $r = r^*$,}\\
       2m,           & \text{for $r < r^*$}\\
\end{array} \right.
\end{equation}
\end{itemize}
\end{theorem}
where $B_i \sim \BER(p(F_i,r,c))$.

\noindent
{\bfseries Proof:}
Part (i) is trivial.
Part (ii) follows from Proposition \ref{prop:gamma-Dnm-r-M} and \ref{prop:gamma-Dnm-r=1-M},
since as $n_i \rightarrow \infty$,
we have
$\X_{[i]} \not= \emptyset$ a.s. for all $i$.

\noindent
Part (iii) follows from Theorem \ref{thm:r and M-asy},
since for $c \in (0,1)$,
it follows that $r > r^* $ implies $r>2$
and as $n_i \rightarrow \infty$,
we have
$\g_{{}_{[i]}}(F_i,r,c) \rightarrow 1$ in probability for all $i$.

\noindent
In part (iv), for $r=2$ and $c \not= 1/2$,
based on Corollary \ref{cor:r=2 and M_c=c},
as $n_i \rightarrow \infty$,
we have
$\g_{{}_{[i]}}(F_i,r,c) \rightarrow 1$ in probability for all $i$.
The result for $r=2$ and $c = 1/2$ is proved in \cite{ceyhan:dom-num-CCCD-NonUnif}.

\noindent
Part (v) follows from Theorem \ref{thm:r and M-asy}. $\blacksquare$

The PICD discussed in this article can be viewed as the one-dimensional
version of proportional-edge proximity catch digraphs introduced in \cite{ceyhan:dom-num-NPE-SPL}
for two-dimensional data.
The extension to higher dimensions $\mathbb R^d$ with $d>2$ is
also provided in \cite{ceyhan:dom-num-NPE-SPL,ceyhan:masa-2007}.

\section{Practical Application: Testing Uniformity with Domination Number of PICDs}
\label{sec:app-test-unif}
Let $X_i$, $i=1,2,\ldots,n$, be iid random variables from a distribution $F$ with finite support.
We will employ domination number of the PICD to test for uniformity
of one-dimensional data in a bounded interval, say $(0,1)$;
i.e., our null hypothesis is
$H_o: F= \U(0,1)$.
For this purpose,
we consider three approaches:
\begin{itemize}
\item[] \textbf{approach (i)} In Theorem \ref{thm:r and M},
we derived the $P(\g_{n,2}(\U,r,c)=2)$ for all $n  \ge 2$, $c \in (0,1)$ and $r \ge 1$
for uniform data on $(0,1)$.
In this approach,
we will use $\g_{n,m}(\U,r,c)$ as an \emph{approximate binomial test statistic}
for testing uniformity of data in $(0,1)$
(by Theorem \ref{thm:scale-inv-NYr},
the results would also be valid for uniform data on any bounded interval
$(a_1,a_2)$ with $-\infty < a_1 < a_2 < \infty$).
Here, the approximation is not the large sample convergence to binomial distribution,
but in estimating the probability of success (i.e., $P(\g_{n,2}(\U,r,c)=2)$)
as we are using the expected number of observations for $n_i$ for each subinterval $i$
under uniformity assumption.

\item[] \textbf{approach (ii)} In Theorem \ref{thm:general-Dnm} in the Supplementary Materials Section,
we have the exact distribution of $\g_{n,m}(F,r^*,c)$.
One could use this distribution in an exact testing procedure,
but for convenience, we estimate the Monte Carlo critical values of $\g_{n,m}(F,r^*,c)$
and use it in our tests.

\item[] \textbf{approach (iii)} In Theorem \ref{thm:r and M-asy},
we have the asymptotic distribution of $\g_{n,m}(F,r^*,c)$.
We will use this distribution in an approximate  testing procedure,
where the asymptotic value of the probability of success
(i.e, $\lim_{n\rightarrow \infty}P(\g_{n,2}(\U,r,c)=2)$)
is used in the binomial test (i.e., large sample
approximation is used for the probability of success).
\end{itemize}

In approaches (i)-(iii),
we divide the interval $(0,1)$ into $m$ subintervals,
and treat the interval endpoints to be the $\Y$ points,
i.e., we set $\Y_m=\{0,1/(m-1),2/(m-1),\ldots,1\}$.
This can be done without loss of generality in this context,
because we are testing uniformity of points from one class in a bounded interval,
and the proximity regions are constructed using arbitrarily chosen $\Y$ points.

In both approaches,
we compute the domination number for each subinterval and use $G_n:=\g_{n,m}(r,c)-m$ as our test statistic.
However in approach (i),
we use an approximate binomial test with $G_n$ approximately having $\BIN(m,p_u(r,c,n_i))$
with $n_i=\lfloor{n/m}\rfloor$.
This is an approximate procedure since $\E[N_i]=n/m$,
i.e., $n_i = n/m$ on the average.
Furthermore,
if $G_n < 0$,
then we set the corresponding $p$-value to 0 for this test,
since this is already evidence of severe deviation from uniformity.
In approach (ii),
we use the exact distribution provided in Theorem \ref{thm:general-Dnm}.
However,
\emph{for convenience,
we estimate the critical value by Monte Carlo simulations}.
In particular,
we generate 10000 samples for each $(r,c)$ combination considered
and compute the domination number $\g_{n,m}(r,c)$ for each sample.
Then for the left-sided (right-sided) alternative,
5th percentile (95th percentile) of the test statistic
constitutes the empirical critical value at $\alpha=0.05$ level.

For comparative purposes,
we employ Kolmogorov-Smirnov (KS) test for uniform distribution
and
Pearson's $\chi^2$ goodness of fit test,
since these are the most well known and commonly used tests for
checking the goodness of distributional fit.
We also consider three recently proposed tests, namely,
a uniformity test based on Too-Lin characterization
of the uniform distribution \citep{milosevic:2018},
and two entropy-based tests, denoted as TB1 and TB2 in \citep{zamanzade:2015}.
The entropy tests due to \cite{zamanzade:2015}
reject the null hypothesis of uniformity for small values of TB1 and TB2.
On the other hand,
the uniformity test denoted as $T_n^{(m)}$ in \citep{milosevic:2018},
uses $m=2$ and $k^{th}$ order statistic Too-Lin characterization
rejects for large absolute values of the test statistic
and we take $k=1$ in $T_n^{(2)}$.
For all these tests TB1, TB2 and $T_n^{(2)}$,
the critical values are obtained by Monte Carlo simulations.

We also compare the performance of PICD domination number test
with that of the arc density of two ICDs:
(i) PICD and (ii) Central ICD (CICD) which is based on
central similarity (CS) proximity region.
For a digraph $D_n=(\V,\A)$ with vertex set $\V$
and arc set $\A$,
the arc density of $D_n$ which is of order $|\V| = n \ge 2$,
denoted $\rho(D_n)$, is defined as
$\rho(D_n) = \frac{|\A|}{n(n-1)}$
where $|\cdot|$ stands for the set cardinality function (\cite{janson:2000}).
So $\rho(D_n)$ is the ratio of the number of arcs
in the digraph $D_n$ to the number of arcs in the complete symmetric digraph of order $n$,
which is $n(n-1)$.
For $n \le 1$, we set $\rho(D_n)=0$.
Arc density of ICDs is shown to be a $U$-statistic,
and hence its asymptotic distribution is a normal distribution,
provided that its asymptotic variance is positive (\cite{ceyhan:metrika-2012}).
Arc density of PICDs is studied in \cite{ceyhan:metrika-2012}
and but not used in testing uniformity before.
Likewise,
CICDs were introduced in \cite{ceyhan:revstat-2016} and
its arc density was employed for testing uniformity in the same article as well.
CS proximity region is defined  as follows (\cite{ceyhan:revstat-2016}):
For $\tau >0$, $c \in (0,1)$ and $x\in \mI_i$
\begin{multline}
\label{eqn:NCSt-general-defn1}
N_{CS}(x,\tau,c)=\\
\begin{cases}
\left(x-\tau\,\left(x-Y_{(i-1)}\right),x+\frac{\tau\,(1-c)}{c}\left(x-Y_{(i-1)}\right)\right) \bigcap \left( Y_{(i-1)},Y_{(i)} \right) & \text{if $x \in (Y_{(i-1)},M_{c,i})$,}
\vspace{0.2cm}\\
\left(x-\frac{c\,\tau\,}{1-c}\left(Y_{(i)}-x\right),x+\tau\,\left(Y_{(i)}-x\right)\right)  \bigcap \left( Y_{(i-1)},Y_{(i)} \right)     & \text{if $x \in \left( M_{c,i},Y_{(i)} \right)$.}
\end{cases}
\end{multline}

\subsection{Empirical Size Analysis}
\label{sec:emp-size}
We perform a size analysis to determine whether the tests have the
appropriate size in testing $H_o: F= \U(0,1)$.
Along this line,
we partition the domain of $p_u(r,c,n)$ for $r$ and $c$ as follows.
We take $c=.01,.02,\ldots,.99$ and $r=1.00,1.01, \ldots, 2.10$,
and consider each $(r,c)$ combination on a $99 \times 210$ grid
with $n=20,50,100$.
For each $(r,c)$ combination,
we generate $N_{mc}=10000$ samples each of size $n$ iid from $\U(0,1)$ distribution.
We also partition the interval $(0,1)$ into $m$ equal subintervals
where $m$ equals $\sqrt{n}$ (rounded to the nearest integer)
whose choice is inspired by the choice of windows size in entropy-based
goodness-of-fit tests (\cite{grzegorzewski:1999}).
This choice is not to justify the use of binomial distribution,
as the distribution of the domination number is available for any $r>1$, $c\in (0,1)$ and finite $n \ge 2$.
That is,
the binomial distribution would hold regardless of the size of $m$,
but it is preferable that it is large enough to give enough resolution for the discrete binomial test.
The reason we use the $(r^*,c)$ combination that renders the asymptotic distribution nondegenerate is that
other choices of $(r,c)$ could make the distribution close to being degenerate for large $n$,
whose rate of convergence to 0 or 1 depends on the values of $r$ and $c$.
Then for each subinterval,
we compute the domination number (which is either 0, 1, or 2),
and sum the domination numbers over the $m$ subintervals
and thus obtain $\g_{n,m}(r,c)$.
We use this summed domination number minus $m$,
i.e., $G_n$,
in an approximate binomial test statistic
(i.e., we follow approach (i) above).
Under $H_o$, $G_n$ approximately has $\BIN(m,p_u(r,c,\lfloor{n/m}\rfloor))$ distribution,
so we compute the $p$-value based on the binomial test with $m$ trials
and probability of success being $p=p_u(r,c,\lfloor{n/m}\rfloor)$ for the two-sided alternative.
For each of the 10000 samples generated,
we also compute the arc density of the ICDs
for the parameters of choice
and appeal to the asymptotic normality of the arc density of these ICDs.
We compute size estimates based on the corresponding normal critical values
for the arc density for each $(r,c)$ (resp. $(\tau,c)$) combination for PICD (resp. CICD).
For each sample,
we also compute KS, $\chi^2$, TB1 and TB2 and $T_n^{(2)}$ tests as well.
In the $\chi^2$ test,
we use the same partition of $(0,1)$ with $m$ subintervals,
and compare the observed and expected frequencies of data points
in these subintervals under uniformity.
Empirical size is estimated as the frequency of number of times $p$-value
is significant at $\alpha=.05$ level divided by $N_{mc}=10000$.
With $N_{mc}=10000$,
empirical size estimates larger than .0536 are deemed liberal,
while those less than .0464 are deemed conservative.
These bounds are also based on binomial test for the proportions for $N_{mc}=10000$ trials
at $.05$ level.
Since the entropy tests TB1 and TB2 and $T_n^{(2)}$ test and PICD domination number test with approach (ii)
are using critical values based on Monte Carlo simulations, we exclude
them in the empirical size comparison, as they, by construction, attain the nominal size.
However, we find the empirical critical values for these tests as the sample $100 \alpha^{th}$ percentile
of the TB1 and TB2 values computed in our simulations,
and
$100 (1-\alpha)^{th}$ percentile
of the $|T_n^{(2)}|$ values computed in our simulations.

\begin{figure}[ht]
\centering
\rotatebox{-90}{ \resizebox{2.12 in}{!}{ \includegraphics{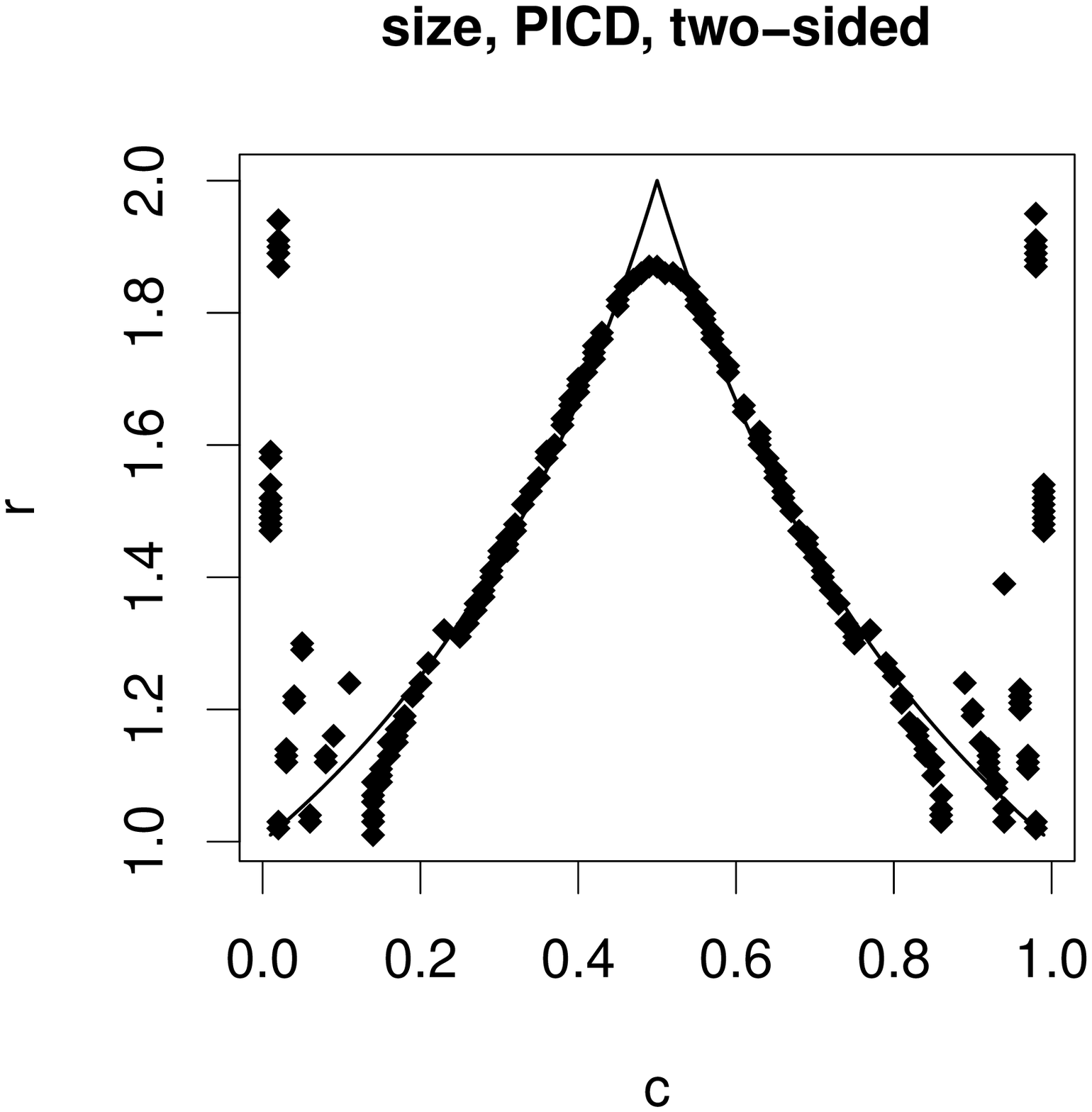}}}
\rotatebox{-90}{ \resizebox{2.12 in}{!}{ \includegraphics{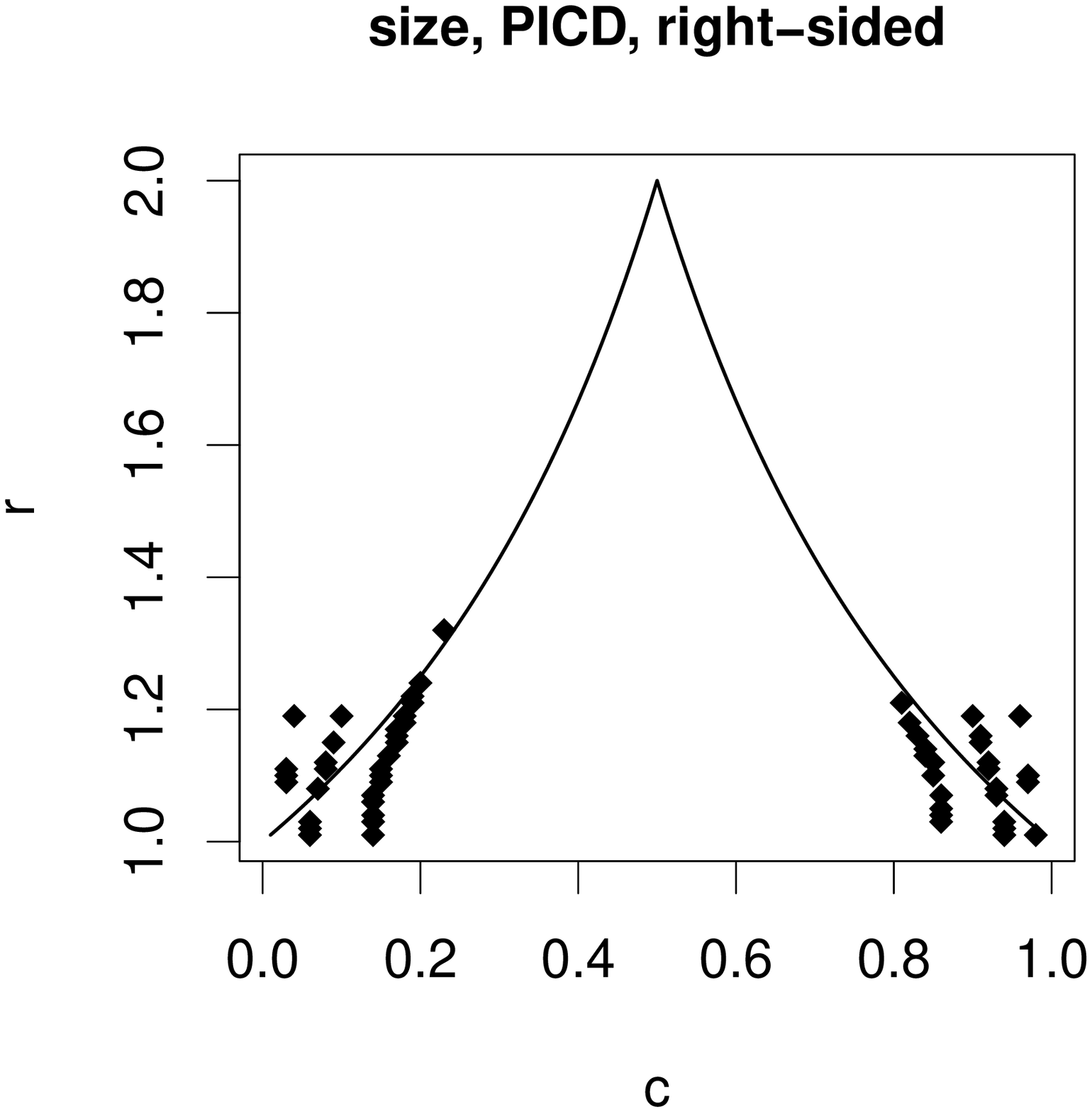}}}
\rotatebox{-90}{ \resizebox{2.12 in}{!}{ \includegraphics{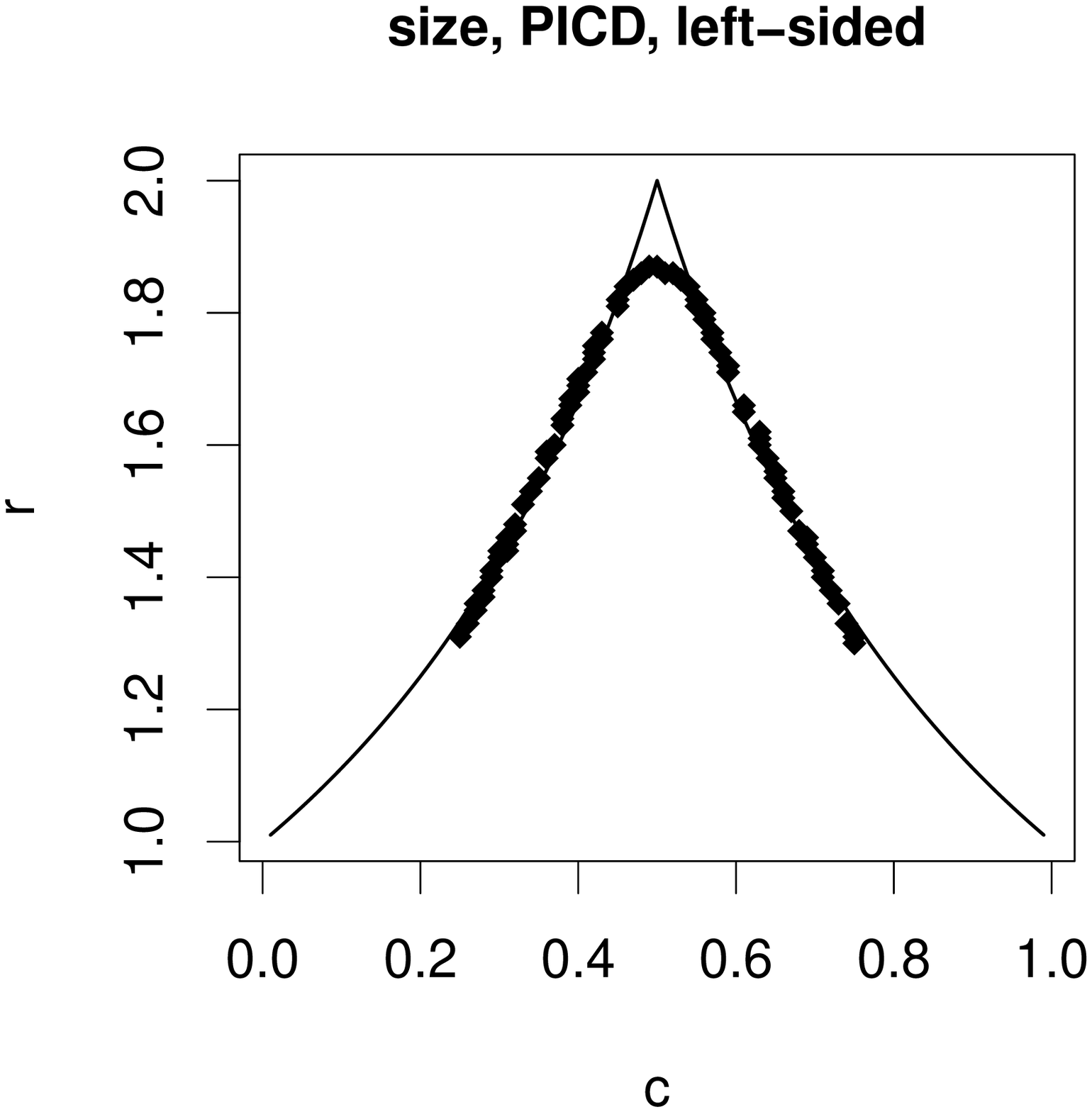}}}
\caption{
\label{fig:PE-Dom-size-image-plots}
The empirical size estimates of the tests based on \textbf{domination number of PICD with approach (i)} for $n=20$ and $N_{mc}=10000$
for $r=1.01,1.02,\ldots,2.10$ and $c=.01,.02,\ldots,.99$ for
the two-sided, right-sided and left-sided alternatives (left to right);
size estimates significantly different from .05 are blanked out,
while size estimates within .0536 and .0464 are plotted as black dots.
The solid lines in the bottom row plots indicate the case of $(r,c)=(r^*,c)$
which yields the asymptotically non-degenerate distribution for the domination number.
}
\end{figure}

We present the empirical size estimates of the tests based on the \emph{domination number of
PICD with approach (i)} as two-level image plots (with empirical sizes not significantly different from 0.05 in black dots,
and others are blanked out in white) with
$n=20$, $c=.01,.02,\ldots,.99$ and $r=1.00,1.01, \ldots, 2.10$
in Figure \ref{fig:PE-Dom-size-image-plots}
(the plots for $n=50$ and $100$ have the similar trend, hence not presented).
Notice that the sizes for the right-sided alternatives are at about the nominal
level for $(r,c)$ around $(1,0)$ or $(1,1)$,
while the sizes for the left-sided alternatives are about the nominal level of 0.05
at the asymptotically non-degenerate $(r,c)=(r^*,c)$ pairs for $c \in (.25,.75)$.
The reason for the asymmetric performance for the left-sided versus right-sided alternatives is that
$p_u(r,c,n)$ values are higher (i.e., close to 1) around $(r,c)=(1,0)$ or $(1,1)$,
and
lower for other values,
but away from 1 or 0 for $(r,c)=(r^*,c)$ pairs.
Therefore,
for the power analysis,
we only consider $(r,c)=(r^*,c)$ pairs,
as empirical size is closer to the nominal level for these parameters in approach (i).

In approach (ii),
by construction the size estimates should be around the nominal level of .05.
But due to the discrete nature of $\g_{n,m}$ with very few atoms for small $n$ and $m$,
the exact test is liberal or conservative depending on whether we include the critical value in our
size estimation.
In particular,
let $\g_{n,m,i}$ be the domination number for sample $i$ and
$\g_{.05}$ be the 5th percentile for the exact distribution of $\g_{n,m}(r,c)$ (as in Theorem \ref{thm:general-Dnm}).
Also let $\alpha_{inc}:=\sum_{i=1}^{N_{mc}} \g_{n,m,i} \le \g_{.05}$
and
$\alpha_{exc}:=\sum_{i=1}^{N_{mc}} \g_{n,m,i} < \g_{.05}$.
Then for testing the left-sided alternative,
$\alpha_{inc}$ tends to be much larger than .05 (implying the procedure is liberal)
and
$\alpha_{exc}$ tends to be much smaller than .05 (implying the procedure is conservative).
In our power computations with approach (ii),
we adjust for this discrepancy.

The size estimates in approach (iii) depend on the sample size $n$, and the parameters $r$ and $c$,
i.e. they tend to be liberal for some values of $(r,c)$, and conservative for others,
especially when $n$ is not large enough.
Our simulations suggest that large sample sizes are needed
(about 30 or more per each subinterval seems to work),
where the required sample size would also depend on $r$ and $c$ as well.
Hence we do not present approach (iii)
except for the large sample simulation cases (in the cases with $n=1000$ here).

We estimate the empirical sizes of the tests based on the arc density of the PICDs and CICDs
for $n=20,50$ and 100 and $c=.01,.02,\ldots,.99$ with $r=1.1,1.2, \ldots, 10.0$ for PICDs
and $\tau=.1,.2, \ldots, 10.0$ for CICDs.
For the one-sided alternatives,
the regions at which size estimates are about the nominal level of 0.05 are somewhat complementary,
in the sense that,
the sizes are appropriate for the parameter combinations in one region for left-sided alternative
and mostly in its complement for the right-sided alternative.
We also observe that
arc density of PICD has appropriate size for the two-sided alternative for more parameter combinations,
and arc density of CICD has appropriate size for the left-sided alternative for more parameter combinations.
See Figure \ref{fig:AD-size-image-plots} in the Supplementary Materials Section for the related image plots of the empirical size estimates.

\subsection{Empirical Power Analysis}
\label{sec:emp-power}
We perform a power analysis to determine which tests have better performance
in detecting deviations from uniformity.
For the alternatives (i.e., deviations from uniformity),
we consider five types of non-uniform distributions with support in $(0,1)$:

\begin{itemize}
  \item[(I)] $f_1(x,\delta)=(2 \delta x+1-\delta)\I(0<x<1)$,
  \item[(II)] $f_2(x,\sigma)=\phi(x,1/2,\sigma)/(\Phi(1,1/2,\sigma)-\Phi(0,1/2,\sigma))\I(0<x<1)$
where $\phi(x,1/2,\sigma)$ is the pdf for normal distribution with mean $\mu=1/2$ and standard deviation $\sigma$,
(i.e., normal distribution with $\mu=1/2$ restricted to $(0,1)$),
  \item[(III)] $f_3(x,\delta)=(\delta\, \left( x-1/2 \right)^{2}+1-\delta/12)\I(0<x<1)$,
  \item[(IV)] $f_4(x,\ve)=(1/(1-2 m \ve))\I(x \in (0,1) \setminus \cup_{i=0}^m (i/m-\ve,i/m+\ve))$, that is, $f_4(x,\ve)$ is a pdf so that
$\ve \times 100$ \% of the regions around the $m$ subinterval end points are prohibited,
and the data is uniform in the remaining regions.
  \item[(V)] $f_5(x,\ve')=(1/2 m \ve')\I(x \in (0,1) \cap (\cup_{i=0}^m (i/m-\ve',i/m+\ve')))$, that is, $f_5(x,\ve')$ is a distribution so that
data is uniform over the $\ve' \times 100$ \% of the regions around the $m$ subinterval end points are prohibited,
and the remaining regions are prohibited.
Notice that the supports of $f_4(x,\ve)$ and $f_5(x,\ve')$ are complimentary in $(0,1)$.
\end{itemize}

That is,
\begin{multline*}
H^I_a:~ f=f_1(x,\delta) \text{ with } \delta \in (0,1)
~~~
H^{II}_a:~ f=f_2(x,\sigma) \text{ with } \sigma > 0
~~~
H^{III}_a:~ f=f_3(x,\delta) \text{ with } \delta \in (0,12]\\
~~~
H^{IV}_a:~ f=f_4(x,\ve) \text{ with } \ve \in (0,1/2)
\text{ and }
H^{V}_a:~ f=f_5(x,\ve') \text{ for } \ve' \in (0,1/2)
\end{multline*}

In type I alternatives,
$\delta=0$ corresponds to $\U(0,1)$ distribution,
and with increasing $\delta>0$, the density of the distribution is more clustered around 1 and less clustered around 0;
in type II alternatives,
with decreasing $\sigma$, the density of the distribution gets more clustered around 1/2
(and less clustered around the end points, 0 and 1);
and
in type III alternatives,
$\delta=0$ corresponds to $\U(0,1)$ distribution,
and with increasing $\delta>0$, the density of the distribution is
more clustered around the end points, 0 and 1, and less clustered around 1/2.
\emph{Types IV and V alternatives are actually motivated from two-class one-dimensional spatial point patterns
called segregation and association.}
Roughly defined,
\emph{segregation} is the pattern in which points from the same class tend to cluster,
while
under \emph{association},
points from one class is clustered around the points from the other class and vice versa.
In one-dimensional case,
the segregation alternative is as in $H^{IV}_a$,
where $X$ points are distributed according to $f_4$ and $Y$ points constitute the end points
of the interval partition of $(0,1)$ (i.e., \{0,1/(m-1),2/(m-1),\ldots,1\}.
Hence, $X$ points tend to stay away from $Y$ points,
which suggests segregation between the classes $X$ and $Y$.
Furthermore,
$\ve=0$ in type IV alternative corresponds to the null case (i.e., uniform distribution).
The association alternative is as in $H^{V}_a$.
The pdf under type I alternative is skewed left for $\delta > 0$,
while pdfs under other alternatives are symmetric around 1/2.
See Figure \ref{fig:pdfs-5-alternatives} in the Supplementary Materials Section for sample plots of the pdfs
with various parameters for alternative types I-III.

\begin{figure}[ht]
\centering
\rotatebox{-90}{ \resizebox{2.12 in}{!}{ \includegraphics{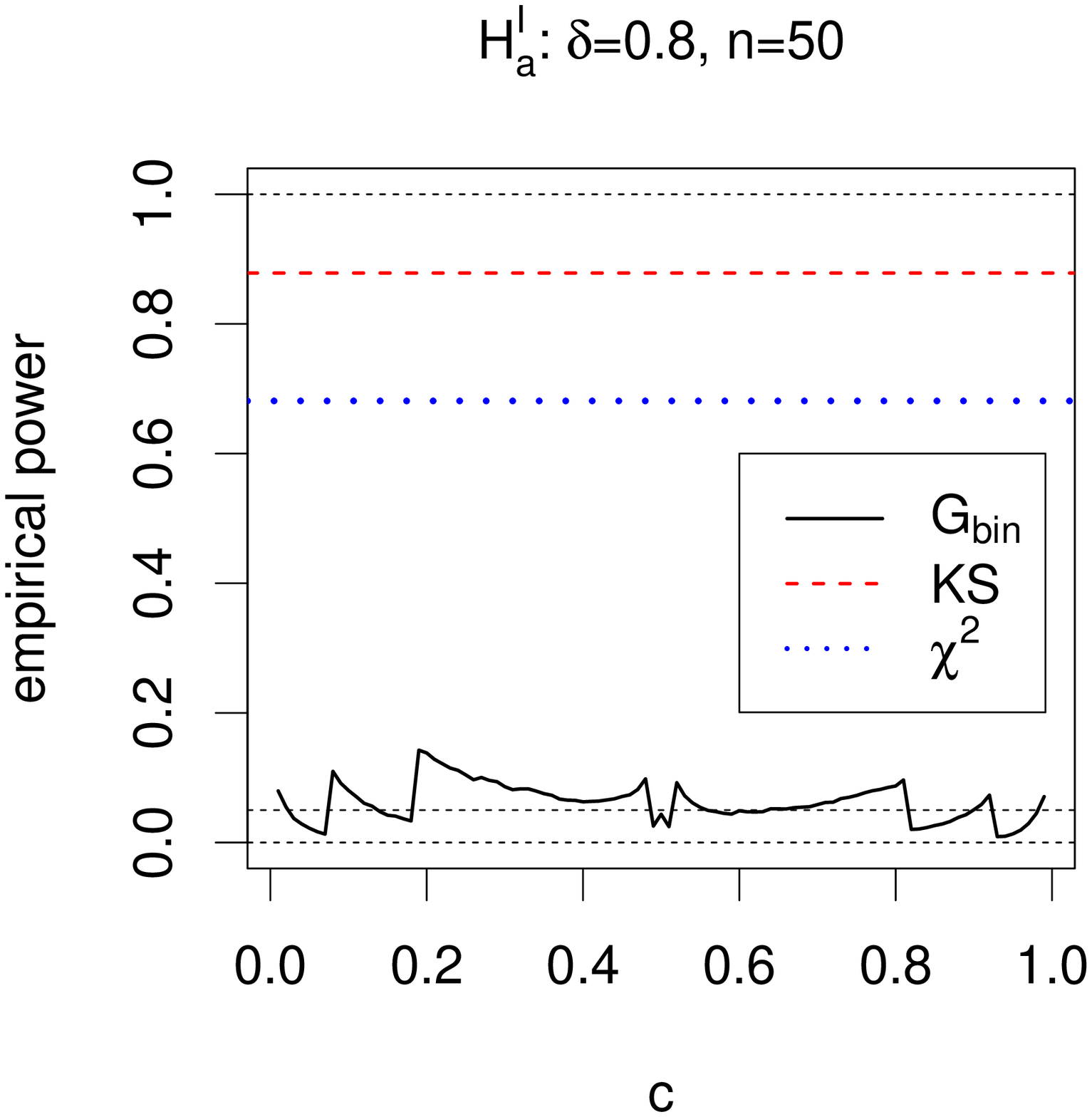}}}
\rotatebox{-90}{ \resizebox{2.12 in}{!}{ \includegraphics{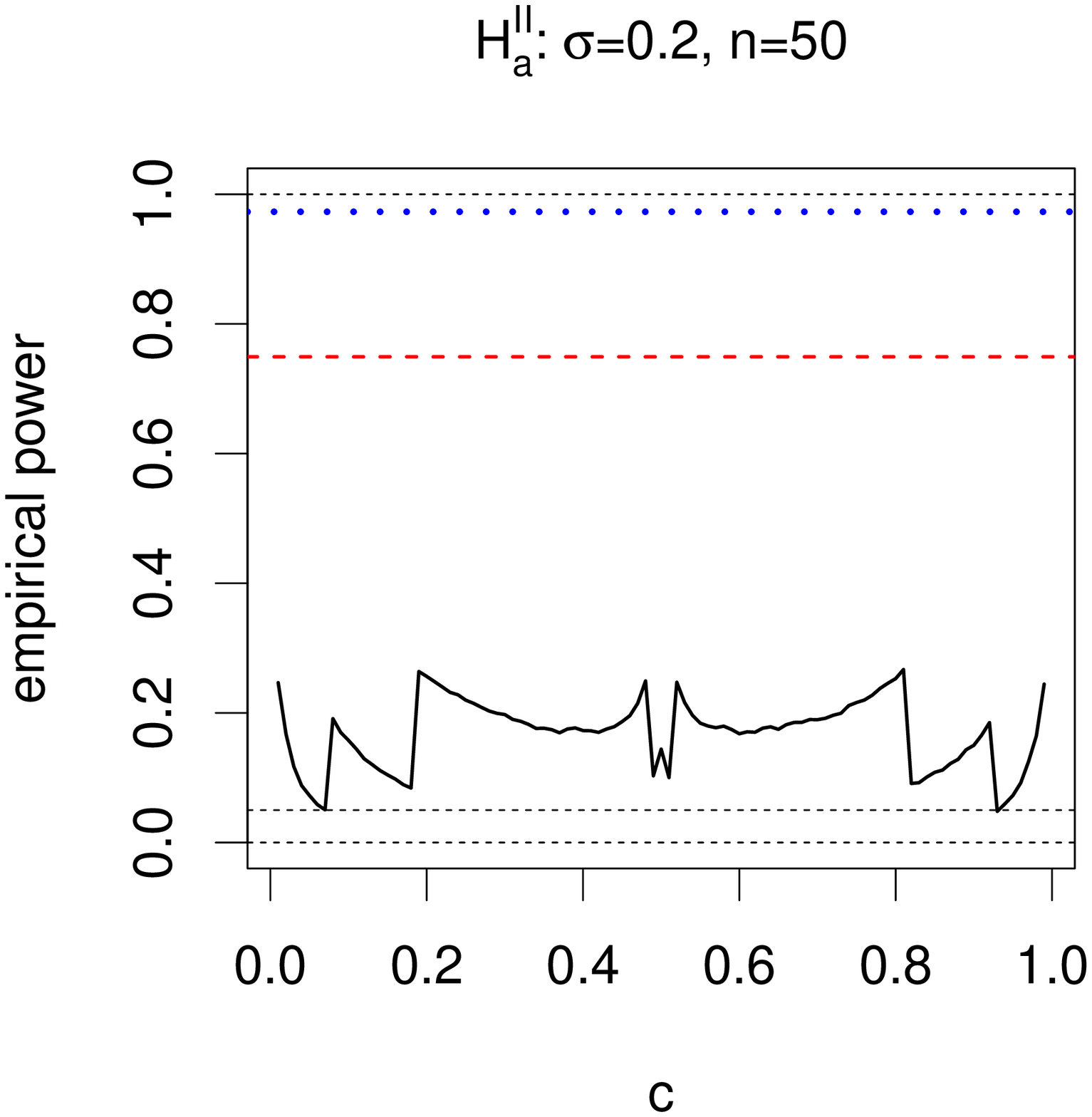}}}
\rotatebox{-90}{ \resizebox{2.12 in}{!}{ \includegraphics{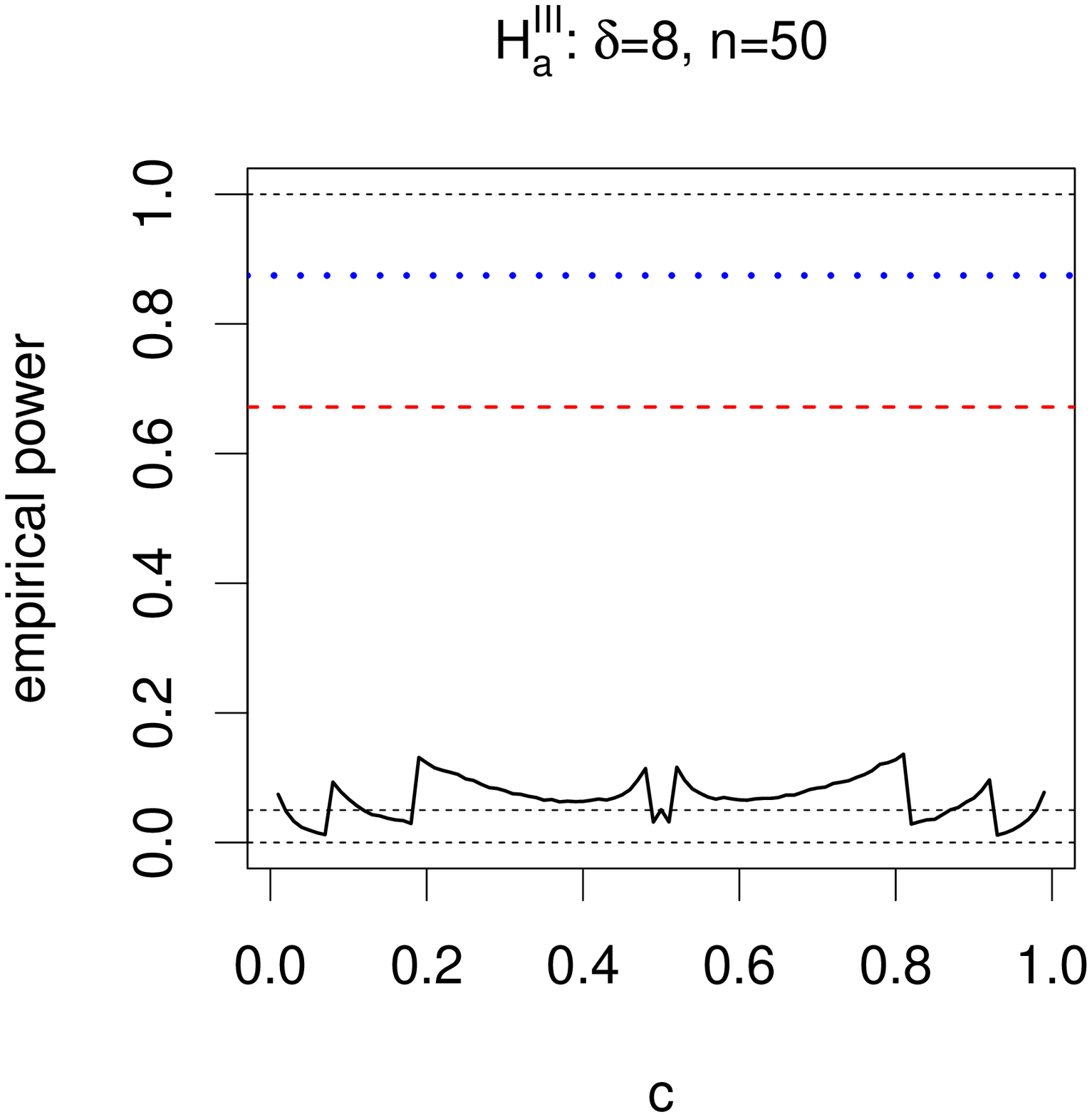}}}
\rotatebox{-90}{ \resizebox{2.12 in}{!}{ \includegraphics{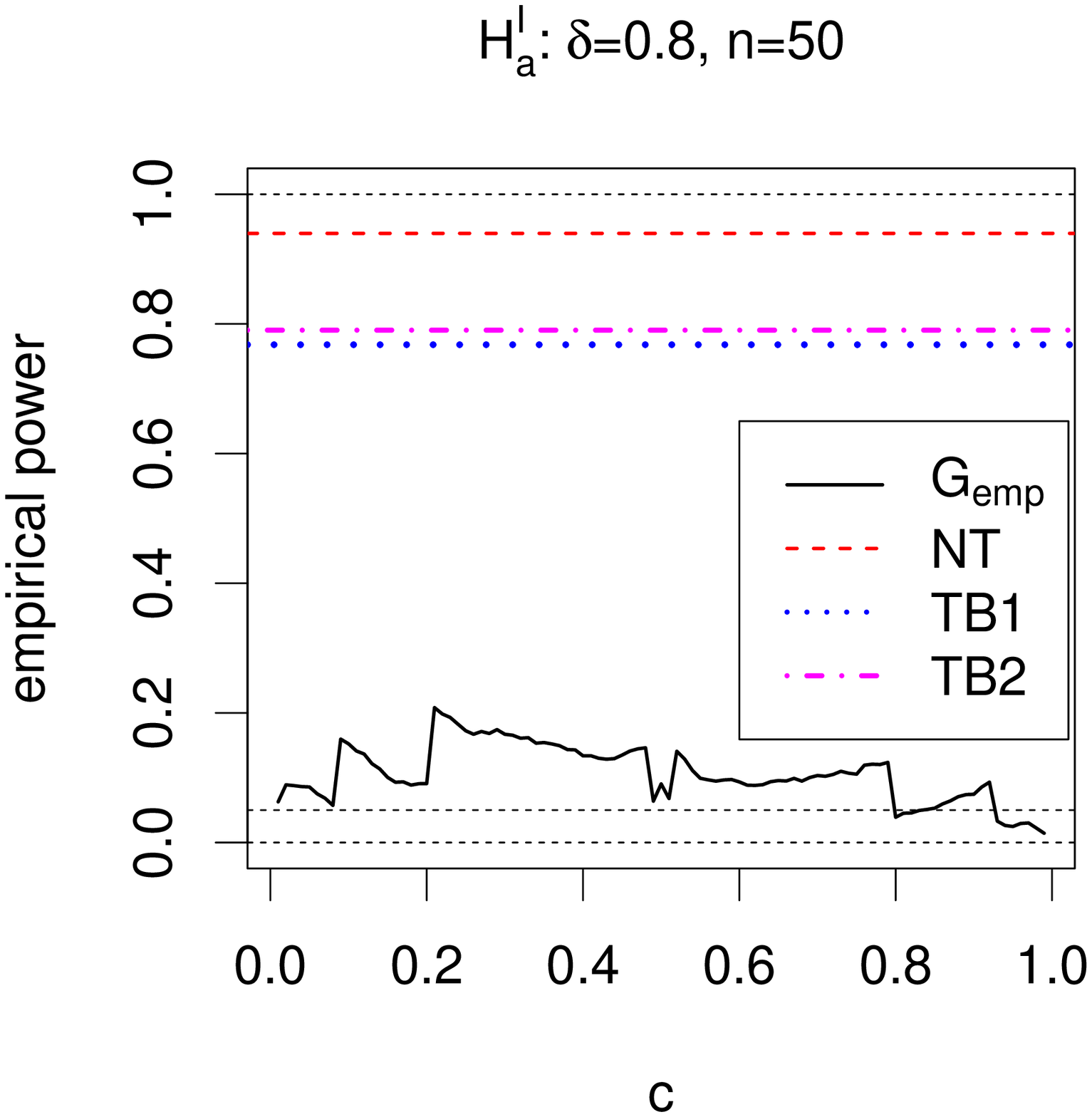}}}
\rotatebox{-90}{ \resizebox{2.12 in}{!}{ \includegraphics{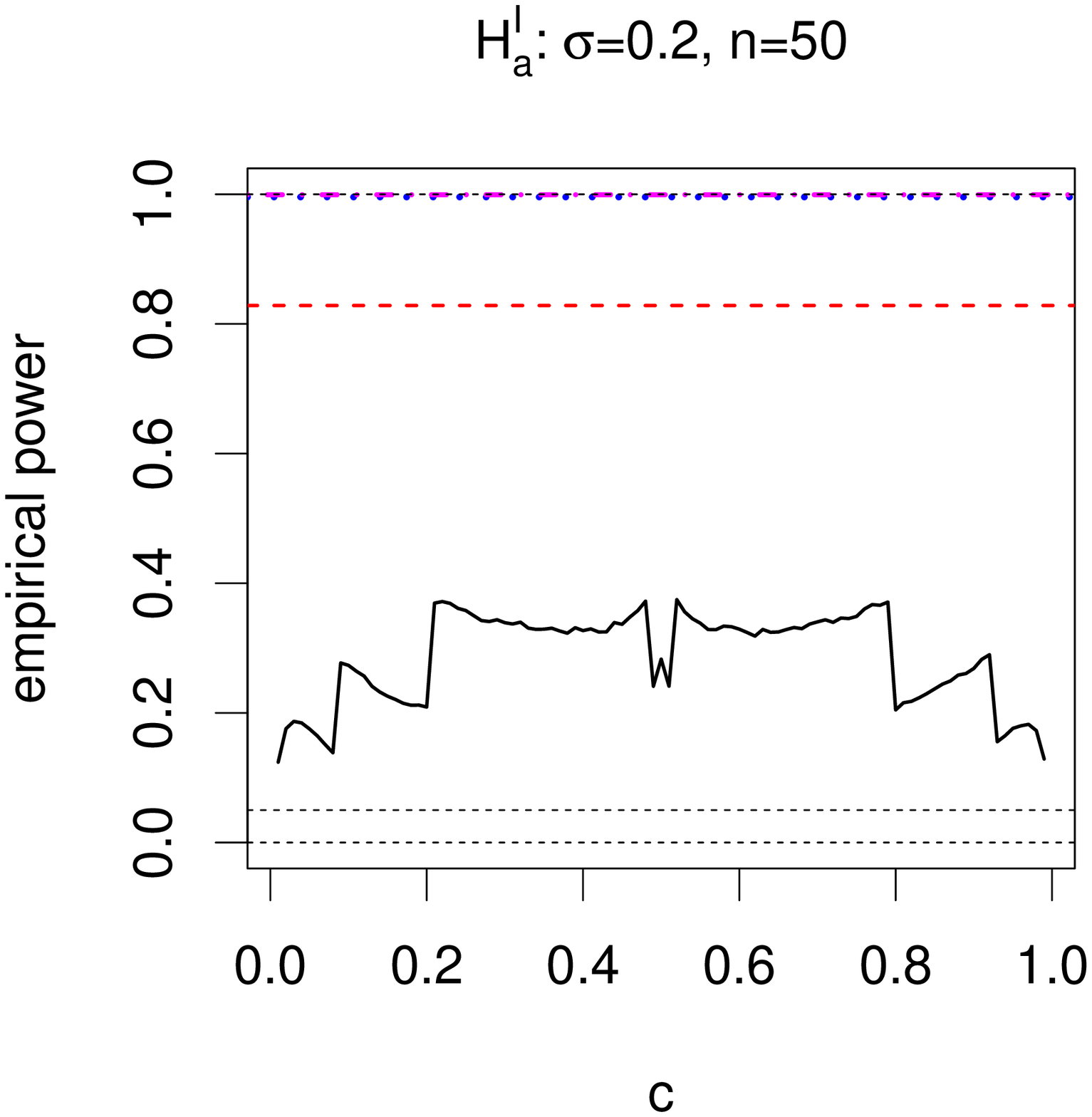}}}
\rotatebox{-90}{ \resizebox{2.12 in}{!}{ \includegraphics{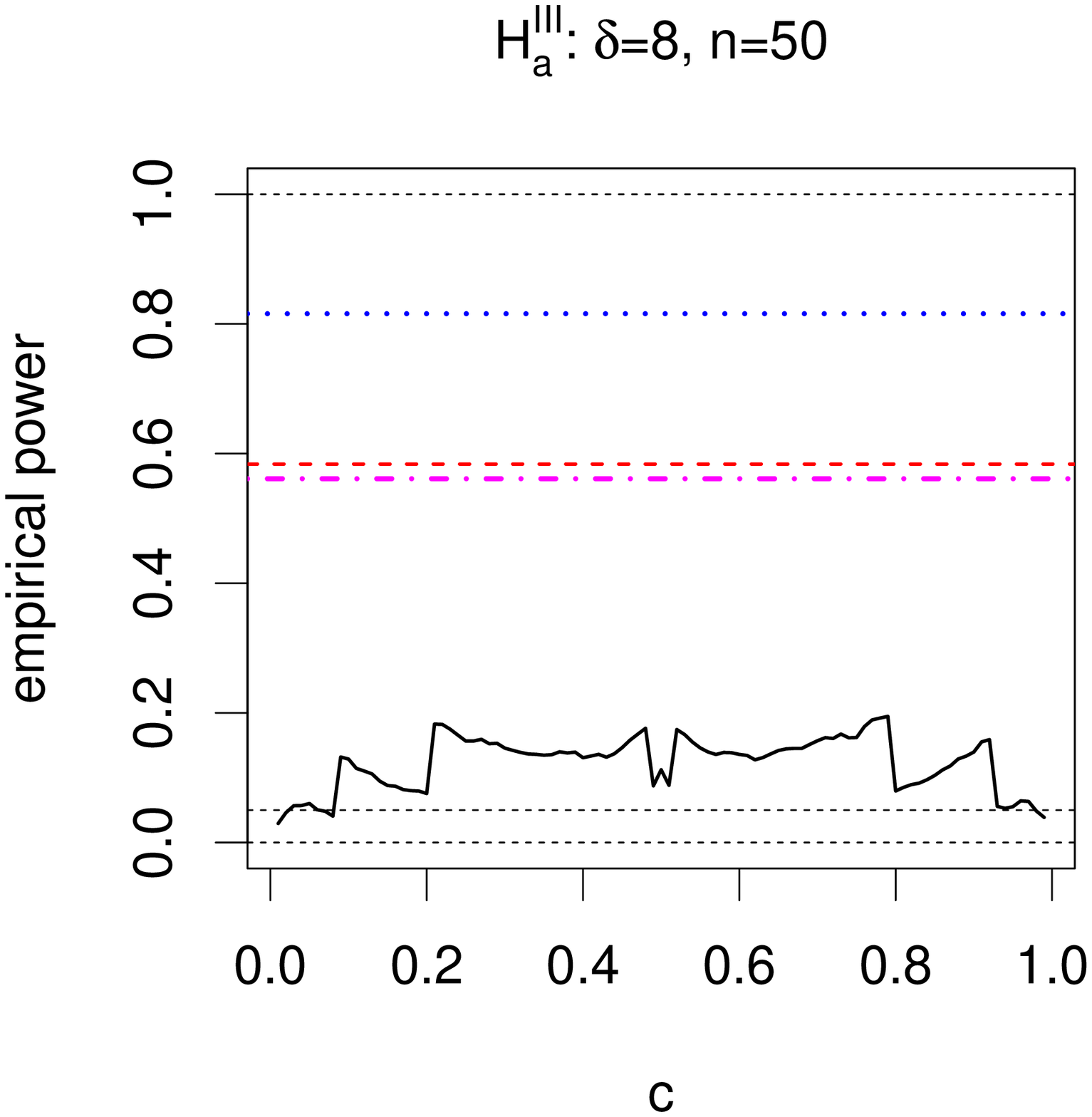}}}
\caption{
\label{fig:sm-power-plots-Ha-I-III}
Power estimates under
under $H^{I}_a:~ F=F_1(x,\delta=.8)$, $H^{II}_a:~ F=F_2(x,\sigma=0.2)$, $H^{III}_a:~ F=F_3(x,\delta=8)$,
with $n=50$ and $N_{mc}=10000$.
$G_{bin}$ and $G_{emp}$: tests based on domination number of PICD with approaches (i) and (ii), respectively,
KS: Kolmogorov-Smirnov test,
$\chi^2$: Chi-square test,
NT: $T_n^{(2)}$ test based on the uniformity characterization,
TB1 and TB2: two versions of the entropy-based tests.
Tests presented in each row are indicated in the legend in that row.
}
\end{figure}

\begin{figure}[ht]
\centering
\rotatebox{-90}{ \resizebox{2.12 in}{!}{ \includegraphics{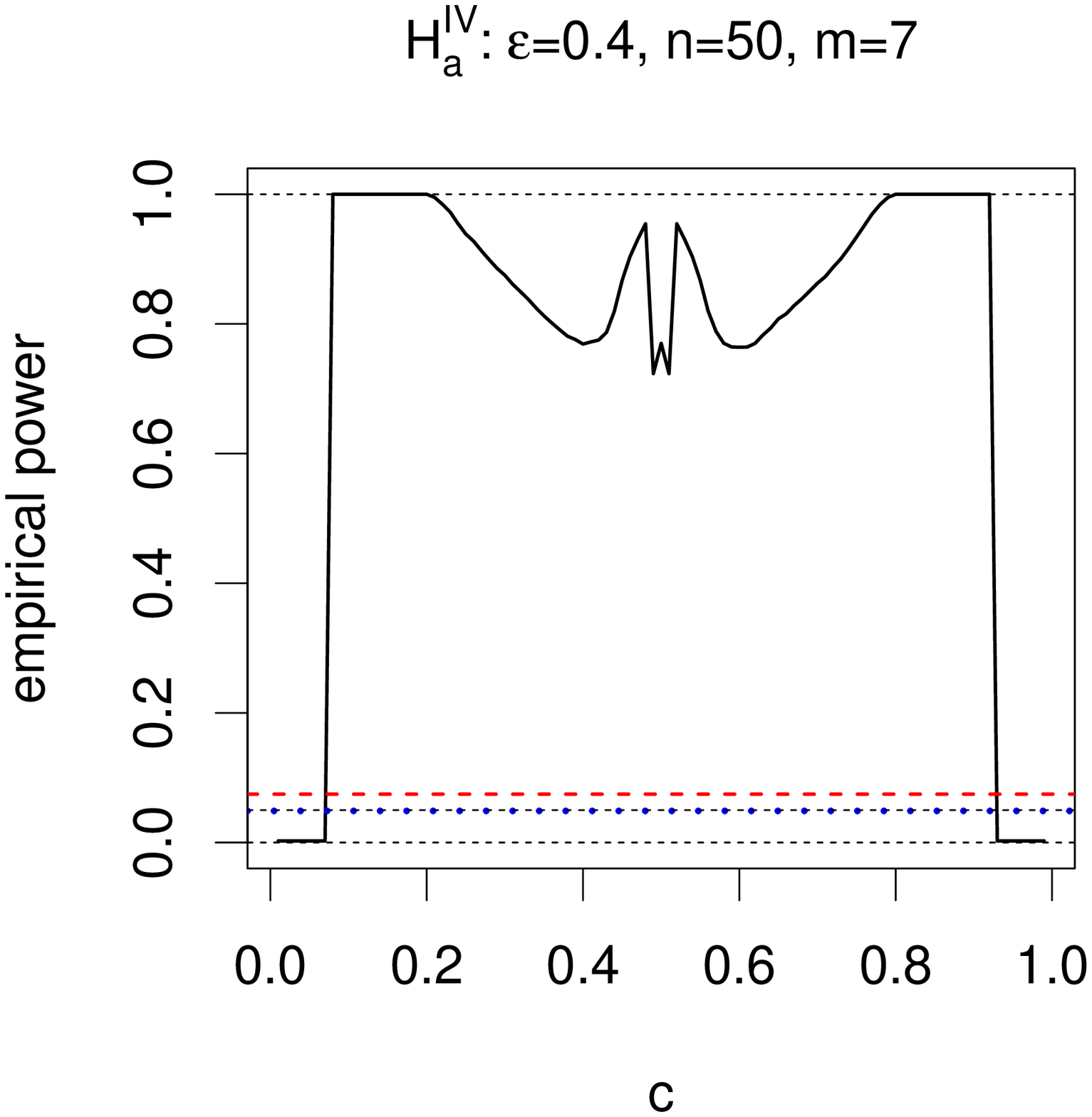}}}
\rotatebox{-90}{ \resizebox{2.12 in}{!}{ \includegraphics{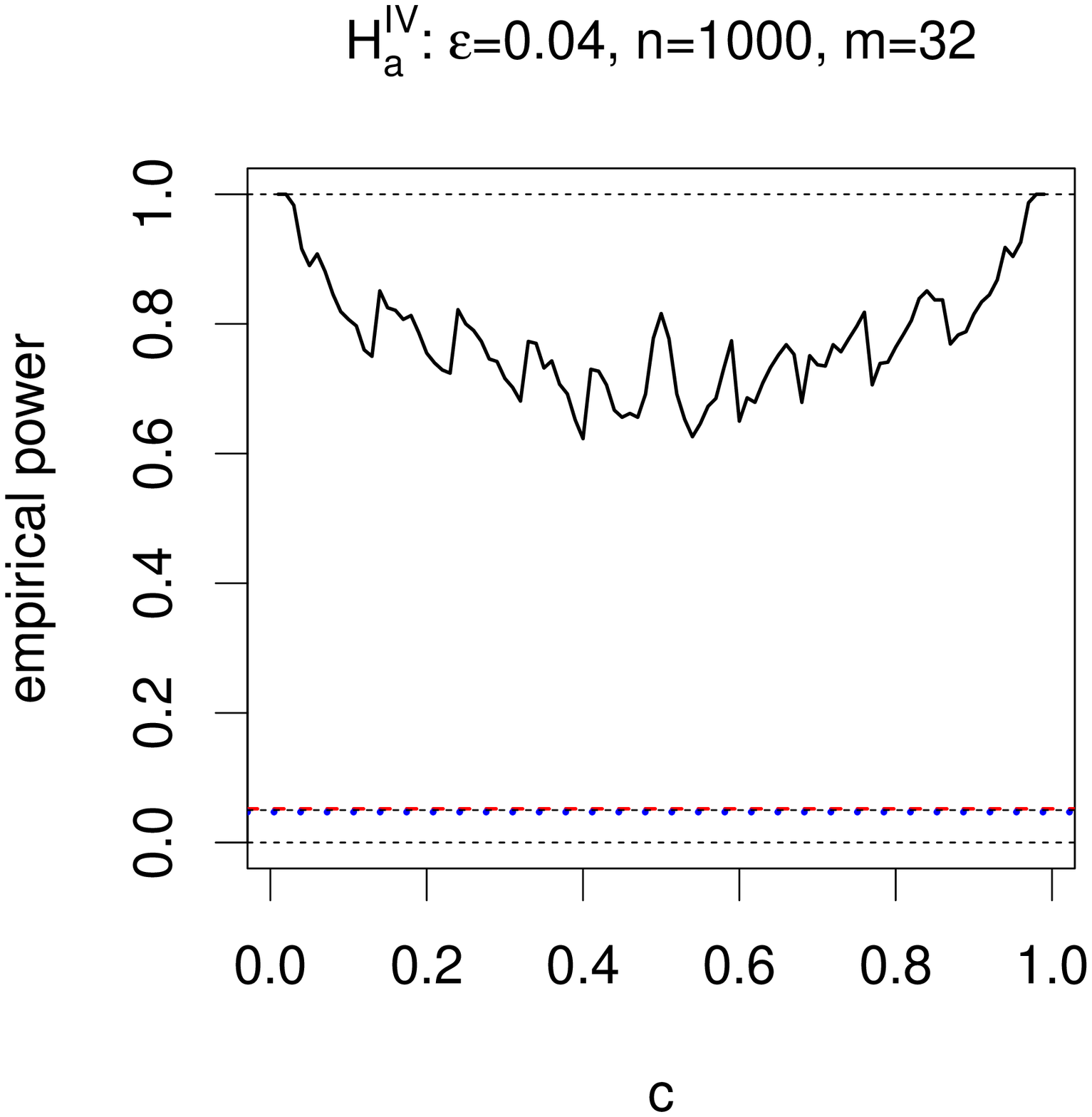}}}
\rotatebox{-90}{ \resizebox{2.12 in}{!}{ \includegraphics{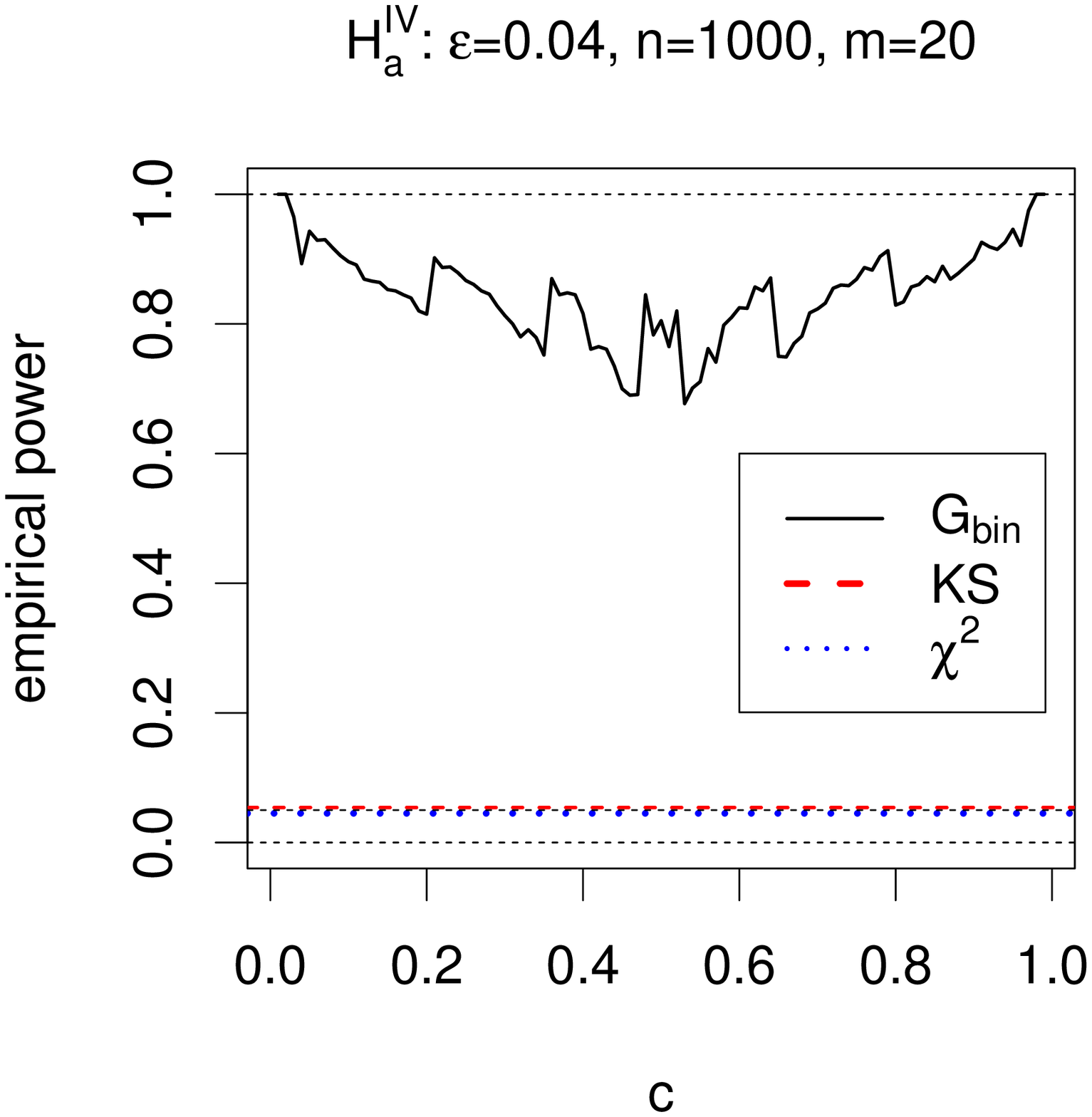}}}
\rotatebox{-90}{ \resizebox{2.12 in}{!}{ \includegraphics{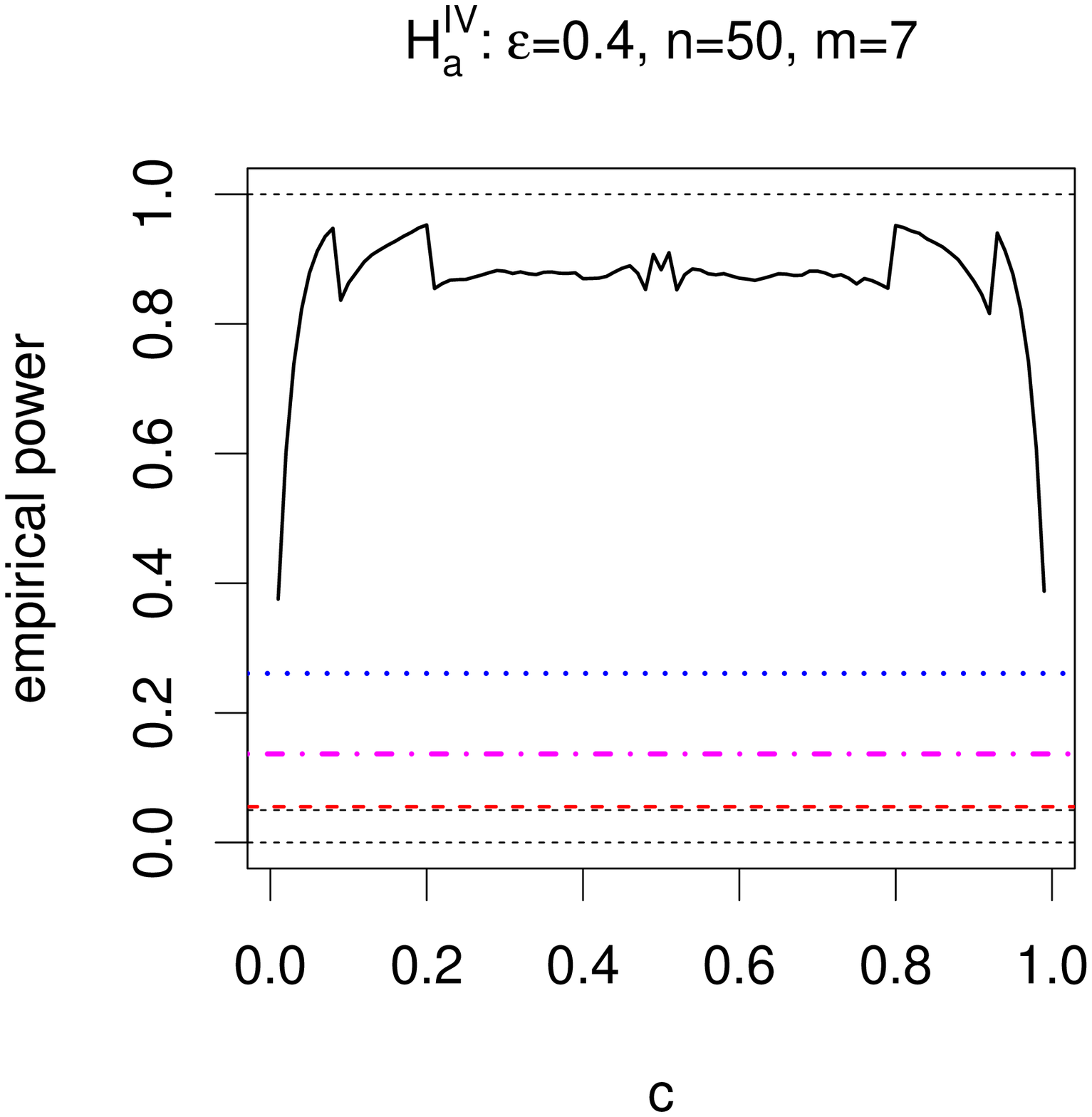}}}
\rotatebox{-90}{ \resizebox{2.12 in}{!}{ \includegraphics{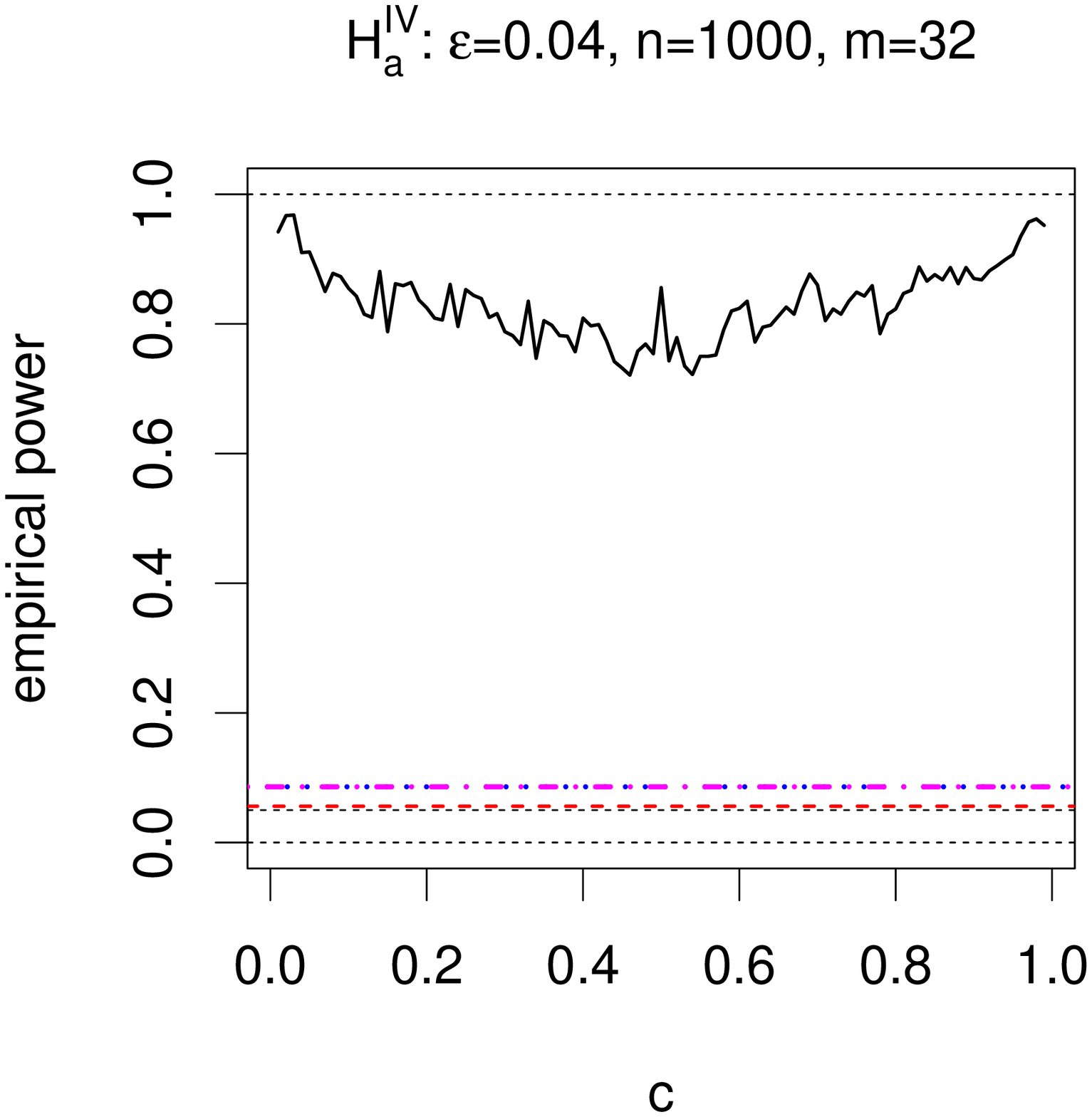}}}
\rotatebox{-90}{ \resizebox{2.12 in}{!}{ \includegraphics{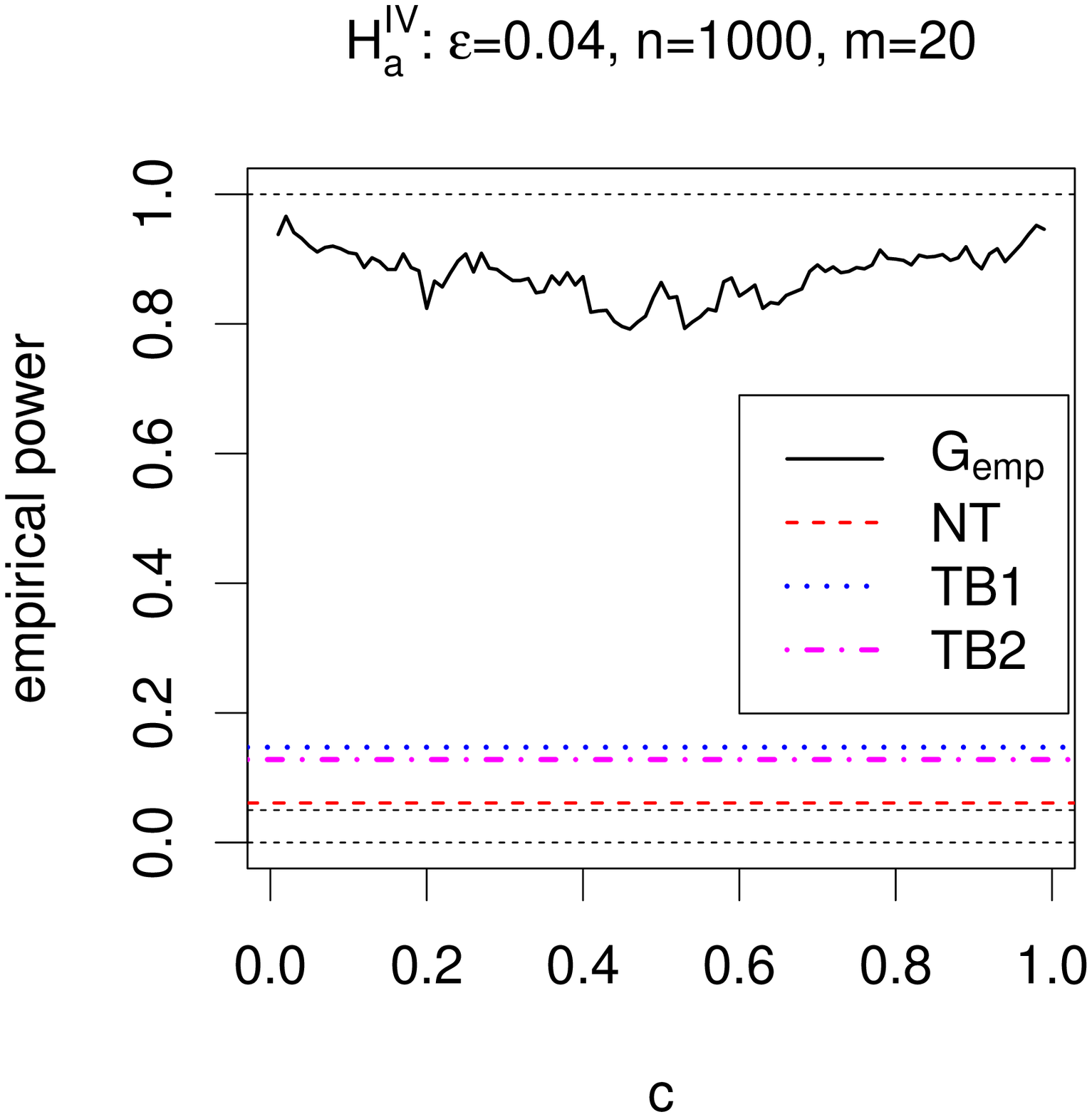}}}
\caption{
\label{fig:sm-power-plots-Ha-IV}
Power estimates under
under $H^{IV}_a:~ F=F_1(x,\ve)$,
case (1) with $\ve=.4, \,n=50, \,m=7$ and $N_{mc}=10000$ (left column),
case (2) $\ve=.04, \,n=1000, \,m=32$ and $N_{mc}=1000$ (middle column),
and
case (3) $\ve=.04, \,n=1000, \,m=20$ and $N_{mc}=1000$ (right column).
Labeling of the tests are as in Figure \ref{fig:sm-power-plots-Ha-I-III}.
Tests presented in each row are indicated in the legend in that row.
}
\end{figure}

Under each alternative,
we generate $n$ points according to the specified alternatives with various parameters.
In particular,
for $H^I_a:~ F=F_1(x,\delta)$, we consider $\delta=.2,.4,.6,.8$,
for $H^{II}_a:~ F=F_2(x,\sigma)$, we consider $\sigma=.1,.2,.3,.4$,
for $H^{III}_a:~ F=F_3(x,\delta)$, we consider $\delta=2,4,6,8$,
and
for $H^{IV}_a:~ F=F_4(x,\ve)$, we consider $\ve=.1,.2,.3,.4$ (also called $H^{IV}_a$-case (1)).
For the domination number of PICDs,
we replicate each case $N_{mc}$ times
for $(r,c)=(r^*,c)$ with $c=.01,.02.\ldots,.99$
(i.e., for $(r,c)$ values that make $\g_{n,2}(\U,r,c)$ non-degenerate in the limit
(see Theorem \ref{thm:r and M-asy})).
We compute the power using the critical values
based on $\BIN(m,p_u(r,c,\lfloor{n/m}\rfloor))$ distribution (i.e., approach (i))
and based on the empirical critical values (i.e., approach (ii)).
For types I-IV alternatives,
we take $n=20, 50, 100$ and $N_{mc}=10000$.
By construction,
our domination number test is more sensitive for segregation/association type alternatives
which also implies the same direction for each subinterval considered
hence, the sum of domination number over the subintervals
detects such deviations from uniformity better.
In fact,
we have consistency results for the domination number test under $H^{IV}_a$ and $H^{V}_a$ type alternatives
(see Section \ref{sec:consistency}).
These consistency results suggest
that domination number test gets very sensitive under very mild forms of
$H^{IV}_a$ and $H^{V}_a$ when $n$ gets large.
Along this line,
we consider two more cases for the type IV alternative
in addition to case (1).
More specifically, we consider $H^{IV}_a$-case (1): $\ve=.1,.2,.3,.4$ $n=50$, $m=7 \approx \sqrt{n}$ and $N_{mc}=10000$,
$H^{IV}_a$-case (2): $n=1000$, $m=32 \approx \sqrt{n}$ and $N_{mc}=1000$;
and
$H^{IV}_a$-case (3): $n=1000$, $m=20$ and $N_{mc}=1000$
where in cases (2) and (3) we take $\ve =.01,.02,.03,.04$.

For the arc density of the ICDs,
we generate $n$ points according to the specified alternatives with various parameters
(where $n$ is taken as in the simulations for the domination number for the null case and each alternative).
With CICDs,
we use $(\tau,c)$ for $\tau=.1,.2,\ldots,10.0$ and $c=.01,.02.\ldots,.99$
and
with PICDs,
we use $(r,c)$ for $r=1.1,1.2,\ldots,10.0$ and $c=.01,.02.\ldots,.99$.
With CICDs, for each $(\tau,c)$ and $\delta$ combination,
and
with PICDs, for each $(r,c)$ and $\delta$ combination,
we replicate the sample generation $N_{mc}$ times.
We compute the power using the asymptotic critical values
based on the normal approximation.
We also keep the parameter combinations ($(r,c)$ for PICDs and $(\tau,c)$ for CICDs)
at which the tests have the appropriate level (of .05),
i.e., if the test is conservative or liberal for the one-sided version in question,
we ignore that parameter combination in our power estimation,
as they would yield unreliable results
which might have a substantial effect on the power values.
We call this procedure the ``size adjustment" for power estimation.
For the arc density of PICDs and CICDs, we only report the maximum power estimates under each alternative.

\begin{table}[ht]
\begin{center}
\begin{tabular}{|c|c|c|c|c|}
\hline
\multicolumn{5}{|c|}{$H^I_a$, $n=50$} \\
\hline
 & $\delta=0.2$ & $\delta=0.4$ & $\delta=0.6$  & $\delta=0.8$\\
\hline
 PICD & .33 & .51 & .76 & .94 \\
    & .15 & .35 & .64 & .88 \\
\hline
 CICD & .19 & .43 & .71 & .92 \\
    & .13 & .33 & .62 & .88 \\
\hline
 $G_{bin}$ & .06 & .08 & .09 & .14 \\
 $G_{emp}$  & .06 & .09 & .12 & .21 \\
\hline
 KS & .11 & .32 & .62 & .88 \\
\hline
 $\chi^2$ & .07 & .17 & .38 & .68 \\
\hline
 $T_n^{(2)}$ & .13 & .37 & .70 & .94 \\
\hline
 TB1 & .08 & .20 & .43 & .77 \\
\hline
 TB2 & .08 & .20 & .43 & .79 \\
\hline
\end{tabular}
\hspace{.05 cm}
\begin{tabular}{|c|c|c|c|c|}
\hline
\multicolumn{5}{|c|}{$H^{II}_a$, $n=50$} \\
\hline
 & $\sigma=0.1$ & $\sigma=0.2$ & $\sigma=0.3$  & $\sigma=0.4$\\
\hline
 PICD & 1.00 & 1.00 & .79 & .39\\
    & 1.00 & 1.00 & .79 & .37\\
\hline
 CICD & 1.00 & 1.00 & .81 & .42\\
    & 1.00 & 1.00 & .81 & .41\\
\hline
 $G_{bin}$ & .99 & .25 & .07 & .06\\
 $G_{emp}$   & .95 & .37 & .11 & .07\\
\hline
 KS & 1.00 & .75 & .17 & .07 \\
\hline
 $\chi^2$ & 1.00 & .97 & .39 & .14 \\
\hline
 $T_n^{(2)}$ & 1.00 & .83 & .17 & .07 \\
\hline
 TB1 & 1.00 & .99 & .54 & .22 \\
\hline
 TB2 & 1.00 & 1.00 & .68 & .31 \\
\hline
\end{tabular}
\vspace{.1 cm}

\begin{tabular}{|c|c|c|c|c|}
\hline
\multicolumn{5}{|c|}{$H^{III}_a$, $n=50$} \\
\hline
 & $\delta=2$ & $\delta=4$ & $\delta=6$  & $\delta=8$\\
\hline
 PICD & .41 & .66 & .92 & .99\\
    & .28 & .66 & .92 & .99\\
\hline
 CICD & .19 & .53 & .86 & .98\\
      & .19 & .52 & .86 & .98\\
\hline
 $G_{bin}$ & .07 & .07 & .09 & .14\\
 $G_{emp}$   & .06 & .07 & .11 & .19\\
\hline
 KS & .09 & .19 & .40 & .67 \\
\hline
 $\chi^2$ & .09 & .27 & .38 & .87\\
\hline
 $T_n^{(2)}$ & .07 & .14 & .31 & .58 \\
\hline
 TB1 & .07 & .19 & .48 & .82 \\
\hline
 TB2 & .03 & .06 & .21 & .56 \\
\hline
\end{tabular}
\hspace{.05 cm}
\begin{tabular}{|c|c|c|c|c|}
\hline
\multicolumn{5}{|c|}{$H^{IV}_a$, $n=50$} \\
\hline
 & $\ve=0.1$ & $\ve=0.2$ & $\ve=0.3$  & $\ve=0.4$\\
\hline
 PICD & .27 & .27 & .34 & .50 \\
    & .06 & .08 & .10 & .14\\
\hline
 CICD & .11 & .15 & .28 & .52 \\
    & .06 & .07 & .08 & .24\\
\hline
 $G_{bin}$ & .28 & 1.00 & 1.00 & 1.00 \\
 $G_{emp}$   & .88 & .95 & .95 & .95 \\
\hline
 KS & .05 & .06 & .06 & .07 \\
\hline
 $\chi^2$ & .05 & .05 & .05 & .05 \\
\hline
 $T_n^{(2)}$ & .05 & .05 & .05 & .06 \\
\hline
 TB1 & .07 & .10 & .15 & .26 \\
\hline
 TB2 & .06 & .07 & .09 & .14 \\
\hline
\end{tabular}
\end{center}
\caption{\label{tab:power-Ha-I-IV}
The power estimates under the alternatives $H_a^{I}$ to $H_a^{IV}$ with all four parameter values considered and $n=50$
and $N_{mc}=10000$ for the tests we employed.
PICD and CICD represent the arc densities of the ICD tests, and for each, top row is without size adjustment and
bottom row is with size adjustment (see the text for the description of size adjustment),
$G_{bin}$ and $G_{emp}$: tests based on domination number of PICD with approaches (i) and (ii),
KS: Kolmogorov-Smirnov test,
$\chi^2$: Chi-square test,
NT: $T_n^{(2)}$ test based on the uniformity characterization,
TB1 and TB2: two versions of the entropy-based tests.
}
\end{table}

\begin{table}[ht]
\begin{center}
\begin{tabular}{|c|c|c|c|c|}
\hline
\multicolumn{5}{|c|}{$H^{IV}_a$, $n=1000$, $m=32$, $N_{mc}=1000$} \\
\hline
 & $\ve=0.01$ & $\ve=0.02$ & $\ve=0.03$  & $\ve=0.04$\\
\hline
 PICD & .08 & .08 & .08 & .08\\
    & .08 & .08 & .08 & .08\\
\hline
 CICD & .08 & .08 & .08 & .08 \\
    & .07 & .08 & .08 & .08 \\
\hline
 DN & .49 & 1.00 & 1.00 & 1.00 \\
    & .66 & .95 & .96 & .97 \\
\hline
 KS & .04 & .04 & .04 & .05 \\
\hline
 $\chi^2$ & .05 & .05 & .06 & .05 \\
\hline
 $T_n^{(2)}$ & .04 & .04 & .04 & .06 \\
\hline
 TB1 & .04 & .07 & .08 & .09 \\
\hline
 TB2 & .04 & .06 & .07 & .09 \\
\hline
\end{tabular}
\hspace{.05 cm}
\begin{tabular}{|c|c|c|c|c|}
\hline
\multicolumn{5}{|c|}{$H^{IV}_a$, $n=1000$, $m=20$, $N_{mc}=1000$} \\
\hline
 & $\ve=0.01$ & $\ve=0.02$ & $\ve=0.03$  & $\ve=0.04$\\
\hline
 PICD & .08 & .08 & .08 & .08 \\
    & .08 & .08 & .08 & .08 \\
\hline
 CICD & .08 & .08 & .08 & .07 \\
    & .08 & .08 & .08 & .07 \\
\hline
 $G_{bin}$ & .40 & 1.00 & 1.00 & 1.00 \\
 $G_{emp}$   & .61 & .95 & .95 & .97 \\
\hline
 KS & .06 & .04 & .05 & .05 \\
\hline
 $\chi^2$ & .05 & .05 & .04 & .05 \\
\hline
 $T_n^{(2)}$ & .06 & .05 & .05 & .06 \\
\hline
 TB1 & .04 & .08 & .09 & .15 \\
\hline
 TB2 & .04 & .07 & .09 & .13 \\
\hline
\end{tabular}
\end{center}
\caption{\label{tab:power-Ha-IV-n1000}
The power estimates under the alternatives $H_a^{IV}$ with all four $\ve$ values considered and $n=1000$
and $N_{mc}=1000$ for the tests we employed.
Labeling of the tests is as in Table \ref{tab:power-Ha-I-IV}.
}
\end{table}

The power comparisons between PICD domination number test, KS, $\chi^2$, TB1, TB2 and $T_n^{(2)}$ tests are presented
in Figure \ref{fig:sm-power-plots-Ha-I-III} for alternatives $H_a^I - H_a^{III}$,
and in Figure \ref{fig:sm-power-plots-Ha-IV} for alternatives $H_a^{IV}$-cases (1)-(3).
The power estimates based on asymptotic critical values of the tests
(i.e., the power estimates for the test based on domination number of PICD with approach (i),
Kolmogorov-Smirnov test, and Chi-square test) are provided in the top row
and those based on Monte Carlo critical values
(for the test based on domination number of PICD with approach (ii),
$T_n^{(2)}$ test based on the uniformity characterization,
two versions of the entropy-based tests) are provided in the bottom row in these figures.
The power estimates under alternatives $H_a^I - H_a^{III}$ and $H_a^{IV}$-case(1) are presented
in Table \ref{tab:power-Ha-I-IV},
and those under alternative $H_a^{IV}$-cases (2) and (3) in Table \ref{tab:power-Ha-IV-n1000};
in both tables the power estimates are rounded to two decimal places.
In Figures \ref{fig:sm-power-plots-Ha-I-III} and \ref{fig:sm-power-plots-Ha-IV},
we do not present the power estimates for ICD arc density tests, due to
the difficulty in presentation since ICD arc density tests depend on two parameters.
For the domination number test,
the power estimates based on asymptotic critical values are provided in the top row
and those based on Monte Carlo critical values are provided in the bottom row in these figures.
\emph{In Tables \ref{tab:power-Ha-I-IV} and \ref{tab:power-Ha-IV-n1000},
we only present the maximum power estimates for the ICD arc density tests for the two-sided alternative
and for the CICD domination number tests.}
Considering Figures \ref{fig:sm-power-plots-Ha-I-III} and \ref{fig:sm-power-plots-Ha-IV}
and Tables \ref{tab:power-Ha-I-IV} and \ref{tab:power-Ha-IV-n1000},
we observe that
power estimates increase as the departure from uniformity gets more severe.
In particular,
power estimate increases as $\delta$ increases in $H_a^{I}$ or $H_a^{II}$,
as $\ve$ increases in $H_a^{IV}$ and
as $\sigma$ decreases in $H_a^{II}$.
Under $H_a^I - H_a^{III}$ and $H_a^{IV}$-case(1),
arc density of PICD and CICD has the highest power estimates,
where PICD arc density tends to perform better (worse) than CICD arc density under $H_a^I - H_a^{III}$
(under $H_a^{IV}$-case (1)).
Under $H_a^I$, ICD arc density tests are followed by $T_n^{(2)}$;
under $H_a^{II}$, ICD arc density tests are followed by TB1 and TB2;
under $H_a^{III}$, ICD arc density tests are followed by $\chi^2$ test;
and
under $H_a^{IV}$-case (1), ICD domination number test is followed by CICD arc density test.
Under $H_a^{IV}$-cases (2) and (3) ICD domination number tests have the highest power estimates,
where under case (1) PICD domination number test with approach (i) and under case (2) domination number test with approach (ii)
has better performance, and power estimates for the other tests are just above .05 or at about .05.
In these large sample cases,
approach (iii) also works, and has higher power estimates than the other two approaches
(corresponding estimates not presented to be consistent with the presentations of the other alternatives).
Moreover,
PICD domination number test performs better when the support is partitioned by $m \approx \sqrt{n}$.
We omit the power performance under $H^{V}_a$
as it is the opposite pattern to the one under $H^{IV}_a$.
More simulation results for the arc density of ICDs are presented in
in the Supplementary Materials Section, where we observe that
the power estimates are symmetric around $c=1/2$ under types II-IV alternatives,
which is in agreement with the symmetry in the corresponding pdfs (around $c=1/2$).

We also considered the power comparisons under $H_a^I-H_a^{III}$ and $H_a^{IV}$ case (1)
at the same alternative parameters with $n=N_{mc}=1000$, to see the effect
of the large samples on the power estimates.
The results are similar to those in the smaller sample cases,
with higher power for each test (hence not presented).
In particular,
under $H_a^I-H_a^{III}$ all tests have much higher power, with most having power virtually 1.00,
but domination number tests with approaches (i) and (ii) exhibit mild improvement,
while under $H_a^{IV}$ case (1), PICD domination number tests attain the highest power estimates, virtually 1.00,
while there is mild improvement in the performance of other tests,
except for TB1 and TB2, which show moderate improvement.
We also observe that in the large sample case,
PICD domination number with approach (iii) attains very high power
under each alternative.

The above methodology can easily be extended for testing non-uniform distributions
(see Remark \ref{rem:test-any-F} in the Supplementary Materials Section).

\section{Consistency of the Tests based on Domination Number of PICDs under $H_a^{IV}$ and $H_a^{V}$}
\label{sec:consistency}

Let $b_{\alpha}$ be the $\alpha \times 100$th percentile of the binomial
distribution $\BIN(m,p_u(r,c,\lfloor{n/m}\rfloor))$.

\begin{theorem}
\label{thm:consistency-I}
\textbf{(Consistency - Type I)}
Let $\g_{n,m}(F,r,c)$ be the domination number under segregation
and association alternatives, $H^{IV}_a$ and $H^{V}_a$, respectively,
in the multiple interval case with $m$ intervals.
The test against segregation with $F = F_4(x,\ve)$
which rejects for $G_n < b_{\alpha}$
and
the test against association with $F = F_5(x,\ve')$
which rejects for $G_n>b_{1-\alpha}$ are consistent.
\end{theorem}

\noindent
{\bf Proof:}
Given $F = F_4(x,\ve)$.
Let $\g_{n,m}(\U,r,c)$ be the domination number for $\X_n$
being a random sample from $\U(0,1)$.
Then $P(\g_{n,m}(F,r,c)=1) \ge P(\g_{n,m}(\U,r,c)=1)$;
and
$P(\g_{n,m}(F,r,c)=2) \le P(\g_{n,m}(\U,r,c)=2)$.
Hence $G_n<m p_u(r,c,\lfloor{n/m}\rfloor)$ with probability 1, as $n \gg m \rightarrow \infty$.
Furthermore,
$\BIN(m,p_u(r,c,\lfloor{n/m}\rfloor))$ distribution converges to normal distribution
with mean $m p_u(r,c,\lfloor{n/m}\rfloor)$ and variance $m p_u(r,c,\lfloor{n/m}\rfloor) (1-p_u(r,c,\lfloor{n/m}\rfloor))$.
Hence consistency follows from the consistency
of tests which have asymptotic normality.
The consistency against the association alternative can be proved similarly.
$\blacksquare$

Below we provide a result which is stronger,
in the sense that it will hold for finite $m$ as $n \rightarrow \infty$.
Let $\overline G_n:=G_n/m$ (i.e., domination number averaged over the number of subintervals)
and $z_{\alpha}$ be the $\alpha \times 100$-th percentile of the standard normal distribution.
\begin{theorem}
\label{thm:consistency-II}
\textbf{(Consistency - Type II)}
Let $\g_{n,m}(F,r,c)$ be the domination numbers under segregation
and association alternatives $H^{IV}_a$ and $H^{V}_a$, respectively, in the multiple interval case with $m$ intervals
where $m < \infty$ is fixed.
Let $m^*(\alpha,\ve):=\left \lceil \Bigl(\frac{\sigma \cdot z_{\alpha}}{\overline G_n(r,c)-\mu} \Bigr)^2 \right \rceil$
where $\lceil \cdot \rceil$ is the ceiling function and
$\ve$-dependence is through $\overline G_{n,m}(r,c)$ under a given alternative.
Then the test against $H^{IV}_a$ which rejects for
$S_{n,m} < z_{\alpha}$ is consistent for all $\ve \in \left( 0,\min(c,1-c)\right)$
and $m \ge m^*(\alpha,\ve)$,
and
the test against $H^{V}_a$ which rejects for $S_{n,m}>z_{1-\alpha}$
is consistent for all $\ve \in \left( 0,\min(c,1-c) \right)$ and $m \ge m^*(1-\alpha,\ve)$.
\end{theorem}

\noindent
{\bf Proof:}
Let $\ve \in \left( 0,\min(c,1-c)\right)$.
Under $H^{IV}_a$, $\g_{n}(F,r,c)$ is
degenerate in the limit as $n \rightarrow \infty$,
which implies $\overline G_n(r,c)$ is a constant a.s.
In particular, for $\ve \in \left( 0,\min(c,1-c)\right)$,
$\overline G_n(r,c)=1$ a.s. as $n \rightarrow \infty$.
Then the test statistic $S_{n,m} = \sqrt{m} (\overline G_n(r,c) - \mu)/\sigma$ is a constant a.s.
and $m \ge m^*(\alpha,\ve)$ implies that $S_{n,m}< z_{\alpha}$ a.s.
Furthermore,
$S_{n,m}\stackrel{\mathcal L}{\rightarrow}N(0,1)$ as $n \to \infty$.
Hence consistency follows for segregation.

Under $H^{V}_a$, as $n \rightarrow \infty$,
$\overline G_n(r,c)=2$ for all $\ve \in \left( 0,\min(c,1-c) \right)$ a.s.
Then $m \ge m^*(1-\alpha,\ve)$ implies that $S_{n,m} > z_{1-\alpha}$ a.s.,
hence consistency follows for association.
$\blacksquare$

Notice that in Theorem \ref{thm:consistency-II}
we actually have more than what consistency requires.
In particular,
we show that the power of the test reaches 1 for $m$ greater than a threshold
as $n \rightarrow \infty$.

\section{Discussion and Conclusions}
\label{sec:disc-conclusions}
In this article,
we derive the distribution of the domination number
of a random digraph family called \emph{parameterized interval catch digraph} (PICD)
which is based on two classes of points, say $\X$ and $\Y$.
Points from one of the classes (say, class $\X$), denoted $\X_n$,
constitute the vertices of the PICDs,
while the points from the other class (say, class $\Y$), denoted $\Y_m$,
are used in the binary relation
that assigns the arcs of the PICDs.
Our PICD is based on a parameterized proximity map
which has an expansion parameter $r$ and a centrality parameter $c$.
We provide the exact and asymptotic distributions of
the domination number of the PICDs
for uniform data
and compute the asymptotic distribution for non-uniform data
for the entire range of $(r,c)$.

We demonstrate an interesting behavior of the domination number of
the PICD for one-dimensional data.
For uniform data or data from a distribution which satisfies some regularity conditions
and fixed finite sample size $n>1$,
the distribution of the domination number restricted to any interval
is a translated form of Bernoulli distribution,
$\BER(p)$, where
$p$ is the probability that the domination number being 2.
In the case of $\Y_2=\{y_1,y_2\}$ with $\U(y_1,y_2)$ data,
for finite $n \ge 1$,
the parameter of the asymptotic distribution of the domination number
of the PICD based on uniform data (i.e. probability of domination number being 2, denoted $p_u(r,c)$) is continuous in $r$ and $c$
for all $r \ge 1$ and $c \in (0,1)$.
For fixed $(r,c) \in [1,\infty) \times (0,1)$,
$p_u(r,c)$ exhibits some discontinuities.
The asymptotic distribution of the domination number is degenerate for the expansion parameter $r > 2$
regardless of the value of $c$.
For $c \in (0,1)$
the asymptotic distribution is nondegenerate when the expansion parameter $r$ equals $r^*=1/\max(c,1-c)$.
For $r=r^*$,
the asymptotic distribution
of the domination number is a translated form of $\BER(p_u(r^*,c))$
where $p_u(r^*,c)$ is continuous in $c$.
For $r > r^*$
the domination number converges in probability to 1,
and for $r < r^*$
the domination number converges in probability to 2.
On the other hand,
at $(r,c)= (2,1/2)$,
the asymptotic distribution is again a translated form of $\BER(p_u(2,1/2))$,
but there is yet another jump at $(r,c)=(2,1/2)$,
as $p_u(2,1/2)=4/9$ while $\lim_{(r,c) \rightarrow (2,1/2)} p_u(r^*,c)=2/3$.
This second jump is due to the symmetry
for the domination number at $c=1/2$
(see the discussion at the end of Section \ref{sec:asy-dist-r-and-c}).

We employ domination number for testing uniformity of one-dimensional data.
In this application,
we have $n$ $\X$ points and we take $m$ $\Y$ points to be the equidistant points in the support of $\X$ points.
For example, if the support of $\X$ points is $(0,1)$,
we take $\Y$ points to be $\Y_m=\{0, 1/(m-1), 2/(m-1), \ldots,1\}$.
Since under $H_o$, $H^{IV}_a$ and $H^{V}_a$
the data is uniform with different support regions,
we can extend the methodology to the random $\Y_m$ case,
but currently the method is only applicable given $\Y_m$ as above.

We compare the size and power performance of PICD domination number
with two well known tests, namely, Kolmogorov-Smirnov (KS) test
and Pearson's $\chi^2$ goodness-of-fit test,
three recently introduced tests, the uniformity test based on Too-Lin characterization,
denoted as $T_n^{(2)}$ \citep{milosevic:2018},
and two entropy-based tests, denoted as TB1 and TB2 in \citep{zamanzade:2015},
and also the arc density of PICDs and of another ICD family called central ICD (CICD),
by Monte Carlo simulations.
Based on the simulation results,
we see that ICDs have better performance than their competitors (in terms of size and power).
Arc density of ICDs perform better than others under most alternatives
for some of the parameter values
and the domination number outperforms others under certain types of alternatives.
In particular,
under the alternatives $H_a^I - H_a^{III}$,
ICD arc density tests outperform other tests,
and under $H_a^{IV}$-cases (1)-(3),
PICD domination number tests outperform other tests.
For the ICD arc density tests,
we use the asymptotic critical values based on normal approximation.
For the PICD domination number test,
we use the binomial critical values with an approximate probability of success (i.e., approach (i))
and also the empirical critical values based on Monte Carlo simulations (i.e., approach (ii)).
For $T_n^{(2)}$, TB1 and TB2 tests,
the critical values are also based on Monte Carlo simulations.

We recommend using the PICD domination number test for uniformity in the following scenario.
If we are testing uniformity of data in multiple intervals (by hypothesis or one can partition the support
of the data),
and the deviation from uniformity is in the same direction at each interval,
then,
by construction,
domination number tends to be more sensitive to detect such alternatives
(even if they are very mild deviations from uniformity).
Among the types of critical value computations,
we recommend the use of the exact distribution provided in Theorem \ref{thm:general-Dnm}
(with Monte Carlo critical values as an approximation in practice), i.e., approach (ii)
for small samples (this approach could be used provided running time is feasible),
and the approximate Binomial test for any $n$, i.e., approach (i)
(see Section \ref{sec:emp-size}).
For large samples,
binomial test with asymptotic probability of success (i.e., approach (iii))
could also be employed.
Our simulations suggest that about 30 or more for each subinterval seems to work
for most $(r^*,c)$ combination,
however, the sample size requirements for approach (iii)
have not been studied thoroughly in this article.
The relevant functions for these tests are \textsf{PEdom1D} and \textsf{TSDomPEBin1D} which are available in the \textsf{R} package
\textsf{pcds} which is available on \textsf{github} and can be installed using the command
\verb|devtools::install_github("elvanceyhan/pcds")| in an \textsf{R} session.
The function \textsf{PEdom1D} computes the domination number when one or two one-dimensional data sets are provided,
and the function \textsf{TSDomPEBin1D} uses the finite sample binomial approximation (i.e. approach (i)) by default
or can use the asymptotic binomial version (i.e., approach (iii)) for very large samples
when \textsf{asy.bin=TRUE} option is employed.
Monte Carlo critical values can also be computed using \textsf{PEdom1D} with sampling from the
uniform distribution of the data sets (i.e., approach (ii)).
See the help pages for \textsf{PEdom1D} and \textsf{TSDomPEBin1D} for more details.
The domination number approach is easily adaptable to testing
nonuniform distributions as well (see Remark \ref{rem:test-any-F} for more detail).
PICDs have other applications, e.g.,
as in \cite{ceyhan:dom-num-NPE-SPL},
we can use the domination number in testing
one-dimensional spatial point patterns
and our results can help make the power comparisons
possible for a large family of distributions
(see, e.g., Section \ref{sec:emp-power}
for a brief treatment of this issue).
PICDs can also be employed in pattern classification as well (see, e.g., \cite{priebe:2003b}
and \cite{ceyhan:jmlr-cccd-2016}).
Furthermore, this article may form the foundation of the generalizations and calculations for uniform and
non-uniform distributions in multiple dimensions.

In our calculations, we extensively make use of the ordering of points in $\R$.
The order statistics of $\Y_m$ partition the support of $X$ points into disjoint intervals.
This nice structure in $\R$ allows us to find a minimum dominating set and hence the domination
number, both in polynomial time.
Furthermore, the components of the digraph restricted to intervals (see Section \ref{sec:special-cases-dom-numb-Dnm})
are not connected to each other, since the defining proximity regions $N(x_i,r,M) \cap N(x_j,r,M)=\emptyset$
for $x_i,\,x_j$ in distinct intervals.
Extension of this approach to higher dimensions is a challenging problem,
since there is no such ordering for point in $\mathbb R^d$ with $d>1$.
However, we can use the Delaunay tessellation based on $\Y_m$ to partition the space
as in \cite{ceyhan:dom-num-NPE-SPL}.
Furthermore,
for most of the article and for all non-trivial results (i.e., for the exact and asymptotic distributions
of the domination number),
we assumed $\Y_m$ is given;
removing the conditioning on $\Y_m$ is a topic of ongoing research
along various directions, namely:
(i) $X$ and $Y$ both have uniform distribution,
(ii) $X$ and $Y$ both have the same (absolutely) continuous distribution,
and
(iii) $X$ is distributed as $F_X$ and $Y$ is distributed as $F_Y$
(where $F_X \ne F_Y$ and both $F_X$ and $F_Y$ are absolutely continuous).

\section*{Acknowledgments}
I would like to thank the anonymous referees, whose constructive
comments and suggestions greatly improved the presentation and flow
of this article.
I also would like to thank Prof B. Milo\v{s}evi\'{c} and Prof E. Zamanzade for
providing the \textsf{R} code for their tests upon request.
This research was supported by the European Commission under the
Marie Curie International Outgoing Fellowship Programme
via Project \# 329370 titled PRinHDD.

\bibliography{References}
\bibliographystyle{apalike}

\newpage
\section*{SUPPLEMENTARY MATERIALS}

\renewcommand{\thepage}{S\arabic{page}}
\renewcommand{\thesection}{S\arabic{section}}
\renewcommand{\thetable}{S\arabic{table}}
\renewcommand{\thefigure}{S\arabic{figure}}
\renewcommand{\theequation}{S\arabic{equation}}
\setcounter{equation}{0}
\setcounter{page}{1}
\setcounter{figure}{0}
\setcounter{section}{1}

\section{Additional and Illustrative Figures}

\subsection{Empirical Size and Power Plots}
\label{sec:emp-size-power}

\begin{figure}[hbp]
\centering
\rotatebox{-90}{ \resizebox{2.12 in}{!}{ \includegraphics{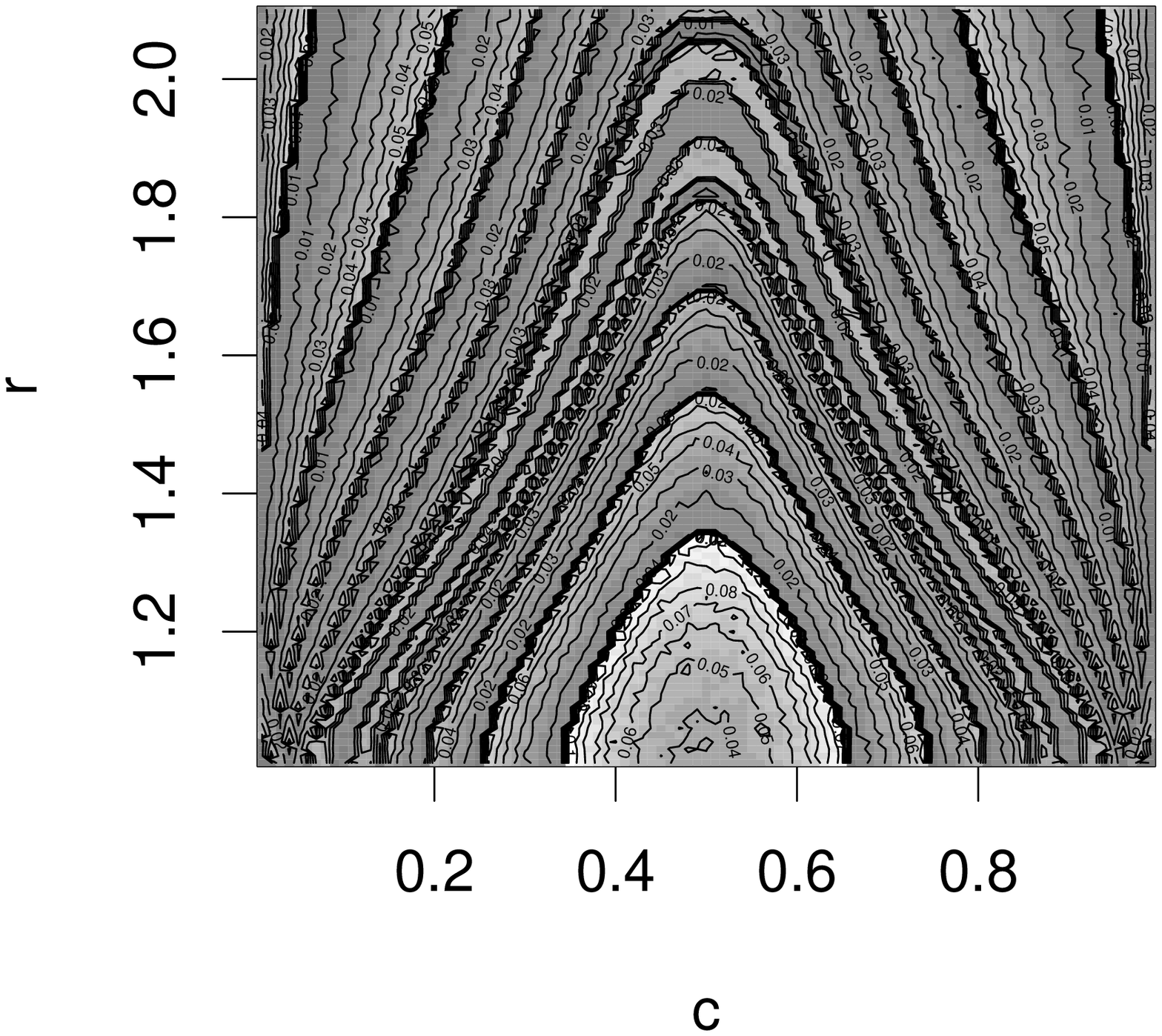}}}
\rotatebox{-90}{ \resizebox{2.12 in}{!}{ \includegraphics{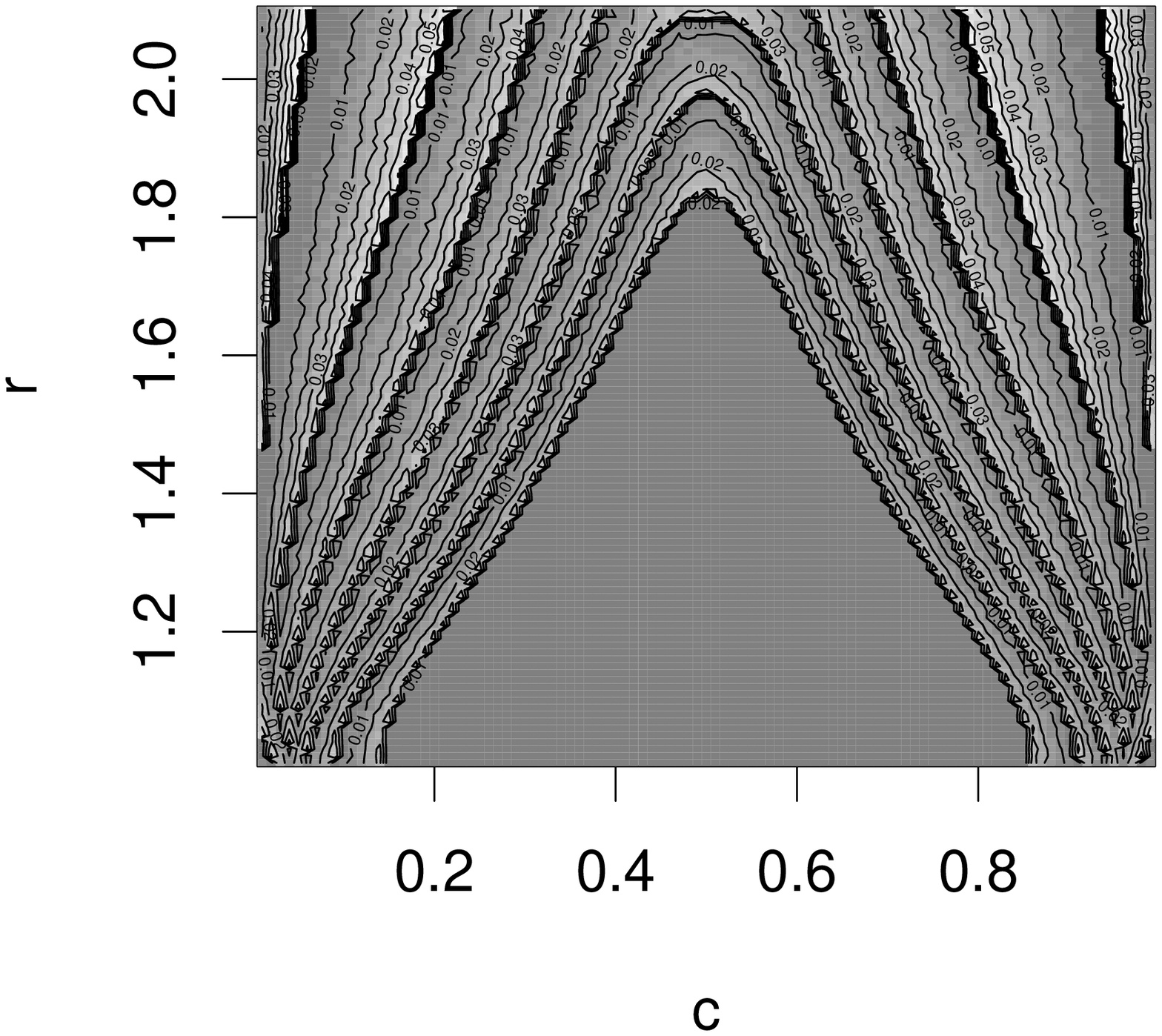}}}
\rotatebox{-90}{ \resizebox{2.12 in}{!}{ \includegraphics{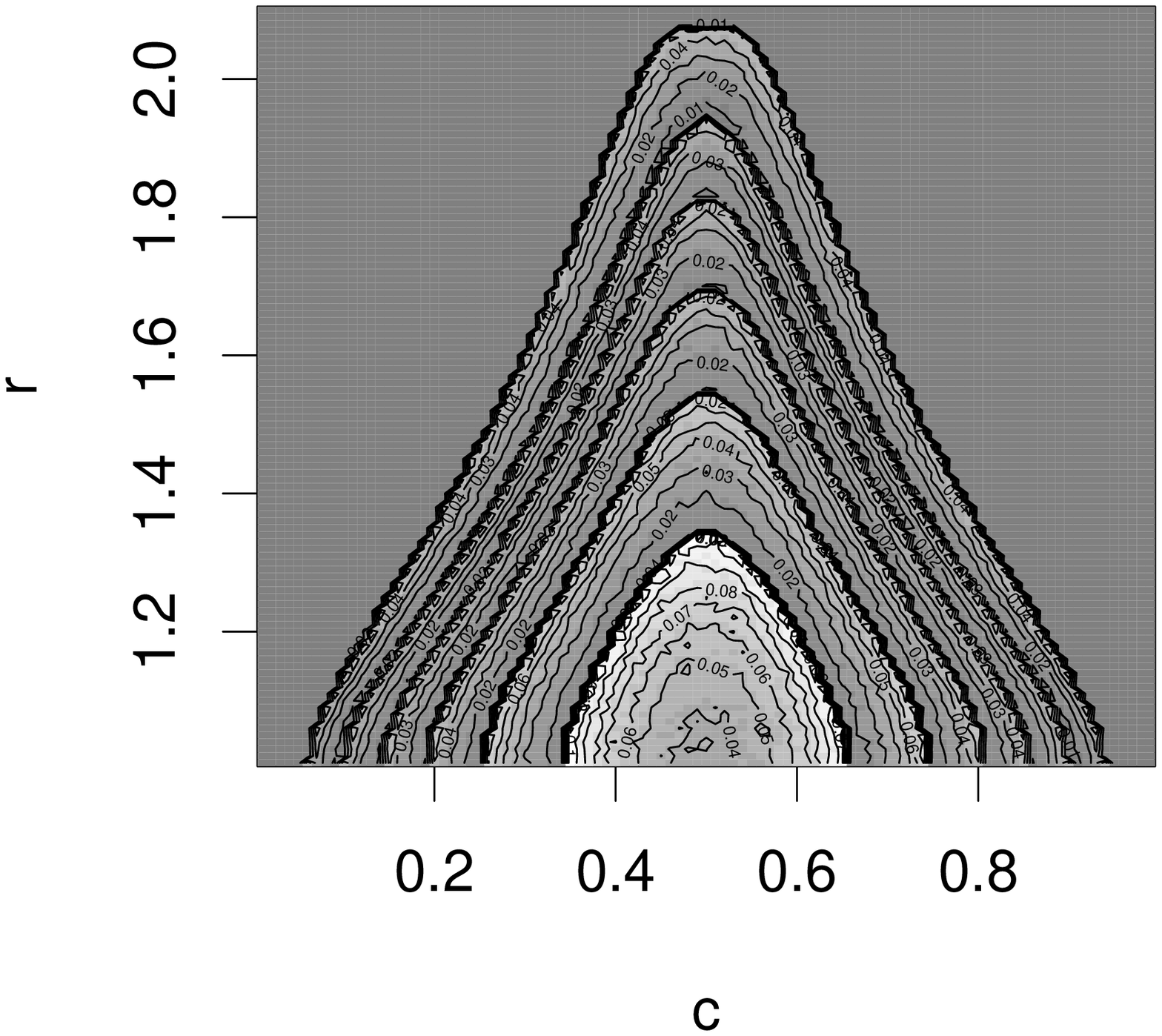}}}
\caption{
\label{fig:sm-size-image-plots-1}
Image plots for the empirical size estimates for \textbf{approach (i)} based on $n=50$
and $N_{mc}=10000$
for $r=1.01,1.02,\ldots,2.10$ and $c=.01,.02,\ldots,.99$ for
the two-sided, right-sided and left-sided alternatives (left to right).
The size estimates are coded in gray-level (as size increases the gray level gets darker).
}
\end{figure}

\begin{figure}[hbp]
\centering
\rotatebox{-90}{ \resizebox{2.12 in}{!}{ \includegraphics{ImagePl4Sizesn20.ps}}}
\rotatebox{-90}{ \resizebox{2.12 in}{!}{ \includegraphics{ImagePl4SizesRSn20.ps}}}
\rotatebox{-90}{ \resizebox{2.12 in}{!}{ \includegraphics{ImagePl4SizesLSn20.ps}}}
\rotatebox{-90}{ \resizebox{2.12 in}{!}{ \includegraphics{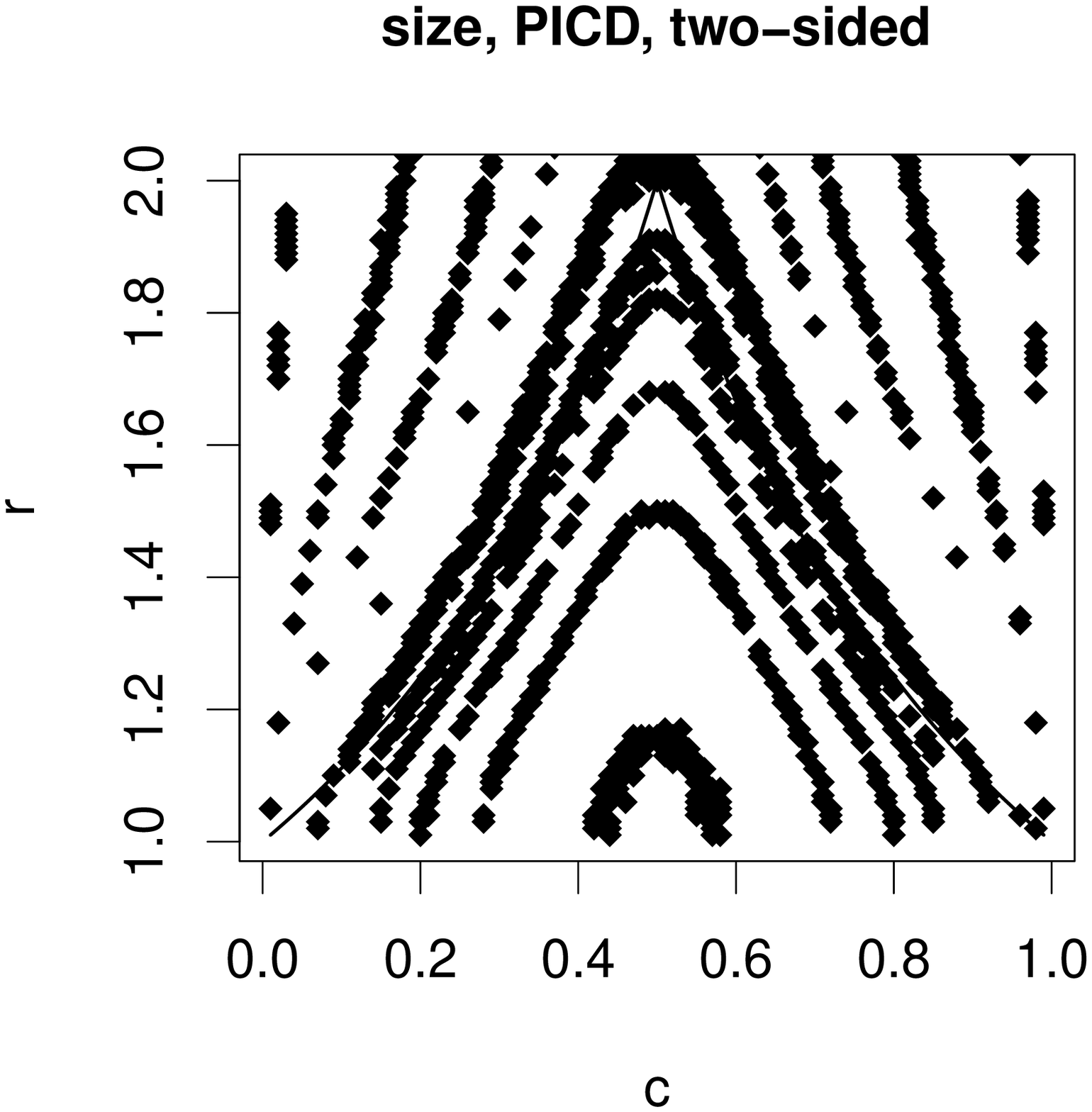}}}
\rotatebox{-90}{ \resizebox{2.12 in}{!}{ \includegraphics{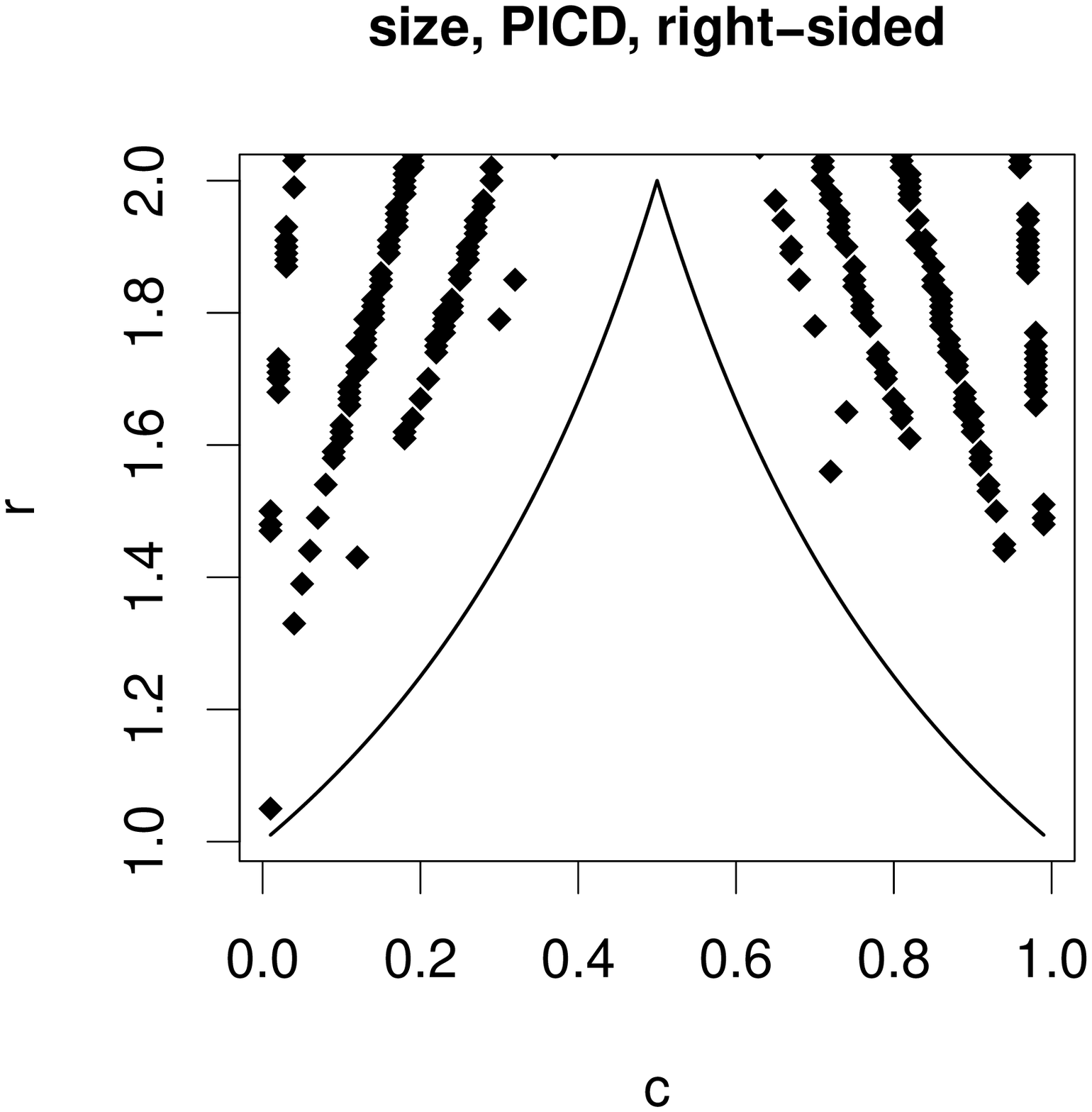}}}
\rotatebox{-90}{ \resizebox{2.12 in}{!}{ \includegraphics{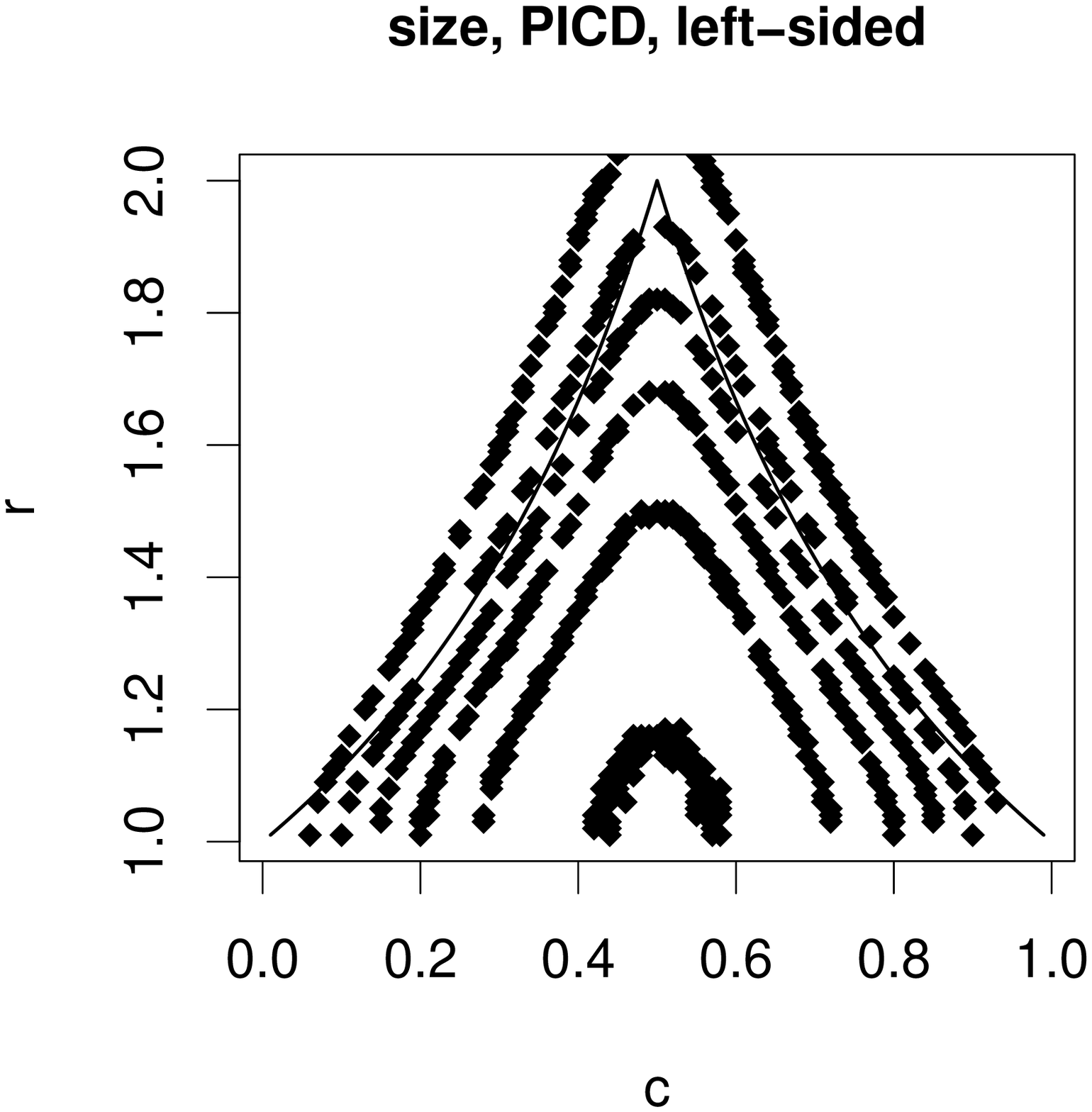}}}
\rotatebox{-90}{ \resizebox{2.12 in}{!}{ \includegraphics{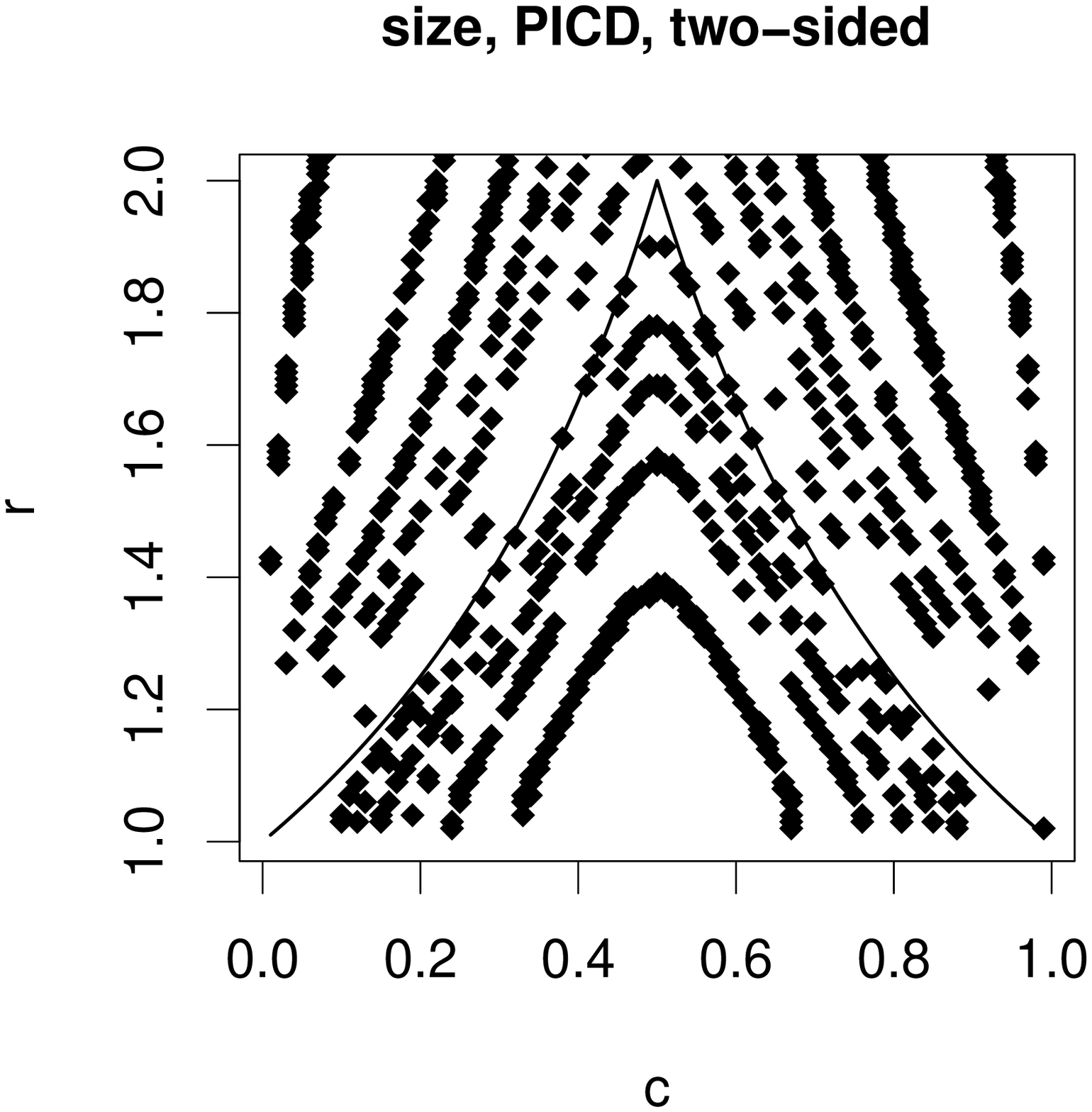}}}
\rotatebox{-90}{ \resizebox{2.12 in}{!}{ \includegraphics{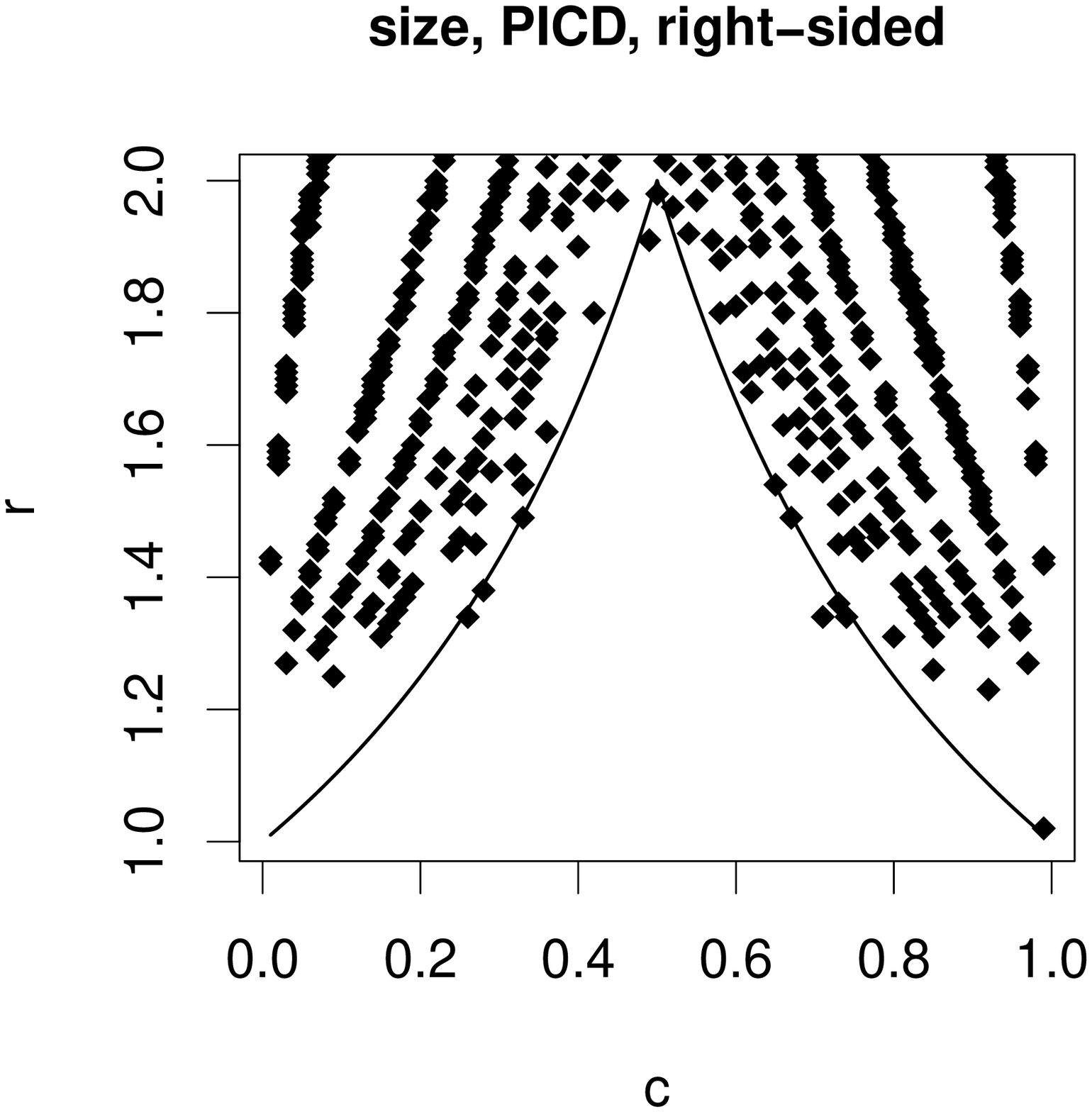}}}
\rotatebox{-90}{ \resizebox{2.12 in}{!}{ \includegraphics{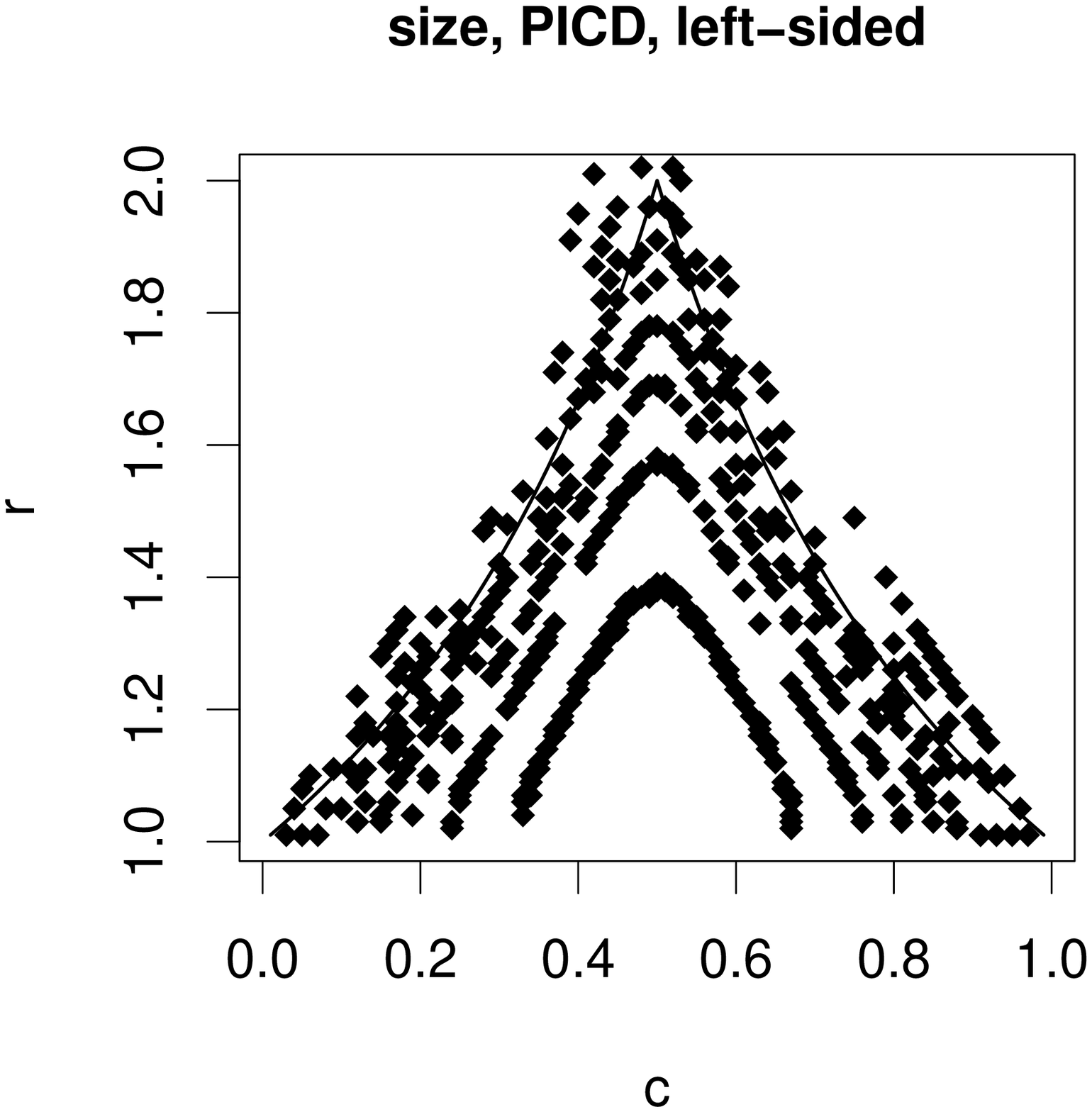}}}
\caption{
\label{fig:size-image-plots}
Image plots for the empirical size estimates for \textbf{approach (i)} based on $n=20$ (top row)
$n=50$ (middle row) and $n=100$ (bottom row) and $N_{mc}=10000$
for $r=1.01,1.02,\ldots,2.10$ and $c=.01,.02,\ldots,.99$ for
the two-sided, right-sided and left-sided alternatives (left to right).
The size estimates significantly different from .05 are blanked out,
while size estimates within .0536 and .0464 are plotted in black.
The solid lines indicates the case of $(r,c) $ (i.e., $(r,c)=(r^*,c)$)
which yields the asymptotically non-degenerate distribution for the domination number.
}
\end{figure}

We present the empirical size estimates of the tests based on the \emph{domination number of
PICD with approach (i)} as gray-scale image plots
for the two-sided, right- and left-sided alternatives with
$n=50$, $c=.01,.02,\ldots,.99$ and $r=1.01, \ldots, 2.10$
in Figure \ref{fig:sm-size-image-plots-1}
(the plots for $n=20$ and $n=100$ have the similar trend, hence not presented).
A similar version of these plots are the
image plots in \ref{fig:size-image-plots} for the empirical size estimates for \textbf{approach (i)}
based on $n=20$ (top row) 50 (middle row) and 100 (bottom row) and $N_{mc}=10000$
for $r=1.01,1.02,\ldots,2.10$ and $c=.01,.02,\ldots,.99$ for
the two-sided, right-sided and left-sided alternatives (left to right).
The size estimates significantly different from .05 are blanked out,
while size estimates within .0536 and .0464 are plotted in black.
The solid lines indicates the case of $(r,c) $ (i.e., $(r,c)=(r^*,c)$)
which yields the asymptotically non-degenerate distribution for the domination number.
Notice that there is symmetry in size estimates around $c=1/2$.

\begin{figure}[hbp]
\centering
\rotatebox{-90}{ \resizebox{2.12 in}{!}{ \includegraphics{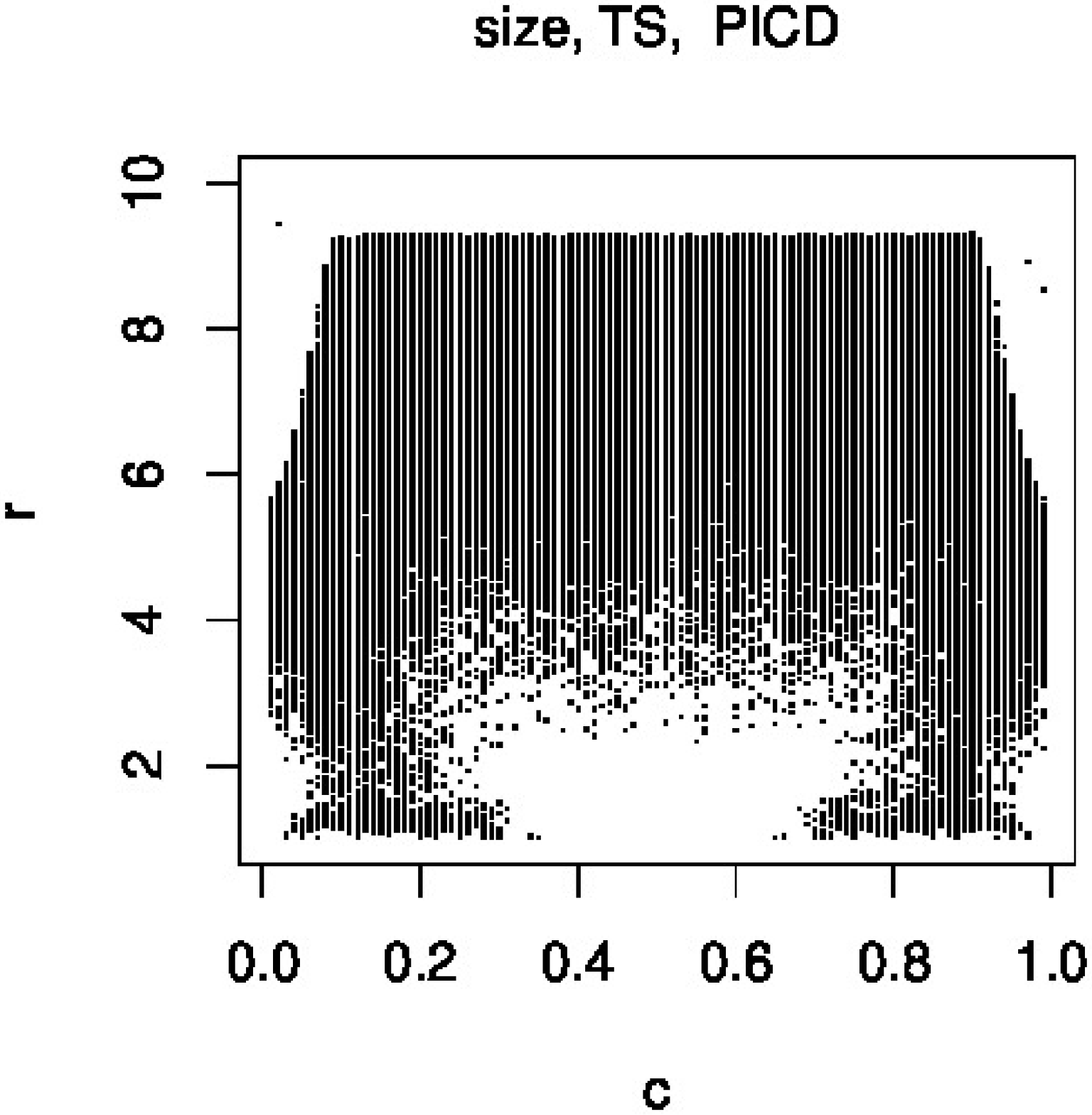}}}
\rotatebox{-90}{ \resizebox{2.12 in}{!}{ \includegraphics{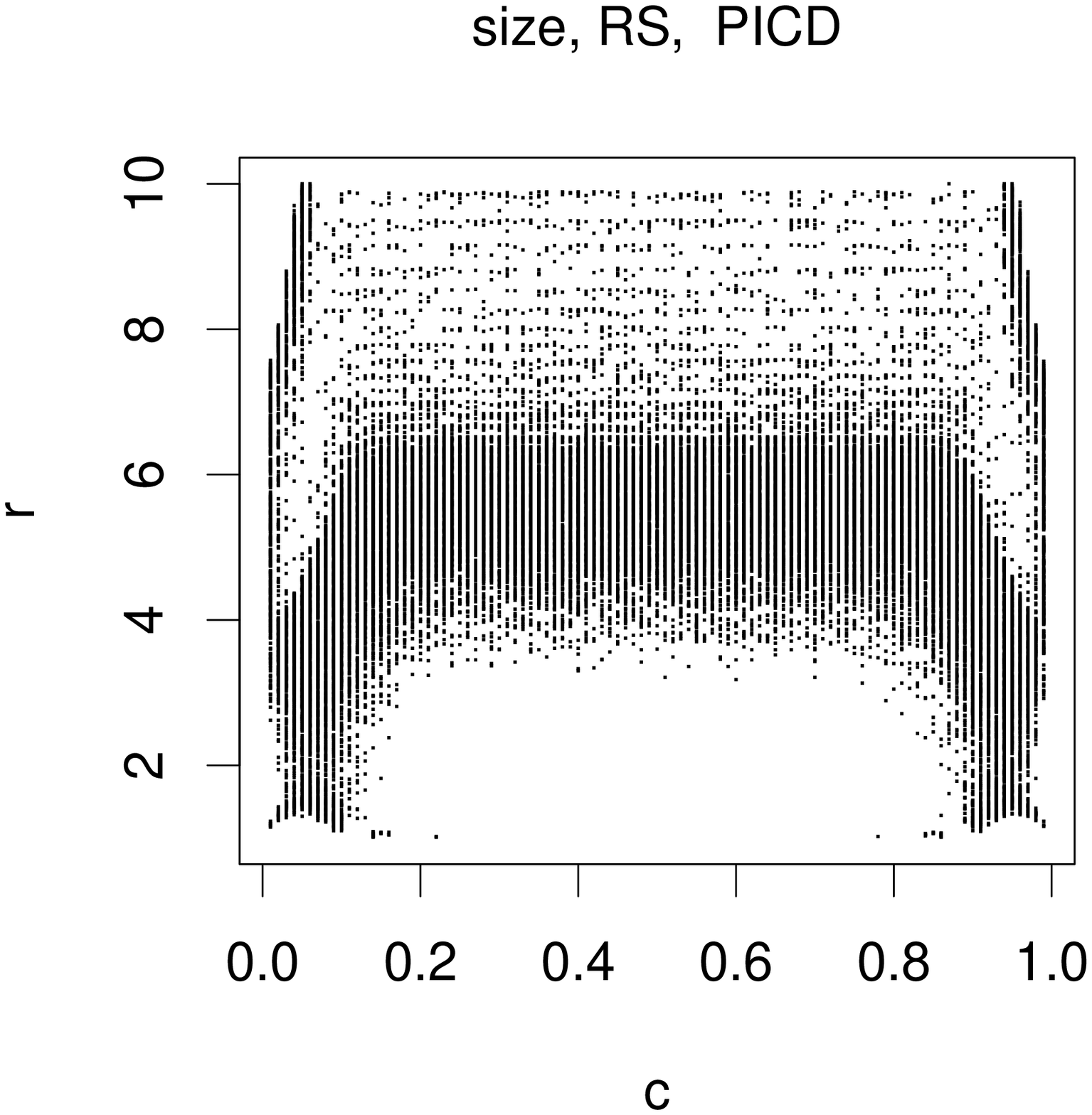}}}
\rotatebox{-90}{ \resizebox{2.12 in}{!}{ \includegraphics{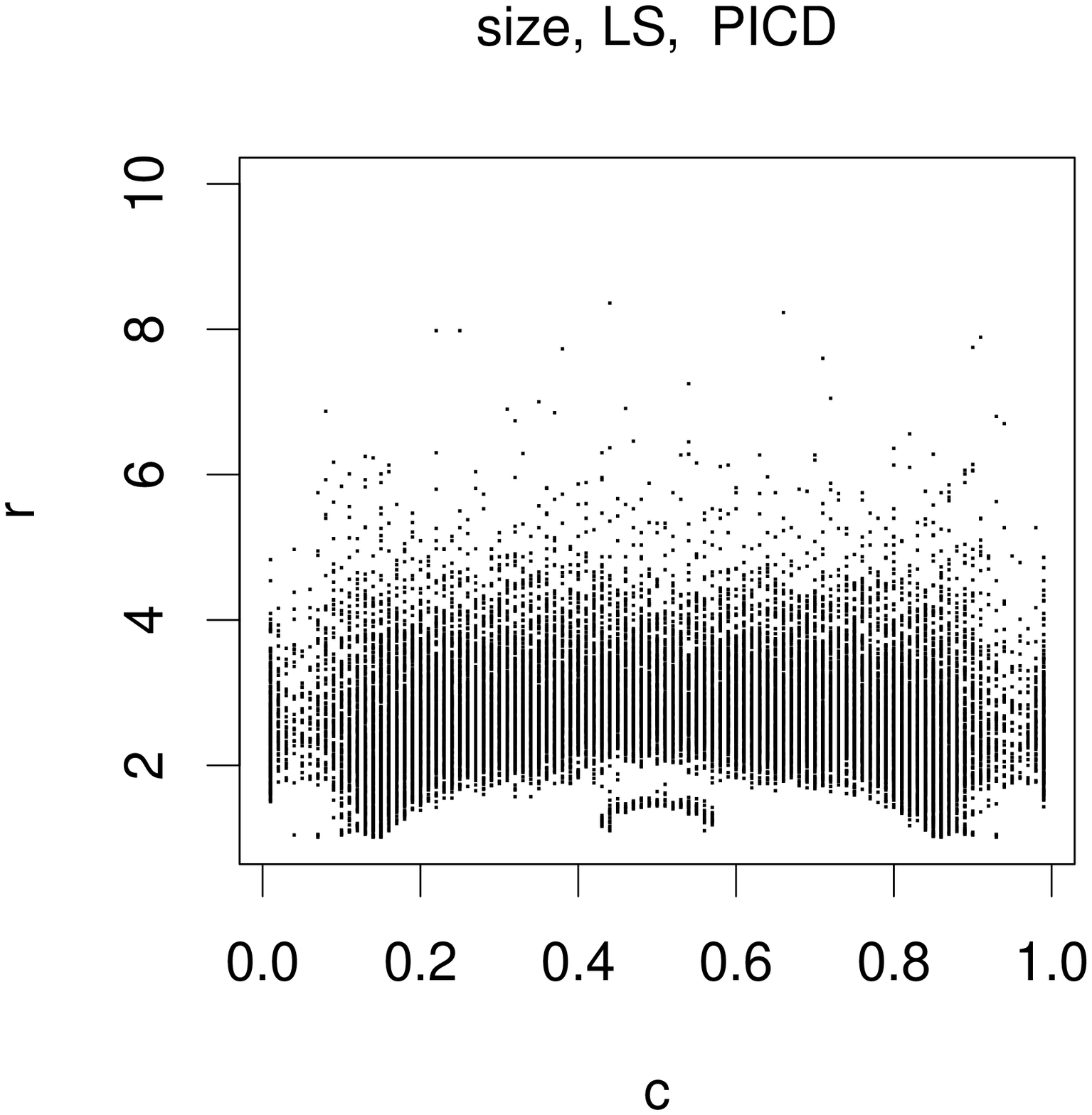}}}
\rotatebox{-90}{ \resizebox{2.12 in}{!}{ \includegraphics{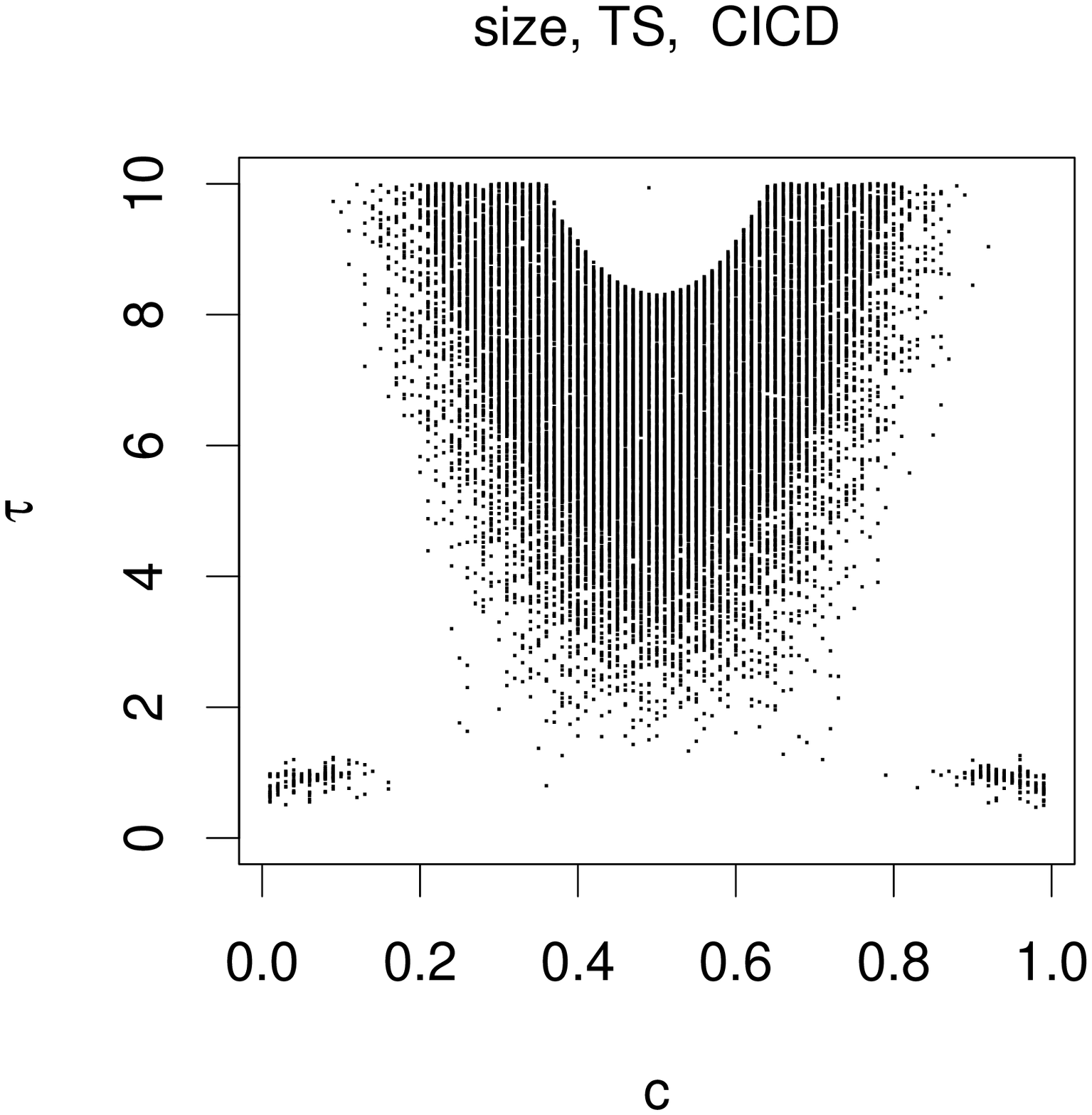}}}
\rotatebox{-90}{ \resizebox{2.12 in}{!}{ \includegraphics{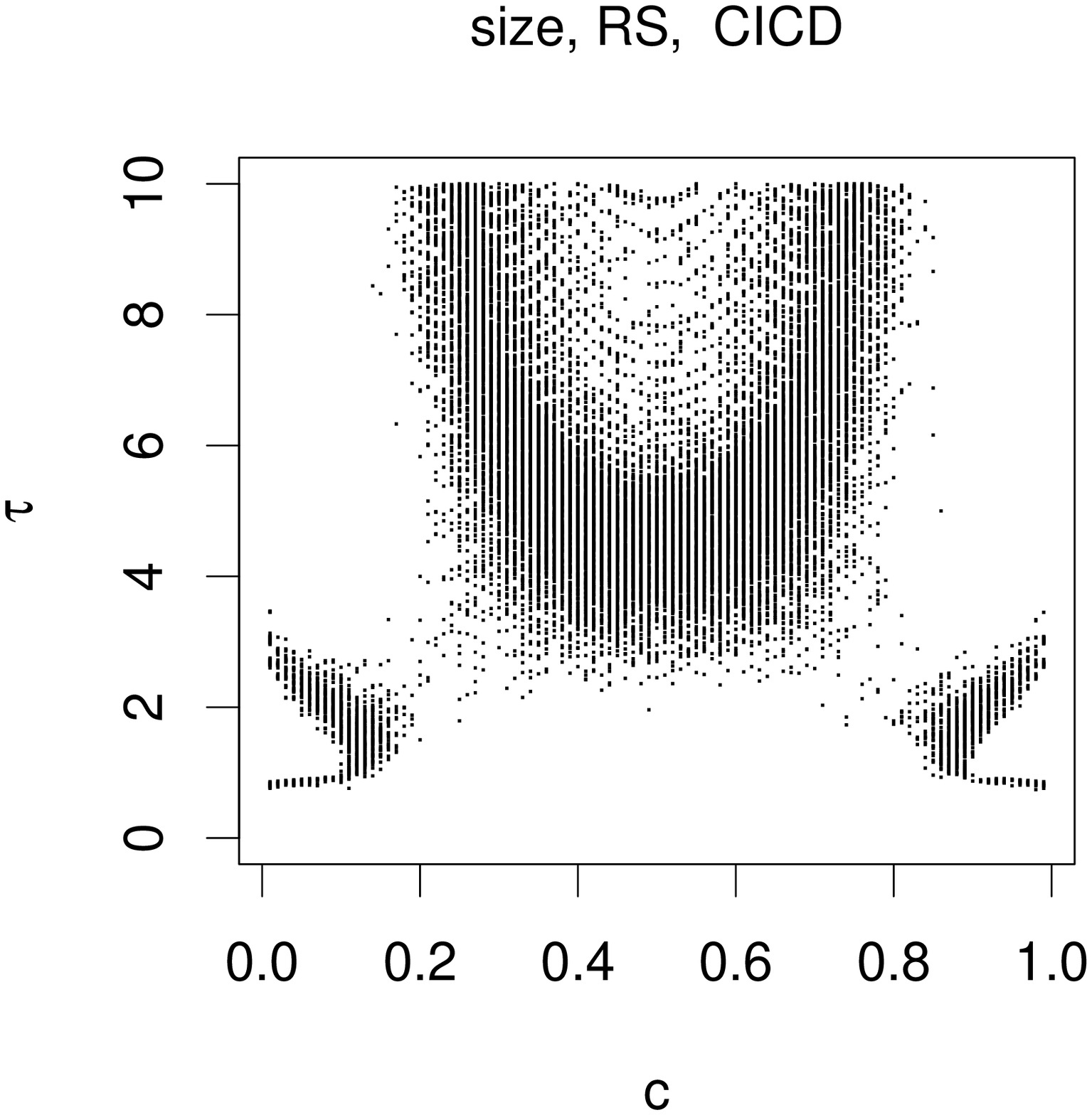}}}
\rotatebox{-90}{ \resizebox{2.12 in}{!}{ \includegraphics{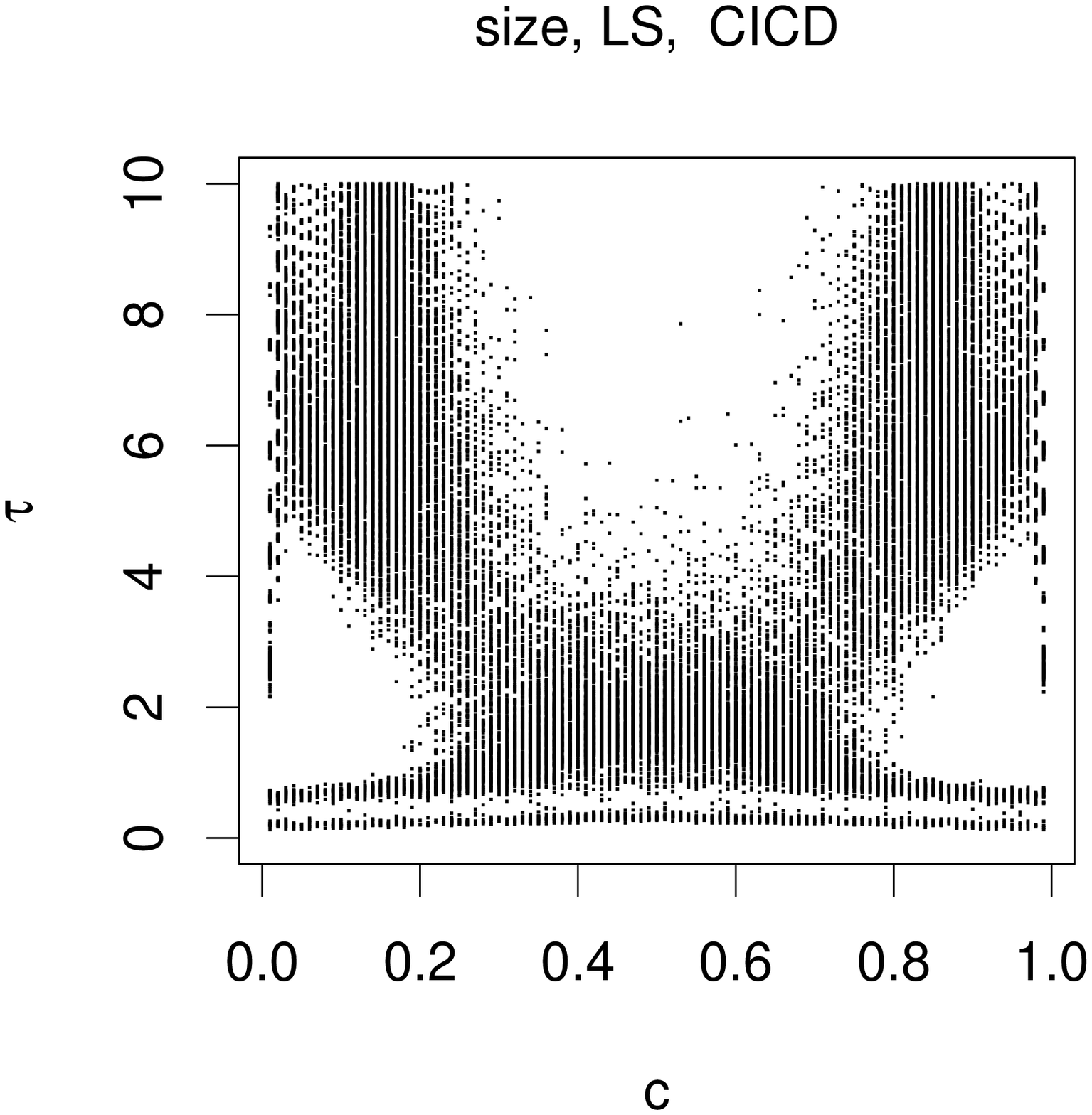}}}
\caption{
\label{fig:AD-size-image-plots}
Two-level (i.e., black and white) image plots for the empirical size estimates
for the arc density of CICD and PICD
based on $n=20$ and $N_{mc}=10000$
the two-sided (TS), right-sided (RS) and left-sided (LS) alternatives.
The empirical sizes not significantly different from 0.05
are represented with black dots, and others are blanked out (i.e., represented with white dots).
For CICD, we use $\tau=.01,.02,\ldots,10.00$ and
for PICD , we use $r=1.01,1.02,\ldots,10.00$ and for both ICDs,
we take $c=.01,.02,\ldots,.99$ with $N_{mc}=10000$ Monte Carlo replications.
}
\end{figure}

We present the empirical size estimates of the tests based on the arc density of the ICDs
in two-level image plots
(with empirical sizes not significantly different from 0.05 in black,
and others blanked out in white)
for the two-sided, right-sided and left-sided alternatives in Figure \ref{fig:AD-size-image-plots}.
The size estimates for PICDs with
$n=20$, $c=.01,.02,\ldots,.99$ and $r=1.01,1.02, \ldots, 10.00$ are plotted in the top row
and those for CICDs with
$n=20$, $c=.01,.02,\ldots,.99$ and $\tau=.01,.02, \ldots, 10.00$ are plotted in bottom row.
The size estimates for $n=50$ and $100$ have similar trends
with sizes closer to nominal level for more parameter combinations
(hence not presented).
Notice the symmetry in size estimates around $c=1/2$.

\subsection{Illustrative Figures}
\label{sec:illustrative-figs}

\begin{figure}[hbp]
\centering
\rotatebox{0}{ \resizebox{2.5 in}{!}{ \includegraphics{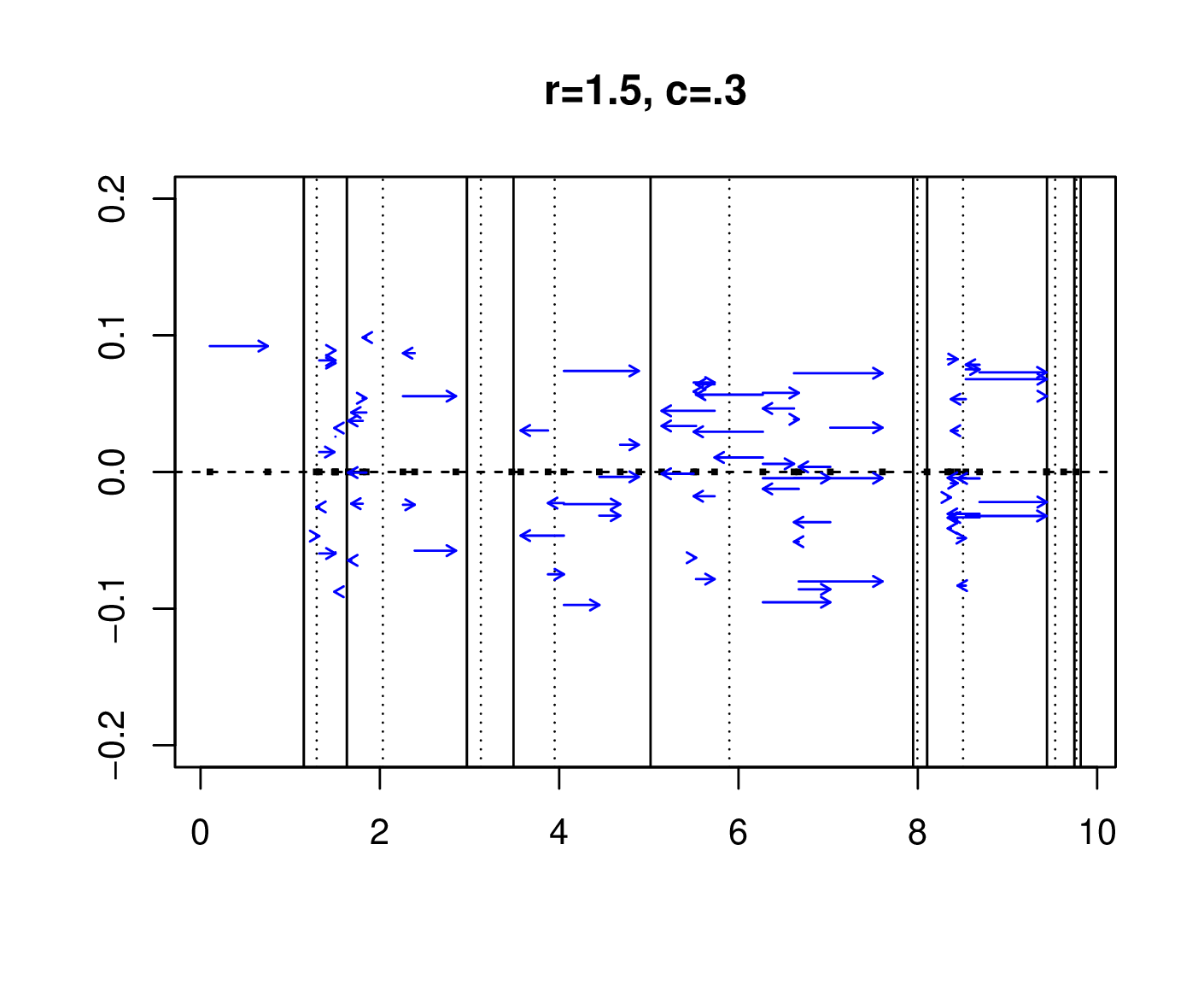}}}
\rotatebox{90}{ \resizebox{2. in}{!}{ \includegraphics{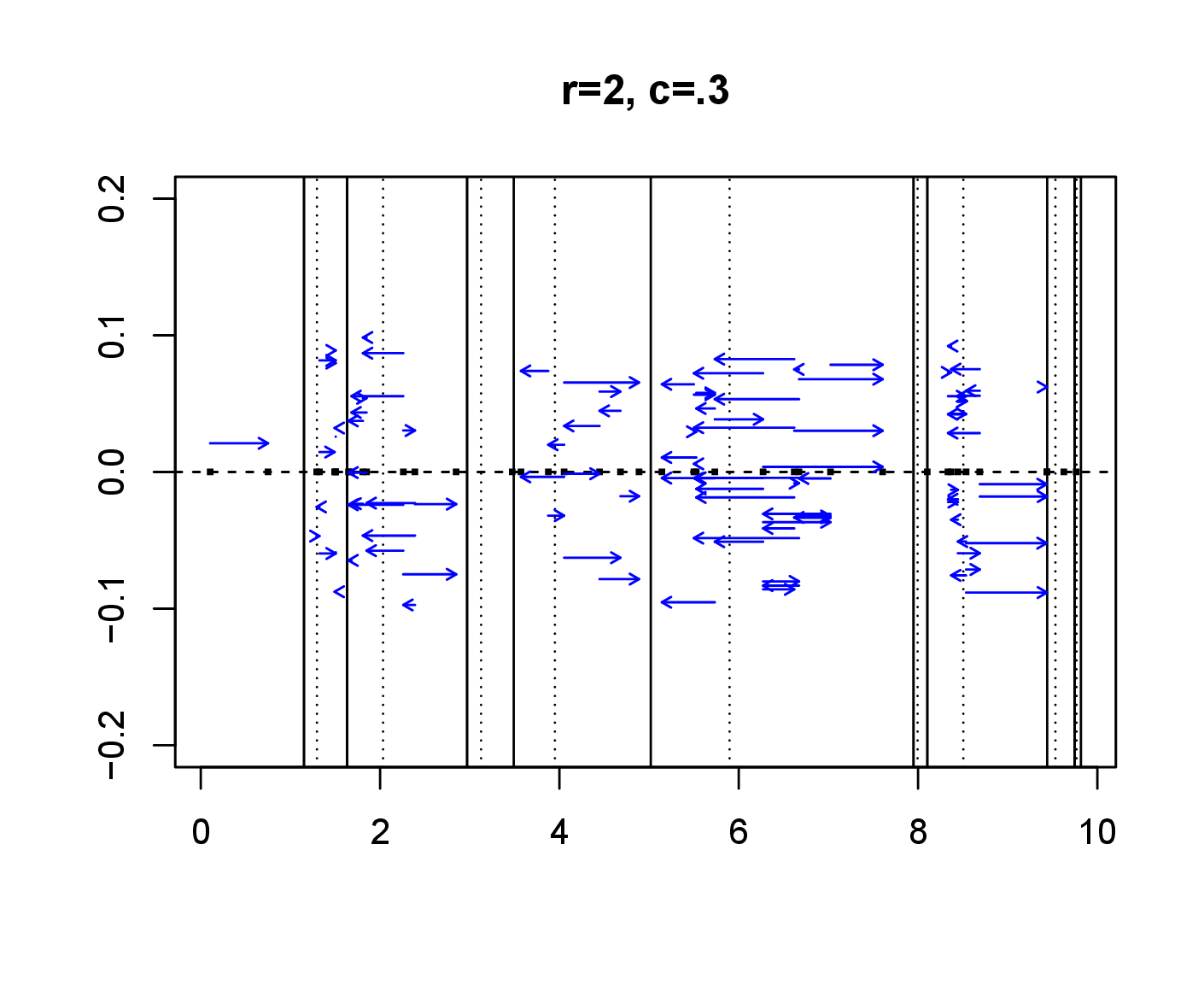}}}
\rotatebox{0}{ \resizebox{2.5 in}{!}{ \includegraphics{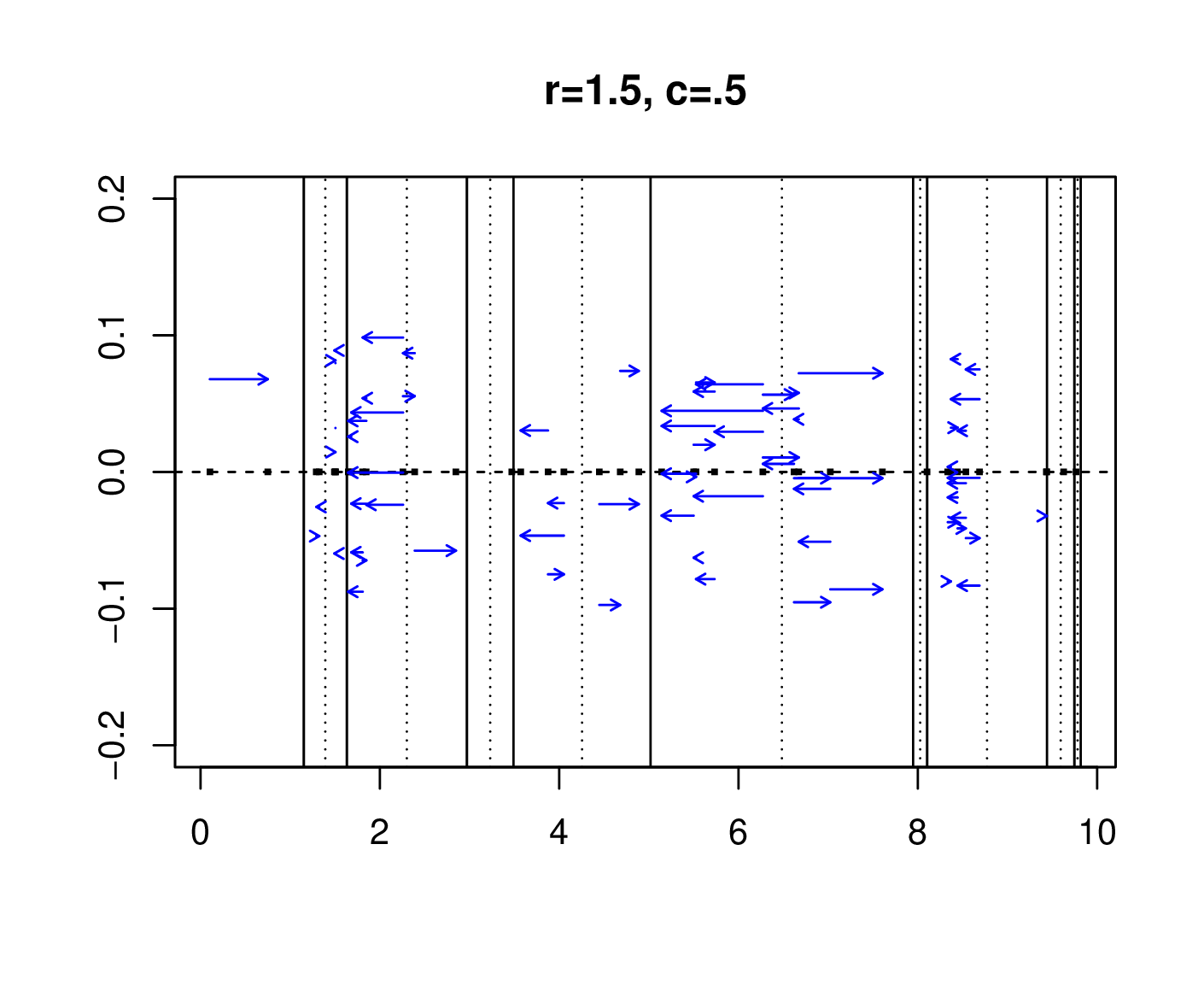}}}
\rotatebox{90}{ \resizebox{2. in}{!}{ \includegraphics{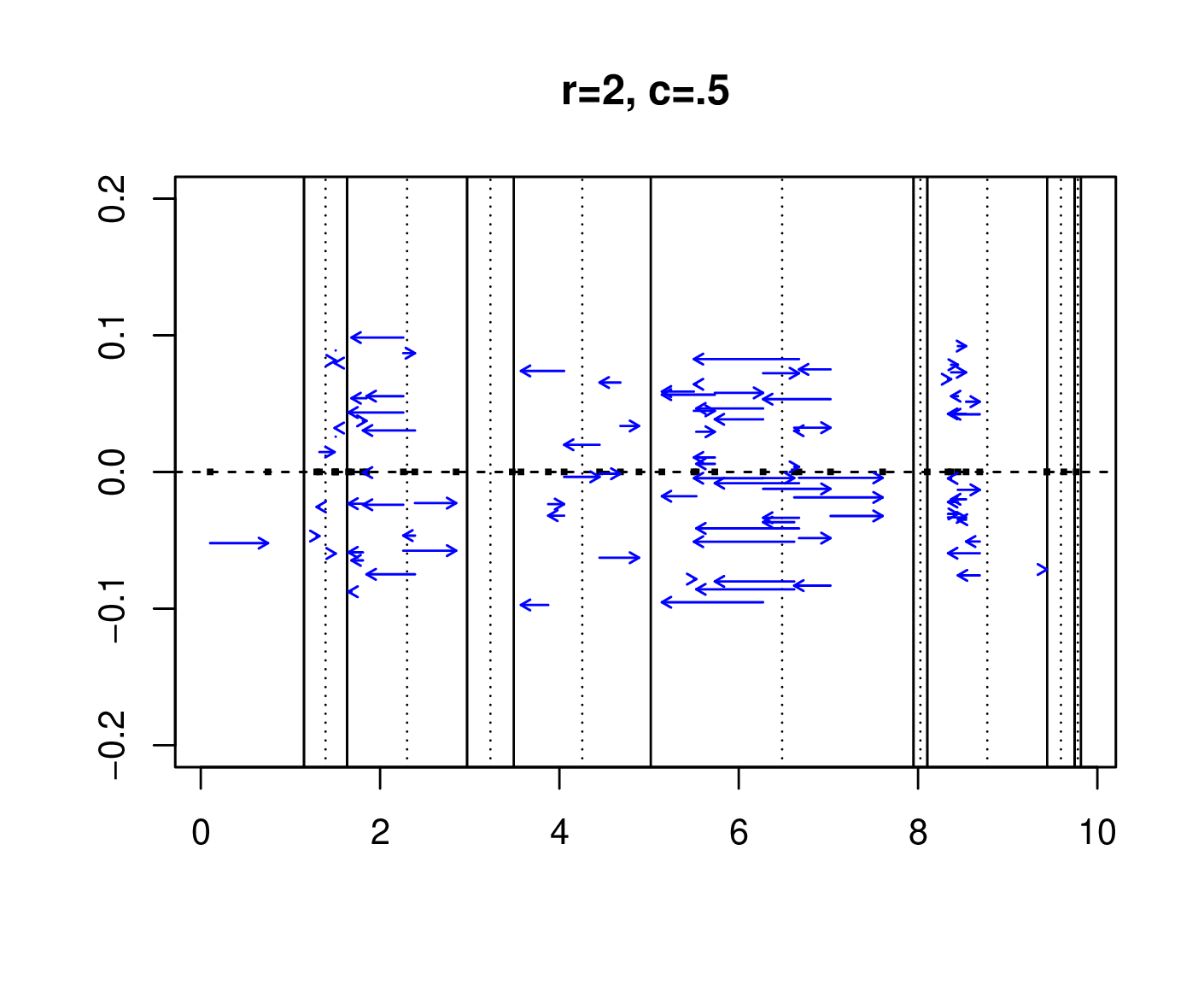}}}
\caption{
\label{fig:arcs-rc}
The arcs for the PICD with $\X_{40}$ and $\Y_{10}$ both of which are uniformly generated in $(0,10)$.
The parameters are provided at the top of each plot, $\Y$ points are represented with solid vertical lines,
and the center values are represented with dotted vertical lines.
Arcs are plotted at jittered locations along the $y$-axis for better visualization.}
\end{figure}

See Figure \ref{fig:arcs-rc} for the arcs of our PICD with $\X_{40}$ and $\Y_{10}$ uniformly generated in $(0,10)$
and $(r,c)=(1.5,.3)$, $(r,c)=(2,.3)$, $(r,c)=(1.5,.5)$ and $(r,c)=(2,.5)$.
This yields a disconnected digraph with subdigraphs
each of which might be null or itself disconnected.
(see, e.g., Figure \ref{fig:arcs-rc} for an illustration).

\begin{figure}[hbp]
\centering
\rotatebox{-90}{ \resizebox{2.5 in}{!}{ \includegraphics{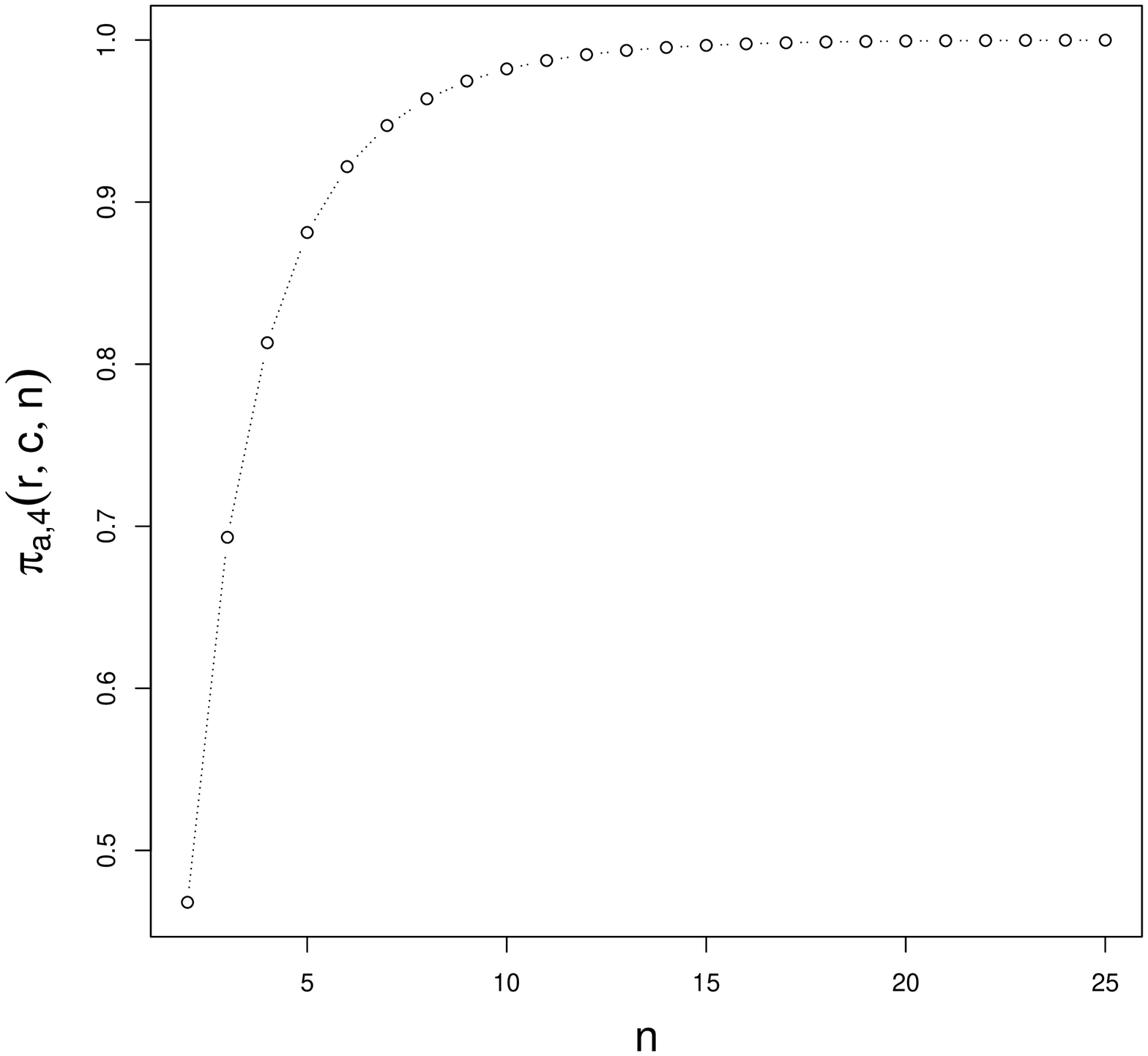}}}
\rotatebox{-90}{ \resizebox{2.5 in}{!}{ \includegraphics{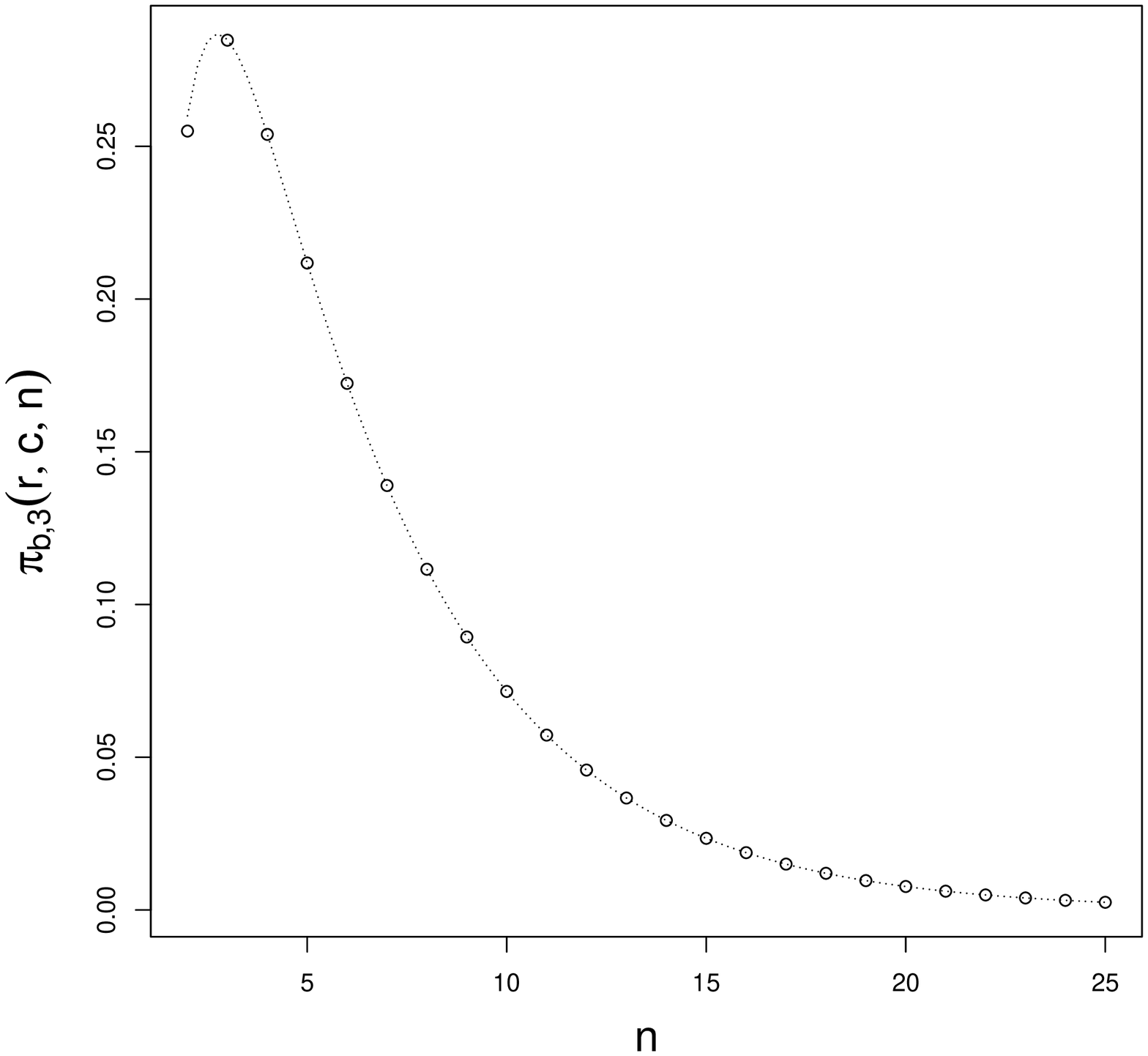}}}
\caption{
\label{fig:pi4-teta3-rc}
The plots of probabilities $\pi_{a,4}(r=1.2,c=0.4,n)$ (left) and $\pi_{b,3}(r=2,c=0.3,n)$ (right)
as a function of number of $\X$ points, $n$, in Main Result 1.}
\end{figure}

We present sample plots for $\pi_{a,4}(r,c,n)$ and $\pi_{b,3}(r,c,n)$ for specific $r$ and $c$ values
as a function of $n$.
As $n$ increases,
$\pi_{a,4}(r,c,n)$ strictly increases towards 1
(see Figure \ref{fig:pi4-teta3-rc} (left)),
and
$\pi_{b,3}(r,c,n)$
decreases (strictly decreases for $n \ge 3$) towards 0
(see Figure \ref{fig:pi4-teta3-rc} (right)).

For the alternatives (i.e., deviations from uniformity),
we consider five families of non-uniform distributions with support in $(0,1)$:
\begin{itemize}
  \item[(I)] $F_1(x,\delta)=(\delta x^2-\delta x+x)\I(0<x<1)+\I(x\ge 1)$,
  \item[(II)] $F_2(x,\sigma)=(\Phi(x,1/2,\sigma)-\Phi(0,1/2,\sigma))/(\Phi(1,1/2,\sigma)-\Phi(0,1/2,\sigma))\I(0<x<1)+\I(x\ge 1)$
where $\Phi(x,1/2,\sigma)$ is the normal distribution function
with mean $\mu=1/2$ and standard deviation $\sigma$, (i.e., normal distribution with $\mu=1/2$ restricted to $(0,1)$),
  \item[(III)] $F_3(x,\delta)=(\delta x^3/3-\delta x^2/2+x+\delta x/6)\I(0<x<1)+\I(x\ge 1)$,
  \item[(IV)] $F_4(x,\ve)$ is a distribution so that
$\ve \times 100$ \% of the regions around the $m$ subinterval end points are prohibited,
and the data is uniform in the remaining regions.
  \item[(V)] $F_5(x,\ve')$ is a distribution so that data is uniform over the
$\ve' \times 100$ \% of the regions around the $m$ subinterval end points are prohibited,
and the remaining regions are prohibited.
Notice that the supports of $F_4(x)$ and $F_5(x)$ are complimentary in $(0,1)$.
\end{itemize}
That is,
\begin{multline*}
H^I_a:~ F=F_1(x,\delta) \text{ with } \delta \in (0,1)
~~~
H^{II}_a:~ F=F_2(x,\sigma) \text{ with } \sigma > 0
~~~
H^{III}_a:~ F=F_3(x,\delta) \text{ with } \delta \in (0,12]\\
~~~
H^{IV}_a:~ F=F_4(x,\ve) \text{ with } \ve \in (0,1/2)
\text{ and }
H^{V}_a:~ F=F_5(x,\ve') \text{ for } \ve' \in (0,1/2)
\end{multline*}

\begin{figure}[hbp]
\begin{center}
\psfrag{f(x)}[cc][][0.8][0]{\put(-7.5,0){\huge{$f_1(x)$}}}
\psfrag{x}{\huge{$x$}}
\psfrag{k=1}{\huge{$\delta=.2$}}
\psfrag{k=2}{\huge{$\delta=.4$}}
\psfrag{k=3}{\huge{$\delta=.6$}}
\psfrag{k=4}{\huge{$\delta=.8$}}
\rotatebox{0}{ \resizebox{2.5 in}{!}{ \includegraphics{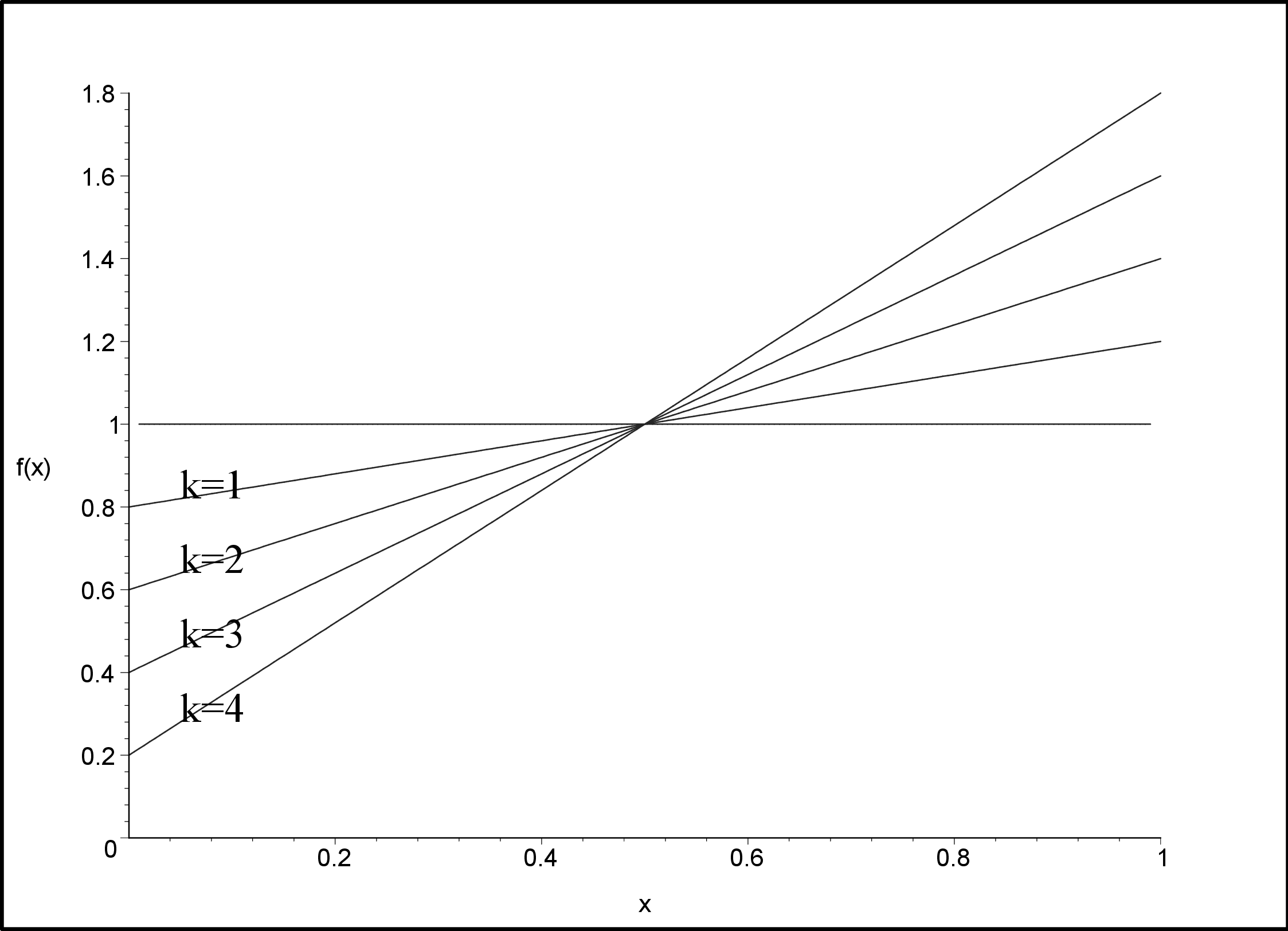}}}
\psfrag{f(x)}[cc][][0.8][0]{\put(-7.5,0){\huge{$f_2(x)$}}}
\psfrag{k=1}{\huge{$\sigma=.1$}}
\psfrag{k=2}{\huge{$\sigma=.2$}}
\psfrag{k=3}{\huge{$\sigma=.3$}}
\psfrag{k=4}{\huge{$\sigma=.4$}}
\rotatebox{0}{ \resizebox{2.5 in}{!}{ \includegraphics{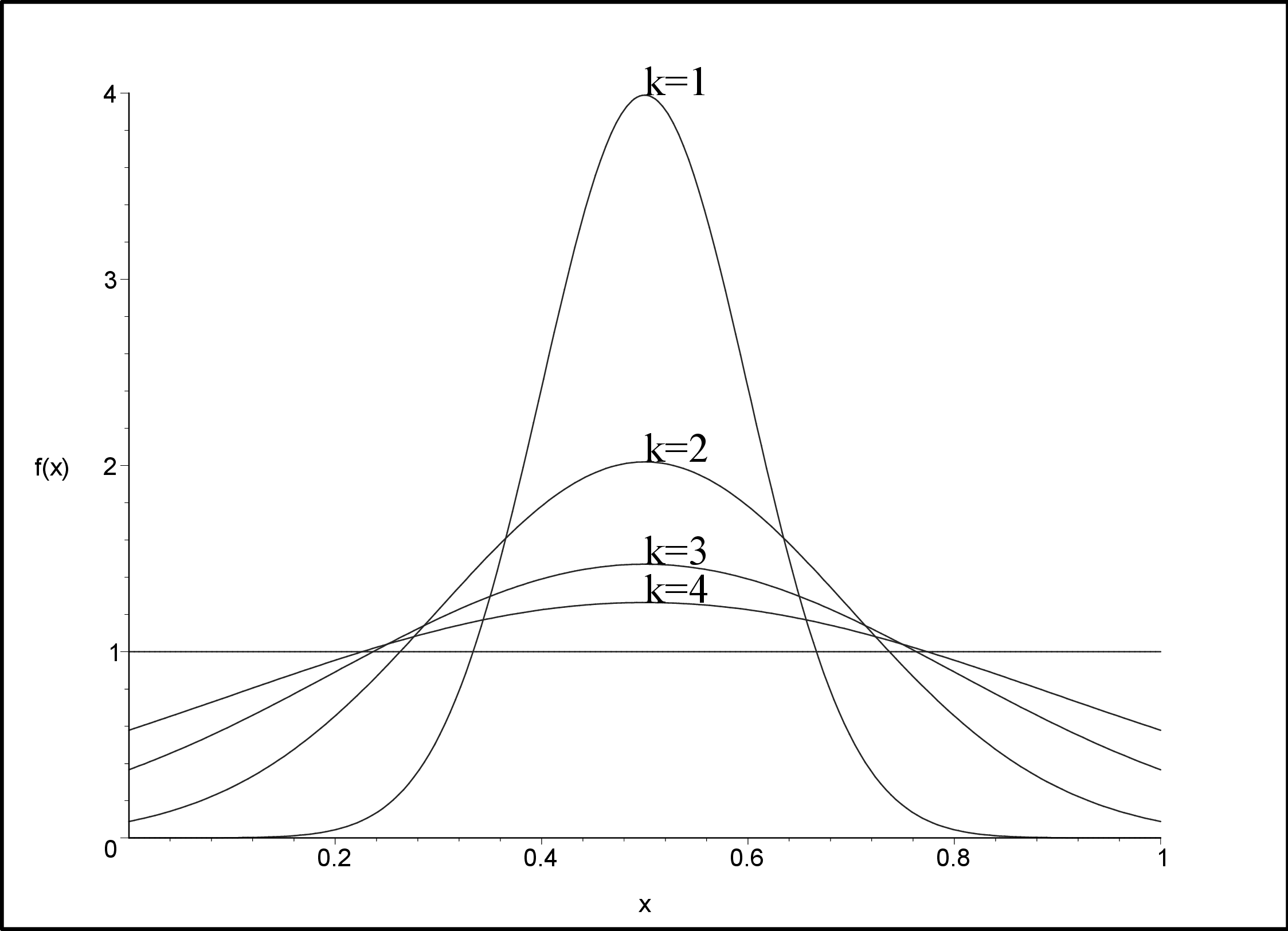}}}
\psfrag{f(x)}[cc][][0.8][0]{\put(-7.5,0){\huge{$f_3(x)$}}}
\psfrag{k=1}{\huge{$\delta=2$}}
\psfrag{k=2}{\huge{$\delta=4$}}
\psfrag{k=3}{\huge{$\delta=6$}}
\psfrag{k=4}{\huge{$\delta=8$}}
\rotatebox{0}{ \resizebox{2.5 in}{!}{ \includegraphics{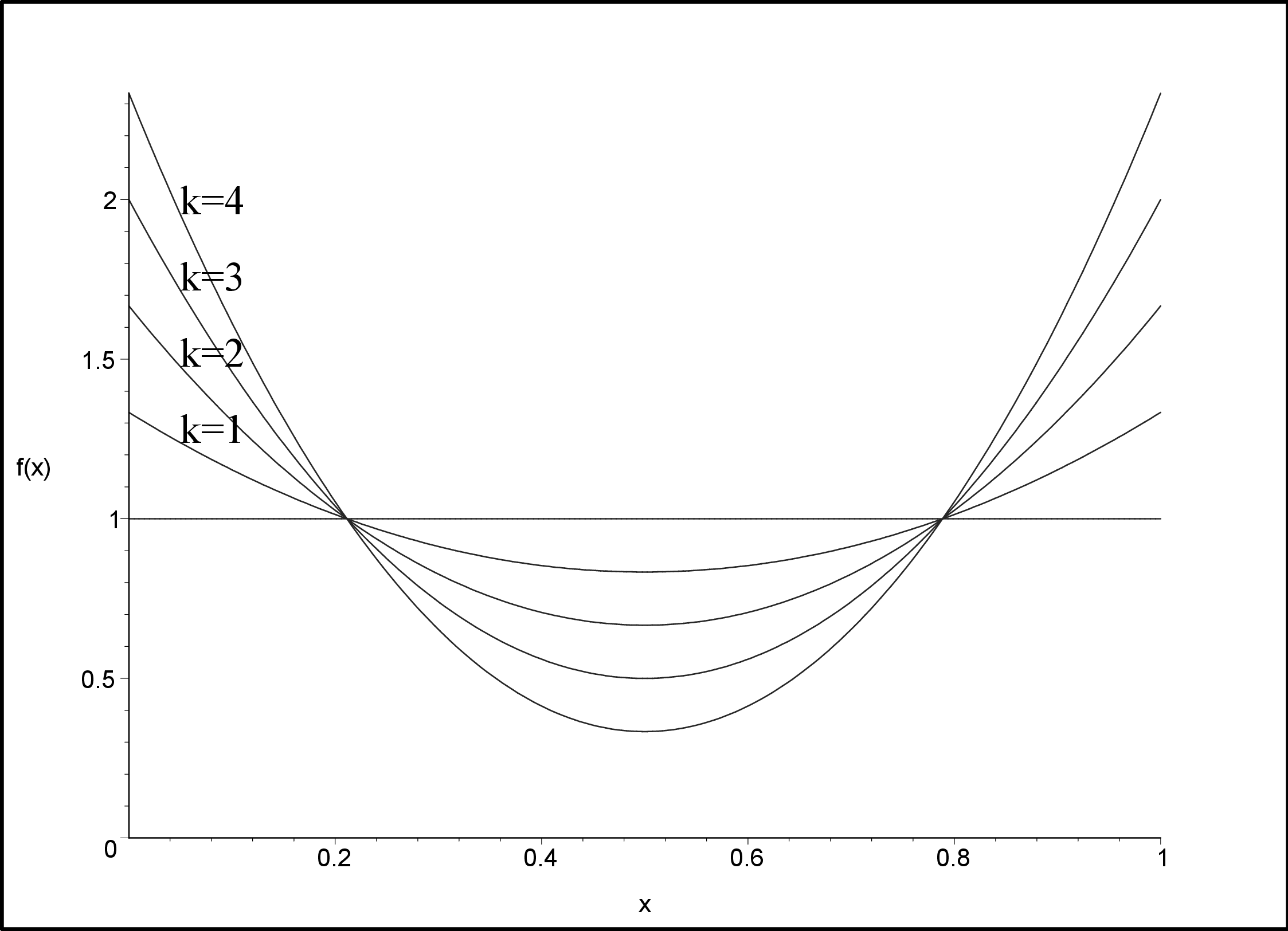}}}
\end{center}
\caption{
\label{fig:pdfs-5-alternatives}
Sample plots for the pdfs of the alternative types I (left), II (middle), and III (right).
We plot pdfs for type I with $\delta=.2,.4,.6,.8$,
for type II with $\sigma=.1,.2,.3,.4$,
and
for type III with $\delta=2,4,6,8$.
The horizontal line at $1$ indicates the pdf for $\U(0,1)$ distribution.
}
\end{figure}

See Figure \ref{fig:pdfs-5-alternatives} for sample plots of the corresponding pdfs
with various parameters for alternative types I-III.

\section{Some Results and Proofs for the Special Cases in Section \ref{sec:special-cases-dom-numb-Dnm}}

As an immediate result of Lemmas \ref{lem:end-intervals} and \ref{lem:gamma 1 or 2},
we have the following upper bound for $\g_{{}_{n,m}}(F_{XY},r,c)$.

\begin{proposition}
\label{prop:gamma-Dnm-r-M}
Let $\mathbf D_{n,m}(F_{XY},r,c)$ be a PICD
and $k_1$, $k_2$, and $k_3$ be three natural numbers defined as
$k_1:=\sum_{i=1}^{m-1} \I(n_i>1)$,
$k_2:=\sum_{i=1}^{m-1} \I(n_i=1)$,
and
$k_3:=\sum_{i \in \{0,m\}} \I(n_i > 0)$.
Then for $n \ge 1,\,m \ge 1$, $r \ge 1$, and $c \in [0,1]$,
we have
$1 \le \g_{{}_{n,m}}(F_{XY},r,c) \le 2\,k_1+k_2+k_3 \le \min(n,2\,m)$.
Furthermore,
$\g_{{}_{1,m}}(F_{XY},r,c)=1$ for all $m \ge 1$, $r \ge 1$, and $c \in [0,1]$;
$\g_{{}_{n,1}}(F_{XY},r,c)=\sum_{i \in \{0,1\}} \I(n_i > 0)$ for all $n \ge 1$ and $r \ge 1$;
$\g_{{}_{1,1}}(F_{XY},r,c)=1$ for all $r \ge 1$;
$\g_{{}_{n,m}}(F_{XY},r,0)=\g_{{}_{n,m}}(F_{XY},r,1)=k_1+k_2+k_3$ for all $m > 1$, $n \ge 1$, and $r \ge 1$;
and
$\g_{{}_{n,m}}(F_{XY},\infty,c)=k_1+k_2+k_3$ for all $m > 1$, $n \ge 1$, and $c \in [0,1]$.
\end{proposition}

For $r=1$, the distribution of $\g_{{}_{[i]}}(F_i,r,c)$ is simpler and
the distribution of $\g_{{}_{n,m}}(F_{XY},r,c)$ has simpler upper bounds.

\begin{proposition}
\label{prop:gamma-Dnm-r=1-M}
Let $\mathbf D_{n,m}(F_{XY},1,c)$ be a PICD,
$k_3$ be defined as in Proposition \ref{prop:gamma-Dnm-r-M},
and $k_4$ be a natural number defined as
$k_4:=\sum_{i=1}^{m-1} \left[\I\left( \left|\X_{[i]} \cap \left( Y_{(i)},M_{c,i} \right)\right|>0 \right)+
\I\left( \left|\X_{[i]} \cap \left( M_{c,i},Y_{(i+1)} \right)\right|>0 \right)\right]$.
Then for $n \ge 1,\,m > 1$, and $c \in [0,1]$,
we have
$1\le \g_{{}_{n,m}}(F_{XY},1,c) = k_3+k_4 \le \min(n,2\,m)$.
\end{proposition}

\noindent {\bf Proof of Lemma \ref{lem:G1-region-in-Ii}:}
By definition,
$\G_1\left( \X_{[i]},r,c \right) = \{x \in \mI_i: \X_{[i]} \subset N(x,r,c)\}$.
Suppose $r \ge 1$ and $c \in [0,1]$.
Then
for $x \in ( Y_{(i)},M_{c,i} ]$,
we have
$\X_{[i]} \subset N(x,r,c)$
iff $Y_{(i)}+r\,(x-Y_{(i)}) > \max\,\left( \X_{[i]} \right)$
iff $x > \frac{\max\,\left( \X_{[i]} \right)+Y_{(i)}(r-1)}{r}$.
Likewise
for $x \in [ M_{c,i},Y_{(i+1)} )$,
we have
$\X_{[i]} \subset N(x,r,c)$
iff $Y_{(i+1)}-r\,(Y_{(i+1)}-x) < \min\,\left( \X_{[i]} \right)$
iff $x < \frac{\min\left( \X_{[i]} \right)+Y_{(i+1)}(r-1)}{r}$.
Hence the desired result follows.
The result for $n_i=1$ is trivial.
$\blacksquare$

\noindent {\bf Proof of Lemma \ref{lem:gamma 1 or 2}:}
Let $X^-_i:=\argmin_{x \in \X_{[i]} \cap \left( Y_{(i)},M_{c,i} \right)}d(x,M_{c,i})$
provided that $\X_{[i]} \cap \left( Y_{(i)},M_{c,i} \right) \not= \emptyset$,
and $X^+_i:=\argmin_{x \in \X_{[i]} \cap \left( M_{c,i},Y_{(i+1)} \right)}d(x,M_{c,i})$
provided that $\X_{[i]} \cap \left( M_{c,i},Y_{(i+1)} \right) \not= \emptyset$.
That is, $X^-_i$ and $X^+_i$ are closest class $\X$ points (if they exist)
to $M_{c,i}$ from left and right, respectively.
Notice that since $n_i > 0$, at least one of $X^-_i$ and $X^+_i$  must exist.
If $\X_{[i]} \cap \left( Y_{(i)},M_{c,i} \right) = \emptyset$,
then $\X_{[i]} \subset N\left( X^+_i,r,c \right)$;
so $\g_{{}_{[i]}}(F_i,r,c)=1$.
Similarly,
if $\X_{[i]} \cap \left(M_{c,i},Y_{(i+1)}\right)= \emptyset$,
then $\X_{[i]} \subset N\left( X^-_i,r,c \right)$;
so $\g_{{}_{[i]}}(F_i,r,c)=1$.
If both of $\X_{[i]} \cap \left( Y_{(i)},M_{c,i} \right)$ and $\X_{[i]} \cap (M_{c,i},Y_{(i+1)})$ are nonempty,
then $\X_{[i]} \subset N\left( X^-_i,r,c \right) \cup N\left( X^+_i,r,c \right)$,
so $\g_{{}_{[i]}}(F_i,r,c) \le 2$.
Since $n_i > 0$, we have $1 \le \g_{{}_{[i]}}(F_i,r,c) \le 2$.
The desired result follows,
since the probabilities $1-p(F_i,r,c))=P(\g_{{}_{[i]}}(F_i,r,c)=1)$ and
$p(F_i,r,c))=P(\g_{{}_{[i]}}(F_i,r,c)=2)$
are both positive.
The special cases in the theorem follow by construction.
$\blacksquare$

\noindent {\bf Proof of Proposition \ref{prop:gamma-Dnm-r-M}:}
Suppose $n \ge 1,\,m \ge 1$, $r \ge 1$, and $c \in [0,1]$.
Then for $i = 1,2,\ldots,(m-1)$,
by Lemma \ref{lem:gamma 1 or 2},
we have $\g_{{}_{[i]}}(F_i,r,c) \in \{1,2\}$ provided that $n_i>1$,
and $\g_{{}_{[i]}}(F_i,r,c) = 1$ for $n_i=1$.
For $i\in \{0,m\}$,
by Lemma \ref{lem:end-intervals},
we have $\g_{{}_{[i]}}(F_i,r,c)= \I(n_i > 0)$.
If $n_i=1$,
then $\g_{{}_{[i]}}(F_i,r,c)=1$
and
if $n_i>1$,
then $\g_{{}_{[i]}}(F_i,r,c)\le 2$.
Since $\g_{{}_{n,m}}(F_{XY},r,c)=\sum_{i=0}^m \g_{{}_{[i]}}(F_i,r,c)\I(n_i>0)$,
the desired result, $\g_{{}_{n,m}}(F_{XY},r,c) \le 2\,k_1+k_2+k_3 \le \min(n,2\,m)$, follows.
The special cases in the theorem follow by construction.
$\blacksquare$

\noindent {\bf Proof of Proposition \ref{prop:gamma-Dnm-r=1-M}:}
Suppose $n \ge 1,\,m > 1$, and $c \in [0,1]$
and let $X^-_i$ and $X^+_i$ be defined as in the proof of Lemma \ref{lem:gamma 1 or 2}.
Then by construction, $\X_{[i]} \cap \left( Y_{(i)},M_{c,i} \right) \subset N\left( X^-_i,1,c \right)$,
but $N\left( X^-_i,1,c \right) \subseteq \left( Y_{(i)},M_{c,i} \right)$.
So $\left[ \X_{[i]} \cap \left( M_{c,i},Y_{(i+1)} \right) \right] \cap N\left( X^-_i,1,c \right) = \emptyset$.
Similarly
$\X_{[i]} \cap \left( M_{c,i},Y_{(i+1)} \right) \subset N\left( X^+_i,1,c \right)$
and $\left[ \X_{[i]} \cap \left( Y_{(i)},M_{c,i} \right) \right] \cap N\left( X^+_i,1,c \right) = \emptyset$.
Then $\g_{{}_{[i]}}(F_i,1,c)=1$,
if $\X_{[i]} \subset \left( Y_{(i)},M_{c,i} \right)$
or
$\X_{[i]} \subset \left( M_{c,i},Y_{(i+1)} \right)$,
and $\g_{{}_{[i]}}(F_i,1,c)=2$,
if $\X_{[i]} \cap \left( Y_{(i)},M_{c,i} \right) \not= \emptyset$
and $\X_{[i]} \cap \left( M_{c,i},Y_{(i+1)} \right) \not= \emptyset$.
Hence for $i=1,2,3,\ldots,(m-1)$,
we have
$\g_{{}_{[i]}}(F_i,1,c) =
\I\left( \left|\X_{[i]} \cap \left( Y_{(i)},M_{c,i} \right)\right|>0 \right)+
\I\left( \left|\X_{[i]} \cap \left( M_{c,i},Y_{(i+1)} \right)\right|>0 \right)$,
and for $i\in \{0,m\}$,
we have $\g_{{}_{[i]}}(F_i,1,c)= \I(n_i > 0)$.
Since $\g_{{}_{n,m}}(F_{XY},1,c)=\sum_{i=0}^m \g_{{}_{[i]}}(F_i,1,c)\I(n_i>0)$,
the desired result follows.
$\blacksquare$

\section{Supplementary Materials for Section \ref{sec:gamma-dist-uniform} }
\subsection{Explicit Forms of $p_{u,a}(r,c,n)$, $p_{u,b}(r,c,n)$, and $p_{u,c}(r,c,n)$ in Theorem \ref{thm:r and M} }
\label{sec:explicit-forms}

\begin{multline*}
p_{u,a}(r,c,n)=\pi_{a,1}(r,c,n) \,\I(r \ge 1/c) +
\pi_{a,2}(r,c,n) \,\I(1/(1-c) \le r < 1/c)+
p_{{}_{a,3}}(r,c,n) \,\I((1-c)/c \le r < 1/(1-c)) +\\
\pi_{a,4}(r,c,n) \,\I( 1 \le r < (1-c)/c),
\end{multline*}
\begin{multline*}
p_{u,b}(r,c,n)=\pi_{b,1}(r,c,n) \,\I(r \ge 1/c) +
\pi_{b,2}(r,c,n) \,\I((1-c)/c \le r < 1/c)+
p_{{}_{b,3}}(r,c,n) \,\I(1/(1-c) \le r < (1-c)/c) +\\
\pi_{b,4}(r,c,n) \,\I( 1 \le r < 1/(1-c)),
\end{multline*}
and
\begin{multline*}
p_{u,c}(r,c,n)(r,c,n)=p_{c,1}(r,c,n) \,\I(r \ge 1/c) +
p_{c,2}(r,c,n) \,\I((1-c)/c \le r < 1/c)+
p_{c,3}(r,c,n) \,\I(\left(1+\sqrt{1-4 c}\right)/(2 c) \le \\
r < (1-c)/c) + p_{c,4}(r,c,n) \,\I( \left(1-\sqrt{1-4 c}\right)/(2 c) \le r < \left(1+\sqrt{1-4 c}\right)/(2 c))+
p_{c,5}(r,c,n) \,\I( 1/(1-c) \le r < \\
\left(1-\sqrt{1-4 c}\right)/(2 c))+ p_{c,6}(r,c,n) \,\I( 1 \le r < 1/(1-c))
\end{multline*}
where
$$\pi_{a,1}(r,c,n)={r}^{2} \left( \left(\frac{2}{r}\right)^{n} (r -1)-
2\,\left( {\frac {r-1}{{r}^{2}}} \right)^{n}r \right) \Big / \left( (r-1)(r+1)^2 \right),$$
\begin{multline*}
\pi_{a,2}(r,c,n)=
r \Bigg( \left( {\frac {c r +1}{r}} \right)^{n}(r+1)-
\left( \frac {(r-1)^{n-1}}{{r}^{2(n-1)}} \right)-
\left( {\frac {1-c}{r}} \right)^{n}(r+1)-
\left( {\frac {c{r}^{2}+c r -r+1}{r}} \right)^{n}- \\
\left( r-1 \right)^{n-1} \left( {\frac { c r +c-1 }{r}} \right)^{n} \Bigg) \Big /
(r+1)^2,
\end{multline*}
\begin{multline*}
p_{{}_{a,3}}(r,c,n)=
\Bigg(  -\left( {\frac {1-c}{r}} \right)^{n}({r}^{3}-r)-
\left( {\frac{c}{r}} \right)^{n}({r}^{3}-r)- \left( r-1 \right)^{n+1}-
\left(\left( 1-c \right) r \right)^{n}({r}^{2}-1)-
\left( c r \right)^{n}({r}^{2}-1)- \\
\left( \frac { r-1 }{r} \right)^{n} r \left(  \left( r - c r -c \right)^n +\left( c r +c-1 \right)^n \right)+
{r}^{3}+{r}^{2}-r-1 \Bigg)  \Big /
\left( (r-1)(r+1)^2 \right),
\end{multline*}
and
\begin{multline*}
\pi_{a,4}(r,c,n)=
\Bigg( \left( r-1 \right)^{n+1}-
\left( {\frac{1-c}{r}} \right)^{n}({r}^{3}-r)-
\left( {\frac{c}{r}} \right)^{n}({r}^{3}-r)-
\left(\left( 1-c \right) r \right)^{n}({r}^{2}-1)-
\left( c r \right)^{n}({r}^{2}-1)-\\
(r-1)^{n}\left( \left(\frac{c r +c -r}{r}\right)^n -(1- c r -c)^n \right)+
r^3+r^2-r-1 \Bigg) \Big /
\left( (r-1)(r+1)^2 \right).
\end{multline*}
Moreover, we have
$\pi_{b,1}(r,c,n)=\pi_{a,1}(r,c,n)$,
$\pi_{b,2}(r,c,n)=\pi_{a,2}(r,c,n)$,
$\pi_{b,4}(r,c,n)=\pi_{a,4}(r,c,n)$,
\begin{multline*}
\pi_{b,3}(r,c,n)=
\Bigg(\left( {\frac {c r +1}{r}} \right)^{n}({r}^2+r)-
r \left( {\frac {r-1}{{r}^{2}}} \right)^{n-1}-
\left( {\frac {1-c}{r}} \right)^{n}({r}^2+r)-
\left( {\frac {c{r}^{2}+c r -r+1}{r}} \right)^{n}r+\\
\left(\left( 1-c r -c \right)  \right)^{n} \Bigg) \Big /
(r+1)^2.
\end{multline*}
Finally,
we have
$p_{c,1}(r,c,n)=\pi_{a,1}(r,c,n)$,
$p_{c,2}(r,c,n)=\pi_{a,2}(r,c,n)$,
$p_{c,3}(r,c,n)=\pi_{b,3}(r,c,n)$,
$p_{c,5}(r,c,n)=\pi_{b,3}(r,c,n)$,
$p_{c,6}(r,c,n)=\pi_{a,4}(r,c,n)$,
and
$$
p_{c,4}(r,c,n)=
\left( r\left( {\frac {c r +1}{r}} \right)^{n}-
r \left( {\frac {1-c}{r}} \right)^{n}-
{c}^{n}(r+1)\right)\Big /
\left(r+1\right).
$$

\subsection{Proof of Theorem \ref{thm:r and M} }
In the proof of Theorem \ref{thm:r and M},
without loss of generality,
we can assume $(y_1,y_2)=(0,1)$
based on Theorem \ref{thm:scale-inv-NYr}.

\noindent
\textbf{\emph{Remark S1:}}
The $\G_1$-region, $\G_1(\X_n,r,c)$, depends on $X_{(1)}$, $X_{(n)}$, $r$, and $c$.
If $\G_1(\X_n,r,c) \not= \emptyset$,
then we have $\G_1(\X_n,r,c)=(\delta_1,\delta_2)$
where at least one of end points $\delta_1$ and $\delta_2$ is a function of $X_{(1)}$ and $X_{(n)}$.
For $\U(0,1)$ data,
given $X_{(1)}=x_1$ and $X_{(n)}=x_n$,
the probability of $p_u(r,c,n)$ is
$\displaystyle \left( 1-(\delta_2-\delta_1)/(x_n-x_1)\right)^{(n-2)}$
provided that $\G_1(\X_n,r,c) \not= \emptyset$;
and $\G_1(\X_n,r,c) = \emptyset$ implies $\g_{{}_{n,2}}(\U,r,c)=2$.
Then $P(\g_{{}_{n,2}}(\U,r,c)=2)=
P(\g_{{}_{n,2}}(\U,r,c)=2,\;\G_1(\X_n,r,c) \not= \emptyset)+
P(\g_{{}_{n,2}}(\U,r,c)=2,\;\G_1(\X_n,r,c) = \emptyset)$ and
\begin{equation}
\label{eqn:Pg2-Unif-int-first}
P(\g_{{}_{n,2}}(\U,r,c)=2,\;\G_1(\X_n,r,c) \not= \emptyset)=
\int\int_{\mS_1}f_{1n}(x_1,x_n)\left(1-\frac{\delta_2-\delta_1}{x_n-x_1}\right)^{(n-2)}\,dx_n dx_1
\end{equation}
where $\mS_1=\{0<x_1<x_n<1:x_1,x_n \not\in \G_1(\X_n,r,c) \text{ and } \G_1(\X_n,r,c) \not= \emptyset\}$
and $f_{1n}(x_1,x_n)=n(n-1)(x_n-x_1)^{(n-2)}\I(0<x_1<x_n<1)$ is the joint density of $X_{(1)},X_{(n)}$.
The integral simplifies to
\begin{equation}
\label{eqn:Pg2-integral-G1nonempty-unif}
P(\g_{{}_{n,2}}(\U,r,c)=2,\;\G_1(\X_n,r,c) \not= \emptyset)=
\int\int_{\mS_1}n(n-1)(x_n-x_1+\delta_1-\delta_2)^{(n-2)}\,dx_n dx_1.
\end{equation}
Since $\G_1(\X_n,r,c) = \emptyset$ implies $\g_{{}_{n,2}}(\U,r,c)=2$,
we have
\begin{equation}
\label{eqn:Pg2-integral-G1empty-unif}
P(\g_{{}_{n,2}}(\U,r,c)=2,\;\G_1(\X_n,r,c)= \emptyset)=
P(\G_1(\X_n,r,c)= \emptyset)=
\int\int_{\mS_2}f_{1n}(x_1,x_n)\,dx_n dx_1
\end{equation}
where $\mS_2=\{0<x_1<x_n<1:\G_1(\X_n,r,c) = \emptyset\}$.
$\square$

\noindent {\bf Proof of Theorem \ref{thm:r and M}:}
Given $X_{(1)}=x_1$ and $X_{(n)}=x_n$,
let $a=x_n/r$ and $b=(x_1+r-1)/r$
and due to symmetry we consider $c \in (0,1/2]$.
There are three cases for $c$, namely,
\noindent \textbf{Case (A)} $c \in \big[\left(3-\sqrt{5}\right)/2,1/2\big]$,
\noindent \textbf{Case (B)} $c \in \big[1/4,\left(3-\sqrt{5}\right)/2\big)$
and
\noindent \textbf{Case (C)} $c \in (0,1/4)$.
Additionally,
For $r \ge 1$ and $c \in (0,1/2]$,
the $\G_1$-region is $\G_1(\X_n,r,c)=(a,c] \cup [c,b)$.
Then there are four cases for $\G_1$-region:
Case (1) $\G_1(\X_n,r,c)=(a,b)$ which occurs when $a<c<b$,
Case (2) $\G_1(\X_n,r,c)=(a,c]$ which occurs when $a<c$ and $b<c$,
Case (3) $\G_1(\X_n,r,c)=[c,b)$ which occurs when $c<b$ and $c<a$, and
Case (4) $\G_1(\X_n,r,c)=\emptyset$ which occurs when $b<c<a$.

Let
\begin{equation*}
p_u(r,c,n) =
\left\lbrace \begin{array}{ll}
       p_{u,a}(r,c,n)  & \text{for $c \in \big[\left(3-\sqrt{5}\right)/2,1/2\big)$,}\\
       p_{u,b}(r,c,n)  & \text{for $c \in \big[1/4,\left(3-\sqrt{5}\right)/2\big)$,}\\
       p_{u,c}(r,c,n)  & \text{for $c \in (0,1/4)$,}\\
\end{array} \right.
\end{equation*}

\noindent \textbf{Case (A) $c \in \big[\left(3-\sqrt{5}\right)/2,1/2\big]$}

\noindent \textbf{Case (A1)}
$\G_1(\X_n,r,c)=(a,b)$, i.e., $a<c<b$:
Moreover,
for $\g_{{}_{n,2}}(\U,2,c)=2$,
$x_1<a$ and $x_n>b$ must hold;
otherwise,
$\g_{{}_{n,2}}(\U,2,c)=1$ would be the case,
since $x_1<b$ and $x_n>a$.
Hence the restrictions for $x_1$ and $x_n$ are
$x_n>\max(rx_1,(x_1+r-1)/r,x_1,0)=\max(rx_1,(x_1+r-1)/r)$,
$x_n < \min(cr,1,x_1+r-1)$,
and
$x_1>\max(0,c r -r+1)$,
$x_1 < \min(c,1)=c$.
Then the region of integration for
$P(\g_{{}_{n,2}}(\U,2,c)=2)$ is
$\mathcal S=\{(x_1,x_n):\max(rx_1,(x_1+r-1)/r) < x_n < \min(cr,1,x_1+r-1), \max(0,c r -r+1) < x_1 < c)\}$.

For $r \ge 1/c$,
$\max(0,c r -r+1)=0$ and $\min(cr,1,x_1+r-1)=1$,
hence we have
\begin{multline}
\label{eqn:r-in-[1/c,infty)-caseA1}
P(\g_{{}_{n,2}}(\U,r,c)=2,\;\G_1(\X_n,r,c)=(a,b))=\\
\left( \int_{0}^{1/(r+1)}\int_{(x_1+r-1)/r}^{1}+\int_{1/(r+1)}^{1/r}\int_{r\,x_1}^{1} \right)
n(n-1)f(x_1)f(x_n)\bigl(F(x_n)-F(x_1)+F(a)-F(b)\bigr)^{(n-2)}\,dx_n dx_1=\\
\frac{2\,r}{(r+1)^2} \left( \left(\frac{2}{r}\right)^{n-1}-
\left(\frac{r-1}{r^2} \right)^{n-1} \right)
\end{multline}

For $1/(1-c) \le r < 1/c$,
$\min(cr,1,x_1+r-1)=cr$,
so we have
\begin{multline}
\label{eqn:r-in-[1/(1-c),1/c)-caseA1}
P(\g_{{}_{n,2}}(\U,r,c)=2,\;\G_1(\X_n,r,c)=(a,b))=\\
\left( \int_{0}^{1/(r+1)}\int_{(x_1+r-1)/r}^{c\,r}+\int_{1/(r+1)}^{c}\int_{r\,x_1}^{c\,r} \right)
n(n-1)f(x_1)f(x_n)\bigl(F(x_n)-F(x_1)+F(a)-F(b)\bigr)^{(n-2)}\,dx_n dx_1=\\
\frac{r^2}{(r+1)^2} \Biggl[ \left(c(r+1)-\frac{r-1}{r}\right)^n
- \left(\frac{r-1}{r}\right)^{n-1} \left((c\,r+c-1)^n+\frac{1}{r^n}\right) \Biggr].
\end{multline}

For $(1-c)/c \le r < 1/(1-c)$,
we have $\max(0,c r -r+1)=c r -r+1$ and $\min(cr,1,x_1+r-1)=cr$,
then
\begin{multline}
\label{eqn:r-in-[(1-c)/c,1/(1-c))-caseA1}
P(\g_{{}_{n,2}}(\U,r,c)=2,\;\G_1(\X_n,r,c)=(a,b))=\\
\left( \int_{c r -r+1}^{1/(r+1)}\int_{(x_1+r-1)/r}^{c\,r}+\int_{1/(r+1)}^{c}\int_{r\,x_1}^{c\,r} \right)
n(n-1)f(x_1)f(x_n)\bigl(F(x_n)-F(x_1)+F(a)-F(b)\bigr)^{(n-2)}\,dx_n dx_1=\\
\frac{r^2(r-1)^{n-1}}{(r+1)^2}\left[ (r-1)-\frac{1}{r^{n-1}}[(r-c\,r-c)^n-(c\,r+c-1)^n] \right].
\end{multline}

For $1 \le r < (1-c)/c$,
we have $\max(0,c r -r+1)=c r -r+1$ and $\min(cr,1,x_1+r-1)=cr$,
and $c r -r+1<cr^2-r+1<c$.
Then
\begin{multline}
\label{eqn:r-in-[1,(1-c)/c)-caseA1}
P(\g_{{}_{n,2}}(\U,r,c)=2,\;\G_1(\X_n,r,c)=(a,b))=\\
\int_{c r -r+1}^{c\,r^2-r+1}\int_{(x_1+r-1)/r}^{c\,r}
n(n-1)f(x_1)f(x_n)\bigl(F(x_n)-F(x_1)+F(a)-F(b)\bigr)^{(n-2)}\,dx_n dx_1=\\
\frac{r^2 (r-1)^{n-1}}{(r+1)^2}
\left[r-1+(1-c\,r-c)^n+ \frac{(r-c\,r-c)^n}{r^{n-1}} \right].
\end{multline}

\noindent \textbf{Case (A2)}
$\G_1(\X_n,r,c)=(a,c]$, i.e., $a<c$ and $b<c$:
Also,
$x_1<a$ and $x_n>c$ must hold,
otherwise $\g_{{}_{n,2}}(\U,r,c)=1$ would be the case.
Then the restrictions on $x_1$ and $x_n$ become
$\max(rx_1,c) < x_n < \min(cr,1)$
and
$0 < x_1 < \min(c r -r+1,c)$.

In this case $r \ge 1/(1-c)$ is not possible, since $c r -r+1>0$.
Hence $r \ge 1/c$ is not possible either.

For $1 \le r < 1/(1-c)$,
we have
\begin{multline}
\label{eqn:r-in-[1,1/(1-c))-caseA2}
P(\g_{{}_{n,2}}(\U,r,c)=2,\;\G_1(\X_n,r,c)=(a,c])=\\
\int_{0}^{c r -r+1}\int_{c}^{c\,r}
n(n-1)f(x_1)f(x_n)(F(x_n)-F(x_1)+F(a)-F(c))^{(n-2)}\,dx_n dx_1=\\
\frac{r}{r+1} \left[c^n\left(r^n-\frac{1}{r^n}\right)-(r-1)^n \left(1-\frac{r-c\,r-c}{r}\right)^n \right].
\end{multline}

\noindent \textbf{Case (A3)}
$\G_1(\X_n,r,c)=[c,b)$, i.e., $c<b$ and $c<a$:
Also,
$x_1<c$ and $x_n>b$ must hold,
otherwise $\g_{{}_{n,2}}(\U,r,c)=1$ would be the case.
Then the restrictions on $x_1$ and $x_n$ become
$\max((x_1+r-1)/r,cr) < x_n < 1$
and
$\max(0,c r -r+1) < x_1 < c$.

In this case
$r \ge 1/c$ is not possible, since $x_n >c\,r$.

For $1/(1-c) \le r < 1/c$,
$\max((x_1+r-1)/r,cr)=cr$
and
$\max(0,c r -r+1)=0$.
Hence
we have
\begin{multline}
\label{eqn:r-in-[1/(1-c),1/c)-caseA3}
P(\g_{{}_{n,2}}(\U,r,c)=2,\;\G_1(\X_n,r,c)=[c,b))=\\
\int_{0}^{c}\int_{c\,r}^{1}
n(n-1)f(x_1)f(x_n)\bigl(F(x_n)-F(x_1)+F(c)-F(b)\bigr)^{(n-2)}\,dx_n dx_1=\\
\frac{1}{(r+1)r^{n-1}} \left[
(r-1)^n (c\,r-1+c)^n
+ (1+c\,r)^n - (c\,r^2+c\,r-r+1)^n - (1-c)^n
\right].
\end{multline}

For $(1-c)/c \le r < 1/(1-c)$,
$\max((x_1+r-1)/r,cr)=cr$
and
$\max(0,c r -r+1)=c r -r+1$.
So we have
\begin{multline}
\label{eqn:r-in-[(1-c)/c,1/(1-c))-caseA3}
P(\g_{{}_{n,2}}(\U,r,c)=2,\;\G_1(\X_n,r,c)=[c,b))=\\
\int_{c r -r+1}^{c}\int_{c\,r}^{1}
n(n-1)f(x_1)f(x_n)\bigl(F(x_n)-F(x_1)+F(c)-F(b)\bigr)^{(n-2)}\,dx_n dx_1=\\
\frac{r}{r+1} \left[
(r-1)^n \left(\left(\frac{c\,r-1+c}{r}\right)^n-1\right)+
(1-c)^n\left(r^n-\frac{1}{r^n}\right)
\right].
\end{multline}

For $1 \le r \le (1-c)/c$,
$\max((x_1+r-1)/r,cr)=cr$
and
$\max(0,c r -r+1)=c r -r+1$ and $c r -r+1<cr^2-r+1<c$.
So we have
\begin{multline}
\label{eqn:r-in-[1,(1-c)/c)-caseA3}
P(\g_{{}_{n,2}}(\U,r,c)=2,\;\G_1(\X_n,r,c)=[c,b))=\\
\left(\int_{c r -r+1}^{cr^2-r+1}\int_{c\,r}^{1}+\int_{cr^2-r+1}^{c}\int_{(x_1+r-1)/r}^{1} \right)
n(n-1)f(x_1)f(x_n)\bigl(F(x_n)-F(x_1)+F(c)-F(b)\bigr)^{(n-2)}\,dx_n dx_1=\\
 \left( r \left( \left( 1-c \right) r \right)^{n}-
\left( r-1 \right)^{n}r-
r \left( {\frac {1-c}{r}} \right)^{n}-
\left( \left( r-1 \right)  \left( 1-c r - c \right)  \right)^{n} \right) \Big /
 \left( r+1 \right).
\end{multline}

\noindent \textbf{Case (A4)}
$\G_1(\X_n,r,c)=\emptyset$, i.e., $b<c<a$:
The restrictions on $x_1$ and $x_n$ are
$\max(x_1+r-1,cr) < x_n < 1$
and
$0 < x_1 < \min(1,c r -r+1)=c r -r+1$.

In this case, $r \ge 1/c$ is not possible, since $x_n>c\,r > 1$;
and $1/(1-c) \le r < 1/c$ is not possible either, since $x_1 < c r -r+1 < 0$.

For $1 \le r < 1/(1-c)$,
we have $\max(x_1+r-1,cr)=cr$ and $c r -r+1 > 0$.
Then
\begin{multline}
\label{eqn:r-in-[1,1/(1-c))-caseA4}
P(\g_{{}_{n,2}}(\U,r,c)=2,\;\G_1(\X_n,r,c)=\emptyset)=\\
\int_{0}^{c r -r+1}\int_{c\,r}^{1}
n(n-1)f(x_1)f(x_n)(F(x_n)-F(x_1))^{(n-2)}\,dx_n dx_1=
1+ (r-1)^n - r^n [c^n + (1-c)^n].
\end{multline}

For $r \ge 1/c$,
the probability $P(\g_{{}_{n,2}}(\U,r,c)=2)$
is the same as in \eqref{eqn:r-in-[1/c,infty)-caseA1};
for $1/(1-c) \le r < 1/c$,
it is the sum of probabilities in \eqref{eqn:r-in-[1/(1-c),1/c)-caseA1}
and \eqref{eqn:r-in-[1/(1-c),1/c)-caseA3};
for $(1-c)/c \le r < 1/(1-c)$,
it is the sum of probabilities in \eqref{eqn:r-in-[(1-c)/c,1/(1-c))-caseA1},
\eqref{eqn:r-in-[1,1/(1-c))-caseA2},
\eqref{eqn:r-in-[(1-c)/c,1/(1-c))-caseA3},
and
\eqref{eqn:r-in-[1,1/(1-c))-caseA4};
for $1 \le r < (1-c)/c$,
it is the sum of probabilities in \eqref{eqn:r-in-[1,(1-c)/c)-caseA1},
\eqref{eqn:r-in-[1,1/(1-c))-caseA2},
\eqref{eqn:r-in-[1,(1-c)/c)-caseA3},
and
\eqref{eqn:r-in-[1,1/(1-c))-caseA4}.

\noindent \textbf{Case (B) $c \in \big[1/4,\left(3-\sqrt{5}\right)/2\big)$: }\\
\noindent \textbf{Case (B1)}:
$\G_1(\X_n,r,c)=(a,b)$, i.e., $a<c<b$:

For $r \ge 1/c$,
the probability
$P(\g_{{}_{n,2}}(\U,r,c)=2,\;\G_1(\X_n,r,c)=(a,b))$ as in
\eqref{eqn:r-in-[1/c,infty)-caseA1}.

For $(1-c)/c \le r < 1/c$,
the probability
$P(\g_{{}_{n,2}}(\U,r,c)=2,\;\G_1(\X_n,r,c)=(a,b))$ is as in
\eqref{eqn:r-in-[1/(1-c),1/c)-caseA1}.

For $1/(1-c) \le r < (1-c)/c$,
we have
\begin{multline}
\label{eqn:r-in-[(1/1-c),(1-c)/c)-caseB1}
P(\g_{{}_{n,2}}(\U,r,c)=2,\;\G_1(\X_n,r,c)=(a,b))=\\
\int_{0}^{c\,r^2-r+1}\int_{(x_1+r-1)/r}^{c\,r}
n(n-1)f(x_1)f(x_n)\bigl(F(x_n)-F(x_1)+F(a)-F(b)\bigr)^{(n-2)}\,dx_n dx_1=\\
\frac{r^2}{(r+1)^2}
\left[
(r-1)^{n-1}\left((1-c\,r-c)^n-\frac{1}{r^{2n-1}}\right)+
\left(\frac{c\,r^2+c\,r-r+1}{r}\right)^n
\right].
\end{multline}

For $1 \le r < 1/(1-c)$,
the probability
$P(\g_{{}_{n,2}}(\U,r,c)=2,\;\G_1(\X_n,r,c)=(a,b))$ is as in
\eqref{eqn:r-in-[1,(1-c)/c)-caseA1}.

\noindent \textbf{Case (B2)}:
$\G_1(\X_n,r,c)=(a,c]$, i.e., $a<c$ and $b<c$:
In this case,
$r \ge 1/(1-c)$ is not possible,
so the cases
$r \ge 1/c$,
$(1-c)/c \le r \le 1/c$,
and
$1/(1-c) \le r \le (1-c)/c$ are not possible either.

For $1 \le r < 1/(1-c)$,
the probability
$P(\g_{{}_{n,2}}(\U,r,c)=2,\;\G_1(\X_n,r,c)=(a,c])$ is as in
\eqref{eqn:r-in-[1,1/(1-c))-caseA2}.

\noindent \textbf{Case (B3)} $\G_1(\X_n,r,c)=[c,b)$, i.e., $c<b$ and $c<a$:
In this case,
$r \ge 1/c$ is not possible.

For $(1-c)/c \le r < 1/c$,
the probability
$P(\g_{{}_{n,2}}(\U,r,c)=2,\;\G_1(\X_n,r,c)=[c,b))$ is as in
\eqref{eqn:r-in-[1/(1-c),1/c)-caseA3}.

For $1/(1-c) \le r < (1-c)/c$,
we have
\begin{multline}
\label{eqn:r-in-[(1/1-c),(1-c)/c)-caseB3}
P(\g_{{}_{n,2}}(\U,r,c)=2,\;\G_1(\X_n,r,c)=(c,b))=\\
\left(\int_{0}^{c\,r^2-r+1}\int_{cr}^{1}+\int_{c\,r^2-r+1}^c\int_{(x_1+r-1)/r}^{1} \right)
n(n-1)f(x_1)f(x_n)\bigl(F(x_n)-F(x_1)+F(c)-F(b)\bigr)^{(n-2)}\,dx_n dx_1=\\
\left(r \left( {\frac {c r +1}{r}} \right)^{n}-
r \left( {\frac {1-c}{r}} \right)^{n}-
\left( {\frac {c{r}^{2}+c r -r+1}{r}} \right)^{n}r-
\left( \left( r-1 \right)  \left( 1-c r -c \right)  \right)^{n} \right) \Big /
\left( r+1 \right).
\end{multline}

For $1 \le r < 1/(1-c)$,
the probability
$P(\g_{{}_{n,2}}(\U,r,c)=2,\;\G_1(\X_n,r,c)=[c,b))$ is as in
\eqref{eqn:r-in-[1,(1-c)/c)-caseA3}.

\noindent \textbf{Case (B4)}
$\G_1(\X_n,r,c)=\emptyset$, i.e., $b<c<a$:
In this case,
$r \ge 1/(1-c)$ is not possible,
so the cases
$r \ge 1/c$,
$(1-c)/c \le r \le 1/c$,
and
$1/(1-c) \le r \le (1-c)/c$ are not possible either.

For $1 \le r < 1/(1-c)$,
the probability
$P(\g_{{}_{n,2}}(\U,r,c)=2,\;\G_1(\X_n,r,c)=(a,c])$ is as in
\eqref{eqn:r-in-[1,1/(1-c))-caseA4}.

For $r \ge 1/c$,
the probability $P(\g_{{}_{n,2}}(\U,r,c)=2)$
is the same as in \eqref{eqn:r-in-[1/c,infty)-caseA1};
for $(1-c)/c \le r < 1/c$,
it is the sum of probabilities in \eqref{eqn:r-in-[1/(1-c),1/c)-caseA1}
and \eqref{eqn:r-in-[1/(1-c),1/c)-caseA3};
for $1/(1-c) \le r < (1-c)/c$,
it is the sum of probabilities in
\eqref{eqn:r-in-[(1/1-c),(1-c)/c)-caseB1}
and
\eqref{eqn:r-in-[(1/1-c),(1-c)/c)-caseB3};
for $1 \le r < 1/(1-c)$
it is the sum of probabilities in \eqref{eqn:r-in-[1,(1-c)/c)-caseA1},
\eqref{eqn:r-in-[1,1/(1-c))-caseA2},
\eqref{eqn:r-in-[1,(1-c)/c)-caseA3},
and
\eqref{eqn:r-in-[1,1/(1-c))-caseA4}.

\noindent \textbf{Case (C) $c \in (0,1/4)$:}\\
\noindent \textbf{Case (C1)} $\G_1(\X_n,r,c)=(a,b)$, i.e., $a<c<b$:\\
For $r \ge 1/c$,
the probability
$P(\g_{{}_{n,2}}(\U,r,c)=2,\;\G_1(\X_n,r,c)=(a,b))$ as in
\eqref{eqn:r-in-[1/c,infty)-caseA1}.

For $(1-c)/c \le r < 1/c$,
the probability
$P(\g_{{}_{n,2}}(\U,r,c)=2,\;\G_1(\X_n,r,c)=(a,b))$ is as in
\eqref{eqn:r-in-[1/(1-c),1/c)-caseA1}.

For $\left(1+\sqrt{1-4 c}\right)/(2c) \le r < (1-c)/c$,
the probability
$P(\g_{{}_{n,2}}(\U,r,c)=2,\;\G_1(\X_n,r,c)=(a,b))$ is as in
\eqref{eqn:r-in-[(1/1-c),(1-c)/c)-caseB1}.

For $\left(1-\sqrt{1-4 c}\right)/(2c) \le r < \left(1+\sqrt{1-4 c}\right)/(2c)$,
the probability
$P(\g_{{}_{n,2}}(\U,r,c)=2,\;\G_1(\X_n,r,c)=(a,b))=0$,
since $x_1<cr^2-r+1<0$ can not hold.

For $1/(1-c) \le r < \left(1-\sqrt{1-4 c}\right)/(2c)$,
the probability
$P(\g_{{}_{n,2}}(\U,r,c)=2,\;\G_1(\X_n,r,c)=(a,b))$ is as in
\eqref{eqn:r-in-[(1/1-c),(1-c)/c)-caseB1}.

For $1 \le r < 1/(1-c)$,
the probability
$P(\g_{{}_{n,2}}(\U,r,c)=2,\;\G_1(\X_n,r,c)=(a,b))$ is as in
\eqref{eqn:r-in-[1,(1-c)/c)-caseA1}.

\noindent \textbf{Case (C2)} $\G_1(\X_n,r,c)=(a,c)$, i.e., $a<c$ and $b<c$:\\
In this case, $r \ge 1/(1-c)$, is not possible.
Hence the cases
$r \ge 1/c$;
$(1-c)/c \le r < 1/c$;
$\left(1+\sqrt{1-4 c}\right)/(2c) \le r < (1-c)/c$;
$\left(1-\sqrt{1-4 c}\right)/(2c) \le r < \left(1+\sqrt{1-4 c}\right)/(2c)$;
and $1/(1-c) \le r < \left(1-\sqrt{1-4 c}\right)/(2c)$ are not possible either.

For $1 \le r < 1/(1-c)$,
the probability
$P(\g_{{}_{n,2}}(\U,r,c)=2,\;\G_1(\X_n,r,c)=(a,b))$ is as in
\eqref{eqn:r-in-[1,1/(1-c))-caseA2}.

\noindent \textbf{Case (C3)} $\G_1(\X_n,r,c)=[c,b)$, i.e., $c<b$ and $c<a$:
In this case, $r \ge 1/c$, is not possible.

For $(1-c)/c \le r < 1/c$,
the probability
$P(\g_{{}_{n,2}}(\U,r,c)=2,\;\G_1(\X_n,r,c)=[c,b))$ is as in
\eqref{eqn:r-in-[1/(1-c),1/c)-caseA3}.

For $\left(1+\sqrt{1-4 c}\right)/(2c) \le r < (1-c)/c$,
the probability
$P(\g_{{}_{n,2}}(\U,r,c)=2,\;\G_1(\X_n,r,c)=[c,b))$ is as in
\eqref{eqn:r-in-[(1/1-c),(1-c)/c)-caseB3}.

For $\left(1-\sqrt{1-4 c}\right)/(2c) \le r < \left(1+\sqrt{1-4 c}\right)/(2c)$,
we have
\begin{multline}
\label{eqn:r-in-[(1-kok(1-4 c))/(2c),(1+kok(1-4 c))/(2c))-caseC3}
P(\g_{{}_{n,2}}(\U,r,c)=2,\;\G_1(\X_n,r,c)=(c,b))=\\
\int_{0}^c\int_{(x_1+r-1)/r}^{1}
n(n-1)f(x_1)f(x_n)\bigl(F(x_n)-F(x_1)+F(c)-F(b)\bigr)^{(n-2)}\,dx_n dx_1=\\
\left(r \left( {\frac {c r +1}{r}} \right)^{n}-
{c}^{n}r-
r \left( {\frac {1-c}{r}} \right)^{n}-
{c}^{n} \right) \Big /
\left( r+1 \right).
\end{multline}

For $1/(1-c) \le r < \left(1-\sqrt{1-4 c}\right)/(2c)$,
the probability
$P(\g_{{}_{n,2}}(\U,r,c)=2,\;\G_1(\X_n,r,c)=[c,b))$ is as in
\eqref{eqn:r-in-[(1/1-c),(1-c)/c)-caseB3}.

For $1 \le r < 1/(1-c)$,
the probability
$P(\g_{{}_{n,2}}(\U,r,c)=2,\;\G_1(\X_n,r,c)=[c,b))$ is as in
\eqref{eqn:r-in-[1,(1-c)/c)-caseA3}.

\noindent \textbf{Case (C4)} $\G_1(\X_n,r,c)=\emptyset$, i.e., $b<c<a$:\\
In this case, $r \ge 1/(1-c)$, is not possible.
Hence the cases
$r \ge 1/c$;
$(1-c)/c \le r < 1/c$;
$\left(1+\sqrt{1-4 c}\right)/(2c) \le r < (1-c)/c$;
$\left(1-\sqrt{1-4 c}\right)/(2c) \le r < \left(1+\sqrt{1-4 c}\right)/(2c)$;
and $1/(1-c) \le r < \left(1-\sqrt{1-4 c}\right)/(2c)$ are not possible either.

For $1 \le r < 1/(1-c)$,
the probability
$P(\g_{{}_{n,2}}(\U,r,c)=2,\;\G_1(\X_n,r,c)=(a,b))$ is as in
\eqref{eqn:r-in-[1,1/(1-c))-caseA4}.

For $r \ge 1/c$,
the probability $P(\g_{{}_{n,2}}(\U,r,c)=2)$
is the same as in \eqref{eqn:r-in-[1/c,infty)-caseA1};
for $(1-c)/c \le r < 1/c$,
it is the sum of probabilities in \eqref{eqn:r-in-[1/(1-c),1/c)-caseA1}
and \eqref{eqn:r-in-[1/(1-c),1/c)-caseA3};
for $\left(1+\sqrt{1-4 c}\right)/(2c) \le r < (1-c)/c$,
it is the sum of probabilities in \eqref{eqn:r-in-[(1/1-c),(1-c)/c)-caseB1}
and \eqref{eqn:r-in-[(1/1-c),(1-c)/c)-caseB3};
for $\left(1-\sqrt{1-4 c}\right)/(2c) \le r < \left(1+\sqrt{1-4 c}\right)/(2c)$,
it is the same as in \eqref{eqn:r-in-[(1-kok(1-4 c))/(2c),(1+kok(1-4 c))/(2c))-caseC3};
for $1/(1-c) \le r < \left(1-\sqrt{1-4 c}\right)/(2c)$,
it is the sum of probabilities in \eqref{eqn:r-in-[(1/1-c),(1-c)/c)-caseB1}
and \eqref{eqn:r-in-[(1/1-c),(1-c)/c)-caseB3};
and
for $1 \le r < 1/(1-c)$,
it is the sum of probabilities in \eqref{eqn:r-in-[1,(1-c)/c)-caseA1},
\eqref{eqn:r-in-[1,1/(1-c))-caseA2},
\eqref{eqn:r-in-[1,(1-c)/c)-caseA3},
and
\eqref{eqn:r-in-[1,1/(1-c))-caseA4}.

By symmetry,
$P(\g_{{}_{n,2}}(\U,r,c)=2)=p_u(r,1-c,n)$
with the understanding that the transformation $c \to 1-c$ is also applied
in the interval endpoints in the piecewise definitions of
$p_{u,a}(r,c,n)$, $p_{u,b}(r,c,n)$ and $p_{u,c}(r,c,n)$.
The special case for $c\in \{0,1\}$ follows trivially by construction.
$\blacksquare$

The proofs of Corollaries \ref{cor:r=2 and M_c=c}
and \ref{cor:r and M_c=1/2} follow,
since $p_u(r,c,n)$ in Theorem \ref{thm:r and M} is continuous in $r$ and $c$ for finite $n \ge 1$.

\subsection{Special Cases for the Exact Distribution of $\g_{n,2}(\U,r,c)$}
\label{sec:special-cases}
Notice that $p_u(r,c,n)$ is continuous in $r$ and $c$ for finite $n \ge 1$.
Hence we provide the following special cases
for the exact distribution of $\g_{n,2}(\U,r,c)$
as corollaries to Theorem \ref{thm:r and M} (Main Result 1):
(I) $r \ge 1$, $c=1/2$,
(II) $r =2$, $c \in (0,1)$,
and
(III) $r=2$, $c=1/2$.

\subsubsection{Special Case I: Exact Distribution of $\g_{n,2}(\U,r,1/2)$}
\label{sec:r-and-M_c=1/2}
For $r \ge 1$, $c=1/2$, and $(y_1,y_2)=(0,1)$,
the $\G_1$-region is $\G_1(\X_n,r,1/2)=(X_{(n)}/r,1/2] \cup [1/2,(X_{(1)}+r-1)/r)$
where
$(X_{(n)}/r,1/2]$ or $[1/2,(X_{(1)}+r-1)/r)$ or both could be empty.
\begin{corollary}
\label{cor:r and M_c=1/2}
Let $\X_n$ be a random sample from $\U(y_1,y_2)$ distribution
with $n \ge 1$ and $r \ge 1$.
Then we have
$\g_{{}_{n,2}}(\U,r,1/2)-1 \sim \BER(p_u(r,1/2,n))$
where
\begin{eqnarray*}
p_u(r,1/2,n)=
\begin{cases}
\frac{2\,r}{(r+1)^2} \left( \left(\frac{2}{r}\right)^{n-1}-\left(\frac{r-1}{r^2} \right)^{n-1} \right) &\text{for} \quad r \ge 2, \\
1-\frac{1+r^{2n-1}}{(2\,r)^{n-1}(r+1)}+\frac{(r-1)^n}{(r+1)^2} \left( 1- \left(\frac{r-1}{2\,r}\right)^{n-1}\right)
&\text{for} \quad 1 \le r < 2.
\end{cases}
\end{eqnarray*}
\end{corollary}

\begin{figure}
\begin{center}
\psfrag{r}{\huge{$r$}}
\psfrag{n}{\huge{$n$}}
\rotatebox{0}{ \resizebox{2.5 in}{!}{ \includegraphics{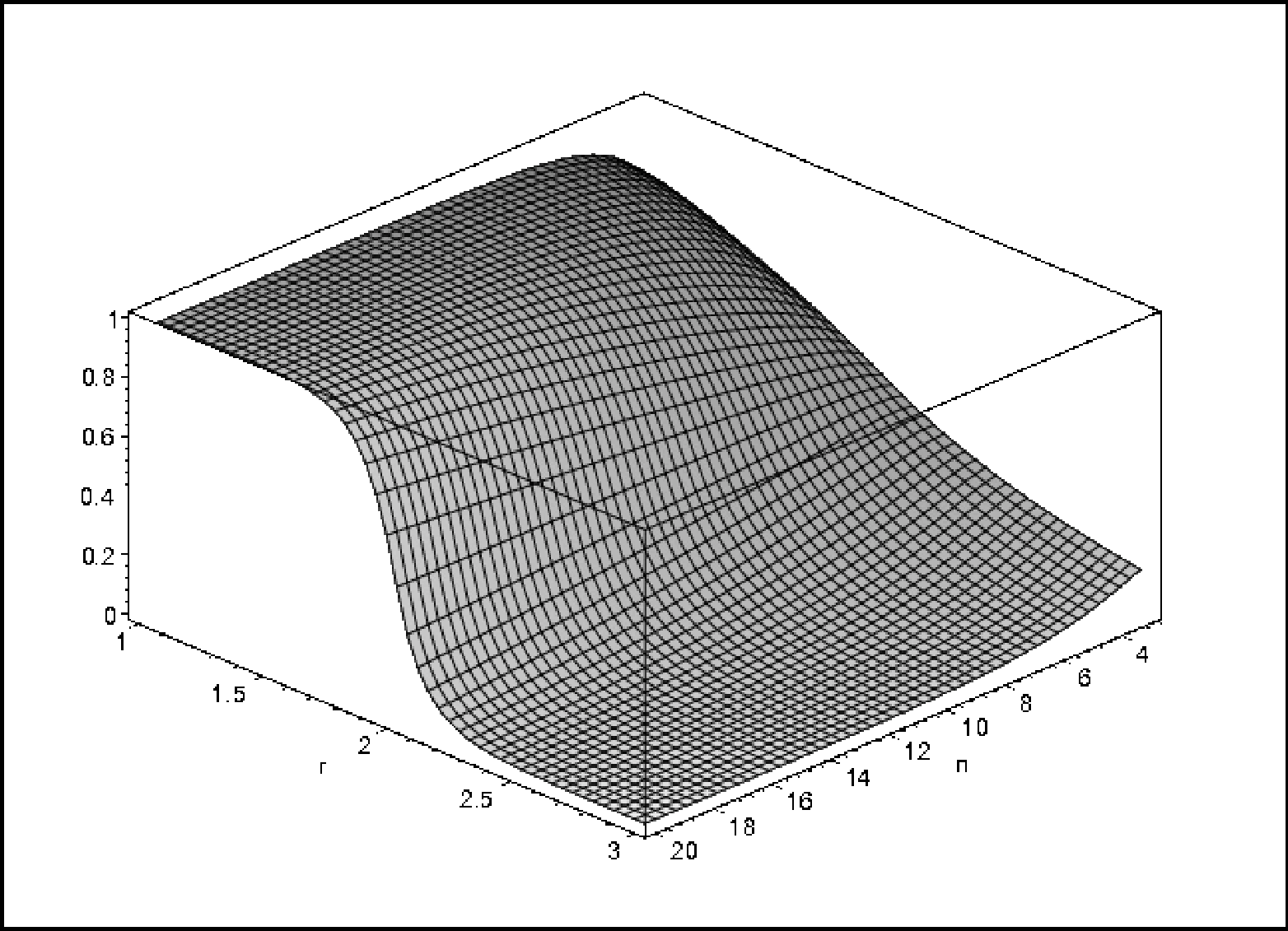}}}
\psfrag{p(r,n)}[cc][][0.8][90]{\put(-25,-15){\huge{$p_u(r,1/2,n)$}}}
\psfrag{n=5}{\huge{$n=5$}}
\psfrag{n=10}[cc][][1.0][0]{\put(-5,-15){\huge{$n=10$}}}
\psfrag{n=20}{\huge{$n=20$}}
\rotatebox{0}{ \resizebox{2.5 in}{!}{ \includegraphics{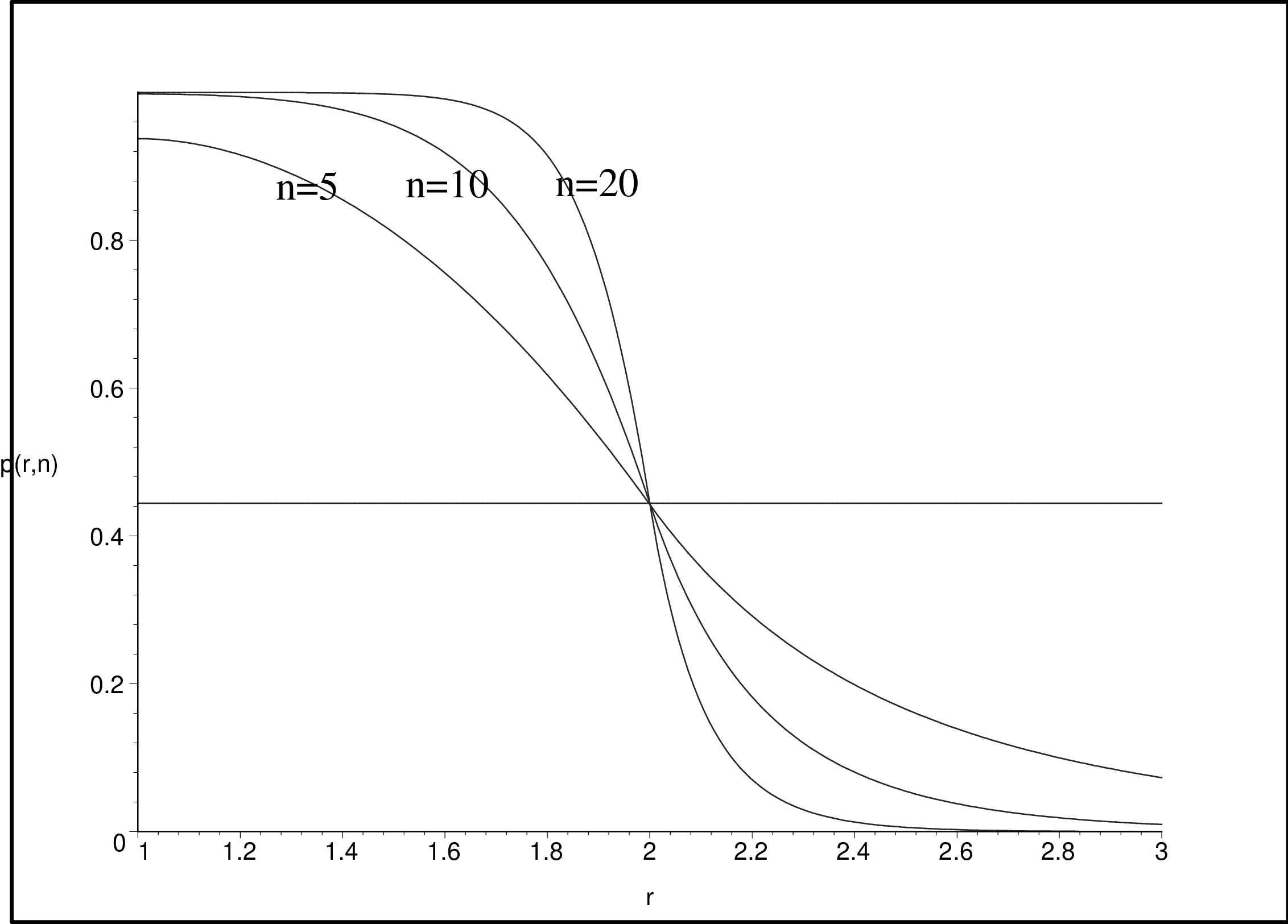}}}
\end{center}
\caption{
\label{fig:Prgam=2-withr}
Three-dimensional surface plots of $p_u(r,1/2,n)$ for $3 \le n \le 20$ and $r \in [1,3]$ (left) and
two-dimensional plots of $p_u(r,1/2,n)$ for $n=5,10,20$ and $r \in [1,3]$ (right).
In the surface plot, although $p_u(r,1/2,n)$ is defined for integer $n$ values,
we plot it as a continuous surface for better visualization.
In the right, the horizontal line is at $4/9$,
which is the limit of $p_u(r,1/2,n)$ at $r=2$ as $n \to \infty$.
}
\end{figure}

We present the three-dimensional surface plot of $p_u(r,1/2,n)$
for $3 \le n \le 20$ and $r \in [1,3]$ in Figure \ref{fig:Prgam=2-withr} (left)
and the two-dimensional plots of $p_u(r,1/2,n)$
for $n=5,10,20$ and $r \in [1,3]$ in Figure \ref{fig:Prgam=2-withr} (right).
Notice that for finite $n \ge 1$,
the probability
$p_u(r,1/2,n)$ is continuous in $r \ge 1$.
For fixed $n$,
$p_u(r,1/2,n)$ is decreasing as $r$ is increasing.
In particular,
for $r=2$,
we have $p_u(2,1/2,n)=4/9-(16/9) \, 4^{-n}$,
hence the distribution of $\g_{{}_{n,2}}(\U,r=2,1/2)$ is same as in \cite{priebe:2001}.
Furthermore,
$\lim_{r \rightarrow 1}p_u(r,1/2,n)=
p_u(1,1/2,n)=1-2^{1-n}$
and
$\lim_{r \rightarrow \infty}p_u(r,1/2,n)=
p_u(\infty,1/2,n)=0$.

In the limit,
as $n \rightarrow \infty$, we have
\begin{equation*}
\g_{{}_{n,2}}(\U,r,1/2)-1 \stackrel{\mathcal L}{\to}
\left\lbrace \begin{array}{ll}
       0           & \text{for $r > 2$,}\\
       \BER(4/9) & \text{for $r = 2$,}\\
       1           & \text{for $1 \le r < 2$.}\\
\end{array} \right.
\end{equation*}
Observe the interesting behavior of the asymptotic distribution
of $\g_{{}_{n,2}}(\U,r,1/2)$ around $r=2$.
The probability
$p_u(r,1/2)$ is continuous (in fact piecewise constant) for $r \in [1,\infty) \setminus \{2\}$.
Hence for $r \not= 2$, the asymptotic distribution is degenerate,
as $p_u(r,1/2) = 0$ for $r>2$
and $p_u(r,1/2) = 1$ for $r<2$
but $p_u(2,1/2) = 4/9$.
That is, for $r=2 \pm \varepsilon$ with arbitrarily small $\varepsilon>0$,
although the exact distribution is non-degenerate,
the asymptotic distribution is degenerate.

\subsubsection{Special Case II: Exact Distribution of $\g_{n,2}(\U,2,c)$}
\label{sec:r=2-and-M_c=c}
For $r=2$, $c \in (0,1)$, and $(y_1,y_2)=(0,1)$,
the $\G_1$-region is $\G_1(\X_n,2,c)=( X_{(n)}/2,c ] \cup [ c,(1+X_{(1)})/2)$.
Notice that $( X_{(n)}/2,c ]$ or $[ c,(1+X_{(1)})/2 )$
could be empty, but not simultaneously.
\begin{corollary}
\label{cor:r=2 and M_c=c}
Let $\X_n$ be a random sample from $\U(y_1,y_2)$ distribution
with $n \ge 1$.
Then we have
$\g_{{}_{n,2}}(\U,2,c)-1 \sim \BER(p_u(2,c,n))$
where
$p_u(2,c,n)=
\nu_{1,n}(c)\I(c \in (0,1/4])+
\nu_{2,n}(c)\I(c \in (1/4,1/3])+
\nu_{3,n}(c)\I(c \in (1/3,1/2])+
\nu_{3,n}(1-c)\I(c \in (1/2,2/3])+
\nu_{2,n}(1-c)\I(c \in (2/3,3/4])+
\nu_{1,n}(1-c)\I(c \in (3/4,1))
$
with

$$\nu_{1,n}(c)= \left( c + \frac{1}{2} \right)^{n-2} \left( \frac{2 c^2}{3}+\frac{2 c}{3}+\frac{1}{6}\right)-
\left( \frac{1-c}{2} \right)^{n-2} \left( \frac{c^2}{6}-\frac{c}{3}+\frac{1}{6}\right)-{c}^{n},$$

$$\nu_{2,n}(c)=  \left( 1 - 3\,c \right)^{n-2} \left( c^2 -\frac{2 c}{3}+\frac{1}{9}\right)+
\left( 3\,c- \frac{1}{2} \right)^{n-2} \left( \frac{2 c}{3}-2 c^2-\frac{1}{18}\right)-
\frac{2}{3}\,\left( \frac{1-c}{2} \right)^{n}+\frac{2}{3}\, \left( \frac{1}{2}+c \right)^{n}-
{\frac {8}{9}}\,{4}^{-n},$$
and
$$\nu_{3,n}(c)=
\frac{2}{3}\, \left( \frac{1}{2}+c \right)^{n}-
\frac{2}{9}\, \left( 3\,c-\frac{1}{2} \right)^{n}-
\frac{2}{9}\, \left( \frac{3 c - 1}{2} \right)^{n}
-\frac{2}{3}\, \left( \frac{1-c}{2} \right)^{n}-
{\frac {8}{9}}\,{4}^{-n}.$$
Furthermore,
$\g_{{}_{n,2}}(\U,2,0)=\g_{{}_{n,2}}(\U,2,1)=1$ for all $n \ge 1$.
\end{corollary}

\begin{figure}[hbp]
\begin{center}
\psfrag{c}{\huge{$c$}}
\psfrag{n}{\huge{$n$}}
\rotatebox{0}{ \resizebox{2.5 in}{!}{ \includegraphics{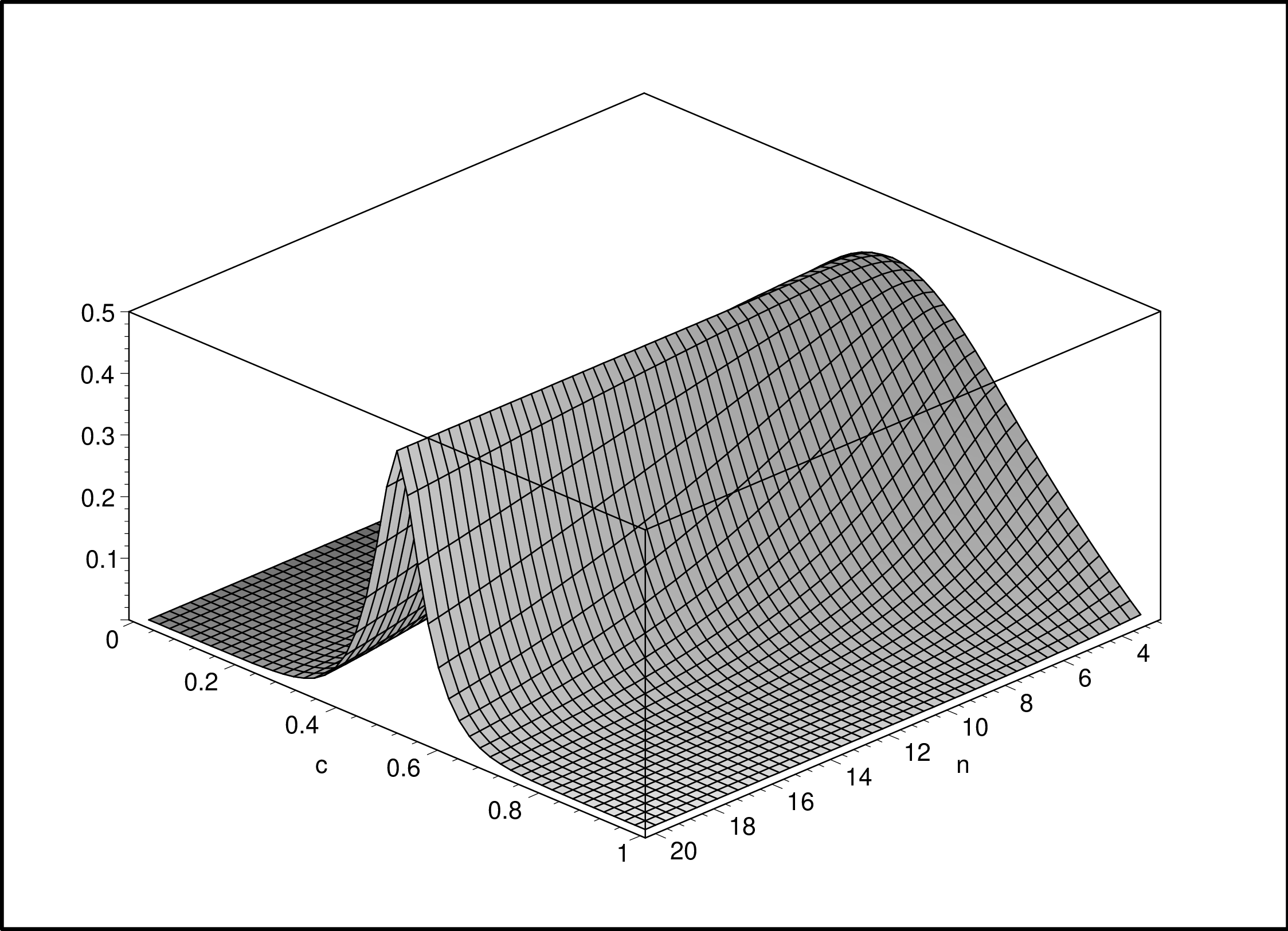}}}
\psfrag{p(c,n)}[cc][][0.8][90]{\put(-25,-15){\huge{$p_u(2,c,n)$}}}
\psfrag{n=5}{\huge{$n=5$}}
\psfrag{n=10}{\huge{$n=10$}}
\psfrag{n=20}{\huge{$n=20$}}
\rotatebox{0}{ \resizebox{2.5 in}{!}{ \includegraphics{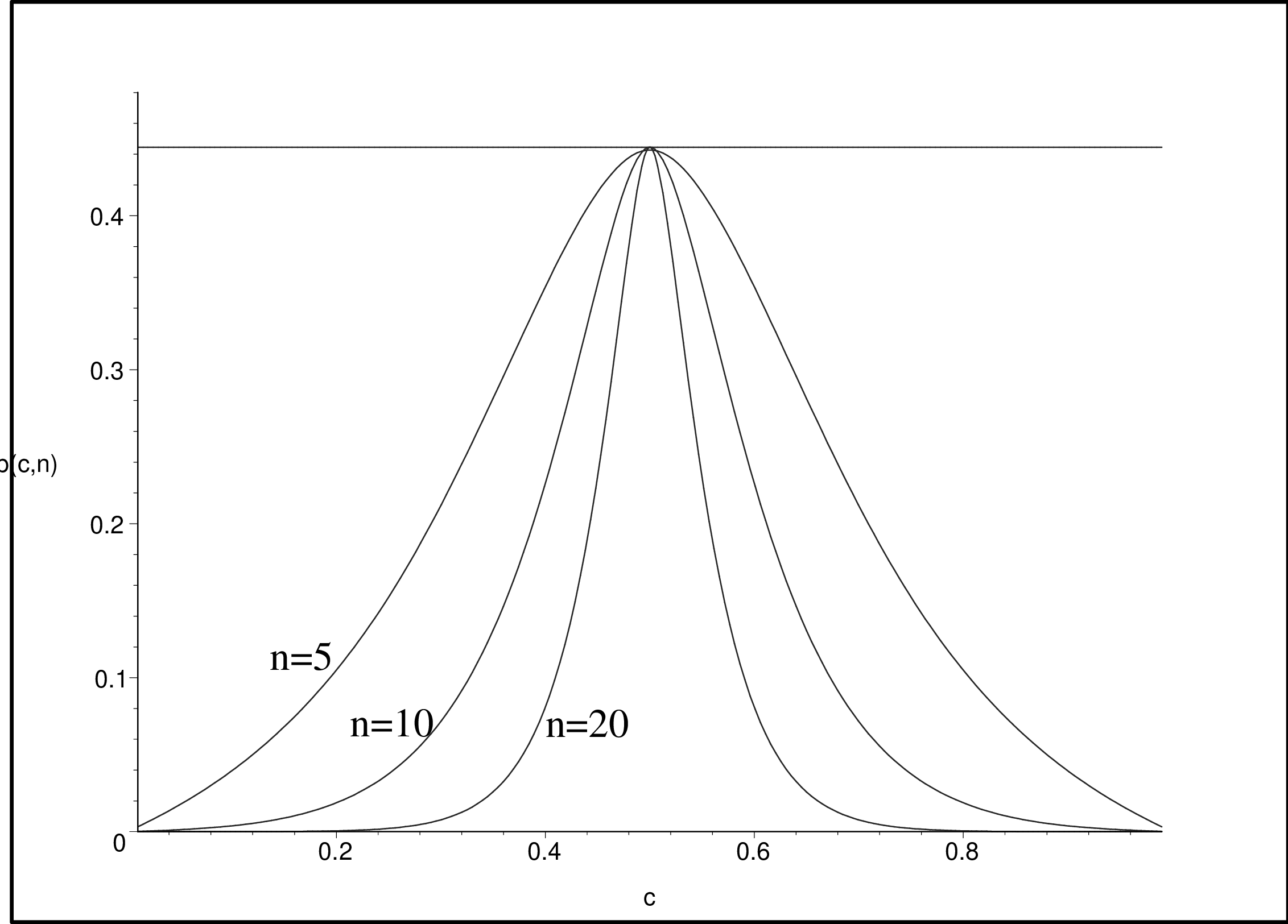}}}
\end{center}
\caption{
\label{fig:Prgam=2-withc}
Three-dimensional surface plot of $p_u(2,c,n)$ for $3 \le n \le 20$ and $c \in (0,1)$ (left) and
two-dimensional plots of $p_u(2,c,n)$ for $n=5,10,20$ and $c \in (0,1)$ (right).
In the surface plot, although $p_u(2,c,n)$ is defined for integer $n$ values,
we plot it as a continuous surface for better visualization.
In the right, the horizontal line is at $4/9$,
which is the limit of $p_u(2,1/2,n)$ as $n \to \infty$.
}
\end{figure}

We present the three-dimensional surface plot of $p_u(2,c,n)$
for $3 \le n \le 20$ and $c \in (0,1)$ in Figure \ref{fig:Prgam=2-withc} (left)
and the two-dimensional plots of $p_u(2,c,n)$
for $n=5,10,20$ and $c \in (0,1)$ in Figure \ref{fig:Prgam=2-withc} (right).
Observe that for finite $n \ge 1$,
the probability $p_u(2,c,n)$ is continuous in $c \in [0,1]$.
For fixed $n$,
$p_u(2,c,n)$ is increasing as $c$ approaches to 1/2.
For $c=1/2$,
we have $p_u(2,c,n)=4/9-(16/9) \, 4^{-n}$,
hence the distribution of $\g_{{}_{n,2}}(\U,2,c=1/2)$ is same as in \cite{priebe:2001}.

In the limit
as $n \rightarrow \infty$,
for $c \in [0,1]$,
we have
\begin{equation*}
\g_{{}_{n,2}}(\U,2,c)-1 \stackrel{\mathcal L}{\to}
\left\lbrace \begin{array}{ll}
       \BER(4/9), & \text{for $c = 1/2$,}\\
       0,           & \text{for $c \not= 1/2$.}\\
\end{array} \right.
\end{equation*}
Observe the interesting behavior of the asymptotic
distribution of $\g_{{}_{n,2}}(\U,2,c)$ around $c=1/2$.
The probability $p(\U,2,c)$ is continuous in $c \in [0,1] \setminus \{1/2\}$
(in fact it is constant),
but there is a jump in $p(\U,2,c)$ at $c=1/2$,
since $p_u(2,1/2)=4/9$
and $p(\U,2,c)=0$ for $c \not= 1/2$.
Hence
the asymptotic distribution is non-degenerate for $c = 1/2$,
and degenerate for $c \not= 1/2$.

\subsubsection{Special Case III: Exact Distribution of $\g_{n,2}(\U,2,1/2)$}
\label{sec:r=2-and-M_c=1/2}
For $r=2$ and $c=1/2$,
we have
$N(x,2,1/2)=B(x,r(x))$ where $r(x)=\min(x,1-x)$ for $x \in (0,1)$.
Hence the PICD based on $N(x,2,1/2)$
is equivalent to the CCCD of \cite{priebe:2001}.
Moreover, $\G_1(\X_n,2,1/2)=\left(X_{(n)}/2,\left(1+X_{(1)}\right)/2\right)$.
It has been shown that $p_u(2,1/2,n)=4/9-(16/9) \, 4^{-n}$
(\cite{priebe:2001}).
Hence, for $\U(y_1,y_2)$ data with $n \ge 1$, we have
\begin{equation}
\label{eqn:finite-sample-unif}
\g_{{}_{n,2}}(\U,2,1/2)=  \left\lbrace \begin{array}{ll}
       1           & \text{w.p. $5/9+(16/9) \, 4^{-n}, $}\\
       2           & \text{w.p. $4/9-(16/9) \, 4^{-n}, $}
\end{array} \right.
\end{equation}
where w.p. stands for ``with probability".
Then as $n \rightarrow \infty$,
$\g_{{}_{n,2}}(\U,2,1/2)-1$ converges in distribution to $\BER(4/9)$.
For $m>2$, \cite{priebe:2001} computed the exact distribution of $\g_{{}_{n,m}}(\U,2,1/2)$ also.
However,
the scale invariance property does not hold for general $F$;
that is, for $X_i \stackrel{iid}{\sim}F$ with support $\mS(F) \subseteq(y_1,y_2)$,
the exact and asymptotic distribution of $\g_{{}_{n,2}}(F,2,1/2)$
depends on $F$ and $\Y_2$ (see \cite{ceyhan:dom-num-CCCD-NonUnif}).

\subsection{Proof of Theorem \ref{thm:r and M-asy} }
Let $c \in (0,1/2]$.
Then $\tau=1/(1-c)$.
We first consider $c \in (0,1/2)$ with the following cases:\\
\noindent \textbf{Case (A)} $c \in \big[\left(3-\sqrt{5}\right)/2,1/2\big)$:\\
In Theorem \ref{thm:r and M},
for $r \ge 1/c > 2$,
it follows that
$\lim_{n \rightarrow \infty}p_u(r,c,n)=\lim_{n \rightarrow \infty}\pi_{a,1}(r,c,n)=0$,
since $2/r <1 $ and $\frac{r-1}{r^2}<1$.

For $1/(1-c) < r < 1/c$,
we have
$\frac{1+c\,r}{r}<1$ (since $r>1/(1-c)$),
$\frac{1-c}{r}<1$,
$\frac{(c\,r^2-r+c\,r+1)}{r}<1$,
$\frac{r-1}{r^2}<1$ (since $r-1<r<r^2$),
and
$\frac{(r-1)(c\,r-1+c)}{r}<1$.
Hence
for $1/(1-c) < r < 1/c$,
$\lim_{n \rightarrow \infty}p_u(r,c,n)=\lim_{n \rightarrow \infty}\pi_{a,2}(r,c,n)=0$

For $(1-c)/c < r < 1/(1-c)$,
we have
$r-1<1$ (since $r<1/(1-c)<2$),
$\frac{(r-1)(c\,r-1+c)}{r}<1$,
$\frac{(r-1)(r-c\,r-c)}{r}<1$,
$c/r < 1$ (since $c<r$),
$(1-c)/r < 1$ (since $1-c<1<r$),
$c\,r<1$ and $(1-c)r<1$ (since $r<1/(1-c)<1/c$).
Hence
$\lim_{n \rightarrow \infty}p_u(r,c,n)=
\lim_{n \rightarrow \infty}p_{{}_{a,3}}(r,c,n)=
\frac{r^3+r^2-r-1}{(r-1)(r+1)^2}=1$.

For $1 \le r < (1-c)/c$,
we have
$r-1<1$ (since $r<1/(1-c)<2$),
$(r-1)(1-c\,r-c)<1$,
$\frac{(r-1)(1-c\,r-c)}{r}<1$,
$\frac{(r-1)(r-c\,r-c)}{r}<1$,
$c/r < 1$ (since $c<r$),
$(1-c)/r < 1$,
$c\,r<1$ and $(1-c)r<1$.
Hence
$\lim_{n \rightarrow \infty}p_u(r,c,n)=
\lim_{n \rightarrow \infty}\pi_{a,4}(r,c,n)=
\frac{r^3+r^2-r-1}{(r-1)(r+1)^2}=1$.

But for $r=1/(1-c)$,
we have
\begin{multline}
p_u(1/(1-c),c,n)=
- \left(
\left( {\frac {\left(c \left( {c}^{2}-3\,c+1 \right)\right)^n }{(c-1)^{n-2}}} \right) +
\left(\left( 1-c \right)^{2n} -1\right) \left(2c -c^2\right)+
\left( {\frac {c^{n+1}}{(1-c)^{n-1}}} \right) ^{n}+
\left( c \left( 1-c \right)  \right) ^{n}\right) \Big /\\
\left({c} \left( c-2 \right)^2\right).
\end{multline}
Letting $n \rightarrow \infty$,
we get $p_u(1/(1-c),c,n) \rightarrow \frac{2c-c^2}{c(2-c)^2}= 1/(2-c)$ for $c \in (0,1/2)$,
since
$\frac{c}{(1-c)}<1$ (as $c<1/2$),
$0<1-c<1$,
$0<c(1-c)<1$,
and
$-1<\frac{c(c^2-3c+1)}{c-1}<1$.

\noindent \textbf{Case (B)} $c \in \big[1/4,\left(3-\sqrt{5}\right)/2\big)$:\\
In Theorem \ref{thm:r and M},
for $r \ge 1/c > 2$,
it follows that
$\lim_{n \rightarrow \infty}p_u(r,c,n)=\lim_{n \rightarrow \infty}\pi_{b,1}(r,c,n)=0$,
since $2/r <1 $ and $\frac{r-1}{r^2}<1$.

For $(1-c)/c < r < 1/c$,
we have
$\frac{1+c\,r}{r}<1$ (since $r>1/(1-c)$),
$\frac{1-c}{r}<1$,
$\frac{(c\,r^2-r+c\,r+1)}{r}<1$,
$\frac{r-1}{r}<1$ (since $r-1<r<r^2$),
and
$\frac{(r-1)(c\,r-1+c)}{r}<1$.
Hence
$\lim_{n \rightarrow \infty}p_u(r,c,n)=\lim_{n \rightarrow \infty}\pi_{b,2}(r,c,n)=0$

For $1/(1-c) < r < (1-c)/c$,
we have
$(r-1)(1-c\,r-c)<1$,
$\frac{(r-1)(1-c\,r-c)}{r}<1$,
$\frac{r-1}{r^2}<1$,
$(1+cr)/r < 1$,
$\frac{c\,r^2-c+c\,r+1}{r}<1$,
$(1-c)/r < 1$ (since $1-c<r$),
$c\,r<1$ and $(1-c)r<1$ (since $r<1/(1-c)<1/c$).
Hence
$\lim_{n \rightarrow \infty}p_u(r,c,n)=\lim_{n \rightarrow \infty}\pi_{b,3}(r,c,n)=0$.

For $1 \le r < 1/(1-c)$,
we have
$r-1<1$ (since $r<1/(1-c)<2$),
$(r-1)(1-c\,r-c)<1$,
$\frac{(r-1)(1-c\,r-c)}{r}<1$,
$c/r < 1$ (since $c<r$),
$(1-c)/r < 1$,
$c\,r<1$ and $(1-c)r<1$.
Hence
$\lim_{n \rightarrow \infty}p_u(r,c,n)=
\lim_{n \rightarrow \infty}\pi_{b,4}(r,c,n)=
\frac{r^3+r^2-r-1}{(r-1)(r+1)^2}=1$.

But for $r=1/(1-c)$,
we have
\begin{multline}
p_u(1/(1-c),c,n)=
\left(
\frac{(c(c^2-3c+1))^n}{(1-c)^{2n-3}}+
(1-(1-c)^{2n})(2c-c^2)-
\frac{c^{n+1}}{(1-c)^{n-1}}-
(c(1-c))^n \right) \Big / \left(c \left( 2-c \right) ^{2}\right).
\end{multline}

Letting $n \rightarrow \infty$,
we get $p_u(1/(1-c),c,n) \rightarrow \frac{2c-c^2}{c(2-c)^2}= 1/(2-c)$ for $c \in (0,1/2)$,
since
$\frac{c}{(1-c)}<1$ (since $c<1/2$),
$0<1-c<1$,
$0<c(1-c)<1$,
and
$-1<\frac{c(c^2-3c+1)}{(c-1)^2}<1$.

\noindent \textbf{Case (C)} $c \in (0,1/4)$:\\
In Theorem \ref{thm:r and M},
for $r \ge 1/c > 2$,
we have
$\lim_{n \rightarrow \infty}p_u(r,c,n)=\lim_{n \rightarrow \infty}p_{c,1}(r,c,n)=0$,
since $2/r <1 $ and $\frac{r-1}{r^2}<1$.

For $(1-c)/c < r < 1/c$,
we have
$\frac{1+c\,r}{r}<1$ (since $r>1/(1-c)$),
$\frac{1-c}{r^2}<1$,
$\frac{1-c}{r}<1$ (since $1-c < r$),
$\frac{(c\,r^2-r+c\,r+1)}{r}<1$,
$\frac{r-1}{r}<1$ (since $r-1<r<r^2$),
$\frac{(r-1)(c\,r-1+c)}{r}<1$,
and
$(r-1)(c\,r-1+c)<1$.
Hence
$\lim_{n \rightarrow \infty}p_u(r,c,n)=\lim_{n \rightarrow \infty}p_{c,2}(r,c,n)=0$

For $\left(1+\sqrt{1-4 c}\right)/(2c) < r < (1-c)/c$,
we have
$\frac{r-1}{r^2}<1$,
$(r-1)(1-c\,r-c)<1$,
$(1+cr)/r < 1$,
$\frac{c\,r^2-c+c\,r+1}{r}<1$,
and
$(1-c)/r < 1$ (since $1-c<r$).
Hence
$\lim_{n \rightarrow \infty}p_u(r,c,n)=\lim_{n \rightarrow \infty}p_{c,3}(r,c,n)=0$.

For $\left(1-\sqrt{1-4 c}\right)/(2c) < r < \left(1+\sqrt{1-4 c}\right)/(2c)$,
we have
$c<1$,
$(1+cr)/r < 1$,
and
$(1-c)/r < 1$ (since $1-c<r$).
Hence
$\lim_{n \rightarrow \infty}p_u(r,c,n)=\lim_{n \rightarrow \infty}p_{c,4}(r,c,n)=0$.

For $1/(1-c) < r < \left(1-\sqrt{1-4 c}\right)/(2c)$,
we have
$\frac{r-1}{r^2}<1$,
$(r-1)(1-c\,r-c)<1$,
$(1+cr)/r < 1$,
$\frac{c\,r^2-c+c\,r+1}{r}<1$,
$\frac{(r-1)(1-c\,r-c)}{r}<1$,
and
$(1-c)/r < 1$ (since $1-c<r$).
Hence
$\lim_{n \rightarrow \infty}p_u(r,c,n)=\lim_{n \rightarrow \infty}p_{c,5}(r,c,n)=0$.

For $1 \le r < 1/(1-c)$,
we have
$r-1<1$ (since $r<1/(1-c)<2$),
$(r-1)(1-c\,r-c)<1$,
$\frac{(r-1)(1-c\,r-c)}{r}<1$,
$c/r < 1$ (since $c<r$),
and
$c\,r<1$ and $(1-c)r<1$.
Hence
$\lim_{n \rightarrow \infty}p_u(r,c,n)=
\lim_{n \rightarrow \infty}p_{c,6}(r,c,n)=
\frac{r^3+r^2-r-1}{(r-1)(r+1)^2}=1$.

But for $r=1/(1-c)$,
we have
\begin{multline}
p_u(1/(1-c),c,n)=
\left(
\frac{(c(c^2-3c+1))^n}{(1-c)^{2n-3}}+
(1-(1-c)^{2n})(2c-c^2)-
\frac{c^{n+1}}{(1-c)^{n-1}}-
(c(1-c))^n \right) \Big / \left(c \left( 2-c \right) ^{2}\right).
\end{multline}

Letting $n \rightarrow \infty$,
we get $p_u(1/(1-c),c,n) \rightarrow \frac{2c-c^2}{c(2-c)^2}= 1/(2-c)$ for $c \in (0,1/2)$,
since
$\frac{c}{(1-c)}<1$ (since $c<1/2$),
$0<1-c<1$,
$0<c(1-c)<1$,
and
$-1<\frac{c(c^2-3c+1)}{(c-1)^2}<1$.

For $c \in (1/2,1)$, we have $\tau=1/c$.
By symmetry, the above results follow with $c$ being replaced by $1-c$
and
as $n \rightarrow \infty$,
we get $p_u(1/c,c,n) \rightarrow 1/(c+1)$.
Hence the desired result follows.

Furthermore,
the result for $c=1/2$ can be derived similarly
by substituting $c=1/2$ in the expressions in Theorem \ref{thm:r and M}
and letting $n$ tend to infinity.
$\blacksquare$

\section{Supplementary Materials for Section \ref{sec:non-uniform} }
\subsection{Proof of Proposition \ref{prop:NF vs NPE} }

Let $U_i:=F(X_i)$ for $i=1,2,\ldots,n$ and $\U_n:=\{U_1,U_2,\ldots,U_n\}$.
Hence, by probability integral transform, $U_i \stackrel{iid}{\sim} \U(0,1)$.
Let $U_{(i)}$ be the $i^{th}$ order statistic of $\U_n$ for $i=1,2,\ldots,n$.
So the image of $N_F(x,r,c)$ under $F$ is
$F(N_F(x,r,c))=N(F(x),r,c)$ for (almost) all $x \in (0,1)$.
Then $F(N_F(X_i,r,c))=N(F(X_i),r,c)=N(U_i,r,c)$ for $i =1,2,\ldots,n$.
Since $U_i \stackrel{iid}{\sim} \U(0,1)$, the distribution of
the domination number of the digraph based on $N(\cdot,r,c)$, $\U_n$, and $\{0,1\}$
is given in Theorem \ref{thm:r and M}.
Observe that for any $j$,
$X_j \in N_F(X_i,r,c)$ iff
$X_j \in F^{-1}(N(F(X_i),r,c))$ iff
$F(X_j) \in N(F(X_i),r,c)$ iff
$U_j \in N(U_i,r,c)$ for $i=1,2,\ldots,n$.
Hence $P(\X_n \subset N_F(X_i,r,c))=P(\U_n \subset N(U_i,r,c))$
for all $i=1,2,\ldots,n$.
Furthermore,
an absolutely continuous $F$ preserves order;
that is,
for $x < y$ with $x,y \in \mathcal S(F)$,
we have $F(x) < F(y)$.
So it follows that $U_{(i)}=F\left(X_{(i)}\right)$.
Therefore, $\X_n \cap \G_1(\X_n,N_F(r,c))=\emptyset$
iff $\U_n \cap \G_1(\U_n,r,c)=\emptyset$.
Hence the desired result follows.
$\blacksquare$

\subsection{Stochastic Ordering between $\g_{{}_{n,2}}(F,r,c)$ and $\g_{{}_{n,2}}(\U,r,c)$ }
\label{sec:stochastic-order}

\begin{proposition}
\label{prop:stoch-order}
Let $\X_n=\{X_1,X_2,\ldots,X_n\}$ be a random sample from
an absolutely continuous distribution $F$ with $\mS(F)\subseteq(0,1)$.
If
\begin{equation}
\label{eqn:stoch-order}
F\bigl( X_{(n)}/r \bigr)< F\left(X_{(n)}\right)/r \text{ and }
F\bigl( X_{(1)} \bigr) < r\,F\left( \left( X_{(1)}+r-1 \right)/r \right)+1-r \text{ hold a.s., }
\end{equation}
then $\g_{{}_{n,2}}(F,r,c)<^{st}\g_{{}_{n,2}}(\U,r,F(c))$
where $<^{st}$ stands for ``stochastically smaller than".
If $<$'s in \eqref{eqn:stoch-order} are replaced with $>$'s,
then $\g_{{}_{n,2}}(\U,r,F(c))<^{st}\g_{{}_{n,2}}(F,r,c)$.
If $<$'s in \eqref{eqn:stoch-order} are replaced with $=$'s,
then $\g_{{}_{n,2}}(F,r,c)\stackrel{d}{=}\g_{{}_{n,2}}(\U,r,F(c))$ where $\stackrel{d}{=}$ stands
for equality in distribution.
\end{proposition}

\noindent {\bfseries Proof:}
Let $\U_n$, $U_i$ and
$U_{(i)}$ be as in Proof of Proposition \ref{prop:NF vs NPE}.
Also,
$F(N(X,r,c))=N(F(X),r,F(c))=N(U,r,F(c))$.
Hence the parameter $c$ for $N(\cdot,r,c)$ with $\X_n$ in $(0,1)$
corresponds to $F(c)$ for $\U_n$.
Then the $\G_1$-region for $\U_n$ based on $N(\cdot,r,F(c))$ is
$\G_1(\U_n,r,F(c))=(U_{(n)}/r,F(c) ] \cup [F(c),\left(U_{(1)}+r-1\right)/r )$;
likewise, $\G_1(\X_n,r,c)=(X_{(n)}/r,c ] \cup [c,\left(X_{(1)}+r-1\right)/r )$.
So,
$F(\G_1(\X_n,r,c))=(F(X_{(n)}/r),F(c) ] \cup [F(c),F\left(\left(X_{(1)}+r-1\right)/r\right) )$.
So the conditions in \eqref{eqn:stoch-order} imply that
$\G_1(\U_n,r,F(c)) \subsetneq F(\G_1(\X_n,r,c))$,
since such an $F$ preserves order.
So $\U_n \cap F(\G_1(\X_n,r,c)) = \emptyset$ implies that
$\U_n \cap \G_1(\U_n,r,F(c)) = \emptyset$ and
$\U_n \cap F(\G_1(\X_n,r,c)) = \emptyset$ iff
$\X_n \cap \G_1(\X_n,r,c) = \emptyset$.
Hence
$$p_{{}_n}(F,r,c)=P(\X_n \cap \G_1(\X_n,r,c) = \emptyset)<
P(\U_n \cap \G_1(\U_n,r,F(c)) = \emptyset)=p_{{}_n}(\U,r,F(c)).$$
Then $\g_{{}_{n,2}}(F,r,c)<^{st}\g_{{}_{n,2}}(\U,r,F(c))$ follows.
The other cases follow similarly.
$\blacksquare$

\begin{remark}
\label{rem:exact-dist-gam}
We can also find the exact distribution of $\g_{{}_{n,2}}(F,r,c)$
for $F$ whose pdf is piecewise constant
with support in $(0,1)$.
Note that the simplest of such distributions is the uniform distribution $\U(0,1)$.
The exact distribution of $\g_{{}_{n,2}}(F,r,c)$ for (piecewise) polynomial
$f(x)$ with at least one piece is of degree 1 or higher and support in $(0,1)$
can be obtained using the multinomial expansion of the term $(\cdot)^{n-2}$
in Equation \eqref{eqn:integrand} with careful bookkeeping.
However, the resulting expression for $p_{{}_n}(F,r,c)$ is extremely lengthy
and not so informative.

Furthermore,
for fixed $n$, one can obtain $p_{{}_n}(F,r,c)$ for $F$
(omitted for the sake of brevity)
by numerical integration of the below expression:
\begin{eqnarray*}
p_{{}_n}(F,r,c)=P\bigl( \g_{{}_{n,2}}(F,r,c)=2 \bigr)&=&
\int\int_{\mS(F)\setminus(\delta_1,\delta_2)}H(x_1,x_n)\,dx_n dx_1,
\end{eqnarray*}
where $H(x_1,x_n)$ is given in Equation $\eqref{eqn:integrand}$. $\square$
\end{remark}

\subsection{Proof of Theorem \ref{thm:kth-order-gen-r,c} }
The asymptotic distribution of  $\g_{{}_{n,2}}(F,r,c)$ for $r=2$ and $c=1/2$
is as follows (see \cite{ceyhan:dom-num-CCCD-NonUnif}).
\begin{theorem}
\label{thm:kth-order-gen}
Let
$\F\bigl(y_1,y_2\bigr):=\Bigl \{F:
(y_1,y_1+\ve) \cup (y_2-\ve,y_2)\cup \bigl( (y_1+y_2)/2-\ve,(y_1+y_2)/2+\ve \bigr)
\subseteq \mS(F)
\subseteq(y_1,y_2) \text{ for some } \ve \in (0,(y_1+y_2)/2) \Bigr\}.
$
Let $\Y_2=\{y_1,y_2\} \subset \R$ with $-\infty < y_1 < y_2<\infty$,
$\X_n=\{X_1,\ldots,X_n\}$ with $X_i \stackrel {iid}{\sim} F \in \F(y_1,y_2)$,
and $D_{n,2}$ be the random $\D_{n,2}$-digraph based on $\X_n$ and $\Y_2$.
\begin{itemize}
\item[(i)]
Then for $n>1$, we have
$\g_{{}_{n,2}}(F,2,1/2)-1 \sim \BER\bigl(p_{{}_n}(F,2,1/2)\bigr)$.
Note also that $\g_{{}_{1,2}}(F,2,1/2)=1$.

\item[(ii)]
Furthermore,
suppose $k \ge 0$ is the smallest integer for which
$F(\cdot)$ has continuous right derivatives up to order $(k+1)$ at $y_1,\,(y_1+y_2)/2$,
$f^{(k)}(y_1^+)+2^{-(k+1)}\,f^{(k)}\left( \left( \frac{y_1+y_2}{2} \right)^+ \right) \not= 0$
and $f^{(i)}(y_1^+)=f^{(i)} \left( \left( \frac{y_1+y_2}{2} \right)^+ \right)=0$ for all $i=0,1,\ldots,k-1$;
and $\ell \ge 0$ is the smallest integer for which
$F(\cdot)$ has continuous left derivatives up to order $(\ell+1)$ at $y_2,\,(y_1+y_2)/2$,
$f^{(\ell)}(y_2^-)+2^{-(\ell+1)}\,f^{(\ell)}\left( \left( \frac{y_1+y_2}{2} \right)^- \right) \not= 0$
and $f^{(i)}(y_2^-)=f^{(i)}\left( \left( \frac{y_1+y_2}{2} \right)^- \right)=0$ for all $i=0,1,\ldots,\ell-1$.
Then for bounded $f^{(k)}(\cdot)$ and $f^{(\ell)}(\cdot)$,
we have the following limit
$$
p(F,2,1/2) =
\lim_{n \rightarrow \infty}p_{{}_n}(F,2,1/2) =
\frac{f^{(k)}(y_1^+)\,f^{(\ell)}(y_2^-)}
{\left[f^{(k)}(y_1^+)+2^{-(k+1)}\,f^{(k)} \left( \left( \frac{y_1+y_2}{2} \right)^+ \right)\right]\,\left[f^{(\ell)}(y_2^-)+
2^{-(\ell+1)}\,f^{(\ell)}\left( \left( \frac{y_1+y_2}{2} \right)^- \right)\right]}.$$
\end{itemize}
\end{theorem}

\noindent {\bf Proof:}
Case (i) follows trivially from Lemma \ref{lem:gamma 1 or 2}.
Also,
the special cases for $n=1$ and $r\in\{1,\infty\}$ follow by construction.

\noindent
Case (ii):
Suppose $(y_1,y_2)=(0,1)$ and $c \in (0,1/2)$ and $r=\tau=1/(1-c)$.
Notice that
$\G_1(\X_n,1/(1-c),c)=\big((1-c) X_{(n)}, c \big] \bigcup \big[c, (1-c) X_{(1)}+c \big ) \subset (0,1)$
and $\g_{{}_{n,2}}(F,1/(1-c),c)=2 \text{ iff } \X_n \cap \G_1(\X_n,1/(1-c),c)=\emptyset$.
Then for finite $n$,
\begin{equation*}
p_{{}_n}(F,1/(1-c),c)=P\bigl( \g_{{}_{n,2}}(F,1/(1-c),c)=2 \bigr)=
\int_{\mS(F) \setminus (\delta_1,\delta_2)} H(x_1,x_n)\,dx_n dx_1,
\end{equation*}
where $(\delta_1,\delta_2)=\G_1(\X_n,1/(1-c),c)$ and $H(x_1,x_n)$ is as in Equation \eqref{eqn:integrand} of the main text.

Let $\ve \in (0,c)$.
Then $P\bigl( X_{(1)} <\ve,\; X_{(n)} > 1-\ve \bigr)\rightarrow 1$
as $n \rightarrow \infty$ with the rate of convergence depending on $F$ and $\ve$.
Moreover, for sufficiently large $n$,
$(1-c)X_{(1)}+c > c$ a.s.;
in fact, $(1-c)X_{(1)}+c \downarrow c$ as $n \rightarrow \infty$ (in probability)
and $(1-c)X_{(n)} > c$ a.s. since $c \in (0,1/2)$.
Then for sufficiently large $n$,
we have $\G_1(\X_n,1/(1-c),c)=[c,(1-c)X_{(1)}+c)$ a.s.
and
\begin{multline}
\label{eqn:asy-g=2-x_1x_n}
p_{{}_n}(F,1/(1-c),c) \approx \int_0^{\ve}\int_{1-\ve}^1 n\,(n-1)f(x_1)f(x_n)
\Bigl[F(x_n)-F(x_1)+F\left(c \right)-F\left( (1-c)x_1+c \right) \Bigr]^{n-2}\,dx_n dx_1\\
=\int_0^{\ve} nf(x_1)
\Biggl(\Bigl[1-F(x_1)+F\left(c \right)-F\left( (1-c)x_1+c \right) \Bigr]^{n-1}-\\
\Bigl[1-\varepsilon-F(x_1)+F\left(c \right)-F\left( (1-c)x_1+c \right) \Bigr]^{n-1}\Biggr)\,dx_1\\
\approx
\int_0^{\ve} nf(x_1)
\Bigl[1-F(x_1)+F\left(c \right)-F\left( (1-c)x_1+c \right) \Bigr]^{n-1}\,dx_1.
\end{multline}
 Let $G(x_1)=1-F(x_1)+F\left(c \right)-F\left( (1-c)x_1+c \right).$
The integral in Equation \eqref{eqn:asy-g=2-x_1x_n} is critical at $x_1=0$,
since $G(0)=1$,
and for $x_1 \in (0,1)$ the
integral converges to 0 as $n \rightarrow \infty$.
Let $\al_i := -\frac{d^{i+1} G(x_1)}{d x_1^{i+1}}\Big|_{(0^+,0^+)}=
f^{(i)}(0^+)+(1-c)^{(i+1)}\,f^{(i)}\left(c^+\right)$.
Then by the hypothesis of the theorem, we have $\al_i = 0$ and
$f^{(i)}\left(c^+\right)=0$ for all $i=0,1,2,\ldots,(k-1)$.
So the Taylor series expansions of $f(x_1)$ around $x_1=0^+$ up to order $k$
and $G(x_1)$ around $0^+$ up to order $(k+1)$
so that $x_1 \in (0,\ve)$ are as follows:
$$f(x_1)=\frac{1}{k!}f^{(k)}(0^+)\,x_1^k+O\left( x_1^{k+1} \right)$$
and
$$G(x_1)=G(0^+)+ \frac{1}{(k+1)!}\left(\frac{d^{k+1}G(0^+)}{d x_1^{k+1}}\right)\,x_1^{k+1}
+O\left( x_1^{k+2} \right) \\
=1-\frac{\al_k}{(k+1)!}\,x_1^{k+1}+O\left(x_1^{k+2}\right).$$
Then substituting these expansions in Equation \eqref{eqn:asy-g=2-x_1x_n},
we obtain
$$
p_{{}_n}(F,1/(1-c),c) \approx \int_0^{\ve}
n\Biggl[\frac{1}{k!}f^{(k)}(0^+)\,x_1^k+O\left(x_1^{k+1}\right)\Biggr]
\Biggl[1-\frac{\al_k}{(k+1)!}\,x_1^{k+1}+O\left(x_1^{k+2}\right)\Biggr]^{n-1}\,d x_1.
$$

Now we let $x_1=w\,n^{-1/(k+1)}$ and get
\begin{multline}
\label{eqn:Pg2-in-(0,1)-(r-1)/r}
p_{{}_n}(F,1/(1-c),c) \approx
\int_0^{\ve\,n^{1/(k+1)}}n\,
\Biggl[\frac{1}{n^{k/(k+1)}\,k!}f^{(k)}(0^+)w^k+O\left(n^{-1}\right)\Biggr]\\
\Biggl[1-\frac{1}{n}\left(\frac{\al_k}{(k+1)!}\,w^{k+1}+
O\left(n^{-(k+2)/(k+1)}\right)\right)\Biggr]^{n-1}\,\left(\frac{1}{n^{1/(k+1)}}\right)\,dw\\
\text{letting $n \rightarrow
\infty,$~~~~~~~~~~~~~~~~~~~~~~~~~~~~~~~~~~~~~~~~~~~~~~~~~~~~~~~~
~~~~~~~~~~~~~~~~~~~~~~~~~~~~~~~~~~~~~~~~~~~~~~~~~~~~}\\
\approx
\int_0^{\infty} \frac{1}{k!} f^{(k)}(0^+)w^k \,
\exp\left[-\frac{\al_k}{(k+1)!}\,w^{k+1}\right]\, dw
= \frac{f^{(k)}(0^+)}{\al_k}
= \frac{f^{(k)}(0^+)}
{f^{(k)}(0^+)+(1-c)^{(k+1)}\,f^{(k)} \left( c^+ \right)},
\end{multline}
as $n \rightarrow \infty$ at rate $O(\kappa_1(f)\cdot n^{-(k+2)/(k+1)})$.

\noindent
Case (iii):
Suppose $(y_1,y_2)=(0,1)$ and $c \in (1/2,1)$ and $r=\tau=1/c$.
Then
$\G_1(\X_n,1/c,c)=\big(c X_{(n)}, c \big] \bigcup \big[c, c X_{(1)}+1-c \big ) \subset (0,1)$
Let $\ve \in (0,c)$.
Then $P\bigl( X_{(1)} <\ve,\; X_{(n)} > 1-\ve \bigr)\rightarrow 1$
as $n \rightarrow \infty$ with the rate of convergence depending on $F$.
Moreover, for sufficiently large $n$,
$c X_{(n)} < c$ a.s.;
in fact, $c X_{(n)} \uparrow c$ as $n \rightarrow \infty$ (in probability)
and $c X_{(1)}+1-c < c$ a.s.
Then for sufficiently large $n$, $\G_1(\X_n,1/c,c)=(c X_{(n)},c]$
a.s.
and
\begin{multline}
\label{eqn:asy-g=2-x_1x_n-1/r}
p_{{}_n}(F,1/c,c) \approx \int_{1-\ve}^1 \int_0^{\ve} n\,(n-1)f(x_1)f(x_n)
\Bigl[F(x_n)-F(x_1)+F\left(c x_n \right)-F\left( c \right) \Bigr]^{n-2}\,dx_1dx_n.\\
=-\int_{1-\ve}^1 n f(x_n)
\Biggl(\Bigl[F(x_n)-F(\varepsilon)+F\left(c x_n \right)-F\left( c \right) \Bigr]^{n-1}-
\Bigl[F(x_n)+F\left(c x_n \right)-F\left( c \right) \Bigr]^{n-1}\Biggr)\,dx_n\\
\approx
\int_{1-\ve}^1 n f(x_n)
\Bigl[F(x_n)+F\left(c x_n \right)-F\left( c \right) \Bigr]^{n-1}\,dx_n.
\end{multline}
 Let $G(x_n)=F(x_n)+F\left(c x_n \right)-F\left( c \right).$
The integral in Equation \eqref{eqn:asy-g=2-x_1x_n-1/r} is critical at $x_n=1$,
since $G(1)=1$,
and for $x_n \in (0,1)$ the
integral converges to 0 as $n \rightarrow \infty$.
So we make the change of variables $z_n=1-x_n$,
then $G(x_n)$ becomes
$$G(z_n)=F(1-z_n)+F(c (1-z_n))-F(c),$$
and Equation \eqref{eqn:asy-g=2-x_1x_n-1/r} becomes
\begin{equation}
\label{eqn:asy-g=2-zn}
p_{{}_n}(F,1/c,c) \approx \int_0^{\ve} n\,f(1-z_n)\left(G(z_n)\right)^{n-1}\,dz_n.
\end{equation}
The new integral is critical at $z_n=0$.
Let $\be_i := (-1)^{i+1}\frac{d^{i+1} G(z_n)}{d z_n^{i+1}}\Big|_{0^+}=
f^{(i)}(1^-)+c^{(i+1)}\,f^{(i)}\left(c^-\right)$.
Then by the hypothesis of the theorem, we have $\be_i = 0$ and
$f^{(i)}\left(c^-\right)=0$ for all $i=0,1,2,\ldots,(\ell-1)$.
So the Taylor series expansions of
$f(1-z_n)$ around $z_n=0^+$ up to $\ell$
and $G(z_n)$ around $0^+$ up to order $(\ell+1)$
so that $z_n \in (0,\ve)$, are as follows:
$$f(1-z_n)=\frac{(-1)^{\ell}}{\ell!}f^{(\ell)}(1^-)\,z_n^{\ell}+O\left(z_n^{\ell+1}\right)$$
$$G(z_n)=G(0^+)+ \frac{1}{(\ell+1)!}\left(\frac{d^{\ell+1}G(0^+)}{d z_n^{\ell+1}}\right)\,z_n^{\ell+1}+
+O\left( z_n^{\ell+2} \right) \\
=1+\frac{(-1)^{\ell+1}\be_{\ell}}{(\ell+1)!}\,z_n^{\ell+1}+O\left(z_n^{\ell+2}\right).$$
Then substituting these expansions in Equation \eqref{eqn:asy-g=2-zn},
we get
$$
p_{{}_n}(F,1/c,c) \approx \int_0^{\ve}
n\Biggl[\frac{(-1)^{\ell}}{\ell!}f^{(\ell)}(1^-)\,z_n^{\ell}+O\left(z_n^{\ell+1}\right)\Biggr]
\Biggl[1-\frac{(-1)^{\ell}\be_{\ell}}{(\ell+1)!}\,z_n^{\ell+1}+O\left(z_n^{\ell+2}\right)\Biggr]^{n-1}\,d z_n.
$$

Now we let $z_n=v\,n^{-1/(\ell+1)}$,
to obtain
\begin{multline}
\label{eqn:Pg2-in-(0,1)-1/r}
p_{{}_n}(F,1/c,c) \approx
\int_0^{\ve\,n^{1/(\ell+1)}}n\,
\Biggl[\frac{(-1)^{\ell}}{n^{\ell/(\ell+1)}\,\ell!}f^{(\ell)}(1^-)v^{\ell}+O\left(n^{-1}\right)\Biggr]\\
\Biggl[1-\frac{1}{n}\left(\frac{(-1)^{\ell}\be_{\ell}}{(\ell+1)!}\,v^{\ell+1}+
O\left(n^{-(\ell+2)/(\ell+1)}\right)\right)\Biggr]^{n-1}\,\left(\frac{1}{n^{1/(\ell+1)}}\right)\,dv\\
\text{letting $n \rightarrow
\infty,$~~~~~~~~~~~~~~~~~~~~~~~~~~~~~~~~~~~~~~~~~~~~~~~~~~~~~~~~
~~~~~~~~~~~~~~~~~~~~~~~~~~~~~~~~~~~~~~~~~~~~~~~~~~~~}\\
\approx
\int_0^{\infty} \frac{(-1)^{\ell}}{\ell!} f^{(\ell)}(1^-)v^{\ell} \,
\exp\left[-\frac{(-1)^{\ell}\be_{\ell}}{(\ell+1)!}\,v^{\ell+1}\right]\, dv
= \frac{f^{(\ell)}(1^-)}{\be_{\ell}}
= \frac{f^{(\ell)}(1^-)}
{f^{(\ell)}(1^-)+c^{(\ell+1)}\,f^{(\ell)} \left( c^- \right)}
\end{multline}
as $n \rightarrow \infty$ at rate $O(\kappa_2(f)\cdot n^{-(\ell+2)/(\ell+1)})$.

For the general case of $\Y=\{y_1,y_2\}$, the transformation
$\phi(x)=(x-y_1)/(y_2-y_1)$ maps $(y_1,y_2)$ to $(0,1)$ and
the transformed random variables $U=\phi(X_i)$ are distributed with
density $g(u)=(y_2-y_1)\,f(y_1+u(y_2-y_1))$ on $(y_1,y_2)$.
Replacing $f(x)$ by $g(x)$ in Equations \eqref{eqn:Pg2-in-(0,1)-(r-1)/r}
and \eqref{eqn:Pg2-in-(0,1)-1/r},
the desired result follows.
$\blacksquare$

Notice the interesting behavior of $p(F,r,c)$ around $(r,c)=(2,1/2)$.
There is a jump in $p(F,r^*,c)$ at $(r,c)=(2,1/2)$.

Note that in Theorem \ref{thm:kth-order-gen-r,c} (ii)
\begin{itemize}
\item
with $(y_1,y_2)=(0,1)$,
we have
$p(F,1/(1-c),c) =
\frac{f^{(k)}(0^+)}
{f^{(k)}(0^+)+(1-c)^{(k+1)}\,f^{(k)}\left( c^+ \right)}$,
\item
if $f^{(k)}(y_1^+)=0$ and
$ f^{(k)} \left( M_c^+ \right)\not=0$,
then $p_{{}_n}(F,1/(1-c),c)\rightarrow 0$ as $n \rightarrow \infty$
at rate $O\bigl( \kappa_1(f)\cdot n^{-(k+2)/(k+1)} \bigr)$
where $\kappa_1(f)$ is a constant depending on $f$
and
\item
if $f^{(k)}(y_1^+)\not=0$ and
$f^{(k)} \left( M_c^+ \right)=0$,
then $p_{{}_n}(F,1/(1-c),c) \rightarrow 1$
as $n \rightarrow \infty$ at rate $O\bigl(\kappa_1(f)\cdot n^{-(k+2)/(k+1)} \bigr)$.
\end{itemize}

Also in Theorem \ref{thm:kth-order-gen-r,c} (iii)
\begin{itemize}
\item
with $(y_1,y_2)=(0,1)$,
we have
$p(F,1/c,c) =
\frac{f^{(\ell)}(1^-)}
{f^{(\ell)}(1^-)+c^{(\ell+1)}\,f^{(\ell)}\left( c^- \right)}$,
\item
if $f^{(\ell)}(y_2^-)=0$ and
$f^{(\ell)} \left( M_c^- \right) \not=0$,
then $p_{{}_n}(F,1/c,c)\rightarrow 0$ as $n \rightarrow \infty$
at rate $O\bigl( \kappa_2(f)\cdot n^{-(\ell+2)/(\ell+1)} \bigr)$
where $\kappa_2(f)$ is a constant depending on $f$
and
\item
if $f^{(\ell)}(y_2^-)\not=0$ and
$f^{(\ell)} \left( M_c^- \right)=0$,
then $p_{{}_n}(F,1/c,c) \rightarrow 1$
as $n \rightarrow \infty$ at rate $O\bigl(\kappa_2(f)\cdot n^{-(\ell+2)/(\ell+1)} \bigr)$.
\end{itemize}

\begin{remark}
\label{rem:unbounded}
In Theorem \ref{thm:kth-order-gen-r,c} parts (ii) and (iii),
we assume that $f^{(k)}(\cdot)$ and $f^{(\ell)}(\cdot)$
are bounded on $(y_1,y_2)$, respectively.
In part (ii),
if $f^{(k)}(\cdot)$ is not bounded on $(y_1,y_2)$ for $k \ge 0$,
in particular at $y_1$, and $M_c$, for example,
$\lim_{x \rightarrow y_1^+}f^{(k)}(x)=\infty$,
then we have
$$p(F,1/(1-c),c) =\lim_{\delta \rightarrow 0^+}\frac{f^{(k)}(y_1+\delta)}
{\left[f^{(k)}(y_1+\delta)+(1-c)^{(k+1)}\,f^{(k)} \left( M_c+\delta \right)\right]}.$$
In part (iii),
if $f^{(\ell)}(\cdot)$ is not bounded on $(y_1,y_2)$ for $\ell \ge 0$,
in particular at $M_c$, and $y_2$, for example,
$\lim_{x \rightarrow y_2^-}f^{(\ell)}(x)=\infty$,
then we have
$$p(F,1/c,c) =\lim_{\delta \rightarrow 0^+}\frac{f^{(\ell)}(y_2-\delta)}
{\left[f^{(\ell)}(y_2-\delta)+c^{(\ell+1)}\,f^{(\ell)} \left( M_c-\delta \right)\right]}.\;\;\square$$
\end{remark}

\begin{remark}
\label{rem:rate-of-conv}
The rates of convergence in  Theorem \ref{thm:kth-order-gen-r,c} parts (ii) and (iii) depend on $f$.
From the proof of Theorem \ref{thm:kth-order-gen-r,c},
it follows that for sufficiently large $n$,
$$p_n(F,1/(1-c),c) \approx p(F,1/(1-c),c) +\frac{\kappa_1(f)}{n^{-(k+2)/(k+1)}}
\text{ and }
p_n(F,1/c,c) \approx p(F,1/c,c) +\frac{\kappa_2(f)}{n^{-(\ell+2)/(\ell+1)}},$$
where
$\kappa_1(f)=\frac{s_1\,s_3^{\frac{1}{k+1}}+s_2\,\G \left(\frac{k+2}{k+1} \right)}
{(k+1)\,s_3^{\frac{k+2}{k+1}} }$
with
$\G(x)=\int_{0}^{\infty} e^{-t}t^{(x-1)}\, dt$,
$s_1=\frac{1}{n^{k+1}k!}\,f^{(k)}(y_1^+)$,
$s_2=\frac{1}{n(k+1)!}\, f^{(k+1)}(y_1^+)$,
and
$s_3=\frac{1}{(k+1)!}p(F,1/(1-c),c)$,
$\kappa_2(f)=\frac{q_1\,\G\left( \frac{\ell+2}{\ell+1} \right)+ q_2\,q_3^{\frac{1}{\ell+1}}}
{(\ell+1)\,q_3^{\frac{\ell+2}{\ell+1}}}$,
$q_1=\frac{(-1)^{\ell+1}}{n(\ell+1)!}\,f^{(\ell+1)}(y_2^-)$,
$q_2=\frac{(-1)^{\ell}}{n^{\ell+1}\ell!}\,f^{(\ell)}(y_2^-)$,
and
$q_3=\frac{(-1)^{\ell+1}}{(\ell+1)!}p(F,1/c,c)$
provided the derivatives exist. $\square$
\end{remark}

\textbf{Examples:}
(S-a) Let $c \in (0,1/2)$.
Then for $F$ with pdf
$f(x)=\frac{\pi}{4 c}\sin(\pi x/c)\I(0 < x \le c)+ g(x)\I(c < x <1 )$,
where $g(x)$ is a nonnegative function such that
$\int_{c}^1 g(t)dt=1/2$,
we have $k=1$, $f'(0^+)=\frac{\pi^2}{4 c^2}$,
and $f'\left( c^+ \right)=\frac{\pi^2}{4 c^2}$ in Theorem \ref{thm:kth-order-gen-r,c} (ii).
Then $p(F,1/(1-c),c)=\frac{1}{c(2-c)}$. $\square$

(S-b)
For the beta distribution with parameters $a,b$,
denoted by $Beta(a,b)$,
where $a,b \ge 1$,
the pdf is given by
$$f(x)= \frac{x^{a-1}(1-x)^{b-1}}{\beta(a,b) } \;
\I(0<x<1) \text{ where } \beta(a,b)=\frac{\G(a)\,\G(b)}{ \G(a +b)}.$$
Then in Theorem \ref{thm:kth-order-gen-r,c} (ii)
we have $k=0$, $f(0^+)=0$,
and $f\left( c^+ \right)=\frac{c^{a-1}(1-c)^{b-1}}{\beta(a,b) }$.
So $p(Beta(a,b),1/(1-c),c)=0$ for $c \ne 1/2$.
As for Theorem \ref{thm:kth-order-gen-r,c} (iii),
we have $\ell=0$, $f(1^-)=0$,
and $f\left( c^- \right)=\frac{c^{a-1}(1-c)^{b-1}}{\beta(a,b) }$.
Then $p(Beta(a,b),1/c,c)=0$ for $c \ne 1/2$.
Moreover,
by Theorem \ref{thm:kth-order-gen},
$p(Beta(a,b),2,1/2)=0$ as well. $\square$

(S-c)
Consider $F$ with pdf $f(x)=\left(\pi \sqrt{x\,(1-x)}\right)^{-1} \;\I(0<x<1)$.
Notice that $f(x)$ is unbounded at $x \in \{0,1\}$.
Using Remark \ref{rem:unbounded},
it follows that $p(F,r^*,c)=1$ for $c \ne 1/2$.
Similarly,
$p(F,2,1/2)=1$ as well.
$\square$

\section{Supplementary Materials for Section \ref{sec:dist-multiple-intervals}}
\subsection{The Exact Distribution of $\g_{{}_{n,m}}(F_{XY},r,c)$ for $\mathbf D_{n,m}(F_{XY},r,c)$, with $F_{XY} \in \mathscr H(\R)$ }
\label{sec:exact-dist-mult-int}
Let $[m]-1:=\bigl\{ 0,1,2,\ldots,m-1 \bigr\}$ and
$\Theta^S_{a,b}:=\bigl\{ (u_1,u_2,\ldots u_b):\;\sum_{i=1}^{b}u_i = a:\; u_i \in S, \;\;\forall i \bigr\}$.
If $Y_i$ have a continuous distribution,
then the order statistics of $\Y_m$ are distinct a.s.
Given $Y_{(i)}=y_{(i)}$ for $i=1,2,\ldots,m$,
let
$\vec{n}$ be the vector of numbers, $n_i$,
$f_{\vec{Y}}(\vec{y})$ be the joint distribution of the order statistics of $\Y_m$,
i.e., $f_{\vec{Y}}(\vec{y})=\frac{1}{m!}\prod_{i=1}^m f(y_i)\,\I(\omega_1<y_1<y_2<\ldots<y_m<\omega_2)$,
and $f_{i,j}(y_i,y_j)$ be the joint distribution of $Y_{(i)},Y_{(j)}$.
Then we have the following theorem.

\begin{theorem}
\label{thm:general-Dnm}
Let $\mathbf D_{n,m}(F_{XY},r,c)$ be the PICD with $F_{XY} \in \mathscr H(\R)$,
$n>1$, $m>1$, $r \in [1,\infty)$ and $c \in (0,1)$.
Then the probability mass function (pmf) of the domination number $\g_{n,m}(F_{XY},r,c)$ is given by
{\small
$$P(\g_{{}_{n,m}}(F_{XY},r^*,c)=q)=\int_{\mathscr S} \sum_{\vec{n} \in \Theta^{[n+1]-1}_{n,(m+1)}}
\sum_{\vec{q}\in \Theta^{[3]-1}_{q,(m+1)}} P(\vec{N}=\vec{n})\,\zeta(q_1,n_1)\,\zeta(q_{m+1},\,n_{m+1})
\prod_{j=2}^{m}\eta(q_i,n_i)f_{\vec{Y}}(\vec{y})\,dy_1 \ldots dy_m$$
}
where
$P(\vec{N}=\vec{n})$ is the joint probability of $n_i$ points
falling into intervals $\mI_i$ for $i=0,1,2,\ldots,m$,
$q_i \in \{0,1,2\}$, $q=\sum_{i=0}^m q_i$ and
\begin{align*}
\zeta(q_i,n_i)&=\max\bigl( \I(n_i=q_i=0),\I(n_i \ge q_i=1) \bigr) \text{ for } i=1,(m+1),
\text{ and }\\
\eta(q_i,n_i)&=\max \bigl( \I(n_i=q_i=0),\I(n_i \ge q_i \ge 1) \bigr)\cdot
p(F_i,r^*,c))^{\I(q_i=2)}\,\bigl( 1-p(F_i,r^*,c) \bigr)^{\I(q_i=1)}\\
& \text{ for $i=1,2,3,\ldots,(m-1),$ and the region of integration is given by}\\
\mathscr S:=\bigl\{&(y_1,y_2,\ldots,y_m)\in
(\omega_1,\omega_2)^2:\,\omega_1<y_1<y_2<\ldots<y_m<\omega_2 \bigr\}.
\end{align*}
The special cases of $n=1$, $m=1$, $r \in \{1,\infty\}$ and $c \in \{0,1\}$ are
as in Proposition \ref{prop:gamma-Dnm-r-M}.
\end{theorem}

Notice that the above theorem might look rather complicated at first glance.
However it is not so and we provide a brief description of this result in words as well:
the probability mass function is obtained by integrating the conditional probability
given $\Y_m$ first and then the number $n_i$ of $\X$ points in each interval $\mI_i$.
When $n_i$ are given,
the domination number is the sum of the domination numbers of the digraphs restricted to $\mI_i$,
and these domination numbers are (conditionally) independent and has the distribution given in Section \ref{sec:non-uniform}.
However the formal proof is omitted as it is very similar to that of Theorem 6.1 in \cite{ceyhan:dom-num-CCCD-NonUnif}.

\begin{corollary}
\label{cor:uniform-Dnm}
Let $\mathbf D_{n,m}(F_{XY},r,c)$ be the PICD with $F_{XY} \in \mathscr U(\R)$
and suppose $n>1$, $m>1$, $r \in [1,\infty)$ and $c \in (0,1)$.
Then the pmf of the domination number of $D$ is given by
$$P(\g_{{}_{n,m}}(\U,r^*,c)=q)=\frac{n!m!}{(n+m)!}
\sum_{\vec{n} \in \Theta^{[n+1]-1}_{n,(m+1)}}
\sum_{\vec{q}\in \Theta^{[3]-1}_{q,(m+1)}}
\zeta(q_1,n_1)\,\zeta(q_{m+1},\,n_{m+1})
\prod_{j=2}^{m}\eta(q_i,n_i).$$
The special cases of $n=1$, $m=1$, $r \in \{1,\infty\}$ and $c \in \{0,1\}$ are
as in Proposition \ref{prop:gamma-Dnm-r-M}.
\end{corollary}
The proof is similar to that of Theorem 2 in \cite{priebe:2001}.
For $n,m < \infty$, the expected value of domination number is
\begin{equation}
\label{eqn:expected-gamma-Dnm}
\E[\g_{{}_{n,m}}(F_{XY},r,c)]=P\left(X_{(1)}<Y_{(1)}\right)+P\left(X_{(n)} > Y_{(m)}\right)+
\sum_{i=1}^{m-1}\sum_{k=1}^n\,P(N_i=k)\,\E[\g_{{}_{[i]}}(F_i,r,c)]
\end{equation}
where
\begin{multline*}
P(N_i=k)=\\
\int_{\omega_1}^{\omega_2}\int_{y_{(i)}}^{\omega_2}
f_{i-1,i}\left(y_{(i)},y_{(i+1)}\right) \Bigl[F_X\left(y_{(i+1)}\right)-
F_X\left(y_{(i)}\right)\Bigr]^k\Bigl[1-\left(F_X\left(y_{(i+1)}\right)-
F_X\left(y_{(i)}\right)\right)\Bigr]^{n-k}\,dy_{(i+1)}dy_{(i)}
\end{multline*}
and $\E[\g_{{}_{[i]}}(F_i,r,c)]=1+p_n(F_i,r,c)$.
Then as in Corollary 6.2 of \cite{ceyhan:dom-num-CCCD-NonUnif},
we have
\begin{corollary}
\label{cor:Egnn goes infty}
For $F_{XY} \in \mathscr H(\R)$ with support $\mS(F_X) \cap \mS(F_Y)$ of positive measure
with $r \in [1,\infty)$ and $c \in (0,1)$,
we have $\lim_{n \rightarrow \infty}\E[\g_{{}_{n,n}}(F_{XY},r,c)] = \infty$.
\end{corollary}

\begin{remark}
\label{rem:test-any-F}
\textbf{Extension of the Methodology to Test Nonuniform Distributions:}
Recall that in Proposition \ref{prop:NF vs NPE},
we have shown that if the defining proximity region for our random digraph is defined as
$N_F(x,r,c):=F^{-1}(N(F(x),r,c))$ where $F$ is an increasing function in $(a,b)$ with $a<b$
the exact (and asymptotic) distribution of the domination number based on the digraph for $N_F$
is the same as $\g_{n,2}(\U,r,c)$.
Hence
we can test whether the distribution of any data set is from $F$ (with disjoint supports with $m$ components) or not
with the above methodology.
For example,
to test a data set is from $H_o:$ ``data is from $F(x)=x^2$ with $\mS(F)=(0,1)$"
(so the inverse is $F^{-1}(x)=\sqrt{x}$ and
the corresponding pdf is $f(x)=2x \I(0<x<1)$),
we need to compute the domination number for the PICD based on
\begin{multline}
\label{eqn:ex-N-F-(0,1)}
N_F\left( x,r,c \right)=F^{-1}\left( N\left( F\left( x \right),r,c \right) \right)=\\
\begin{cases}
F^{-1}\left(  \left( 0, \min\left( 1,r\,x^2 \right) \right)  \right)= \left( 0,\min\left( 1,\sqrt{r}\,x \right) \right) & \text{if $x \in \left( 0,\sqrt{c} \right)$,}\\
F^{-1}\left(  \left( \max\left( 0,1-r\left( 1-x^2 \right) \right),1 \right)  \right)=\left( \max\left( 0,\sqrt{1-r\left( 1-x^2 \right)} \right),1 \right)     & \text{if $x \in \left( \sqrt{c},1 \right)$.}
\end{cases}
\end{multline}
Then the domination number will have the same distribution as $\g_{n,m}(\U,r,c)$
and hence can be used for testing data is from $F$ or not
with similar procedures outlined above.
$\square$
\end{remark}

\end{document}